\documentclass[longauth,traditabstract]{aa} 
\pdfoutput=1
\usepackage{graphicx}
\usepackage{amsmath,amsfonts,amssymb}
\usepackage{txfonts}
\usepackage[breaklinks,colorlinks,citecolor=blue]{hyperref}
\usepackage{color}
\usepackage{fixltx2e}
\usepackage{natbib}
\usepackage{float}
\usepackage[usenames,dvipsnames,svgnames,table]{xcolor}
\usepackage[flushleft]{threeparttable}
\bibpunct{(}{)}{;}{a}{}{,} 
\usepackage[switch]{lineno}
\usepackage{ifthen}
\usepackage{lscape}

\def\setsymbol#1#2{\expandafter\def\csname #1\endcsname{#2}}
\def\getsymbol#1{\csname #1\endcsname}

\def\Planck{\textit{Planck}}

\def\alltwentythirteenresultspapers{\nocite{planck2013-p01, planck2013-p02, planck2013-p02a, planck2013-p02d, planck2013-p02b, planck2013-p03, planck2013-p03c, planck2013-p03f, planck2013-p03d, planck2013-p03e, planck2013-p01a, planck2013-p06, planck2013-p03a, planck2013-pip88, planck2013-p08, planck2013-p11, planck2013-p12, planck2013-p13, planck2013-p14, planck2013-p15, planck2013-p05b, planck2013-p17, planck2013-p09, planck2013-p09a, planck2013-p20, planck2013-p19, planck2013-pipaberration, planck2013-p05, planck2013-p05a, planck2013-pip56, planck2013-p06b, planck2013-p01a}}

\def\alltwentyfifteenresultspapers{\nocite{planck2014-a01, planck2014-a03, planck2014-a04, planck2014-a05, planck2014-a06, planck2014-a07, planck2014-a08, planck2014-a09, planck2014-a11, planck2014-a12, planck2014-a13, planck2014-a14, planck2014-a15, planck2014-a16, planck2014-a17, planck2014-a18, planck2014-a19, planck2014-a20, planck2014-a22, planck2014-a24, planck2014-a26, planck2014-a28, planck2014-a29, planck2014-a30, planck2014-a31, planck2014-a35, planck2014-a36, planck2014-a37, planck2014-ES}}

\newbox\tablebox    \newdimen\tablewidth
\def\leaderfil{\leaders\hbox to 5pt{\hss.\hss}\hfil}
\def\endPlancktable{\tablewidth=\columnwidth 
    $$\hss\copy\tablebox\hss$$
    \vskip-\lastskip\vskip -2pt}
\def\endPlancktablewide{\tablewidth=\textwidth 
    $$\hss\copy\tablebox\hss$$
    \vskip-\lastskip\vskip -2pt}
\def\tablenote#1 #2\par{\begingroup \parindent=0.8em
    \abovedisplayshortskip=0pt\belowdisplayshortskip=0pt
    \noindent
    $$\hss\vbox{\hsize\tablewidth \hangindent=\parindent \hangafter=1 \noindent
    \hbox to \parindent{$^#1$\hss}\strut#2\strut\par}\hss$$
    \endgroup}
\def\doubleline{\vskip 3pt\hrule \vskip 1.5pt \hrule \vskip 5pt}

\def\L2{\ifmmode L_2\else $L_2$\fi}

\def\DeltaT{\ifmmode \Delta T\else $\Delta T$\fi}
\def\deltat{\ifmmode \Delta t\else $\Delta t$\fi}
\def\fknee{\ifmmode f_{\rm knee}\else $f_{\rm knee}$\fi}
\def\Fmax{\ifmmode F_{\rm max}\else $F_{\rm max}$\fi}
\def\solar{\ifmmode{\rm M}_{\mathord\odot}\else${\rm M}_{\mathord\odot}$\fi}
\def\Msolar{\ifmmode{\rm M}_{\mathord\odot}\else${\rm M}_{\mathord\odot}$\fi}
\def\Lsolar{\ifmmode{\rm L}_{\mathord\odot}\else${\rm L}_{\mathord\odot}$\fi}
\def\inv{\ifmmode^{-1}\else$^{-1}$\fi}
\def\mo{\ifmmode^{-1}\else$^{-1}$\fi}
\def\sup#1{\ifmmode ^{\rm #1}\else $^{\rm #1}$\fi}
\def\expo#1{\ifmmode \times 10^{#1}\else $\times 10^{#1}$\fi}
\def\,{\thinspace}
\def\lsim{\mathrel{\raise .4ex\hbox{\rlap{$<$}\lower 1.2ex\hbox{$\sim$}}}}
\def\gsim{\mathrel{\raise .4ex\hbox{\rlap{$>$}\lower 1.2ex\hbox{$\sim$}}}}

\def\simprop{\mathrel{\raise .4ex\hbox{\rlap{$\propto$}\lower 1.2ex\hbox{$\sim$}}}}
\def\deg{\ifmmode^\circ\else$^\circ$\fi}
\def\pdeg{\ifmmode $\setbox0=\hbox{$^{\circ}$}\rlap{\hskip.11\wd0 .}$^{\circ}
          \else \setbox0=\hbox{$^{\circ}$}\rlap{\hskip.11\wd0 .}$^{\circ}$\fi}
\def\arcs{\ifmmode {^{\scriptstyle\prime\prime}}
          \else $^{\scriptstyle\prime\prime}$\fi}
\def\arcm{\ifmmode {^{\scriptstyle\prime}}
          \else $^{\scriptstyle\prime}$\fi}
\newdimen\sa  \newdimen\sb
\def\parcs{\sa=.07em \sb=.03em
     \ifmmode \hbox{\rlap{.}}^{\scriptstyle\prime\kern -\sb\prime}\hbox{\kern -\sa}
     \else \rlap{.}$^{\scriptstyle\prime\kern -\sb\prime}$\kern -\sa\fi}
\def\parcm{\sa=.08em \sb=.03em
     \ifmmode \hbox{\rlap{.}\kern\sa}^{\scriptstyle\prime}\hbox{\kern-\sb}
     \else \rlap{.}\kern\sa$^{\scriptstyle\prime}$\kern-\sb\fi}
\def\ra[#1 #2 #3.#4]{#1\sup{h}#2\sup{m}#3\sup{s}\llap.#4}
\def\dec[#1 #2 #3.#4]{#1\deg#2\arcm#3\arcs\llap.#4}
\def\deco[#1 #2 #3]{#1\deg#2\arcm#3\arcs}
\def\rra[#1 #2]{#1\sup{h}#2\sup{m}}

\def\dots{\relax\ifmmode \ldots\else $\ldots$\fi}
\def\WHzsr{\ifmmode $W\,Hz\mo\,sr\mo$\else W\,Hz\mo\,sr\mo\fi}
\def\mHz{\ifmmode $\,mHz$\else \,mHz\fi}
\def\GHz{\ifmmode $\,GHz$\else \,GHz\fi}
\def\mKs{\ifmmode $\,mK\,s$^{1/2}\else \,mK\,s$^{1/2}$\fi}
\def\muKs{\ifmmode \,\mu$K\,s$^{1/2}\else \,$\mu$K\,s$^{1/2}$\fi}
\def\muKRJs{\ifmmode \,\mu$K$_{\rm RJ}$\,s$^{1/2}\else \,$\mu$K$_{\rm RJ}$\,s$^{1/2}$\fi}
\def\muKHz{\ifmmode \,\mu$K\,Hz$^{-1/2}\else \,$\mu$K\,Hz$^{-1/2}$\fi}
\def\MJysr{\ifmmode \,$MJy\,sr\mo$\else \,MJy\,sr\mo\fi}
\def\MJysrmK{\ifmmode \,$MJy\,sr\mo$\,mK$_{\rm CMB}\mo\else \,MJy\,sr\mo\,mK$_{\rm CMB}\mo$\fi}
\def\microns{\ifmmode \,\mu$m$\else \,$\mu$m\fi}
\def\micron{\microns}
\def\muK{\ifmmode \,\mu$K$\else \,$\mu$\hbox{K}\fi}
\def\microK{\ifmmode \,\mu$K$\else \,$\mu$\hbox{K}\fi}
\def\muW{\ifmmode \,\mu$W$\else \,$\mu$\hbox{W}\fi}
\def\kms{\ifmmode $\,km\,s$^{-1}\else \,km\,s$^{-1}$\fi}
\def\kmsMpc{\ifmmode $\,\kms\,Mpc\mo$\else \,\kms\,Mpc\mo\fi}
\providecommand{\sorthelp}[1]{}

\newcommand{\extended}{``extended''}
\newcommand{\healpix}{{\tt HEALPix}}
\newcommand{\lastcat}{PCCS}
\newcommand{\plccs}{PCCS2}
\newcommand{\ecat}{PCCS2E}
\newcommand{\snr}{S/N}
\newcommand{\Herschel}{\textit{Herschel}}

\newcommand{\lcv}{\left(\begin{array}{r}}
\newcommand{\rcv}{\end{array}\right)}
\def\erf{\mathop{\rm erf}}
\begin{document}

\author{\small
Planck Collaboration: P.~A.~R.~Ade\inst{101}
\and
N.~Aghanim\inst{68}
\and
F.~Arg\"{u}eso\inst{24}
\and
M.~Arnaud\inst{84}
\and
M.~Ashdown\inst{80, 7}
\and
J.~Aumont\inst{68}
\and
C.~Baccigalupi\inst{99}
\and
A.~J.~Banday\inst{111, 12}
\and
R.~B.~Barreiro\inst{75}
\and
N.~Bartolo\inst{35, 76}
% ...
}
\institute{\small
APC, AstroParticule et Cosmologie, Universit\'{e} Paris Diderot, CNRS/IN2P3, CEA/lrfu, Observatoire de Paris, Sorbonne Paris Cit\'{e}, 10, rue Alice Domon et L\'{e}onie Duquet, 75205 Paris Cedex 13, France\goodbreak
\and
Aalto University Mets\"{a}hovi Radio Observatory and Dept of Radio Science and Engineering, P.O. Box 13000, FI-00076 AALTO, Finland\goodbreak
\and
Aalto University Mets\"{a}hovi Radio Observatory, P.O. Box 13000, FI-00076 AALTO, Finland\goodbreak
\and
African Institute for Mathematical Sciences, 6-8 Melrose Road, Muizenberg, Cape Town, South Africa\goodbreak
\and
Agenzia Spaziale Italiana Science Data Center, Via del Politecnico snc, 00133, Roma, Italy\goodbreak
\and
Aix Marseille Universit\'{e}, CNRS, LAM (Laboratoire d'Astrophysique de Marseille) UMR 7326, 13388, Marseille, France\goodbreak
\and
Astrophysics Group, Cavendish Laboratory, University of Cambridge, J J Thomson Avenue, Cambridge CB3 0HE, U.K.\goodbreak
\and
Astrophysics \& Cosmology Research Unit, School of Mathematics, Statistics \& Computer Science, University of KwaZulu-Natal, Westville Campus, Private Bag X54001, Durban 4000, South Africa\goodbreak
\and
Atacama Large Millimeter/submillimeter Array, ALMA Santiago Central Offices, Alonso de Cordova 3107, Vitacura, Casilla 763 0355, Santiago, Chile\goodbreak
\and
CGEE, SCS Qd 9, Lote C, Torre C, 4$^{\circ}$ andar, Ed. Parque Cidade Corporate, CEP 70308-200, Bras\'{i}lia, DF, Brazil\goodbreak
\and
CITA, University of Toronto, 60 St. George St., Toronto, ON M5S 3H8, Canada\goodbreak
\and
CNRS, IRAP, 9 Av. colonel Roche, BP 44346, F-31028 Toulouse cedex 4, France\goodbreak
\and
CRANN, Trinity College, Dublin, Ireland\goodbreak
\and
California Institute of Technology, Pasadena, California, U.S.A.\goodbreak
\and
Centre for Theoretical Cosmology, DAMTP, University of Cambridge, Wilberforce Road, Cambridge CB3 0WA, U.K.\goodbreak
\and
Centro de Estudios de F\'{i}sica del Cosmos de Arag\'{o}n (CEFCA), Plaza San Juan, 1, planta 2, E-44001, Teruel, Spain\goodbreak
\and
Computational Cosmology Center, Lawrence Berkeley National Laboratory, Berkeley, California, U.S.A.\goodbreak
\and
Consejo Superior de Investigaciones Cient\'{\i}ficas (CSIC), Madrid, Spain\goodbreak
\and
DSM/Irfu/SPP, CEA-Saclay, F-91191 Gif-sur-Yvette Cedex, France\goodbreak
\and
DTU Space, National Space Institute, Technical University of Denmark, Elektrovej 327, DK-2800 Kgs. Lyngby, Denmark\goodbreak
\and
D\'{e}partement de Physique Th\'{e}orique, Universit\'{e} de Gen\`{e}ve, 24, Quai E. Ansermet,1211 Gen\`{e}ve 4, Switzerland\goodbreak
\and
Departamento de Astrof\'{i}sica, Universidad de La Laguna (ULL), E-38206 La Laguna, Tenerife, Spain\goodbreak
\and
Departamento de F\'{\i}sica, Universidad de Oviedo, Avda. Calvo Sotelo s/n, Oviedo, Spain\goodbreak
\and
Departamento de Matem\'{a}ticas, Universidad de Oviedo, Avda. Calvo Sotelo s/n, Oviedo, Spain\goodbreak
\and
Department of Astronomy and Astrophysics, University of Toronto, 50 Saint George Street, Toronto, Ontario, Canada\goodbreak
\and
Department of Astrophysics/IMAPP, Radboud University Nijmegen, P.O. Box 9010, 6500 GL Nijmegen, The Netherlands\goodbreak
\and
Department of Physics \& Astronomy, University of British Columbia, 6224 Agricultural Road, Vancouver, British Columbia, Canada\goodbreak
\and
Department of Physics and Astronomy, Dana and David Dornsife College of Letter, Arts and Sciences, University of Southern California, Los Angeles, CA 90089, U.S.A.\goodbreak
\and
Department of Physics and Astronomy, University College London, London WC1E 6BT, U.K.\goodbreak
\and
Department of Physics, Florida State University, Keen Physics Building, 77 Chieftan Way, Tallahassee, Florida, U.S.A.\goodbreak
\and
Department of Physics, Gustaf H\"{a}llstr\"{o}min katu 2a, University of Helsinki, Helsinki, Finland\goodbreak
\and
Department of Physics, Princeton University, Princeton, New Jersey, U.S.A.\goodbreak
\and
Department of Physics, University of California, Santa Barbara, California, U.S.A.\goodbreak
\and
Department of Physics, University of Illinois at Urbana-Champaign, 1110 West Green Street, Urbana, Illinois, U.S.A.\goodbreak
\and
Dipartimento di Fisica e Astronomia G. Galilei, Universit\`{a} degli Studi di Padova, via Marzolo 8, 35131 Padova, Italy\goodbreak
\and
Dipartimento di Fisica e Scienze della Terra, Universit\`{a} di Ferrara, Via Saragat 1, 44122 Ferrara, Italy\goodbreak
\and
Dipartimento di Fisica, Universit\`{a} La Sapienza, P. le A. Moro 2, Roma, Italy\goodbreak
\and
Dipartimento di Fisica, Universit\`{a} degli Studi di Milano, Via Celoria, 16, Milano, Italy\goodbreak
\and
Dipartimento di Fisica, Universit\`{a} degli Studi di Trieste, via A. Valerio 2, Trieste, Italy\goodbreak
\and
Dipartimento di Matematica, Universit\`{a} di Roma Tor Vergata, Via della Ricerca Scientifica, 1, Roma, Italy\goodbreak
\and
Discovery Center, Niels Bohr Institute, Blegdamsvej 17, Copenhagen, Denmark\goodbreak
\and
Discovery Center, Niels Bohr Institute, Copenhagen University, Blegdamsvej 17, Copenhagen, Denmark\goodbreak
\and
European Southern Observatory, ESO Vitacura, Alonso de Cordova 3107, Vitacura, Casilla 19001, Santiago, Chile\goodbreak
\and
European Space Agency, ESAC, Planck Science Office, Camino bajo del Castillo, s/n, Urbanizaci\'{o}n Villafranca del Castillo, Villanueva de la Ca\~{n}ada, Madrid, Spain\goodbreak
\and
European Space Agency, ESTEC, Keplerlaan 1, 2201 AZ Noordwijk, The Netherlands\goodbreak
\and
Finnish Centre for Astronomy with ESO (FINCA), University of Turku, V\"{a}is\"{a}l\"{a}ntie 20, FIN-21500, Piikki\"{o}, Finland\goodbreak
\and
Gran Sasso Science Institute, INFN, viale F. Crispi 7, 67100 L'Aquila, Italy\goodbreak
\and
HGSFP and University of Heidelberg, Theoretical Physics Department, Philosophenweg 16, 69120, Heidelberg, Germany\goodbreak
\and
Haverford College Astronomy Department, 370 Lancaster Avenue, Haverford, Pennsylvania, U.S.A.\goodbreak
\and
Helsinki Institute of Physics, Gustaf H\"{a}llstr\"{o}min katu 2, University of Helsinki, Helsinki, Finland\goodbreak
\and
INAF - Osservatorio Astrofisico di Catania, Via S. Sofia 78, Catania, Italy\goodbreak
\and
INAF - Osservatorio Astronomico di Padova, Vicolo dell'Osservatorio 5, Padova, Italy\goodbreak
\and
INAF - Osservatorio Astronomico di Roma, via di Frascati 33, Monte Porzio Catone, Italy\goodbreak
\and
INAF - Osservatorio Astronomico di Trieste, Via G.B. Tiepolo 11, Trieste, Italy\goodbreak
\and
INAF/IASF Bologna, Via Gobetti 101, Bologna, Italy\goodbreak
\and
INAF/IASF Milano, Via E. Bassini 15, Milano, Italy\goodbreak
\and
INFN, Sezione di Bologna, Via Irnerio 46, I-40126, Bologna, Italy\goodbreak
\and
INFN, Sezione di Roma 1, Universit\`{a} di Roma Sapienza, Piazzale Aldo Moro 2, 00185, Roma, Italy\goodbreak
\and
INFN, Sezione di Roma 2, Universit\`{a} di Roma Tor Vergata, Via della Ricerca Scientifica, 1, Roma, Italy\goodbreak
\and
INFN/National Institute for Nuclear Physics, Via Valerio 2, I-34127 Trieste, Italy\goodbreak
\and
IPAG: Institut de Plan\'{e}tologie et d'Astrophysique de Grenoble, Universit\'{e} Grenoble Alpes, IPAG, F-38000 Grenoble, France, CNRS, IPAG, F-38000 Grenoble, France\goodbreak
\and
ISDC, Department of Astronomy, University of Geneva, ch. d'Ecogia 16, 1290 Versoix, Switzerland\goodbreak
\and
IUCAA, Post Bag 4, Ganeshkhind, Pune University Campus, Pune 411 007, India\goodbreak
\and
Imperial College London, Astrophysics group, Blackett Laboratory, Prince Consort Road, London, SW7 2AZ, U.K.\goodbreak
\and
Infrared Processing and Analysis Center, California Institute of Technology, Pasadena, CA 91125, U.S.A.\goodbreak
\and
Institut N\'{e}el, CNRS, Universit\'{e} Joseph Fourier Grenoble I, 25 rue des Martyrs, Grenoble, France\goodbreak
\and
Institut Universitaire de France, 103, bd Saint-Michel, 75005, Paris, France\goodbreak
\and
Institut d'Astrophysique Spatiale, CNRS, Univ. Paris-Sud, Universit\'{e} Paris-Saclay, B\^{a}t. 121, 91405 Orsay cedex, France\goodbreak
\and
Institut d'Astrophysique de Paris, CNRS (UMR7095), 98 bis Boulevard Arago, F-75014, Paris, France\goodbreak
\and
Institut f\"ur Theoretische Teilchenphysik und Kosmologie, RWTH Aachen University, D-52056 Aachen, Germany\goodbreak
\and
Institute of Astronomy, University of Cambridge, Madingley Road, Cambridge CB3 0HA, U.K.\goodbreak
\and
Institute of Theoretical Astrophysics, University of Oslo, Blindern, Oslo, Norway\goodbreak
\and
Instituto Nacional de Astrof\'{\i}sica, \'{O}ptica y Electr\'{o}nica (INAOE), Apartado Postal 51 y 216, 72000 Puebla, M\'{e}xico\goodbreak
\and
Instituto de Astrof\'{\i}sica de Canarias, C/V\'{\i}a L\'{a}ctea s/n, La Laguna, Tenerife, Spain\goodbreak
\and
Instituto de F\'{\i}sica de Cantabria (CSIC-Universidad de Cantabria), Avda. de los Castros s/n, Santander, Spain\goodbreak
\and
Istituto Nazionale di Fisica Nucleare, Sezione di Padova, via Marzolo 8, I-35131 Padova, Italy\goodbreak
\and
Jet Propulsion Laboratory, California Institute of Technology, 4800 Oak Grove Drive, Pasadena, California, U.S.A.\goodbreak
\and
Jodrell Bank Centre for Astrophysics, Alan Turing Building, School of Physics and Astronomy, The University of Manchester, Oxford Road, Manchester, M13 9PL, U.K.\goodbreak
\and
Kavli Institute for Cosmological Physics, University of Chicago, Chicago, IL 60637, USA\goodbreak
\and
Kavli Institute for Cosmology Cambridge, Madingley Road, Cambridge, CB3 0HA, U.K.\goodbreak
\and
Kazan Federal University, 18 Kremlyovskaya St., Kazan, 420008, Russia\goodbreak
\and
LAL, Universit\'{e} Paris-Sud, CNRS/IN2P3, Orsay, France\goodbreak
\and
LERMA, CNRS, Observatoire de Paris, 61 Avenue de l'Observatoire, Paris, France\goodbreak
\and
Laboratoire AIM, IRFU/Service d'Astrophysique - CEA/DSM - CNRS - Universit\'{e} Paris Diderot, B\^{a}t. 709, CEA-Saclay, F-91191 Gif-sur-Yvette Cedex, France\goodbreak
\and
Laboratoire Traitement et Communication de l'Information, CNRS (UMR 5141) and T\'{e}l\'{e}com ParisTech, 46 rue Barrault F-75634 Paris Cedex 13, France\goodbreak
\and
Laboratoire de Physique Subatomique et Cosmologie, Universit\'{e} Grenoble-Alpes, CNRS/IN2P3, 53, rue des Martyrs, 38026 Grenoble Cedex, France\goodbreak
\and
Laboratoire de Physique Th\'{e}orique, Universit\'{e} Paris-Sud 11 \& CNRS, B\^{a}timent 210, 91405 Orsay, France\goodbreak
\and
Lawrence Berkeley National Laboratory, Berkeley, California, U.S.A.\goodbreak
\and
Lebedev Physical Institute of the Russian Academy of Sciences, Astro Space Centre, 84/32 Profsoyuznaya st., Moscow, GSP-7, 117997, Russia\goodbreak
\and
Max-Planck-Institut f\"{u}r Astrophysik, Karl-Schwarzschild-Str. 1, 85741 Garching, Germany\goodbreak
\and
Max-Planck-Institut f\"{u}r Extraterrestrische Physik, Giessenbachstra{\ss}e, 85748 Garching, Germany\goodbreak
\and
McGill Physics, Ernest Rutherford Physics Building, McGill University, 3600 rue University, Montr\'{e}al, QC, H3A 2T8, Canada\goodbreak
\and
National University of Ireland, Department of Experimental Physics, Maynooth, Co. Kildare, Ireland\goodbreak
\and
Nicolaus Copernicus Astronomical Center, Bartycka 18, 00-716 Warsaw, Poland\goodbreak
\and
Niels Bohr Institute, Blegdamsvej 17, Copenhagen, Denmark\goodbreak
\and
Niels Bohr Institute, Copenhagen University, Blegdamsvej 17, Copenhagen, Denmark\goodbreak
\and
Nordita (Nordic Institute for Theoretical Physics), Roslagstullsbacken 23, SE-106 91 Stockholm, Sweden\goodbreak
\and
Optical Science Laboratory, University College London, Gower Street, London, U.K.\goodbreak
\and
SISSA, Astrophysics Sector, via Bonomea 265, 34136, Trieste, Italy\goodbreak
\and
SMARTEST Research Centre, Universit\`{a} degli Studi e-Campus, Via Isimbardi 10, Novedrate (CO), 22060, Italy\goodbreak
\and
School of Physics and Astronomy, Cardiff University, Queens Buildings, The Parade, Cardiff, CF24 3AA, U.K.\goodbreak
\and
School of Physics and Astronomy, University of Nottingham, Nottingham NG7 2RD, U.K.\goodbreak
\and
Sorbonne Universit\'{e}-UPMC, UMR7095, Institut d'Astrophysique de Paris, 98 bis Boulevard Arago, F-75014, Paris, France\goodbreak
\and
Space Research Institute (IKI), Russian Academy of Sciences, Profsoyuznaya Str, 84/32, Moscow, 117997, Russia\goodbreak
\and
Space Sciences Laboratory, University of California, Berkeley, California, U.S.A.\goodbreak
\and
Special Astrophysical Observatory, Russian Academy of Sciences, Nizhnij Arkhyz, Zelenchukskiy region, Karachai-Cherkessian Republic, 369167, Russia\goodbreak
\and
Sub-Department of Astrophysics, University of Oxford, Keble Road, Oxford OX1 3RH, U.K.\goodbreak
\and
The Oskar Klein Centre for Cosmoparticle Physics, Department of Physics,Stockholm University, AlbaNova, SE-106 91 Stockholm, Sweden\goodbreak
\and
Theory Division, PH-TH, CERN, CH-1211, Geneva 23, Switzerland\goodbreak
\and
UPMC Univ Paris 06, UMR7095, 98 bis Boulevard Arago, F-75014, Paris, France\goodbreak
\and
Universit\'{e} de Toulouse, UPS-OMP, IRAP, F-31028 Toulouse cedex 4, France\goodbreak
% ...
}

\title{\Planck\ 2015 results. XXVI. The Second Planck Catalogue of Compact Sources}
\authorrunning{Planck Collaboration}
\titlerunning{Second Planck Catalogue of Compact Sources}

\providecommand{\sorthelp}[1]{} 


\abstract{The Second Planck Catalogue of Compact Sources is a list of discrete objects detected in single-frequency maps from the full duration of the \Planck\ mission and supersedes previous versions. It consists of compact sources, both Galactic and extragalactic, detected over the entire sky.  Compact sources detected in the lower frequency channels are assigned to the \plccs, while at higher frequencies they are assigned to one of two subcatalogues, the \plccs\ or \ecat, depending on their location on the sky.  The first of these (\plccs) covers most of the sky and allows the user to produce subsamples at higher reliabilities than the target 80\,\% integral reliability of the catalogue. The second (\ecat) contains sources detected in sky regions where the diffuse emission makes it difficult to quantify the reliability of the detections.
Both the \plccs\ and \ecat\ include polarization measurements, in the form of polarized flux densities, or upper limits, and orientation angles for all seven polarization-sensitive \Planck\ channels.
The improved data-processing of the full-mission maps and their reduced noise levels allow us to increase the number of objects in the catalogue, improving its completeness for the target 80\,\% reliability as compared with the previous versions, the \lastcat\ and the Early Release Compact Source Catalogue (ERCSC).
 }


\maketitle

\alltwentyfifteenresultspapers
\alltwentythirteenresultspapers

\section{Introduction}
\label{sec:intro}
This paper, one of a set associated with the 2015 release of data from the \Planck\ mission,\footnote{\Planck\ (\url{http://www.esa.int/Planck}) is a project of the European Space Agency  (ESA) with instruments provided by two scientific consortia funded by ESA member states and led by Principal Investigators from France and Italy, telescope reflectors provided through a collaboration between ESA and a scientific consortium led and funded by Denmark, and additional contributions from NASA (USA).} describes the second release of the Planck Catalogue of Compact Sources (\plccs)\footnote{\url{http://archives.esac.esa.int/pla}}.
It outlines the construction of the single-frequency catalogues from an analysis of each of the nine \Planck\ frequency-channel, full-mission maps.
The construction of these catalogues builds on much of the same infrastructure and methodology as the first incarnation of the Planck Catalogue of Compact Sources (\lastcat), and the reader is referred to \cite{planck2013-p05} for a full description of the catalogue construction procedures, which are only summarized here. 
Table~\ref{tab:summary} lists some basic properties of the nine frequency subcatalogues of the current release. 
\begin{table}
\begingroup
\newdimen\tblskip \tblskip=5pt
\caption{Characteristics of the \plccs\ and \ecat\ catalogues: flux density at 90\% completeness in total intensity;  number of sources detected in each catalogue in total intensity; and number of sources that have a  polarized signal is measured above a 99.99\% confidence level.  See Tables~\ref{tab:all_pccs_stats} and \ref{tab:pccs_pol} for more details. }
\label{tab:summary}
\nointerlineskip
\vskip -3mm
\footnotesize
\setbox\tablebox=\vbox{
   \newdimen\digitwidth
   \setbox0=\hbox{\rm 0}
   \digitwidth=\wd0
   \catcode`*=\active
   \def*{\kern\digitwidth}
\def\leaderfill{\leaders \hbox to 5pt{\hss.\hss}\hfil}
   \newdimen\signwidth
   \setbox0=\hbox{+}
   \signwidth=\wd0
   \catcode`!=\active
   \def!{\kern\signwidth}
\halign{\hbox to 0.5in{#\leaderfill}\tabskip=0.7em &\tabskip=1em\hfil#\hfil&\hfil#\hfil &\hfil#\hfil &\hfil#\hfil&\hfil#\tabskip=0em\hfil\cr
\noalign{\doubleline}
\omit&\omit&\multispan{2}\hfil No. of sources\hfil&\multispan{2}\hfil Polarized sources\hfil\cr
\noalign{\vskip-6pt}
\omit&Flux density 90\%&\multispan{2}\hrulefill&\multispan{2}\hrulefill\cr
\omit\hfil Channel\hfil&completeness&\plccs&\ecat&\plccs&\ecat\cr
\omit&[mJy]& &  & &\cr
\noalign{\vskip 3pt\hrule\vskip 3pt}
*30&427&1560&\dots&122& \dots\cr
*44&692&*934&\dots&*30& \dots\cr
*70&501&1296&\dots&*34& \dots\cr
100&269&1742&*2487&*20&*43\cr
143&177&2160&*4139&*25&111\cr
217&152&2135&16842&*11&325\cr
353&304&1344&22665&**1&666\cr
545&555&1694&31068&\dots&\dots\cr
857&791&4891&43290&\dots&\dots\cr
\noalign{\vskip 3pt\hrule\vskip 3pt}}}
\endPlancktable
\endgroup
\end{table}

One of the primary differences of this release from the \lastcat\ is the division of the six highest frequency catalogues into two subcatalogues, the \plccs\ and the \ecat. This division separates sources for which the reliability (the fraction of sources above a given \snr\ which are real) can be quantified (\plccs) from those of unknown reliability (\ecat).  This separation is primarily based on the Galactic coordinates of the source, as described in Sect.~\ref{sec:selection}.
The target integral reliability of the the entire catalogue, as in the \lastcat, is $80\,\%$ or greater. The advantage of setting the reliability target this low is that it improves the odds of discovering interesting sources with unusual properties, which might otherwise have been rejected by restrictive selection criteria. 
On the other hand, a highly reliable catalogue is desirable for follow-up observations. That is the aim of the \plccs. To this end, we have provided additional information in the catalogue which will allow the user to select  a subset of highly reliable sources from the \plccs.
This takes the form of an additional flag per source that indicates the highest reliability catalogue to which that source belongs, allowing the user to perform a cut on the \plccs\ to reduce it to the desired percentage reliability subset.  To assist users, we also flag those sources identified in other catalogues, mainly at radio wavelengths. 

\begin{figure}
\begin{center}
\includegraphics[width=0.5\textwidth]{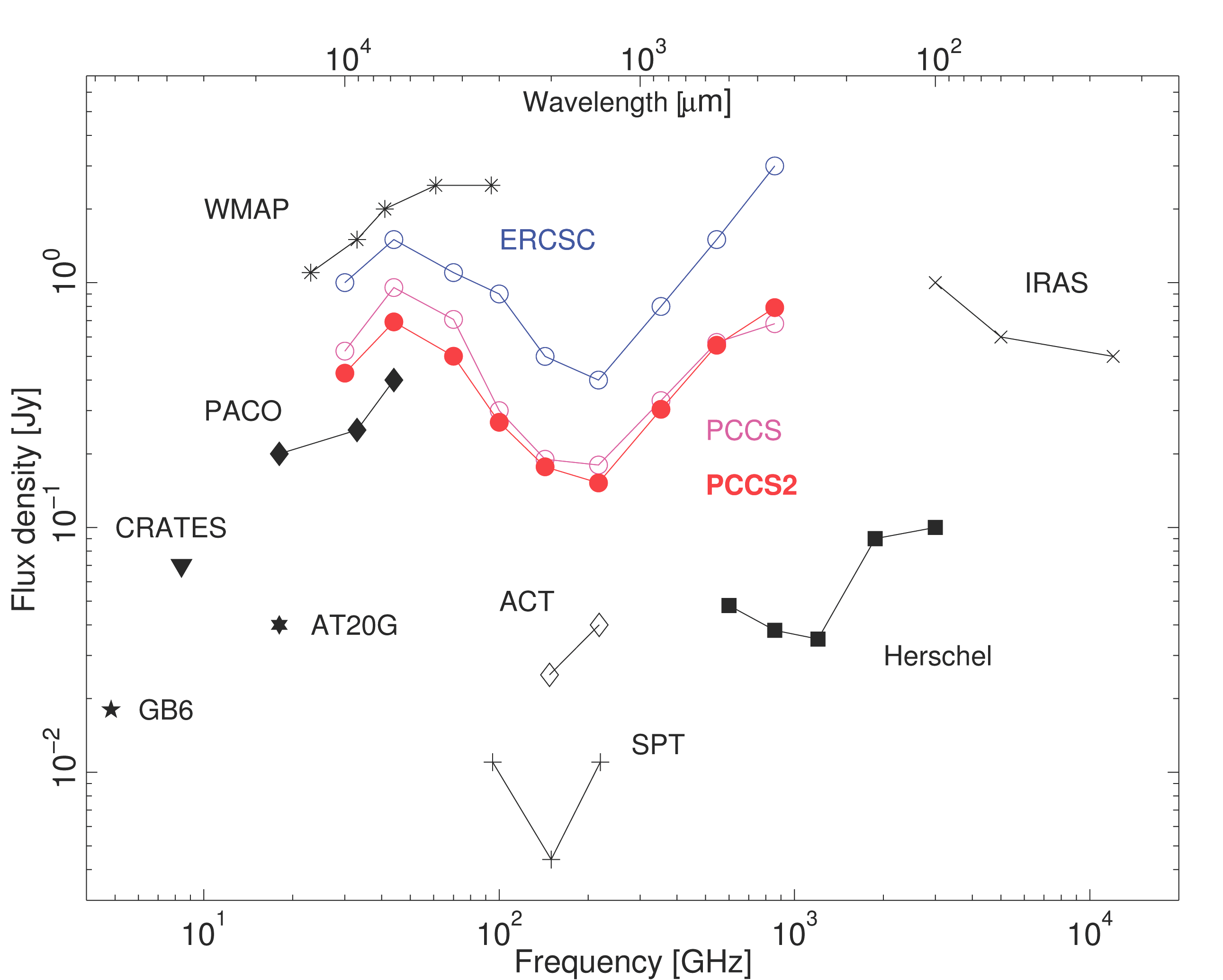}
\caption{Sensitivity (the flux density at 90\,\% completeness) of the \plccs, compared with \lastcat, ERCSC, WMAP and others as described in text.
The sensitivities displayed for the LFI channels are for the full sky.
For the HFI channels, the 90\,\% completeness limits plotted for the \lastcat\ were evaluated in the extragalactic zone, as defined in \cite{planck2013-p05}. The regions of sky to which the 90\,\% completeness limits apply are therefore similar (but not identical) to those of the \plccs. These comparisons are discussed further in Sect.~\ref{sec:characteristics}.
\label{fig:sensitivity}}
\end{center}
\end{figure}

The principal data-driven difference between the \lastcat\ and the \plccs\ and \ecat\ catalogues is the additional data from the \extended\ mission and the inclusion of polarization measurements  in seven out of the nine frequency channels. The 545 and 857\,GHz channels are not sensitive to polarized signals, so the polarization measurements span the range from 30 to 353\,GHz.  The polarization measurements provided are based on the positions of the compact sources discovered in the temperature maps; there is no independent search for compact sources in polarization. The additional data, together with improved data processing, have the effect of reducing the noise and hence improving the completeness of the \plccs\ and \ecat\ catalogues over that of the \lastcat, as demonstrated in Fig.~\ref{fig:sensitivity}. In this figure we compare the sensitivity of the PCCS2 with that of the PCCS \citep{planck2013-p05}, the Early Release Compact Source Catalogue (see \citealt{planck2011-6.1} for the 30--70\,GHz channels and \citealt{planck2012-VII} for the others), WMAP (see \citealt{gnuevo08} for the  channels at 41\,GHz and below, and \citealt{lanz13} for 61 and 94\,GHz). The sensitivity levels for the \Herschel\ SPIRE and PACS instruments are from \cite{clements10} and \cite{rigby11}, respectively. The other wide-area surveys shown as a comparison are: the Green Bank 6\,cm Survey, GB6 \citep{gregory96}, the Combined Radio All-Sky Targeted Eight GHz Survey, CRATES \citep{healey07},  the Australia Telescope 20\,GHz Survey, AT20G \citep{murphy10}, the Planck--ATCA Co-eval Observation Project, PACO \citep{bonavera11}, the South Pole Telescope, SPT \citep{Moc13}, the Atacama Cosmology Telescope, ACT \citep{Mar13}, and the Infrared Astronomical Satellite catalogue, IRAS \citep{beichman88}.

In Sect.~\ref{sec:pccs} we describe the construction of the catalogues, including the criteria used to place a source into the \plccs\ or the \ecat, and the methods used to measure the flux  densities and linear polarization parameters. 
Section~\ref{sec:validation} discusses the validation and quality assessment of the catalogues. Here the internal and external consistency tests are described, as are the completeness (the fraction of the total number of sources above a given flux density that are present in the catalogue) and reliability of the catalogues.
The overall characteristics of the \plccs\ and \ecat\ are presented in detail in Sect.~\ref{sec:characteristics}; they are also compared with the characteristics of the \lastcat.
The released product, which is composed of the catalogues and their associated maps, is described in Sect.~\ref{sec:content}.
We summarize our conclusions in Sect.~\ref{sec:conclusions}.
Details of the estimators used for photometry and for polarization measurements are given in the appendices.

\section{The Second Planck Catalogue of Compact Sources}
\label{sec:pccs}
\subsection{Data}
\label{sec:data}
After four years of operations, the data from the full \Planck\ mission have been transformed into full-sky \healpix\footnote{\url{http://healpix.sourceforge.net}} maps \citep{gorski2005} by the data processing centres (DPCs) \citep{planck2014-a07, planck2014-a09}. The Low Frequency Instrument (LFI) DPC produced the 30, 44, and 70\,GHz maps after the completion of eight full surveys (spanning the period 12 August 2009 to 3 August 2013). In addition, special LFI maps covering the period 1 April 2013 to 30 June 2013 were produced in order to compare the \Planck\ flux-density scales with those of the Very Large Array and the Australia Telescope Compact Array, by performing simultaneous observations of a sample of sources over that period. The High Frequency Instrument (HFI) DPC produced the 100, 143, 217, 353, 545, and 857\,GHz maps after five full surveys (2009 August 12 to 2012 January 11).  The flux densities of all sources measured in the full-mission maps are an average of several observations. Single-survey maps are available from the Planck Legacy Archive,\footnote{\url{http://archives.esac.esa.int/pla}} but single-survey flux densities are not provided in this catalogue.  It is important to note that even single-survey maps may include more than one observation of an individual source, and hence extracting flux densities from the single-survey maps does not guarantee a single-epoch observation for a given source. However, in the Planck Legacy Archive the time-ordered data are available and users can produce maps from arbitrary time intervals.
 Table~\ref{tab:beam_data} gives the parameters of the \Planck\ beams used in this paper; further details may be found in \cite{planck2014-a03} and~\cite{planck2014-a08}.

\begin{table}
\begingroup
\newdimen\tblskip \tblskip=5pt
\caption{Parameters of the beams used in the construction of the \plccs\ and \ecat.}
\label{tab:beam_data}
\nointerlineskip
\vskip -3mm
\footnotesize
\setbox\tablebox=\vbox{
   \newdimen\digitwidth
   \setbox0=\hbox{\rm 0}
   \digitwidth=\wd0
   \catcode`*=\active
   \def*{\kern\digitwidth}
\def\leaderfill{\leaders \hbox to 5pt{\hss.\hss}\hfil}
   \newdimen\signwidth
   \setbox0=\hbox{+}
   \signwidth=\wd0
   \catcode`!=\active
   \def!{\kern\signwidth}
\halign{\hbox to 0.5in{#\leaderfill}\tabskip=1em&\hfil#\hfil&\hfil#\hfil &\hfil#\hfil &\hfil#\hfil&\tabskip=0pt\hfil#\hfil\cr
\noalign{\doubleline}
\omit&\multispan2{\hfil FWHM\hfil}&\multispan{3}{\hfil Beam area\hfil}\cr
\noalign{\vskip-6pt}
\omit&\multispan{2}\hrulefill&\multispan{3}\hrulefill\cr
\omit Channel&Fitted&Effective&$\Omega_{\rm beam}$&$\Omega_{\rm beam_1}$&$\Omega_{\rm beam_2}$\cr
\omit&[arcmin]&[arcmin]&[arcmin$^2$]&[arcmin$^2$]&[arcmin$^2$]\cr
\noalign{\vskip 3pt\hrule\vskip 5pt}
*30&32.29&32.41&1190.06&1117.30&1188.93\cr
*44&27.00&27.10&*832.00&*758.00&*832.00\cr
*70&13.21&13.32&*200.90&*186.10&*200.59\cr
100&*9.66&*9.69&*106.22&*100.78&*106.03\cr
143&*7.22&*7.30&**60.44&**56.97&**60.21\cr	
217&*4.90&*5.02&**28.57&**26.46&**28.46\cr
353&*4.92&*4.94&**27.69&**25.32&**27.53\cr
545&*4.68&*4.83&**26.44&**24.06&**26.09\cr
857&*4.22&*4.64&**24.37&**22.58&**23.93\cr
\noalign{\vskip 3pt\hrule\vskip 3pt}}}
\endPlancktable
\tablefoot{Two FWHM values are given: one from an elliptical Gaussian fit to the beam, and another that is the FWHM of a Gaussian with the same solid angle as the main beam, $\Omega_{\rm beam}$. The FWHM found from the main beam solid angle is used to evaluate $\Omega_{\rm beam_1}$ and $\Omega_{\rm beam_2}$, which are the beam solid angles within a radius equal to this FWHM and twice this FWHM, respectively.}
\endgroup
\end{table}

\subsection{Catalogue construction}
\label{sec:construction}
\begin{figure*}
\begin{center}
\includegraphics[width=0.95\textwidth]{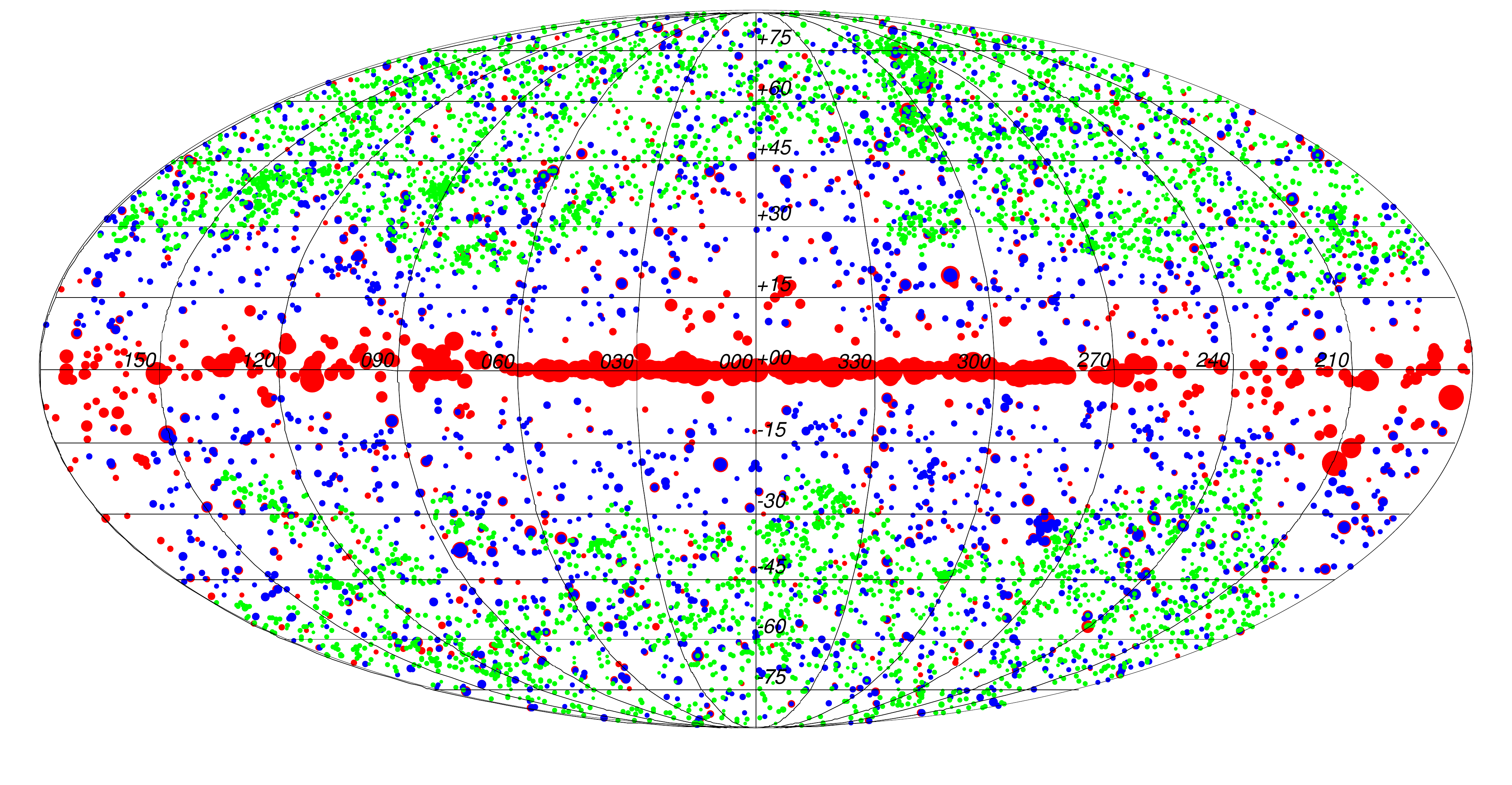}
\includegraphics[width=0.95\textwidth]{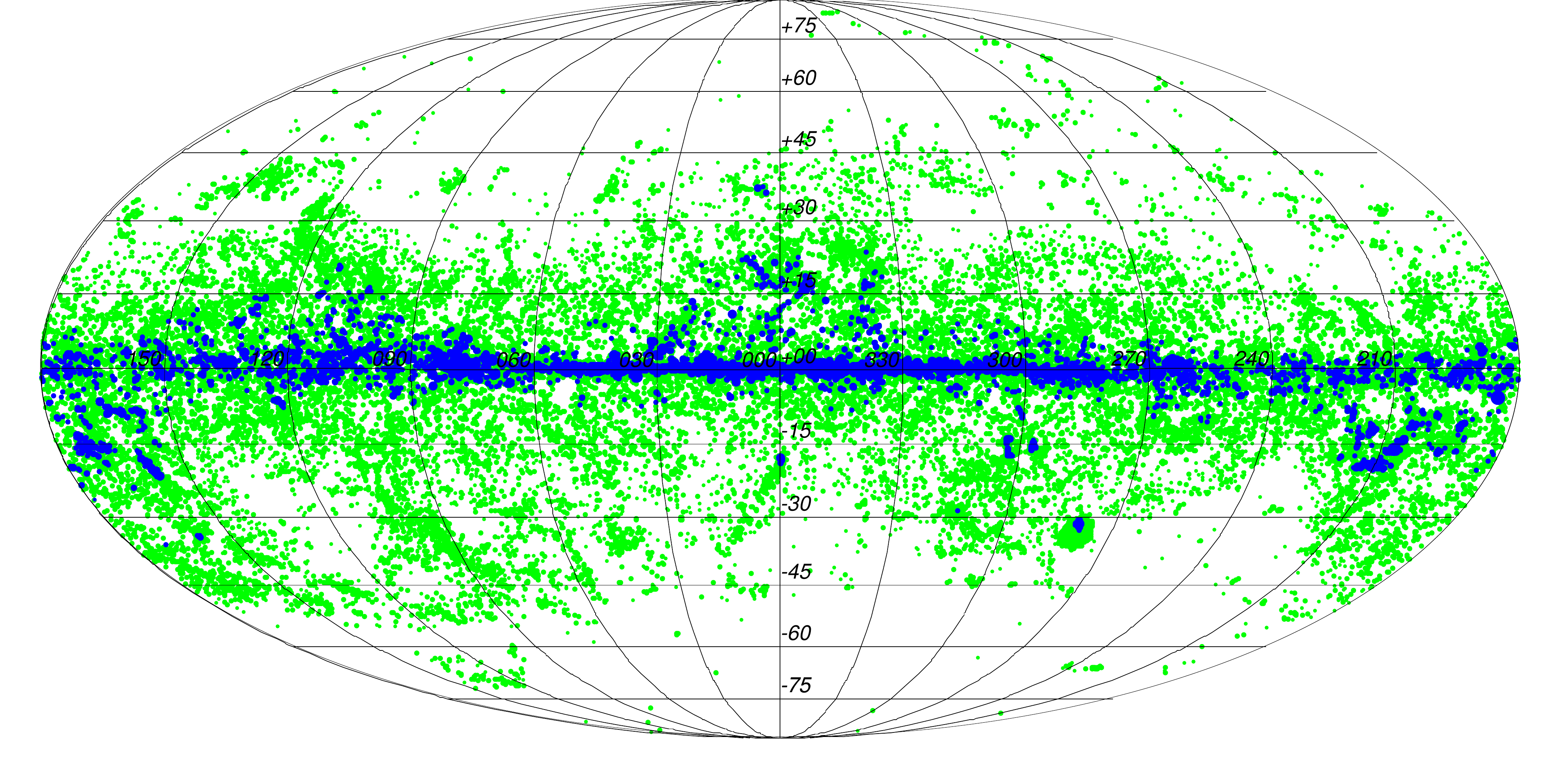}
\caption{\textit{Top}: distribution of the validated sources from the \plccs. Red, blue, and green circles show sources from the 30, 143, and 857\,GHz  catalogues, respectively. \textit{Bottom}: locations of the sources in the 143 and 857 \ecat, shown by blue and green circles, respectively.  The figure is a full-sky Mollweide projection with the Galactic equator horizontal; longitude increases to the left with the Galactic centre in the centre of the map. The size of the filled circles gives an idea of the relative flux densities of the sources per frequency, where the larger circles correspond to larger flux densities. Note that a different size range for each channel was necessary for visualization purposes.}
\label{fig:bothcat_positions_temperature}
\end{center}
\end{figure*}
\begin{figure*}
\begin{center}
\includegraphics[width=0.95\textwidth]{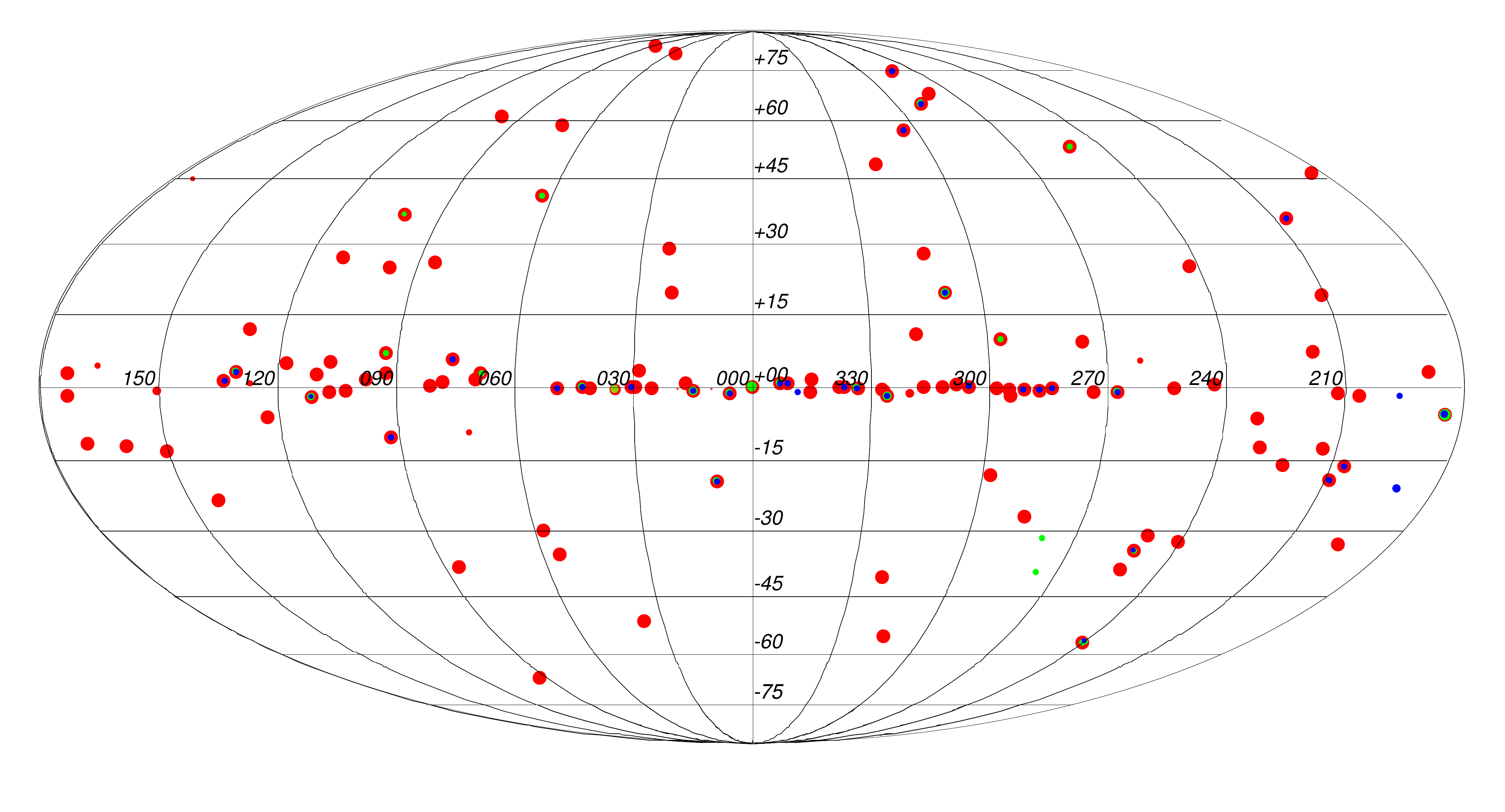}
\caption{Distribution of the polarized sources in the lowest channels of the \plccs.  Red, green, and blue circles show sources from the 30, 44, and 70\,GHz  catalogues, respectively. As in Fig.~\ref{fig:bothcat_positions_temperature}, the size of the circle gives a qualitative idea of the relative polarized flux density of the source.} 
\label{fig:lficat_positions_pol}
\end{center}
\end{figure*}
\begin{figure*}
\begin{center}
\includegraphics[width=0.95\textwidth]{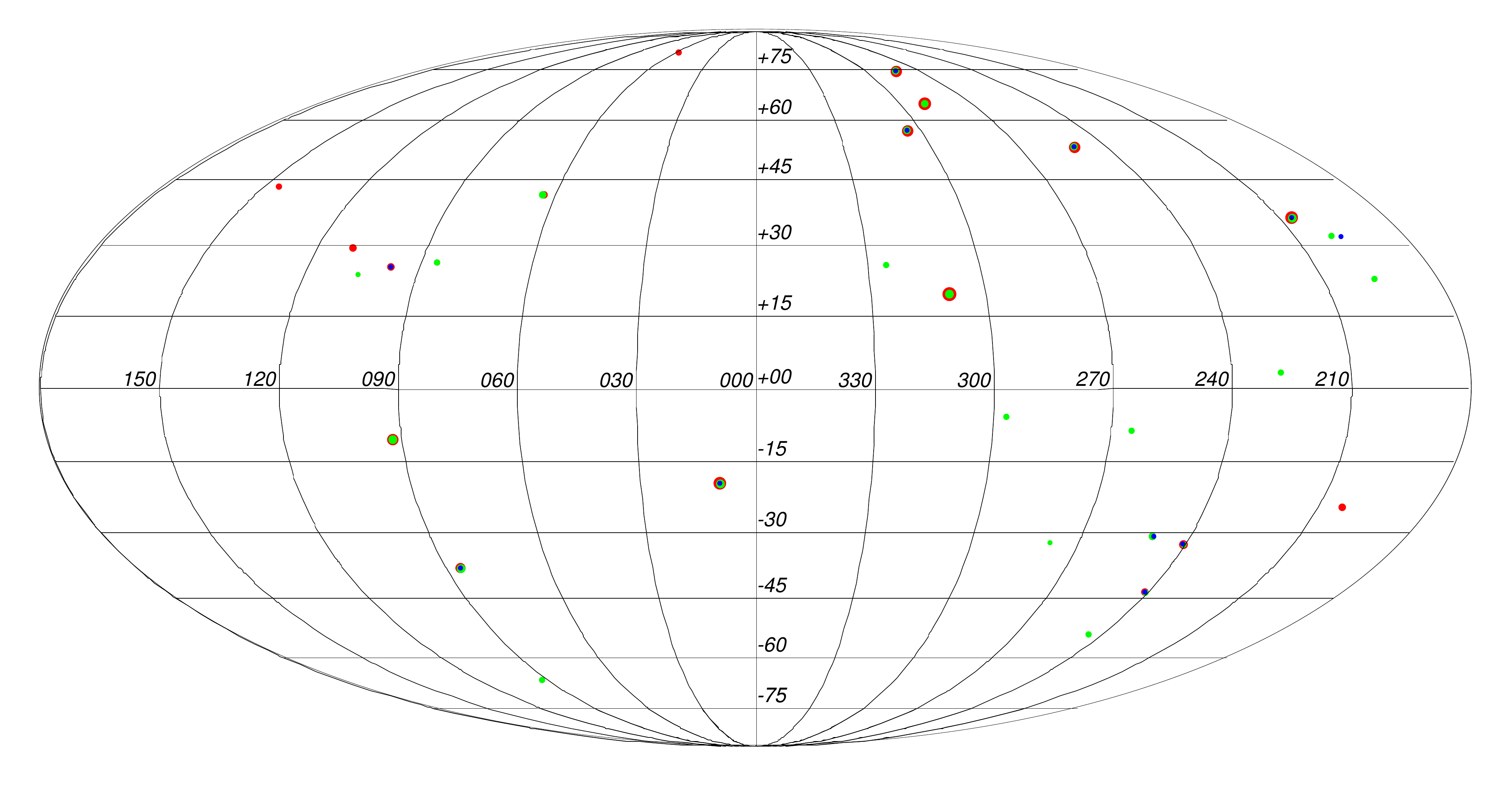}
\includegraphics[width=0.95\textwidth]{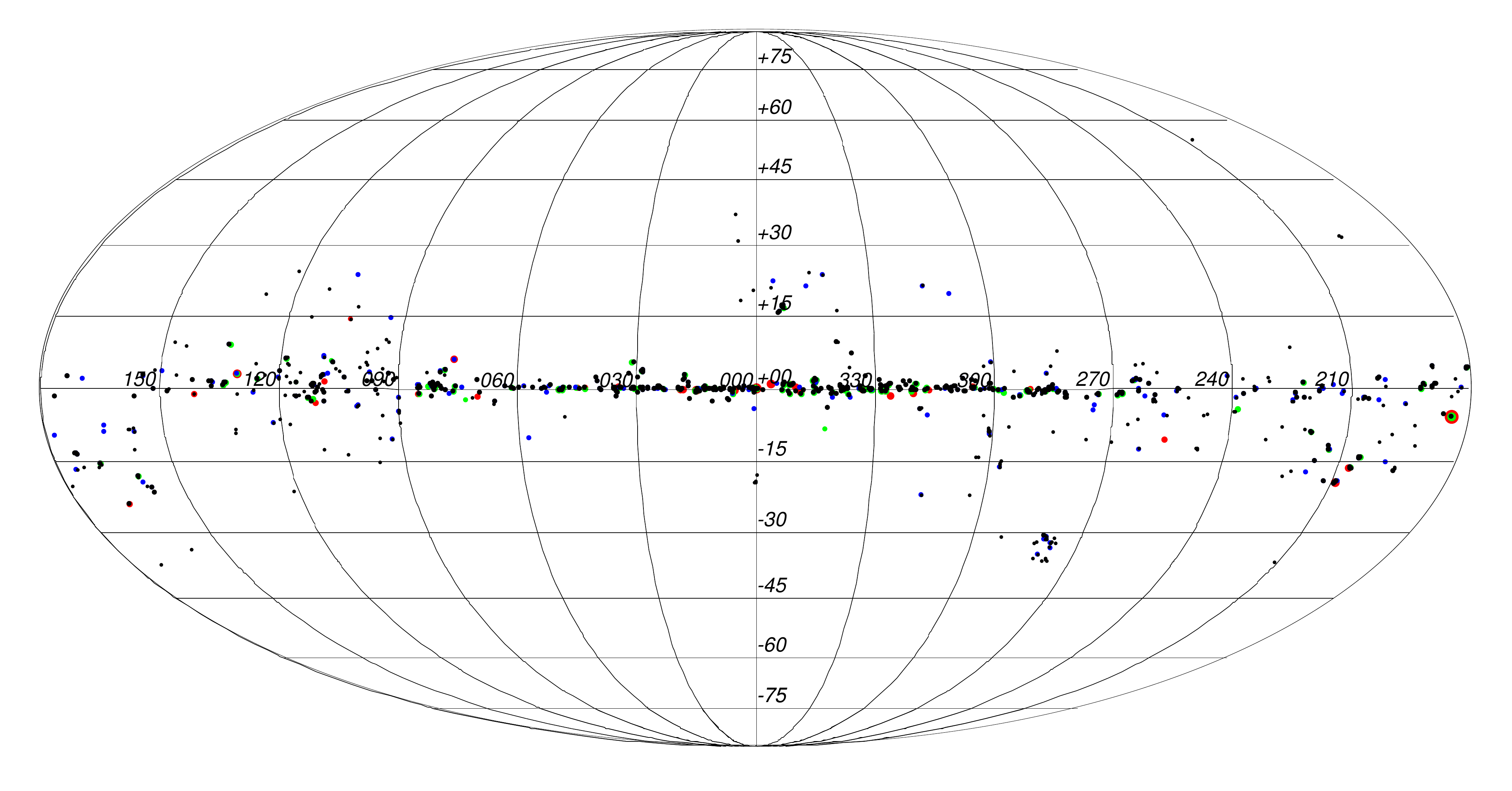}
\caption{Distribution of the  polarized sources from the \plccs\ (top) and the \ecat\ (bottom).  Red, blue, green, and black circles show sources from the 100, 143, 217, and 353\,GHz  catalogues, respectively. As in previous figures, the size of the filled circles gives a qualitative idea of the relative flux densities of the sources at each frequency, where the larger dots correspond to larger flux densities. Note that a different circle size range for each channel was necessary for visualization purposes.}
\label{fig:hficat_positions_pol}
\end{center}
\end{figure*}
The compact sources in the catalogue were detected at each frequency independently using improved versions of the detection pipelines used to create the \lastcat. These pipelines are based on the Mexican Hat Wavelet~2 algorithm \citep[MHW2;][]{gnuevo06,caniego06}.
This is a cleaning and denoising algorithm used to convolve the maps, preserving the amplitude of the sources while greatly reducing the large-scale structures visible at
 these frequencies (e.g. diffuse Galactic emission) and small scale fluctuations (e.g. instrumental noise) in the vicinity of the sources. The LFI and HFI DPCs have different implementations of the MHW2 algorithm, but consistency checks between them have been performed.
The differences between the implementations are described in \cite{planck2013-p05}, and are due to the different characteristics of the maps at LFI and HFI frequencies, requiring alternative methods to reduce the numbers of spurious detections.
Both implementations project the full-sky maps onto square patches where the filtering and detection is performed. The sizes of the patches and the overlap between patches have been chosen in such a way that the full sky is effectively covered.  Sources above a fixed signal-to-noise ratio (\snr) threshold are selected and their positions are translated from patch to spherical sky coordinates. Because the patches overlap, multiple detections of the same object can occur; these must be found and removed,  keeping the detection with the highest \snr\ for inclusion in the catalogue.

\subsection{Defining the \plccs\ and \ecat}
\label{sec:selection}

For reasons explained in Sect.~\ref{sec:intro}, for this release we provide two subcatalogues for some frequency channels, which we call \plccs\ and \ecat.  
We also provide a new parameter for each source that gives the highest reliability catalogue to which the source belongs. However, it is not possible to evaluate this parameter for every source. 
It is this consideration which separates the sources into the \plccs\ and the \ecat; those sources without this field are placed in the \ecat.
Figure \ref{fig:bothcat_positions_temperature} shows the distribution of the \plccs\ and \ecat\ sources across the sky for three of the \Planck\ frequency channels (30, 143, and 857\,GHz).

\paragraph{LFI:}
One measure of the reliability of the sources detected in the three LFI channels is based on a comparison with existing catalogues of radio sources.
First, each single-frequency catalogue is compared with the appropriate external radio catalogues. Next, all the remaining unidentified sources outside a $|b|>20^{\circ}$ Galactic cut are examined on a source-by-source basis by performing a manual search in archival repositories such as the NASA/IPAC Extragalactic Database (NED).
Second, a multifrequency analysis is used to assess whether or not a source is present in more than one \Planck\ channel between 30 and 100\,GHz.
All sources with plausible identifications in external catalogues are assigned to the \plccs\ with a high degree of reliability. Additionally, all sources that are detected in two or more \Planck\ channels
 in the multifrequency analysis are also placed in the \plccs, albeit with a lower degree of reliability. 
Given the small number of remaining sources, we do not create a \ecat\ for the LFI bands; instead, we flag the least reliable sources.
Further details can be found in Sect.~\ref{sec:lfi_reliability}, which describes our assessment of the reliability of the \plccs\ at the lower frequency channels.

\paragraph{HFI:}
There are no external full-sky catalogues at HFI frequencies; hence the reliability cannot be evaluated following the same procedure as for LFI.
Instead, we follow a similar procedure to the one used for the \lastcat\; as described in \cite{planck2013-p05}.  In that paper, we used two measures of reliability for the HFI catalogues.
The first measure, which we called \emph{simulation reliability}, is determined from source injection into simulated maps and is defined as the fraction of detected sources that matched the positions of injected sources.  If the simulations are accurate, such that the spurious and real detection number counts mirror the real catalogue, this reliability is exact.  
The second measure, which we called \emph{injection reliability}, is determined from source injection into the real maps, and in this case we are only simulating the real source component.  
The total source counts are composed of real and spurious components. In order to understand the reliability, we need to understand the spurious component.
However, if we have a knowledge of the real component we can infer the spurious component from the total source counts.
This is the motivation for this second approach, where injection reliability is defined as the fraction of total extracted sources that correspond to injected sources recovered by PCCS2 detection methods.
This method therefore provides an approximate assessment of the reliability. 
Naturally, the first method is preferred over the second, but the second approach becomes necessary when the simulations are not a sufficiently accurate representation of the sky signal. 
In order to establish whether the simulations are of sufficient quality, we require that the simulated catalogues pass two internal consistency tests: that they have the same injected source completeness as the real catalogues; and that they have total detected number counts, as a function of \snr, that are consistent with those in the real data. (The intrinsic number counts are assumed to be power law functions of flux density and are fitted to the detection counts at higher flux densities, where the catalogues are reliable and complete, and extrapolated to lower flux densities).  
In order to achieve this internal consistency we need to exclude the Galactic plane region from the analysis, due to discrepancies arising from deficiencies in the simulation of diffuse dust emission near the beam scale (e.g. Galactic cirrus) and uncertainties in defining an input source model for the Galactic sources.
The region excluded increases with frequency and is not a simple Galactic latitude cut,  but is based on the level of the dust emission.  The Galactic masks used at each of the six HFI frequencies are described in Sect.~\ref{sec:qa_reliability}.
In addition to these Galactic plane regions, for the highest four frequency channels we also exclude the region of sky inside a ``filament mask'' from the reliability assessment. Note that there is a different filament mask for each of these channels.
These filament masks describe the areas of the sky in which residual structures, not related to sources, are present in the MHW2 filtered maps. The creation of these masks is explained further in Sect.~\ref{sec:filament_masks}.
The union of the filament mask and the Galactic plane region then defines the area of the sky in which sources are assigned to the \ecat.
Thus, whether a source is assigned to the \plccs\ or \ecat\ is determined solely by its location on the sky.
The HFI reliability assessment for the \plccs\ is described further in Sect.~\ref{sec:hfi_reliability} (recall there is no reliability assessment for the \ecat).

\subsection{Photometry}
\label{sec:photometry}

As in the \lastcat, we provide four different measures of the flux density for each source. They are determined by the source detection algorithm (DETFLUX), aperture photometry (APERFLUX), point spread function fitting (PSFFLUX), and Gaussian fitting (GAUFLUX). Only the first is obtained from the filtered maps; the other measures are estimated from the full-sky maps at the positions of the sources. The source detection algorithm photometry, the aperture photometry, and the point spread function (PSF) fitting use the \Planck\ band-average effective beams, calculated with {\tt FEBeCoP} (Fast Effective Beam Convolution in Pixel space; \citealt{mitra2010, planck2014-a05, planck2014-a08}). Note that only the PSF fitting algorithm takes into account the variation of the PSF with position on the sky.
The \plccs\ has been produced from the \Planck\ full-mission maps (eight sky surveys in the LFI and five sky surveys in the HFI), and therefore  supersedes the previous catalogues (for the \lastcat\ only 1.5 surveys were analysed). It also includes the latest calibration and beam information, and  we have improved some of the algorithms used to measure the photometry of the sources. In order to assess the differences between the photometry in the \lastcat\ and \plccs\ we have compared both sets of catalogues at all \Planck\ bands.

A major change is in the Gaussian fitting photometry. We have implemented a new version of the algorithm that produces more robust measures, particularly for extended objects where the difference between the flux densities in the \lastcat\ and \plccs\ can be as large as $100\%$. In the previous version of the algorithm, for some sources the fitting code was not converging properly and this issue has been addressed by using a new fitting approach; see Appendix~\ref{sec:gauflux} for a description of the method and its validation.  In addition, the photometry from the detection pipeline has changed at some frequencies by several percent, because it now takes into account the latest information about the effective beam FWHM and corrects for the biases listed in Table \ref{tab:qa}, which range from $1\%$ to $12\%$, depending on the frequency. The other two techniques, aperture photometry and PSF fitting, produce similar results in both catalogues. In the first case, the algorithm has not changed, while in the second, although the algorithm has been changed to improve the positional accuracy, this does not affect our measurements because we use the coordinates from the detection pipeline as the reference for all photometric measures. In both cases the differences are always at the percent level.
Moreover, flux densities extracted from the publicly released \Planck\ maps at 30, 44, and 70\,GHz require a small correction for beam efficiency, since a small amount of power lies outside the main beam \citep{planck2014-a03}.  These small multiplicative corrections are 1.00808, 1.00117, and 1.00646 at 30, 44, and 70\,GHz, respectively.  The flux densities provided for sources in the \plccs\ catalogues have been corrected accordingly.
Uncertainties are provided for all four flux-density measures. In Table \ref{tab:all_pccs_stats} we show the uncertainties associated with the  flux densities of the faintest sources in the extragalactic zone of each catalogue, after excluding the faintest $10\,\%$ of sources as obtained with the source detection algorithm. These uncertainties range from $90$ to $130$\,mJy for the 30--70\,GHz catalogues, and from $30$ to $270$\,mJy for the 100--857\,GHz catalogues. This gives an idea of the sensitivity of the catalogue and the associated uncertainties. However, the uncertainties depend not only on the flux density of the sources but also on their position in the sky, so we provide noise maps that can be used to estimate the expected uncertainty in the flux density of a source at any position in the sky.

\paragraph{Detection pipeline photometry ({DETFLUX}).}  The detection pipelines assume that sources are point-like. The amplitude of a detected source is converted to flux density using the solid angle of the effective beam (from Table~\ref{tab:beam_data}), and the conversion from map units into intensity units. The uncertainty in the flux density for each source is measured as the local noise in an annulus around the source in the MHW2-filtered map, where bright pixels belonging to other compact sources in the vicinity, if any, are excluded from the calculation. If a source is resolved by \Planck\ its flux density will be underestimated. In this case it may be better to use the GAUFLUX estimation.  The estimation of the flux density provided by the HFI detection pipeline has been improved since the \lastcat\ release, by removing a bias that lowered the recovered flux densities in  the higher frequency channels \citep[see][]{planck2013-p05}. The photometric performance of the  \plccs\ detection pipeline is assessed in  Sect.~\ref{sec:qa_photometry}.

\paragraph{Aperture photometry ({APERFLUX}).}  The flux density is estimated by integrating the data in a circular aperture centred at the position of the source. An annulus around the aperture is used to evaluate the level of the background. The annulus is also used to make a local estimate of the noise to calculate the uncertainty in the flux density. The flux density is corrected for the fraction of the beam solid angle falling outside the aperture and for the fraction of the beam solid angle falling in the annulus. The aperture photometry was computed using an aperture with a radius equal to the average FWHM of the effective beam (the effective FWHM in Table~\ref{tab:beam_data}), and an annulus with an inner radius of 1~FWHM and an outer radius of 2~FWHM. The effective beams, also given in Table~\ref{tab:beam_data},  were used to compute the beam solid angle corrections. For details see the \lastcat\ paper \citep{planck2013-p05}.

\paragraph{PSF fit photometry ({PSFFLUX}).}
The flux density and its uncertainty are obtained by fitting a model of the PSF at the position of the source. The model has four free parameters: the amplitude of the source; a background offset; and two coordinates for the location of the source. The PSF is obtained from the effective beam by means of a bicubic spline interpolation for source positions that are different from the centre of a pixel. Note that the PSF fitting now includes subpixel positioning, which is a new feature introduced after the production of the \lastcat. For details see Appendix~\ref{sec:psfflux}.

\paragraph{Gaussian fit photometry ({GAUFLUX}).}
The approach to Gaussian fitting has been completely revised since the \lastcat.  The algorithm now allows the position of the source to vary as the best fit is found. The same parameters are returned for each source: its flux density;  the major and minor semi-axes; and an orientation angle. Additionally, as in the \lastcat, the semi-axis values are used in the construction of the flag for extended sources.
The new method uses a downhill simplex method in multidimensions, the Nelder-Mead method,  to find the best-fit values in the full parameter space of position, flux density, and elliptical Gaussian parameters. The method has been shown to be robust and stable \citep{Press:1992:NRC:148286}.  Optimization is based on the reduced log-likelihood with prior regularization for the size of the source defined by the effective beam at each frequency. The downhill simplex methods does not produce estimates of the flux density uncertainties. For this purpose a Markov Chain Monte Carlo has been used. 
For details see Appendix~\ref{sec:gauflux}.

\subsection{Polarization}
\label{sec:polarization}
\begin{figure}
\begin{center}
\includegraphics[width=0.52\textwidth]{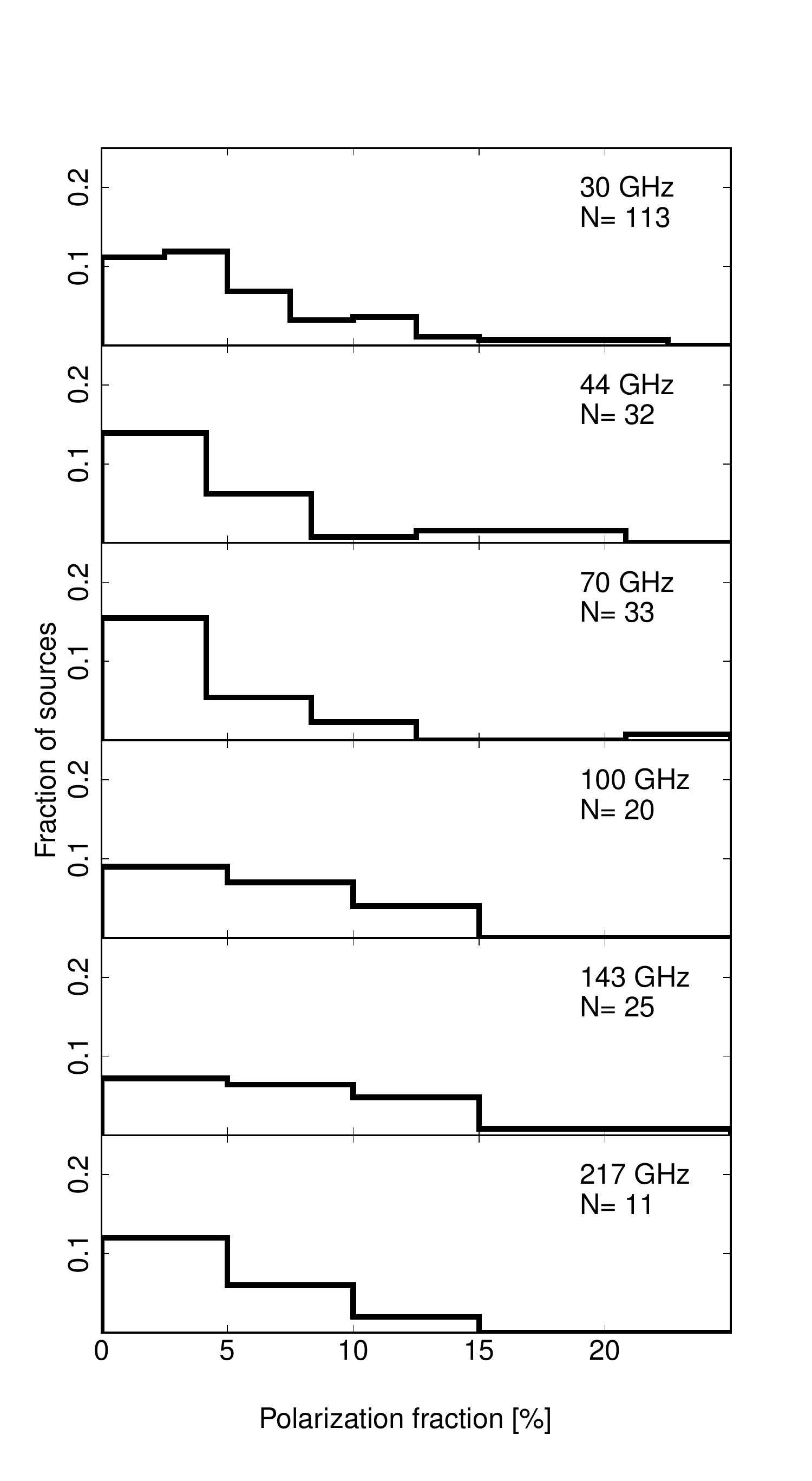}
\caption{Normalized histograms of the recovered polarization fraction from the \plccs\ catalogue from 30 to 217\,GHz (the 353\,GHz channel is not shown because the catalogue contains only one source). The number of sources in each histogram is indicated.
\label{fig:HFI_DX11c_POLFRAC}}
\end{center}
\end{figure}
In the \Planck\ polarization maps, the polarized sources are embedded in a background that is the combination of instrumental noise and  diffuse emission. The nature of the diffuse emission depends on the observation frequency; for example, polarized synchrotron emission in the lower frequency channels and infrared emission in the higher frequency channels.
In both regimes the polarization fraction of the compact sources (the ratio between their polarized flux densities and total intensity) is typically lower than  1--2\,\%. This presents a challenge in terms of disentangling the true polarized flux density of a source from the background.
In order to tackle this problem, a two-step process has been proposed \citep{LopezCaniego09}. First, a maximum-likelihood filter is applied, reducing the noise and enhancing the \snr\ of the sources embedded in the $Q$ and $U$ maps \citep{Argueso09}. Second, the significance of each detection is assessed based on the statistics of the local background in the vicinity of the source.
Several significance levels were investigated and we concluded that, for the typical polarization backgrounds present in the \Planck\ polarization maps, a significance threshold of 99.99\,\% successfully distinguishes the polarized emission of a compact source from a peak in the background.
This approach has been used in the present catalogues to attempt to measure the polarized flux densities and uncertainties of all sources found in the temperature maps. Polarization measurements are provided for all sources where the significance of the detected polarized signal reaches or exceeds the limit of 99.99\,\%; for the remaining sources we provide the 99\,\% upper limit. 
Figures~\ref{fig:lficat_positions_pol}~and~\ref{fig:hficat_positions_pol} show the distributions of the significantly polarized sources in the LFI and HFI polarized frequency channels.
Normalized histograms of the polarization fraction for the population of significantly polarized sources in the \plccs\ catalogues are shown in Fig.~\ref{fig:HFI_DX11c_POLFRAC}.

If a source is not strictly point-like, the filtering procedure used to reduce the noise will also remove signal. For this reason we also provide aperture photometry measurements for polarization, which, although noisier,  do not remove as much signal from compact (but not point-like) sources as filtering. The aperture photometry package is common to both LFI and HFI, whereas there are two different implementations of the maximum-likelihood estimator, which will allow us to assess the robustness of the methods by comparing their results in a common set of simulations. 
The results of the HFI and LFI polarization pipelines are compared in Sect.~\ref{sec:qa_polarization}.
The polarized flux density of a source, $P$,  is evaluated using
\begin{equation}
\label{eqn:P_def}
P = \sqrt{  Q^2+U^2 },
\end{equation}
where $Q$ and $U$ are the flux densities in the Stokes $Q$ and ${U}$ maps, measured at the position of the source detected in the $I$ map.
We follow the IAU/IEEE convention \citep{IAUdef} for defining the angle of polarization of a source: polarization angles are taken as increasing anticlockwise (north through east). In this paper, position angle zero is taken as the direction of the north Galactic pole.
The polarization angle is defined by
\begin{equation}
\label{eqn:Pangle_def}
\psi=\frac{1}{2} \arctan(-U/Q).
\end{equation}
The minus sign is necessary to correct from the \texttt{HEALPix} convention for position angles used in
the \Planck\ Stokes parameter maps, in which position angle increases clockwise. As noted in \cite{planck2014-a01}, the convention used for Planck polarization maps is the one usual in CMB studies and is used in WMAP papers, whereas the IAU/IEEE convention adopted here is standard for astronomical sources. Polarization angles
are given in degrees in the range $-90\deg$ to $90\deg$.
The estimate of $P$ acquired using Eq.~(\ref{eqn:P_def}) is biased, because the errors in the $Q$ and $U$ measurements, on average, contribute positively to the measurement of $P$ (see, e.g. \citealt{Montier15}).
However, in our significance regime we can use the approximation
\begin{equation}
\label{eqn:P_debiased}
P_{\rm debiased} = \sqrt{ P^2 -\sigma^2_P  }
\end{equation}
to debias our estimate of $P$, where $\sigma_P$ is the error in $P$ and is calculated by propagating the errors in $Q$ and $U$, where $\sigma_{Q,U}$ are calculated as the local rms in an annulus around the source in the maximum-likelihood filtered $Q$ and $U$ maps, under the assumption of no correlation:
\begin{equation}
\label{eqn:P_err}
\sigma_P = \sqrt{ \frac{1}{  Q^2+U^2  }\left( Q^2 \times \sigma_Q^2 + U^2 \times \sigma_U^2 \right) } \, .
\end{equation}
The polarization angle error is obtained by propagating the errors in $Q$ and $U$:
\begin{equation}
\label{eqn:P_ang_err}
\sigma_{\psi} = \frac{1}{ 2\left( Q^2+U^2  \right )} \sqrt{  Q^2 \times \sigma_U^2 + U^2 \times \sigma_Q^2  } \, .
\end{equation}

As shown in Table \ref{tab:pccs_pol} the typical uncertainty in the polarized flux density is 45--90\,mJy between 30 and 70\,GHz and 30--180\,mJy between 100 and 353\,GHz.

\subsubsection{Corrections for bandpass mismatch}
\label{subsec:bandpass_mismatch}
Mismatch between the bandpass shapes of the two orthogonally-polarized detectors in each feed horn causes
leakage of total intensity into the polarization signal for any emission whose spectrum differs from that of the primary calibrator, namely the CMB dipole; therefore all foreground emission including that from compact sources suffers from temperature-to-polarization leakage. Correction requires a model of the spectrum of the source,  as well as a model for the spectral response of each detector or bolometer.  Since the detecting elements used in the two instruments are different, LFI and HFI treated bandpass mismatch differently.  The magnitude of the correction can be very different from one source to another. In the lower \textit{Planck} frequencies, the correction can vary from a fraction of a percent up to $100\,\%$. In the higher frequency channels this correction is always below the percent level. The details are presented in Appendix \ref{sec:mismatch}. 

\subsubsection{Evaluation of marginal polarization measurements}
At four HFI frequencies (100, 143, 217, and 353\,GHz) we present an additional set of polarized flux-density and polarization angle estimates for sources detected only marginally in polarization. 
These are derived using the Bayesian ``PowellSnakes'' algorithm \citep[see][]{planck2013-p05}.  The aim is to disentangle the sources that have some polarized emission from those that are consistent with no polarized signal.  
This allows us to probe fainter polarization signals, and thus to provide deeper and more complete polarization catalogues without any loss of reliability (as we show in Sect.~\ref{sec:qa_polarization}). The details of the method are presented in Appendix~\ref{sec:marginal_pol}.

\section{Validation of the \plccs}
\label{sec:validation}
The contents of the \plccs\ and the four different flux-density estimates have been validated by simulations (internal validation) and comparison with other astrophysical data (external validation), as was done for the \lastcat\ \citep{planck2013-p05}. The validation of the low-frequency sources can be performed in part by using the large number of existing catalogues. Detections identified with known sources have been flagged as such in the catalogues.
In contrast, the validation of sources at higher frequencies must be done using simulations, specifically through a Monte Carlo quality assessment process in which artificial sources are injected into both real and simulated maps.
In the following subsections, we discuss tests on the completeness, reliability, astrometry, and photometry of the single-frequency catalogues, as well as comparisons between different \Planck\ bands.  We also describe internal and external validation of the polarization measurements.

\subsection{Completeness}
\label{sec:qa_completeness}
\begin{figure}
\begin{center}
\includegraphics[width=0.5\textwidth]{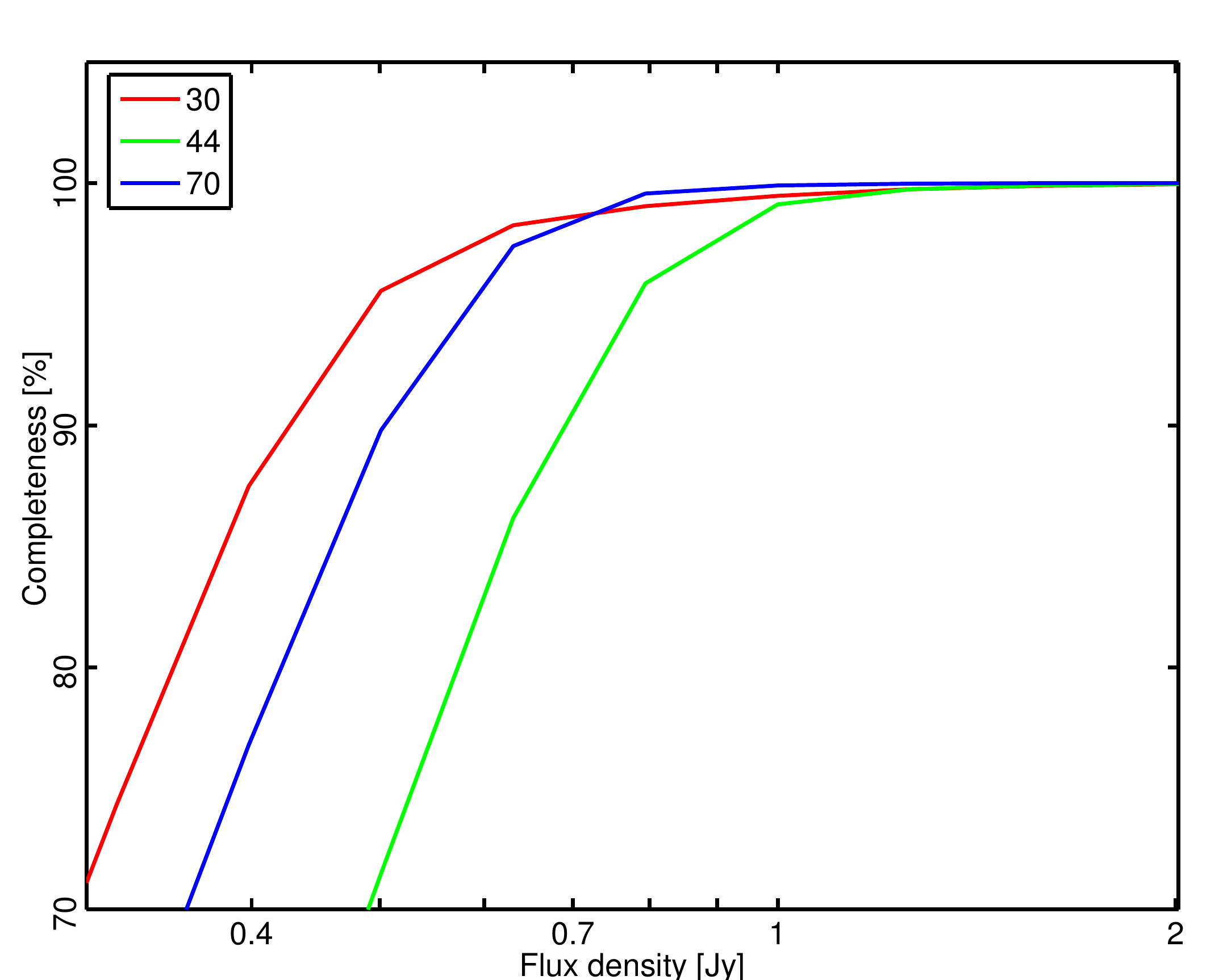}
\includegraphics[width=0.5\textwidth]{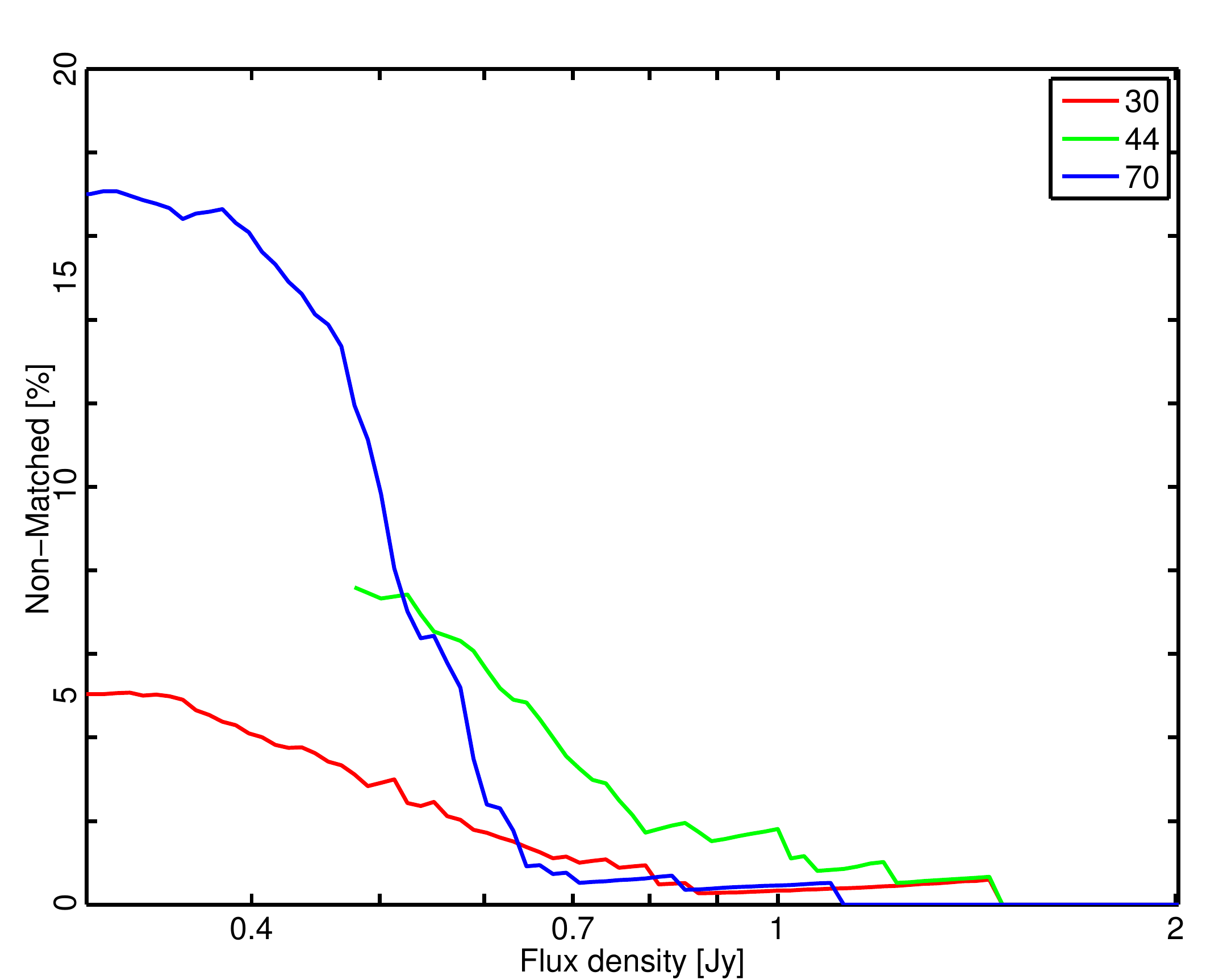}
\caption{Validation of the sources in the 30, 44, and 70\,GHz channels. The top panel shows the cumulative completeness per flux-density bin of the \plccs\ catalogue. The lower panel shows the number of unidentified sources per flux-density bin. These sources have no clear counterpart in any of the external catalogues used for validation. However, many of these source have been detected in more than one \Planck\ channel and this implies, first, that they could be real sources and not spurious, and one can consider the number of non-matched sources presented here as an upper limit. Second, the fact that they are not in the external catalogues suggests that, if real, they could be potentially interesting objects, maybe going through a flaring phase. 
\label{fig:LFI_comp_reliab}}
\end{center}
\end{figure}
\subsubsection{LFI}
\label{sec:lfi_completeness}
In the case of the three lowest frequencies,
we compared the \Planck\ compact source detections to external catalogues of radio sources.  We began by constructing a band-merged catalogue based on positional coincidence and included all the sources detected above the initial $4\sigma$ detection threshold at any of the LFI frequencies. The catalogue contained 2039 sources, many of which were detected at only one frequency. We then used this catalogue to make a position-based search for identifications with three external catalogues of radio sources using a search radius of $1.5 \times \sigma_{\rm b}$, where $\sigma_{\rm b} = {\rm FWHM} / (2\sqrt{2\,\ln 2})$ can be evaluated from the fitted FWHM in Table~\ref{tab:beam_data}. These catalogues are: (1) in the southern hemisphere, the Australia Telescope 20 GHz Survey (AT20G; \citealt{murphy10}), a catalogue of sources brighter than 40\,mJy that covers the whole southern sky $(|b|<0\deg)$; (2) in the northern hemisphere, where no large-area survey at similar frequencies to AT20G is available, the 8.4\,GHz Combined Radio All-sky Targeted Eight GHz Survey (CRATES; \citealt{healey07}), a compilation of flat-spectrum $(\alpha \, \textgreater -0.5)$ radio sources with nearly uniform extragalactic $(|b|>10\deg)$ coverage for sources brighter than 65\,mJy at 4.8\,GHz; (3) the full-sky New WMAP Point Source Catalogue (NEWPS; \citealt{caniego07,massardi09}) covering the frequency range 23--61\,GHz, that include sources brighter than 700\,mJy at 23\,GHz and is complete above 2\,Jy. These catalogues have a similar source density, so there should not be multiple associations of AT20G, CRATES, or NEWPS sources within a \Planck\ beam, and their frequencies range from 8 to 61\,GHz. The NEWPS catalogue  was produced by analysing the WMAP maps, and this dataset is very similar to that of \Planck\ in terms of format (all-sky Healpix maps), background characteristics, and angular resolution ($\sim$13--33\arcmin\ in \Planck\ LFI, vs. $\sim$14--56\arcmin\ in WMAP), which makes it a very good dataset for validation. In addition, the two catalogues, \plccs\ and NEWPS, were produced using the same source-extraction tool (the Mexican Hat wavelet). For the search radius, we use $1.5 \times \sigma_{\rm b}$, which includes $\sim$87\,\% of the area of the \Planck\ beam, as a compromise between the arcsecond resolution of CRATES and AT20G, and the $\sim$ degree resolution of the 23 GHz WMAP channel. In any case, given the similar source density of the catalogues, one could use a slightly larger search radius and the results would not change. As in the \lastcat, the \plccs\ includes $94 \%$ of the sources in NEWPS when using a $1.5 \times \sigma_b$ search radius, so in order to study the completeness of the catalogue deeper samples like CRATES and AT20G are needed. However, the frequencies of these two surveys, 8.4 and 20\,GHz, are lower than the lowest \Planck\ frequency, and variability and spectral effects could push some of the sources below the \plccs\ detection thresholds. Thus, the completeness that we estimate by comparing the \plccs\ against these three catalogues is a lower limit. For this reason we used an alternative completeness estimate that can be derived from knowledge of the noise in the maps when the completeness is greater than $50\%$. If the flux density estimates \textit{S} are subject to Gaussian errors with amplitude given by the noise of the filtered patches, the cumulative completeness per patch should be
\begin{equation}
C(S) = \frac{1}{2} + \frac{1}{2} \erf\left(\frac{S - q\sigma_S(\theta,\phi)}{\sigma_S(\theta,\phi)}\right),
\label{eq:erf}
\end{equation}
where $\sigma_{S}^2(\theta,\phi)$ is the variance of the filtered patch located at $(\theta,\phi)$, $q$ is the \snr\ threshold and $\erf(x) = \frac{2}{\sqrt{\pi}}\int_0^x e^{-t^2} dt$ is the standard error function. The true completeness will depart from this limit when the simplifying assumptions of non-Gaussian noise and uniform Gaussian beams are broken.
Using this expression, the cumulative completeness of each LFI band is derived by making use of a model of the source counts $N(S)$ \citep{deZotti05} that accounts for various source populations (flat-spectrum radio quasars, BL Lac objects, steep-spectrum sources, GPS sources, early phase gamma-ray after glows, etc.). The true completeness will depart from this limit when the simplifying assumptions of non-Gaussian noise and uniform Gaussian beams are broken. 

Figure~\ref{fig:LFI_comp_reliab} demonstrates that the catalogues at 30, 44, and 70\,GHz are essentially 100\,\% complete above flux densities of 1\,Jy, and for 30 and 70\,GHz are still more than 95\,\% complete down to around $0.6$\,Jy. The flux densities that correspond to the 90\,\% completeness level are shown in Table~\ref{tab:all_pccs_stats}.
\subsubsection{HFI}
\label{sec:hfi_completeness}
\begin{figure*}
\begin{center}
\includegraphics[width=0.49\textwidth]{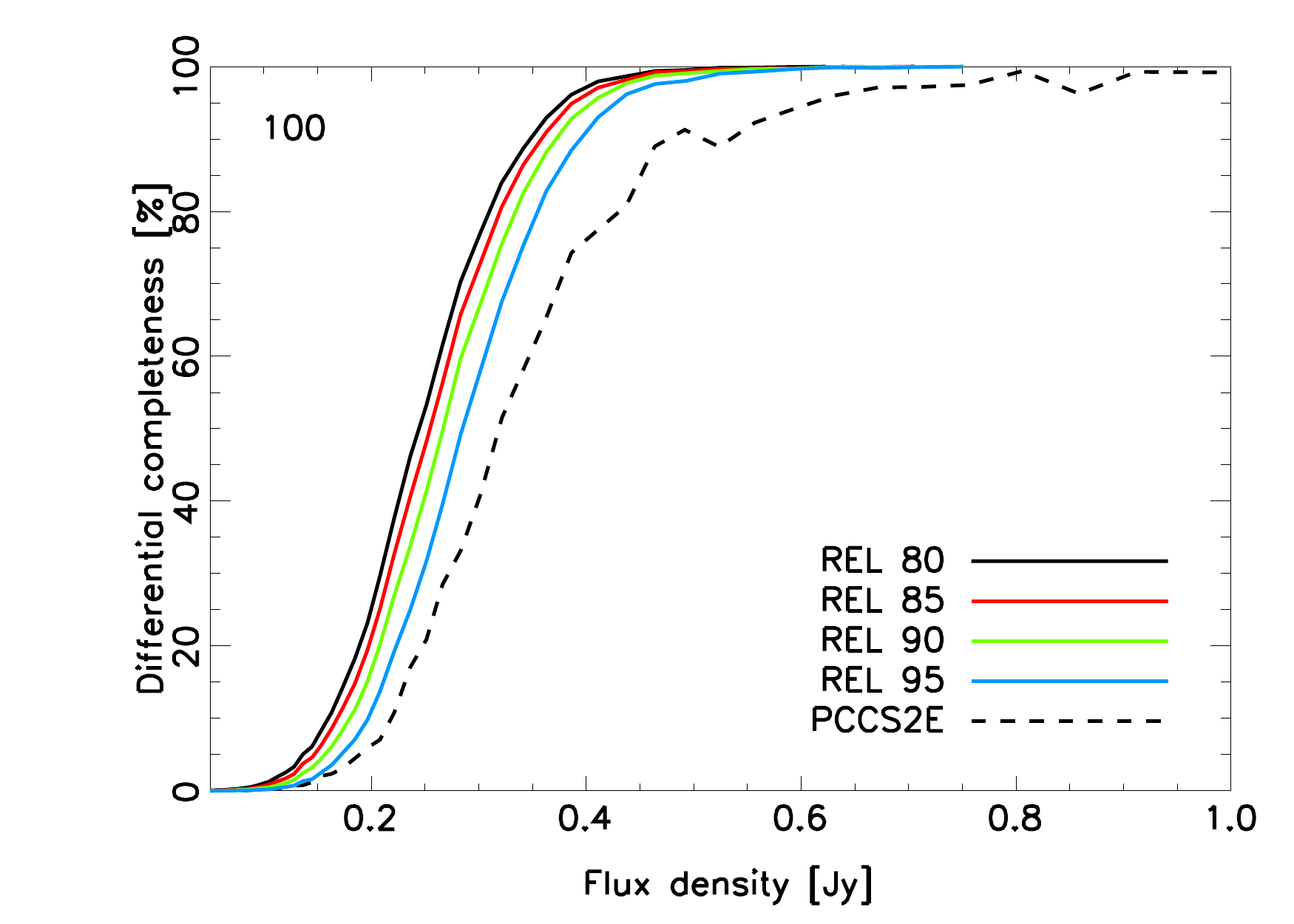}
\includegraphics[width=0.49\textwidth]{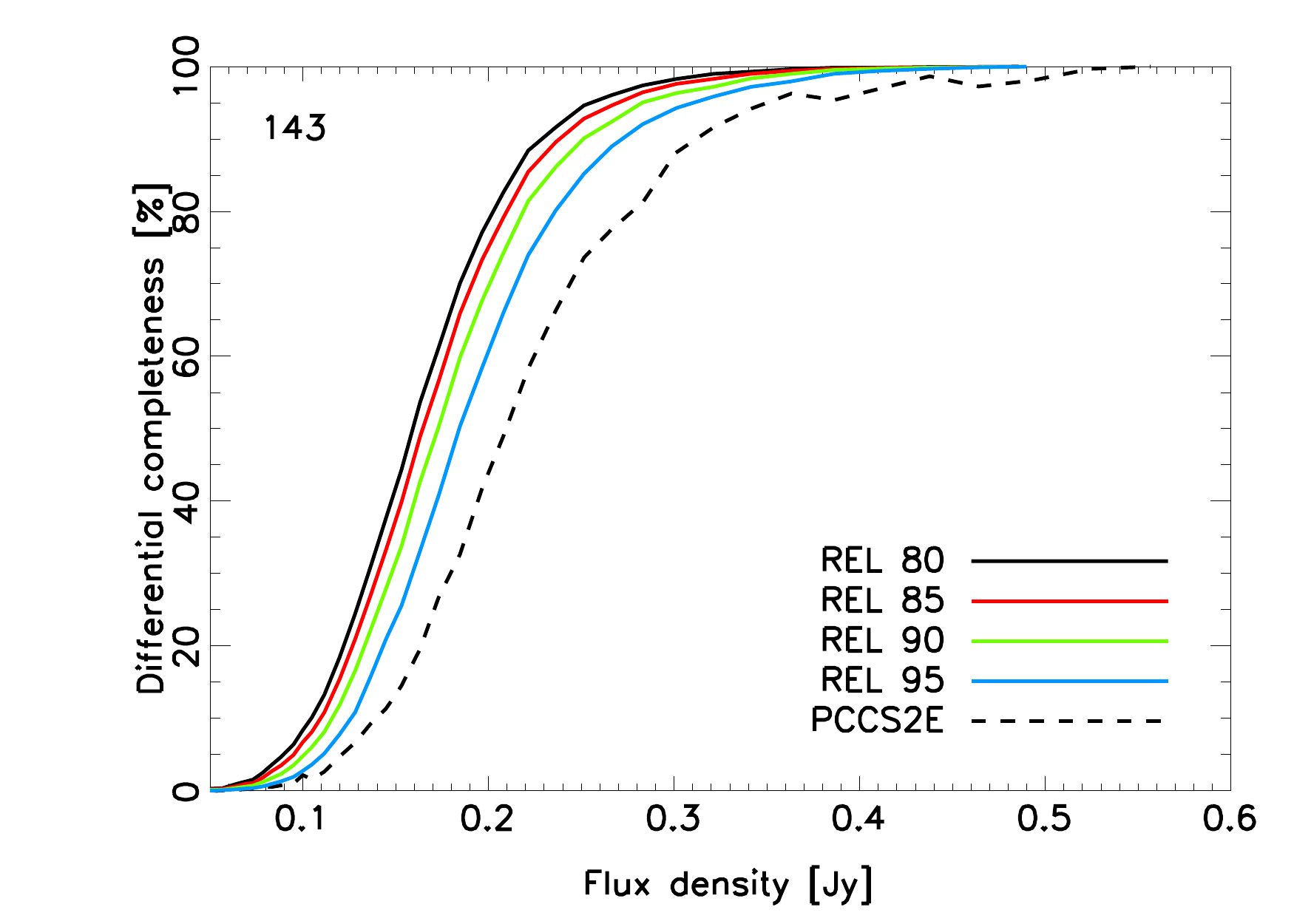}\\
\includegraphics[width=0.49\textwidth]{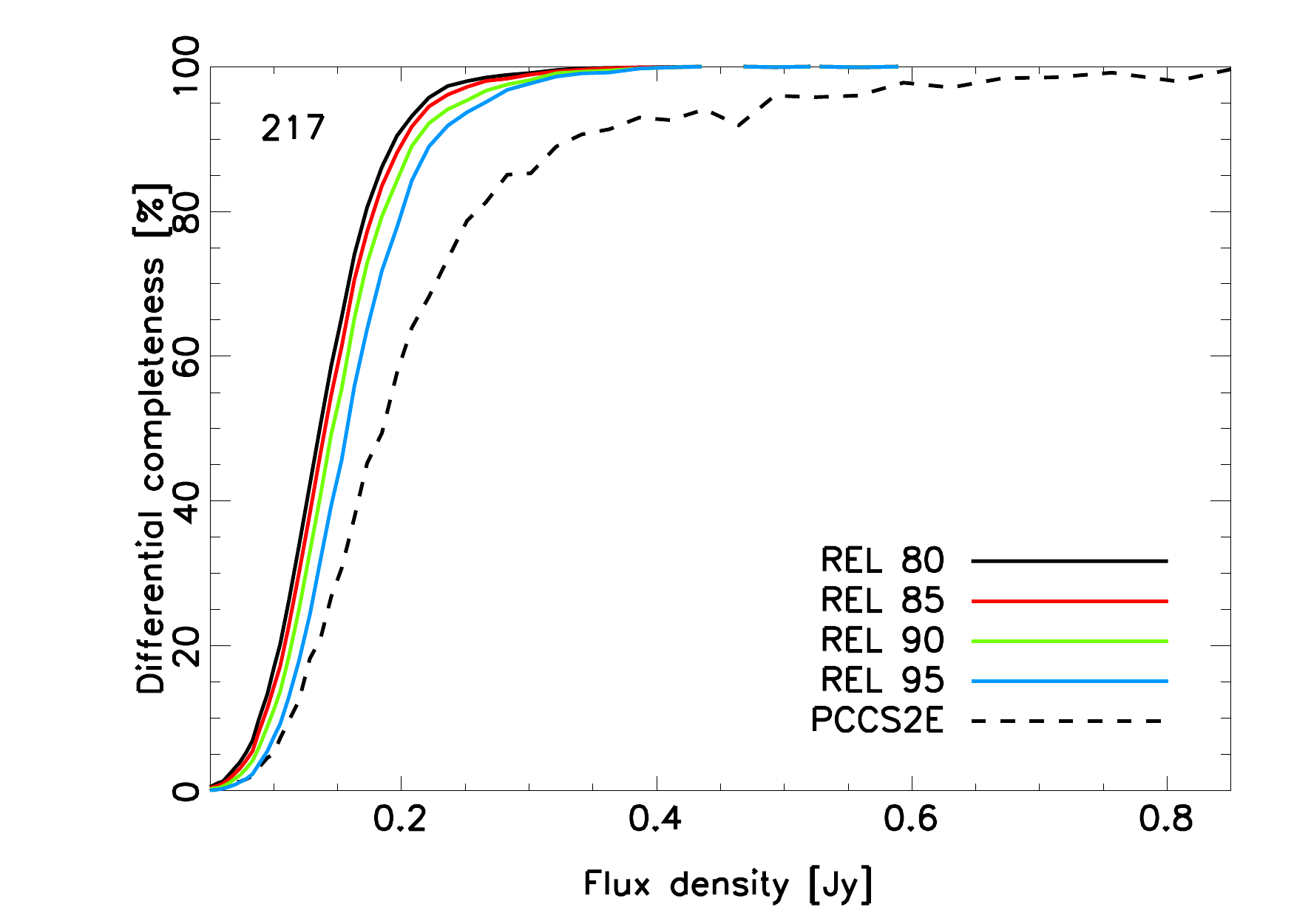}
\includegraphics[width=0.49\textwidth]{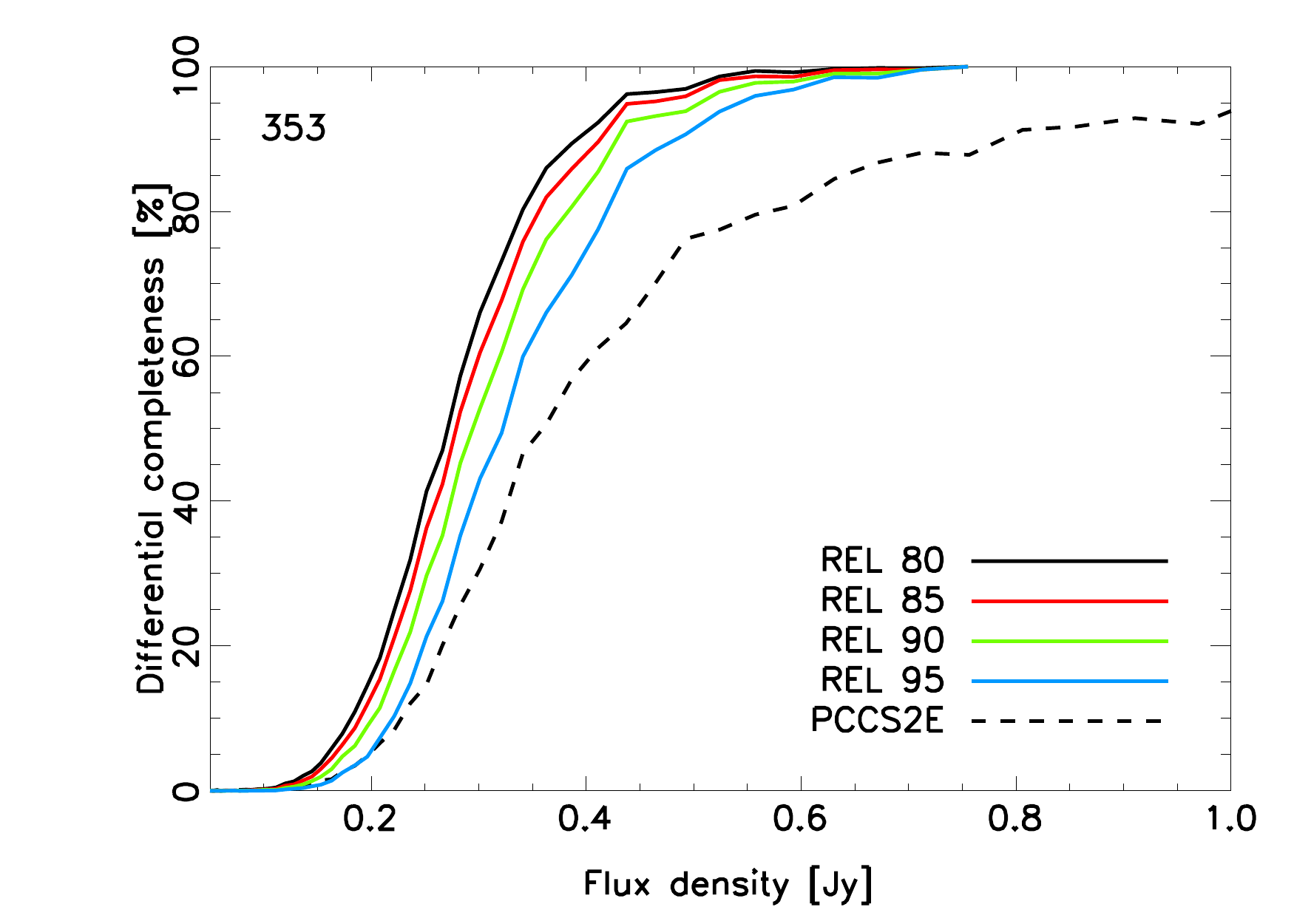}\\
\includegraphics[width=0.49\textwidth]{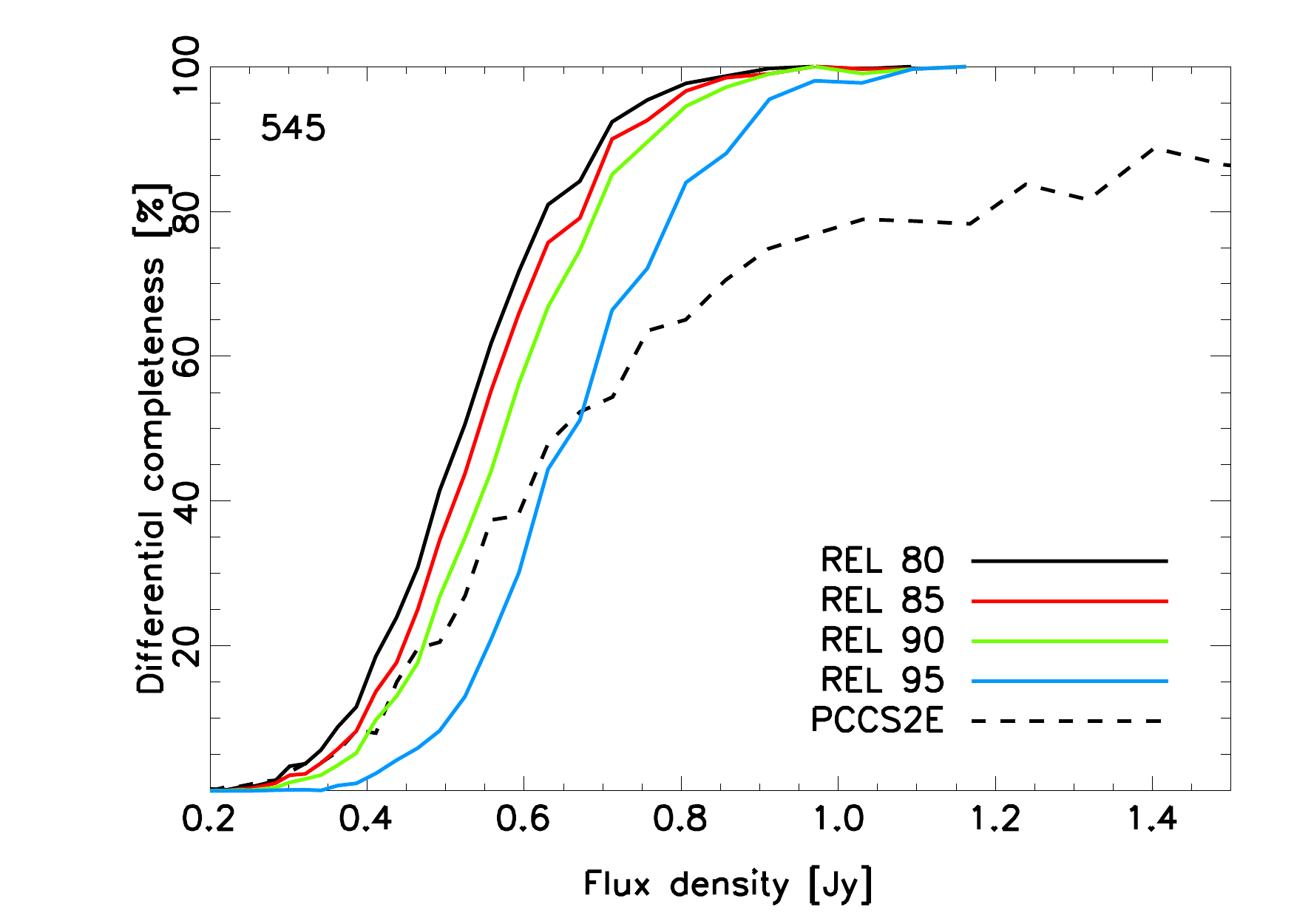}
\includegraphics[width=0.49\textwidth]{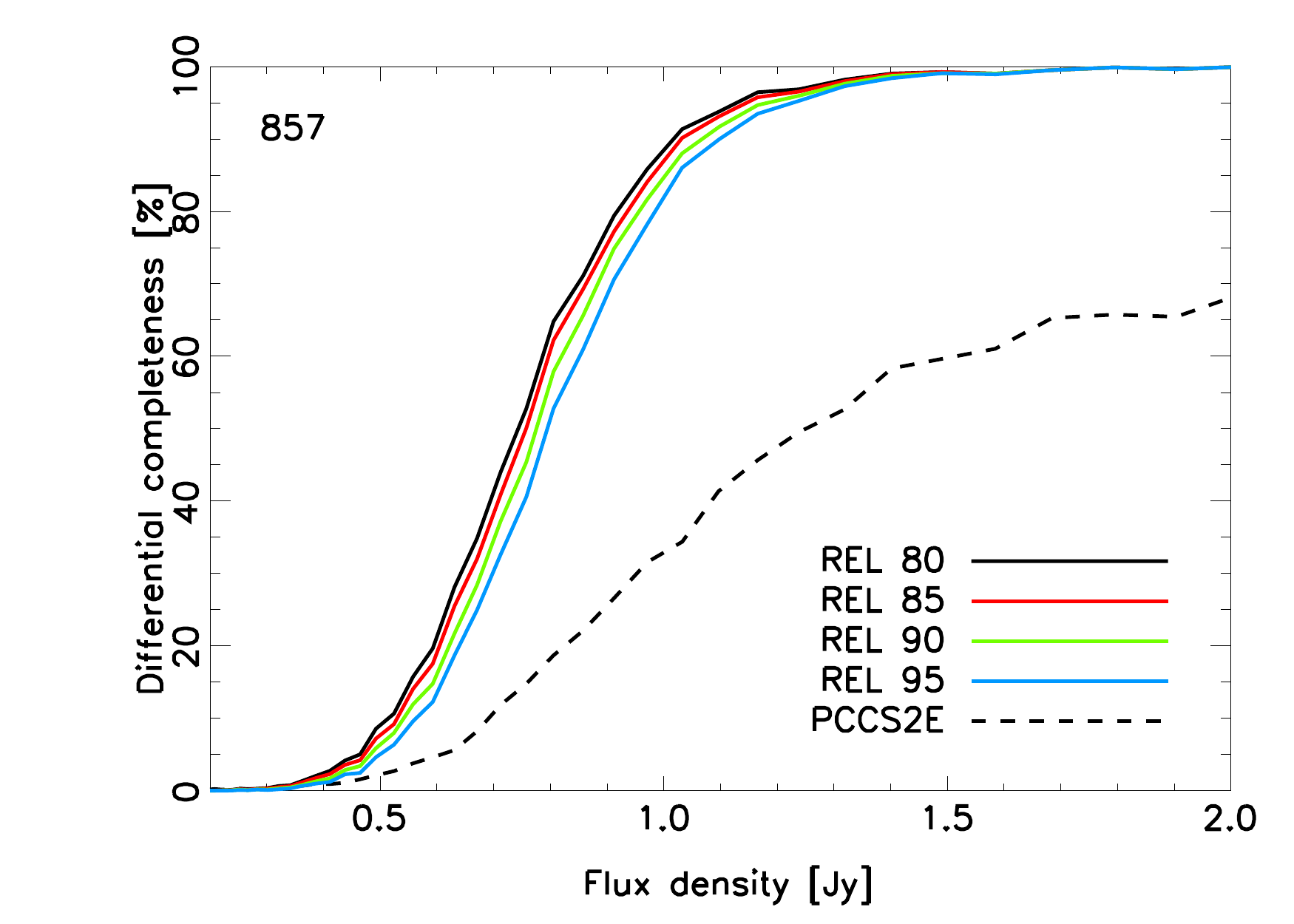}\\
\caption{HFI completeness results from the Monte Carlo quality assessment.  Completeness is shown per channel for the 80, 85, 90, and 95\,\% reliability catalogues, and for the \ecat.
\label{fig:HFI_compl_MCQA}}
\end{center}
\end{figure*}
\begin{figure*}
\begin{center}

\begin{tabular}{p{1.5cm}ccc}
\vspace{-2.7cm}{100\,GHz}&
\hspace{-0.9cm} \includegraphics[width=0.3\textwidth]{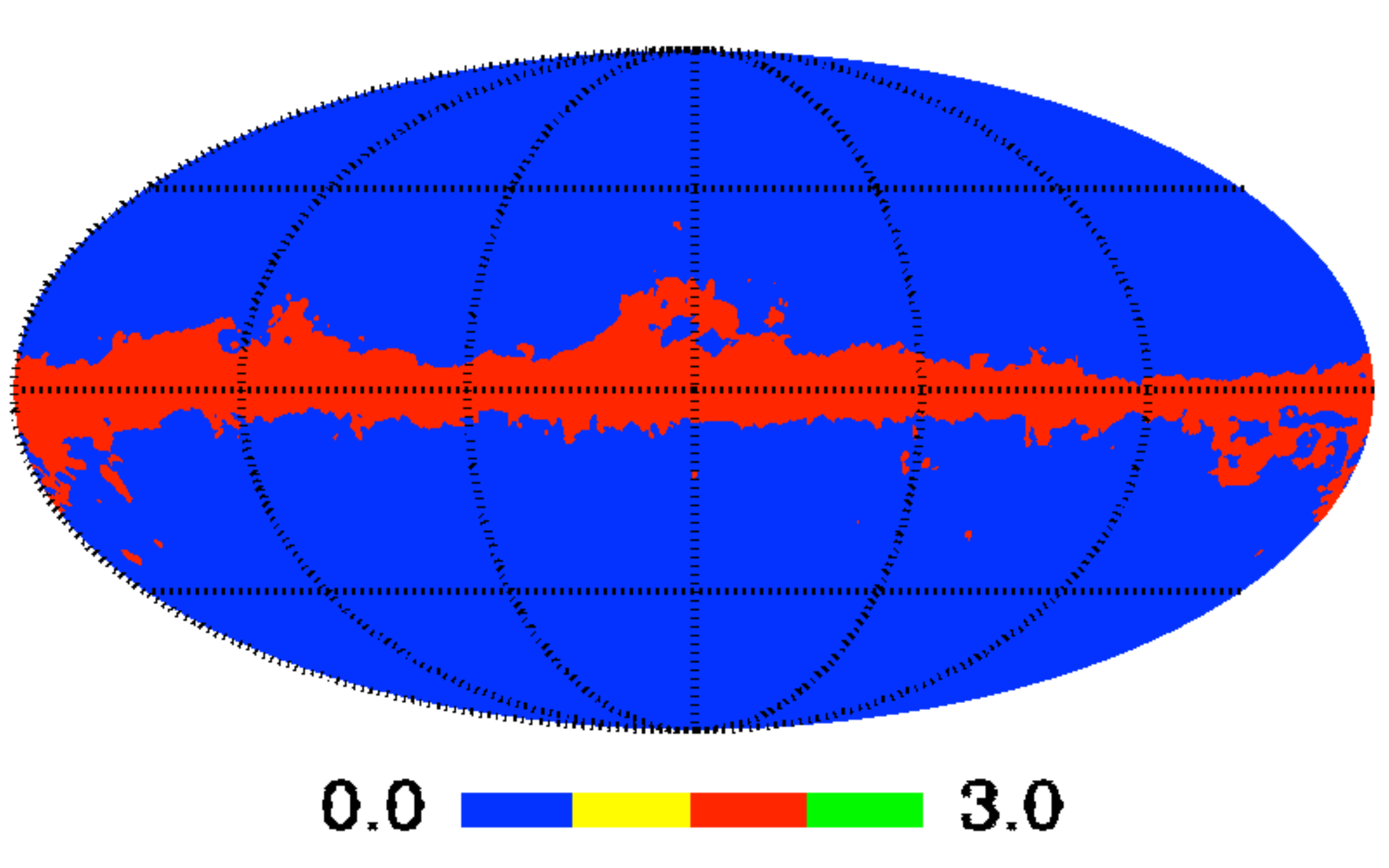} &
 \hspace{-0.5cm}{ \includegraphics[width=0.3\textwidth]{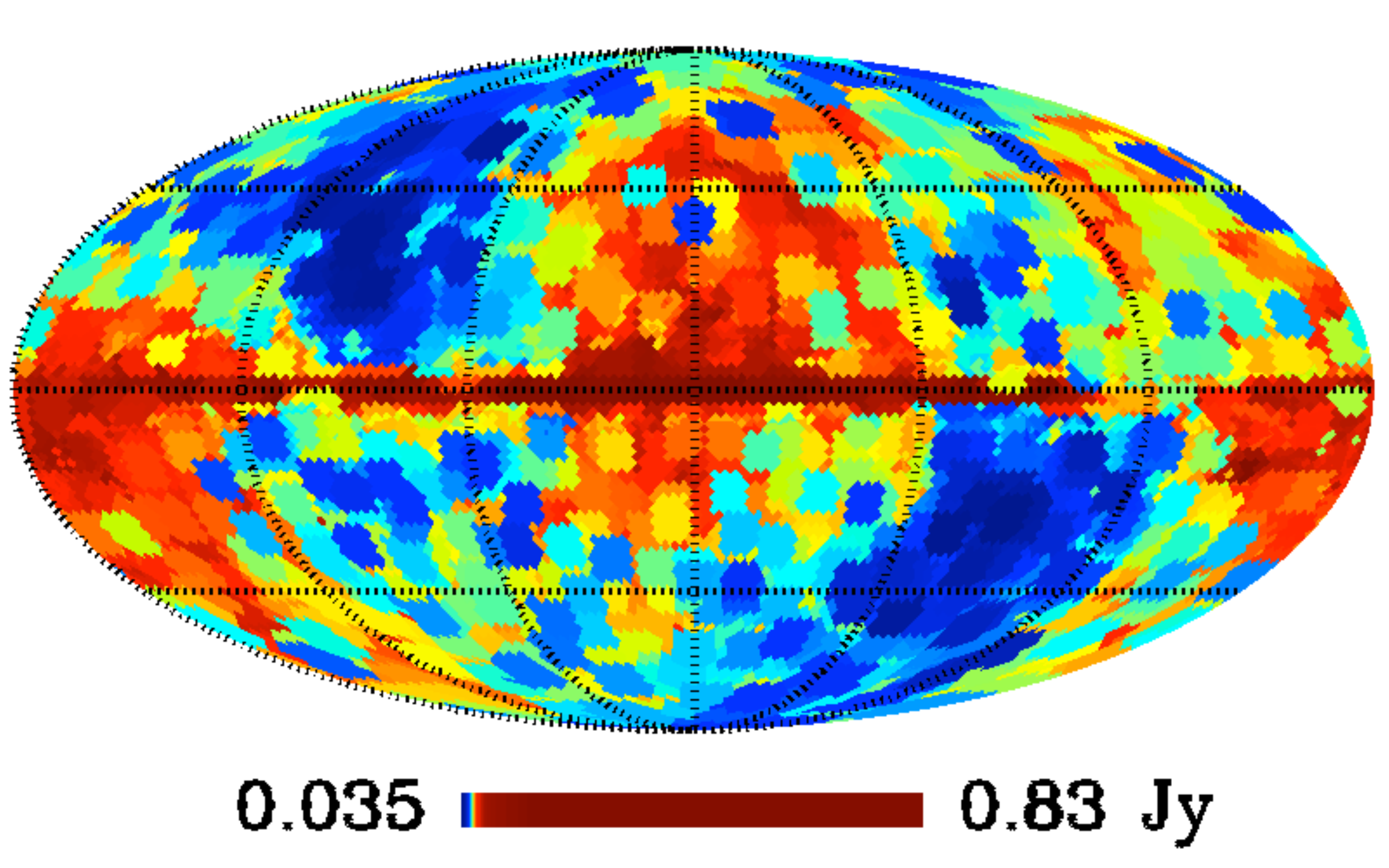} }&
\hspace{-0.5cm} \includegraphics[width=0.3\textwidth]{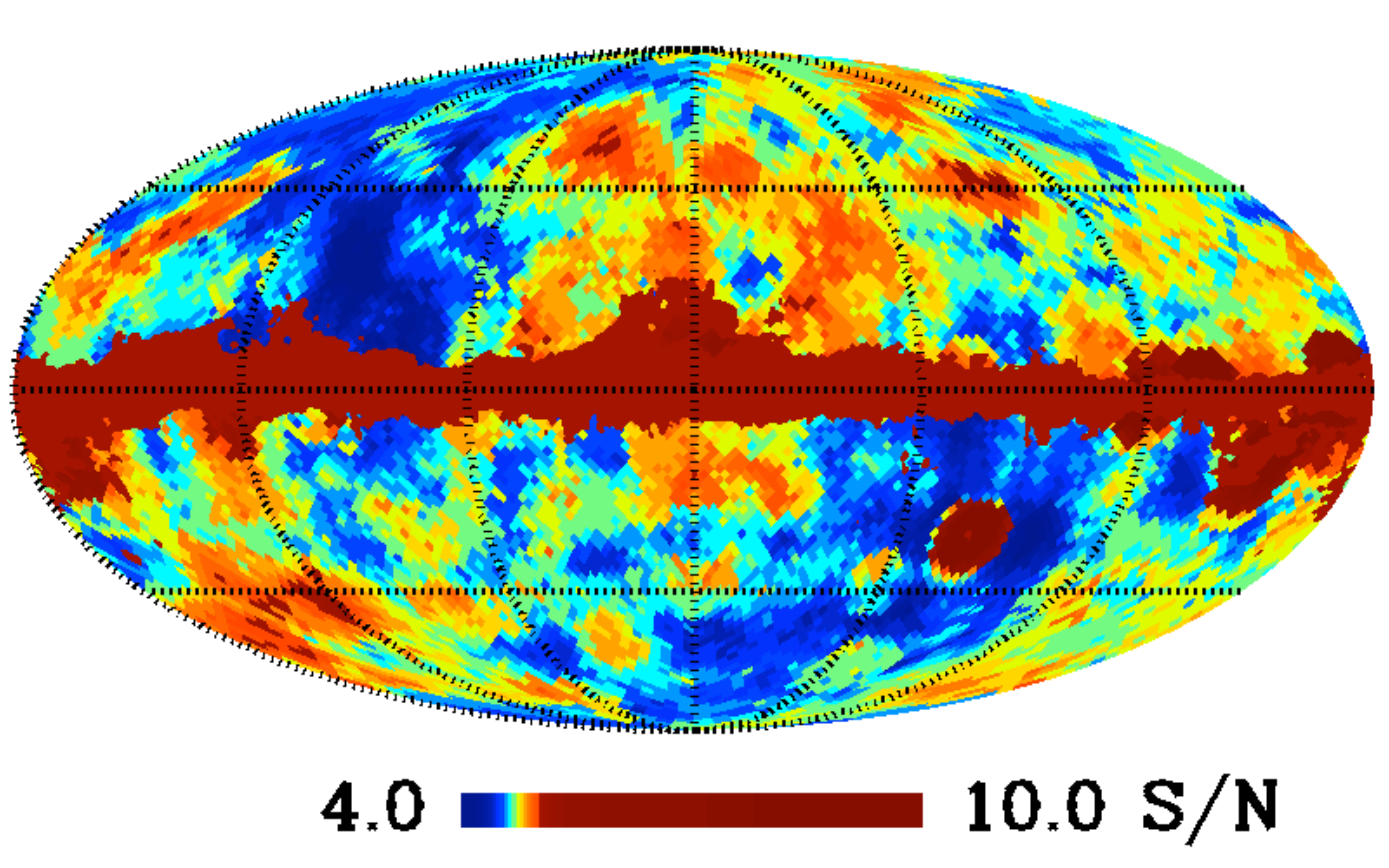} \\
\vspace{-2.7cm}{143\,GHz}&
\hspace{-0.9cm} \includegraphics[width=0.3\textwidth]{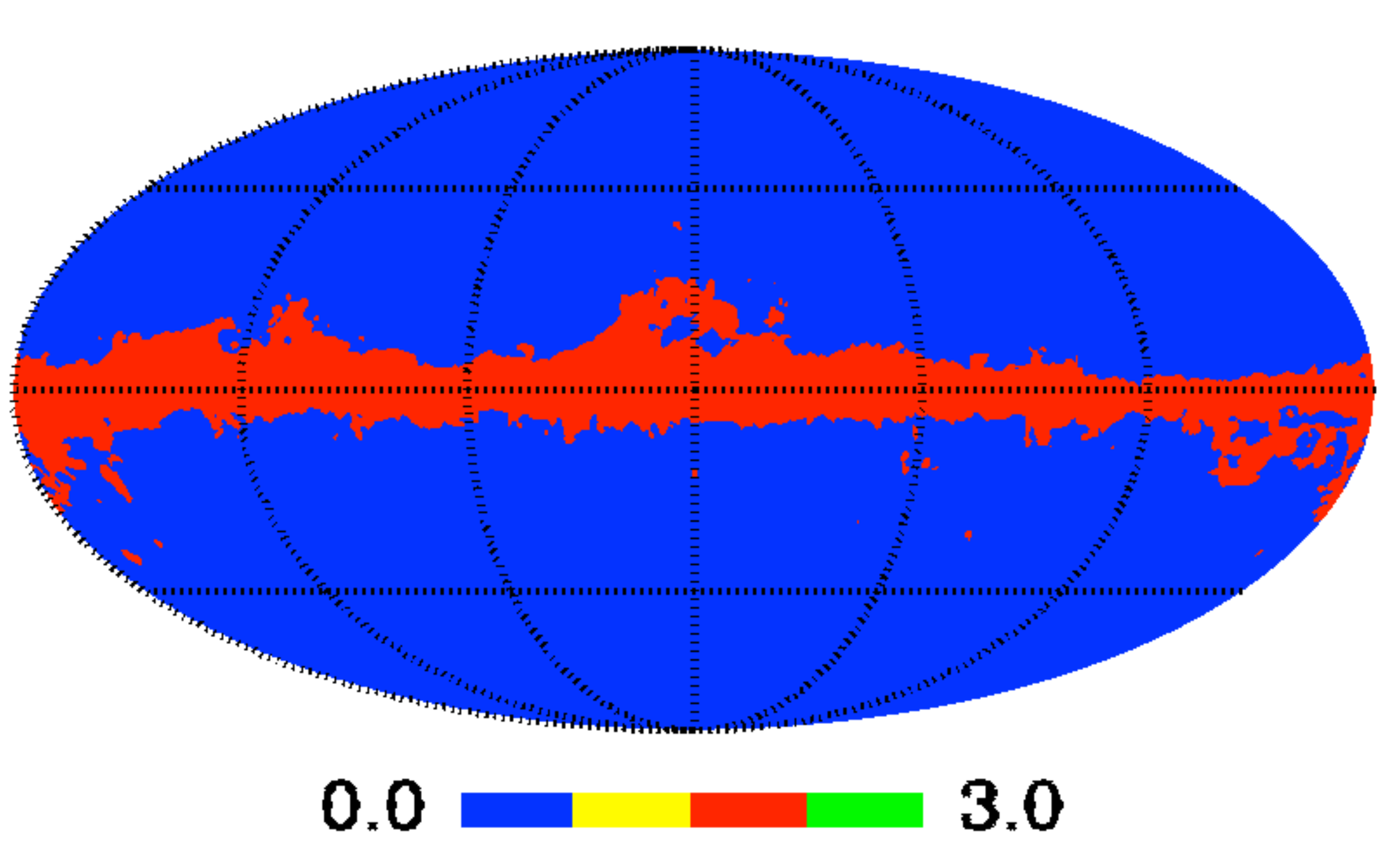} &
\hspace{-0.5cm}{\includegraphics[width=0.3\textwidth]{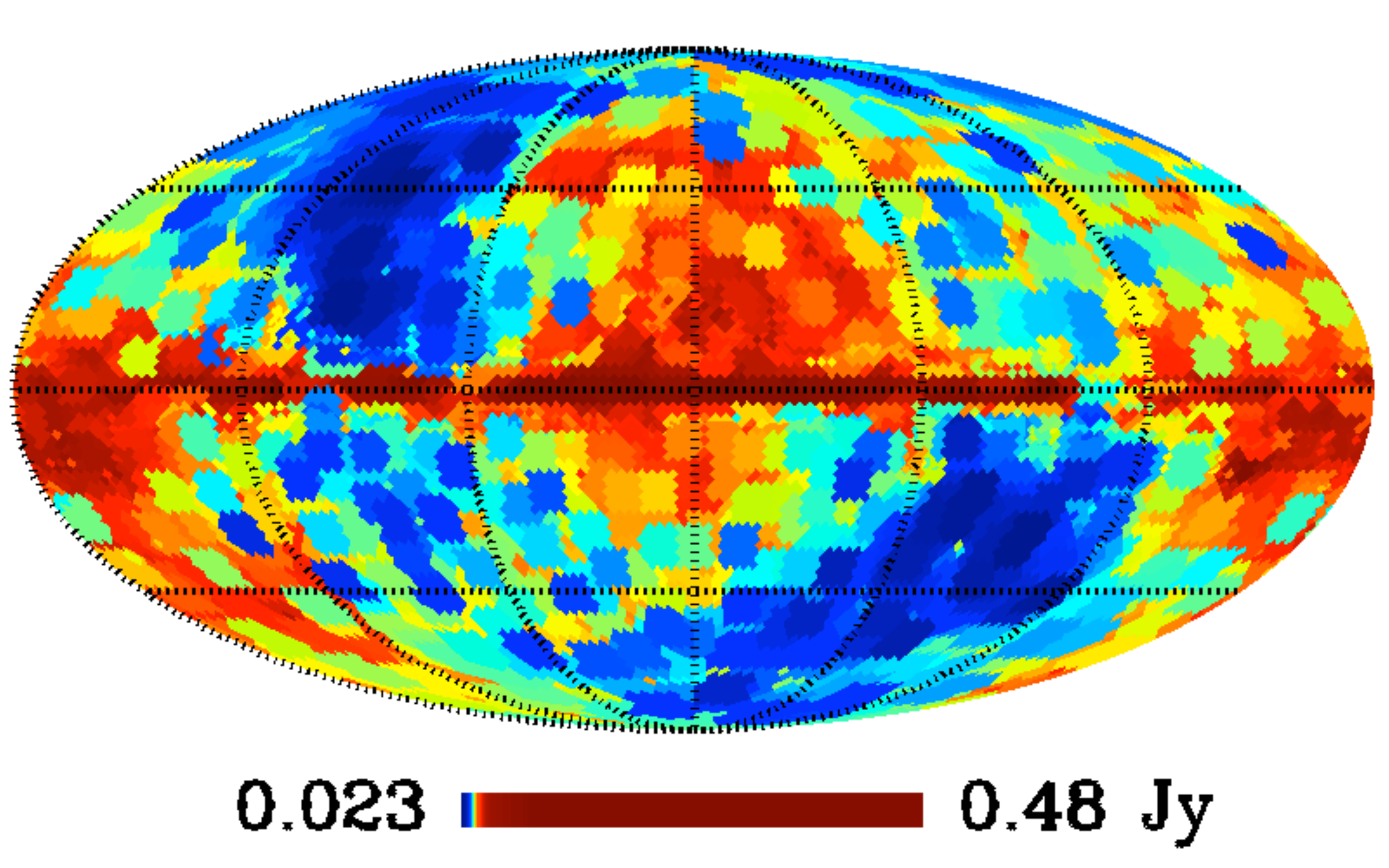} }&
\hspace{-0.5cm} \includegraphics[width=0.3\textwidth]{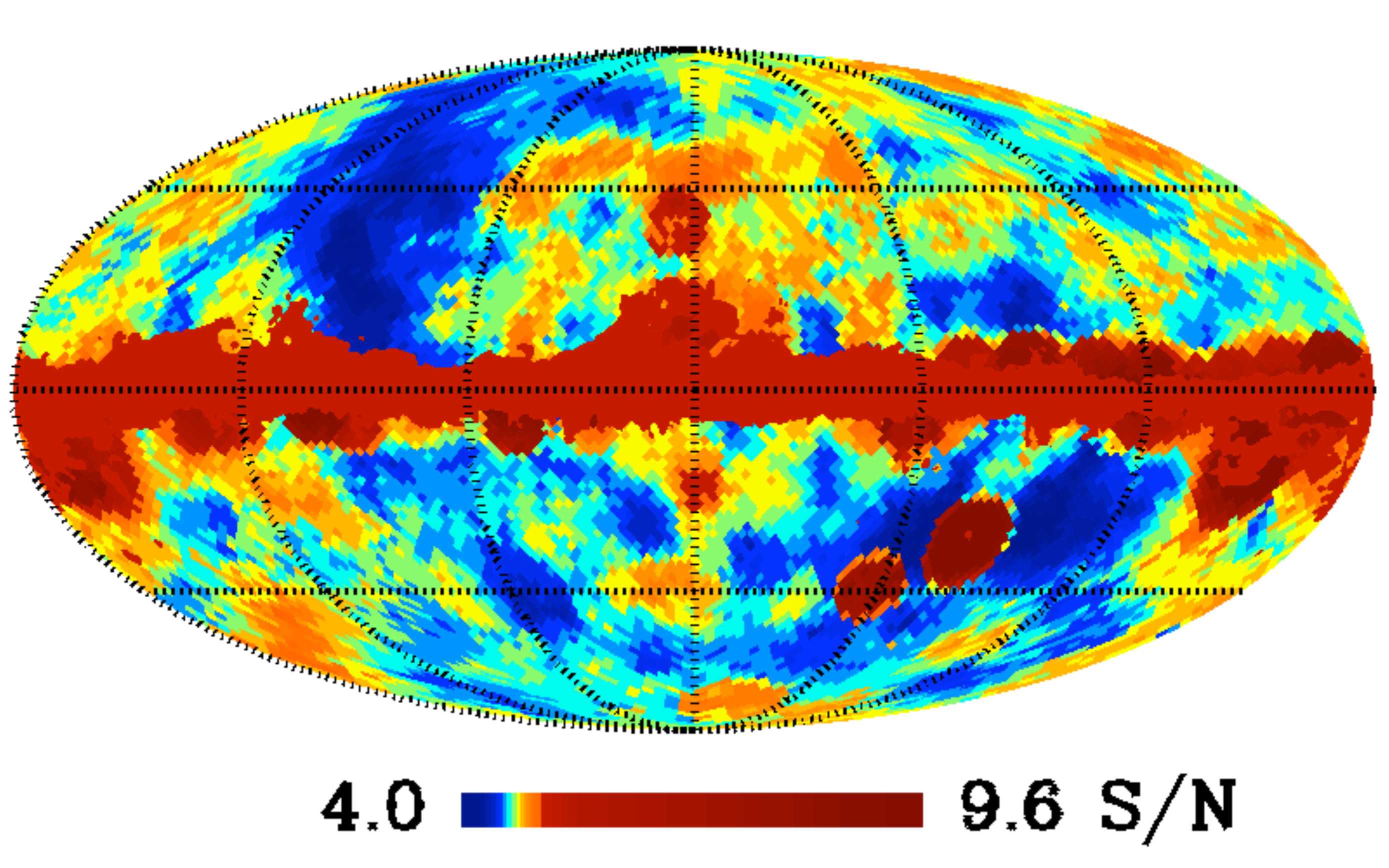} \\
\vspace{-2.7cm}{217\,GHz}&
\hspace{-0.9cm} \includegraphics[width=0.3\textwidth]{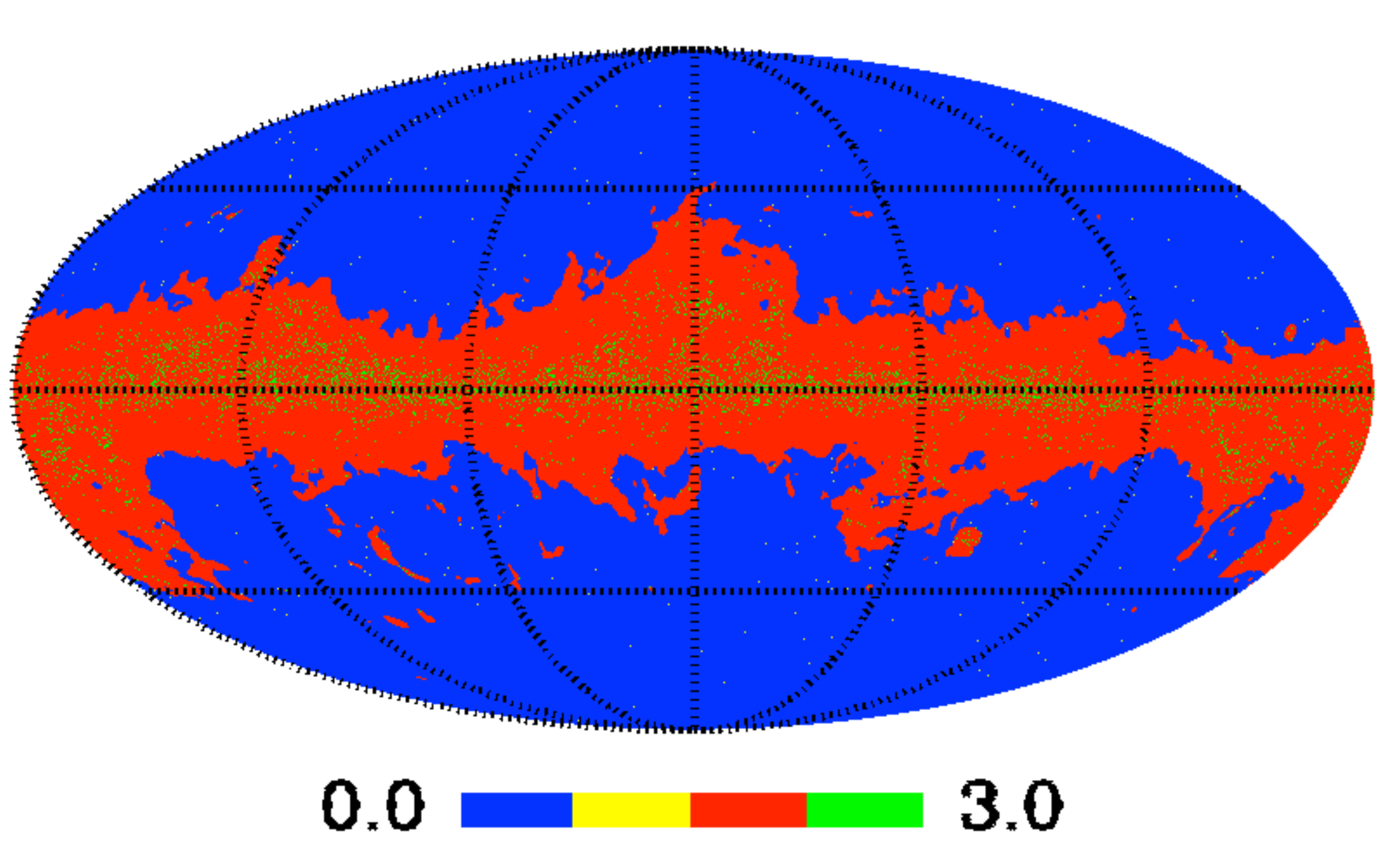} &
 \hspace{-0.5cm} {\includegraphics[width=0.3\textwidth]{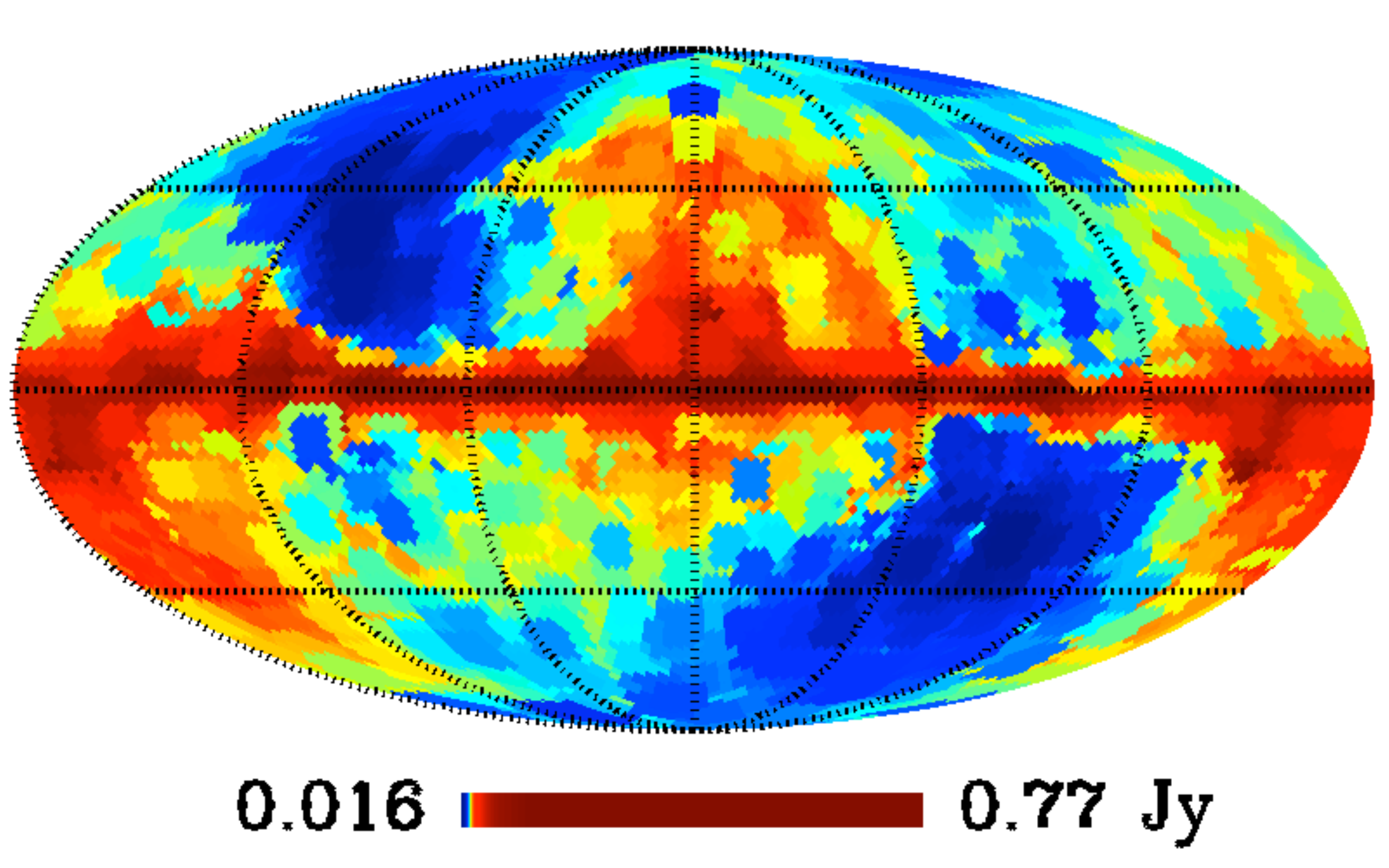} }&
\hspace{-0.5cm} \includegraphics[width=0.3\textwidth]{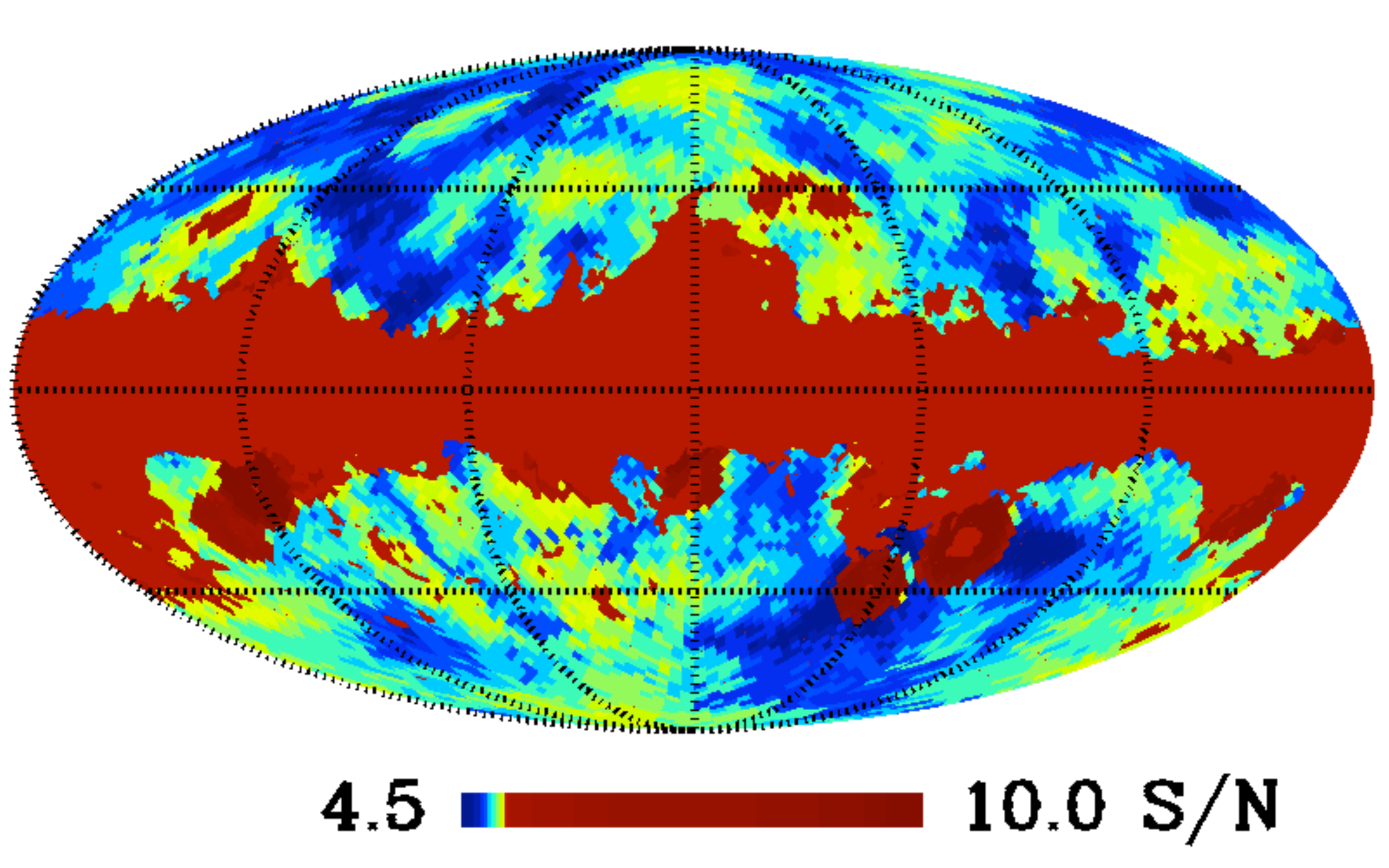}\\
\vspace{-2.7cm}{353\,GHz}&
\hspace{-0.9cm} \includegraphics[width=0.3\textwidth]{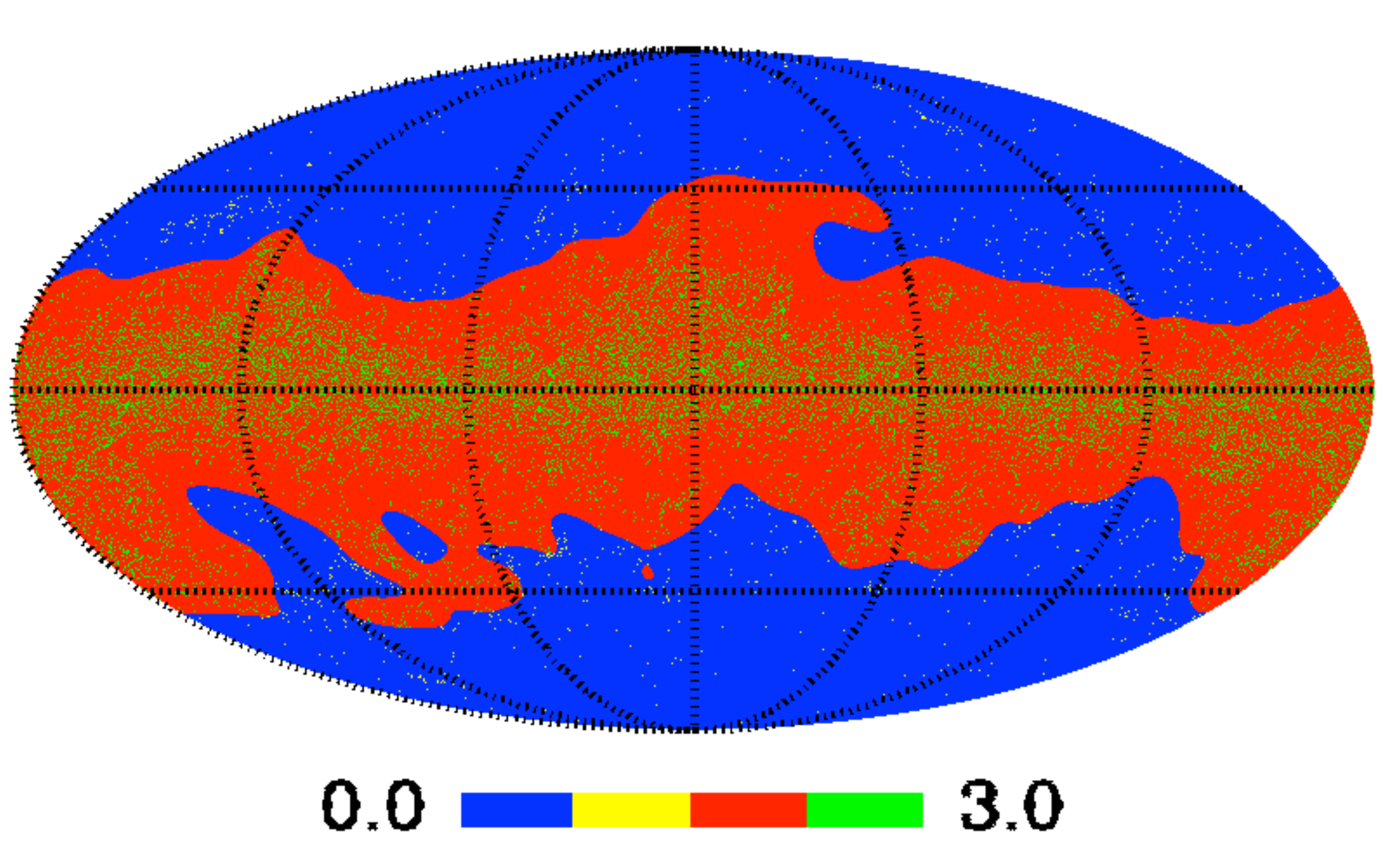} &
 \hspace{-0.5cm}{ \includegraphics[width=0.3\textwidth]{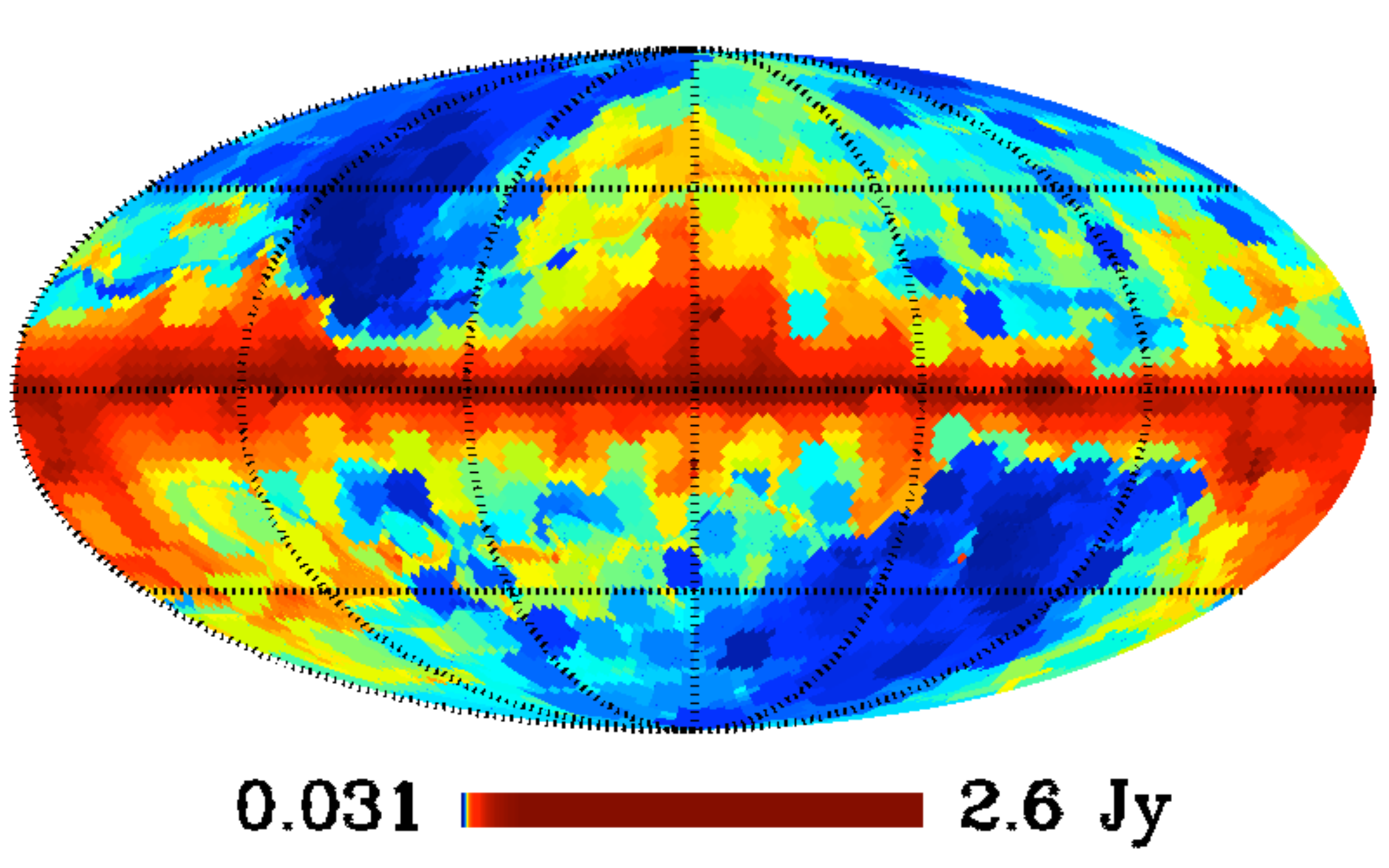} }&
\hspace{-0.5cm} \includegraphics[width=0.3\textwidth]{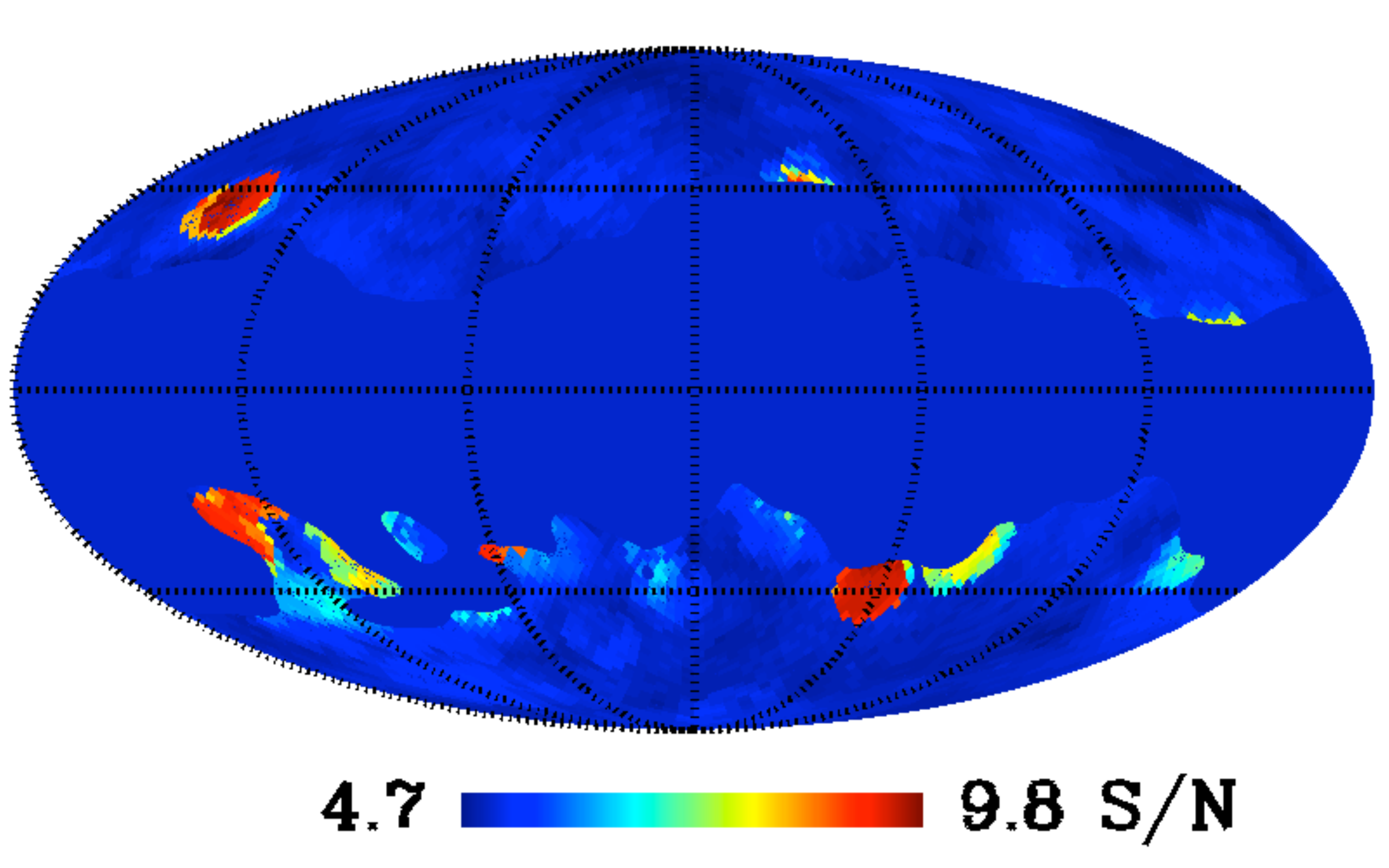} \\
\vspace{-2.7cm}{545\,GHz}&
\hspace{-0.9cm} \includegraphics[width=0.3\textwidth]{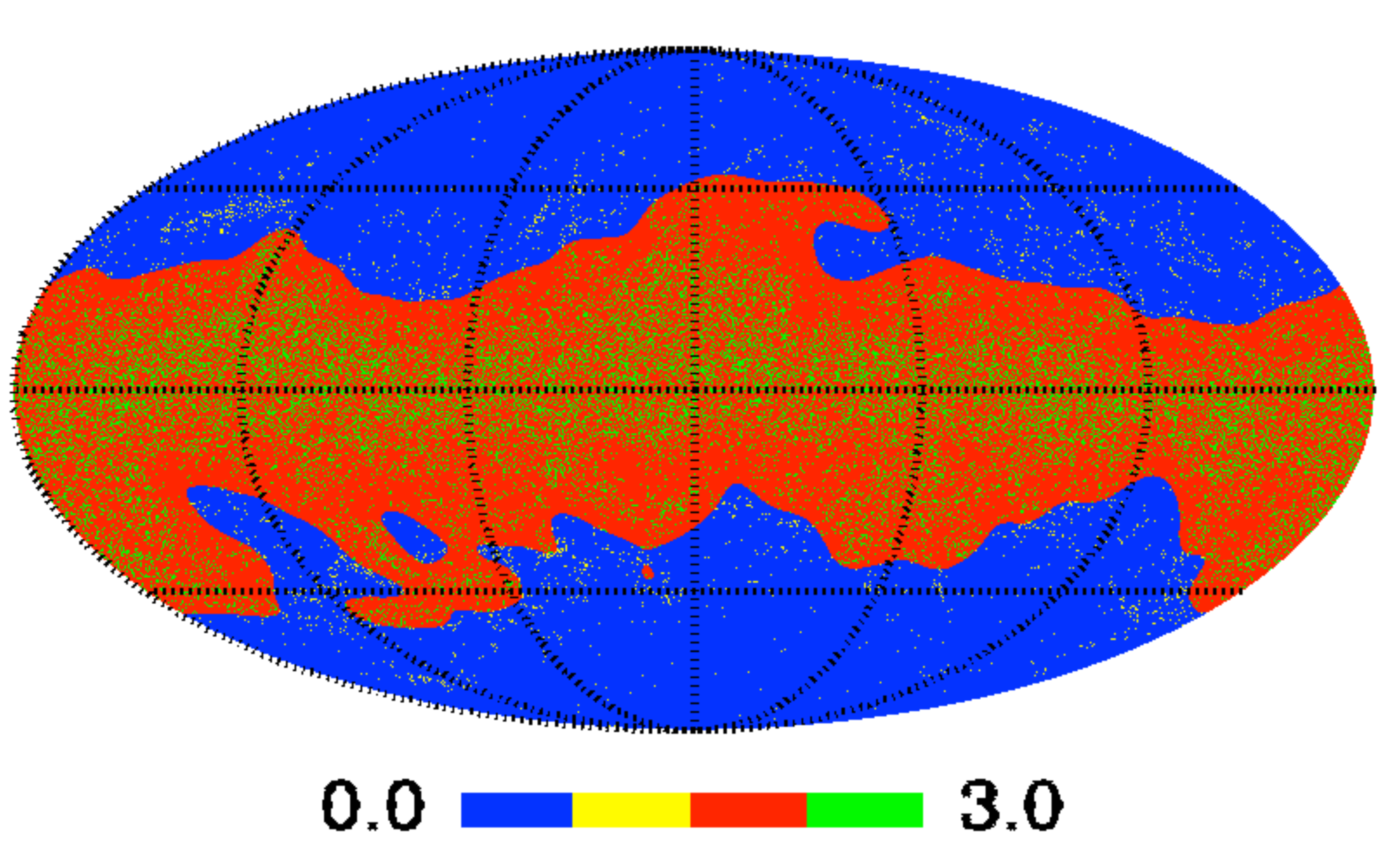} &
 \hspace{-0.5cm} { \includegraphics[width=0.3\textwidth]{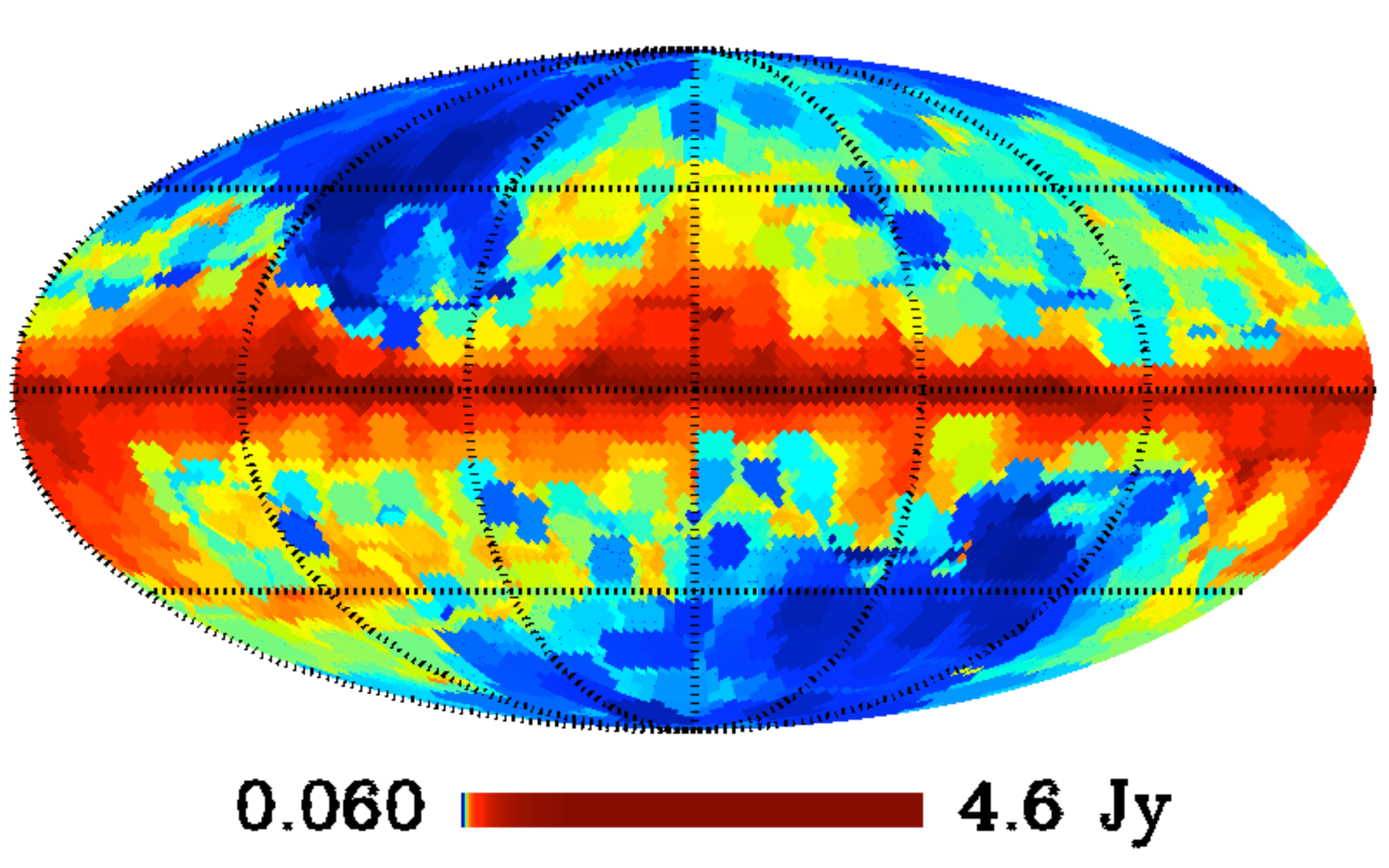} }&
\hspace{-0.5cm} \includegraphics[width=0.3\textwidth]{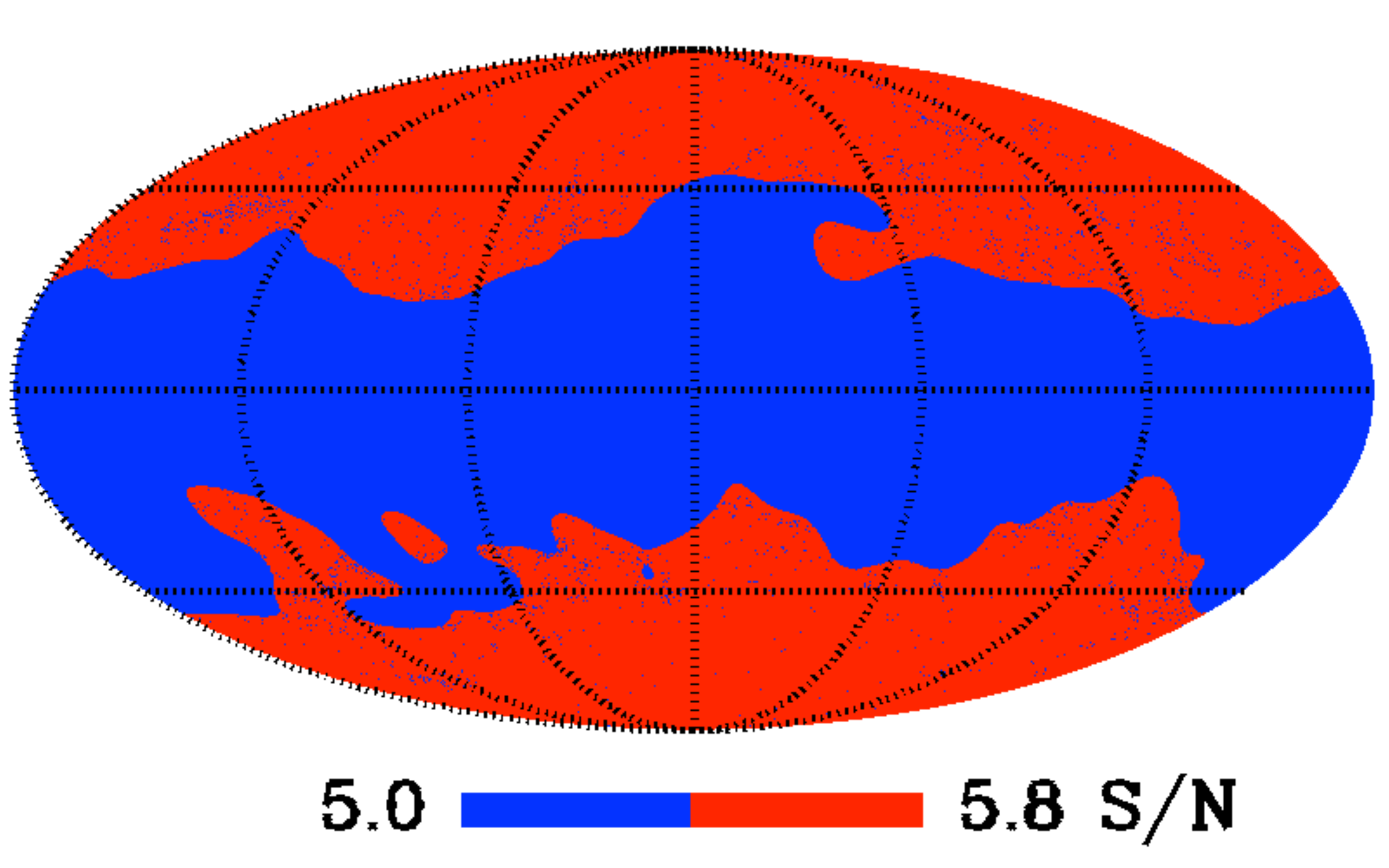} \\
\vspace{-2.7cm}{857\,GHz}&
\hspace{-0.9cm} \includegraphics[width=0.3\textwidth]{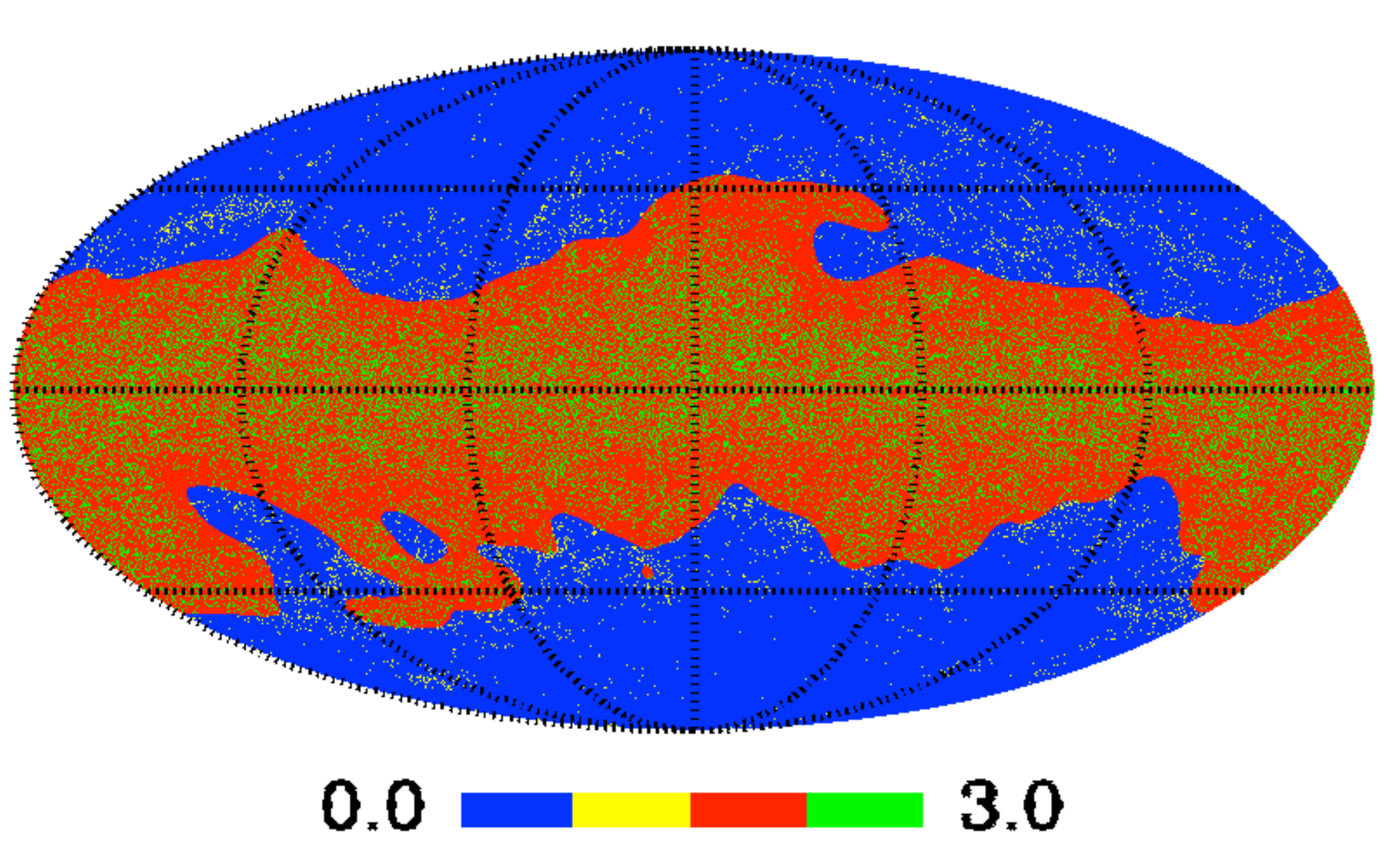} &
 \hspace{-0.5cm} { \includegraphics[width=0.3\textwidth]{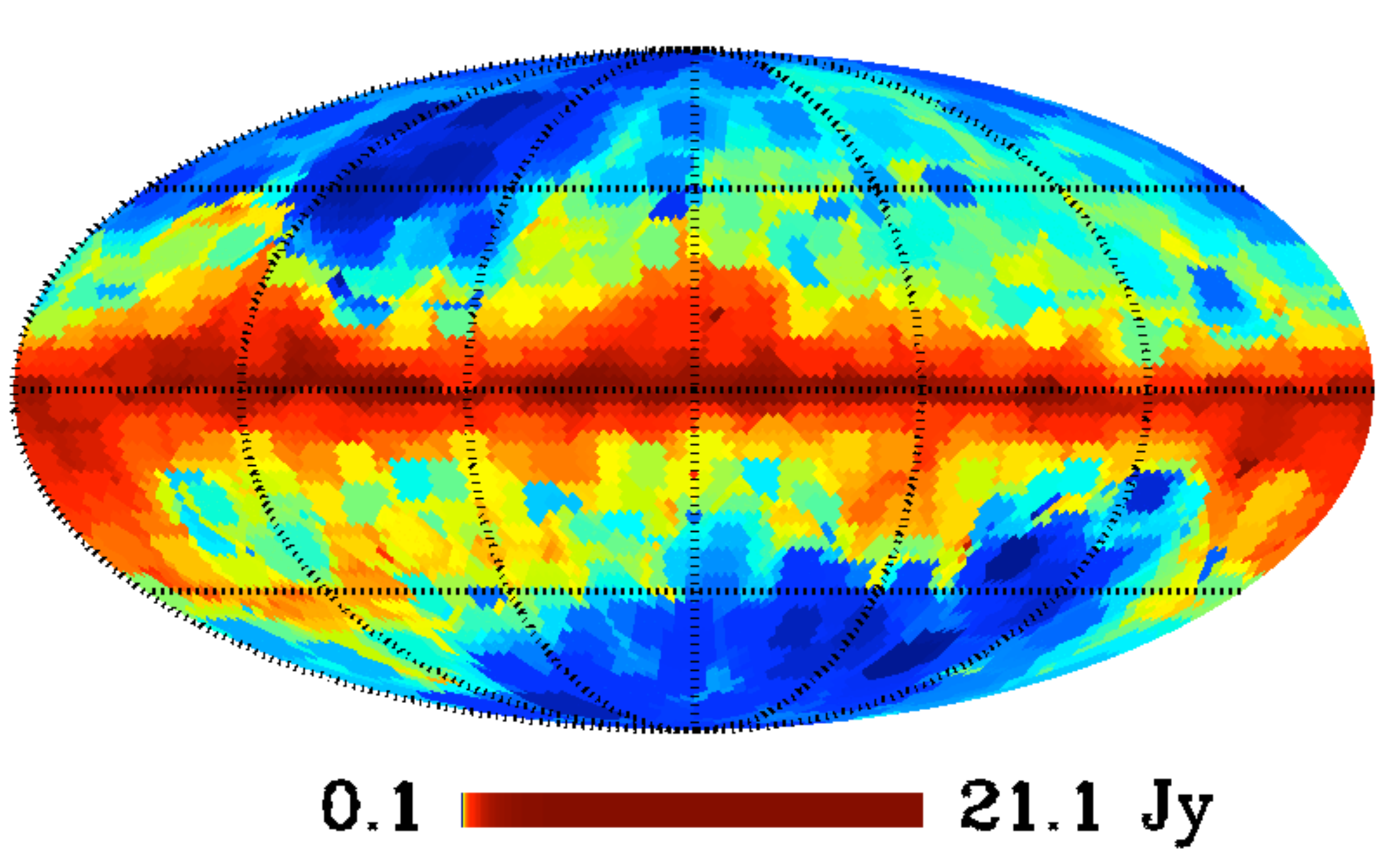} }&
\hspace{-0.5cm} \includegraphics[width=0.3\textwidth]{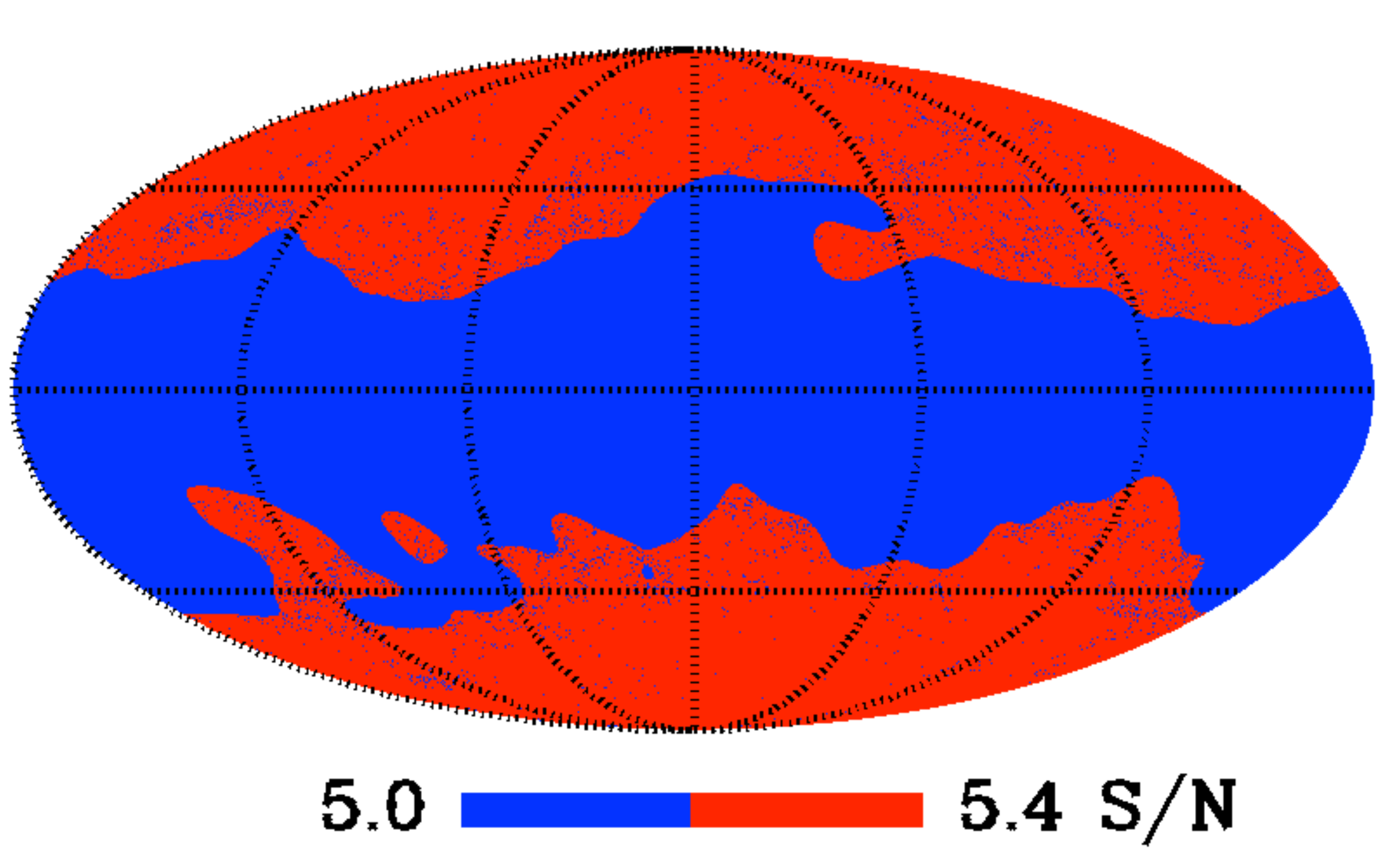}
\end{tabular}
\end{center}
\caption{\textit{Left}: zone masks constructed from the filament masks and the Galactic masks. The area covered by the \plccs\ is given by the zero (blue) values in the mask, and that covered by the \ecat\ is given by the non-zero values. The Galactic region is traced in red, and the filament mask is yellow outside the Galactic region and green inside it.
\textit{Centre}: rms noise level as determined by the HFI MHW2 code. 
\textit{Right}:  \snr\ thresholds applied to the raw catalogue; a  flat \snr\ cut is applied in the region of the \ecat\ and for the \plccs\ at 545 and 857\,GHz.}
\label{fig:maps}
\end{figure*}
\begin{figure}
\begin{center}
\includegraphics[width=0.5\textwidth]{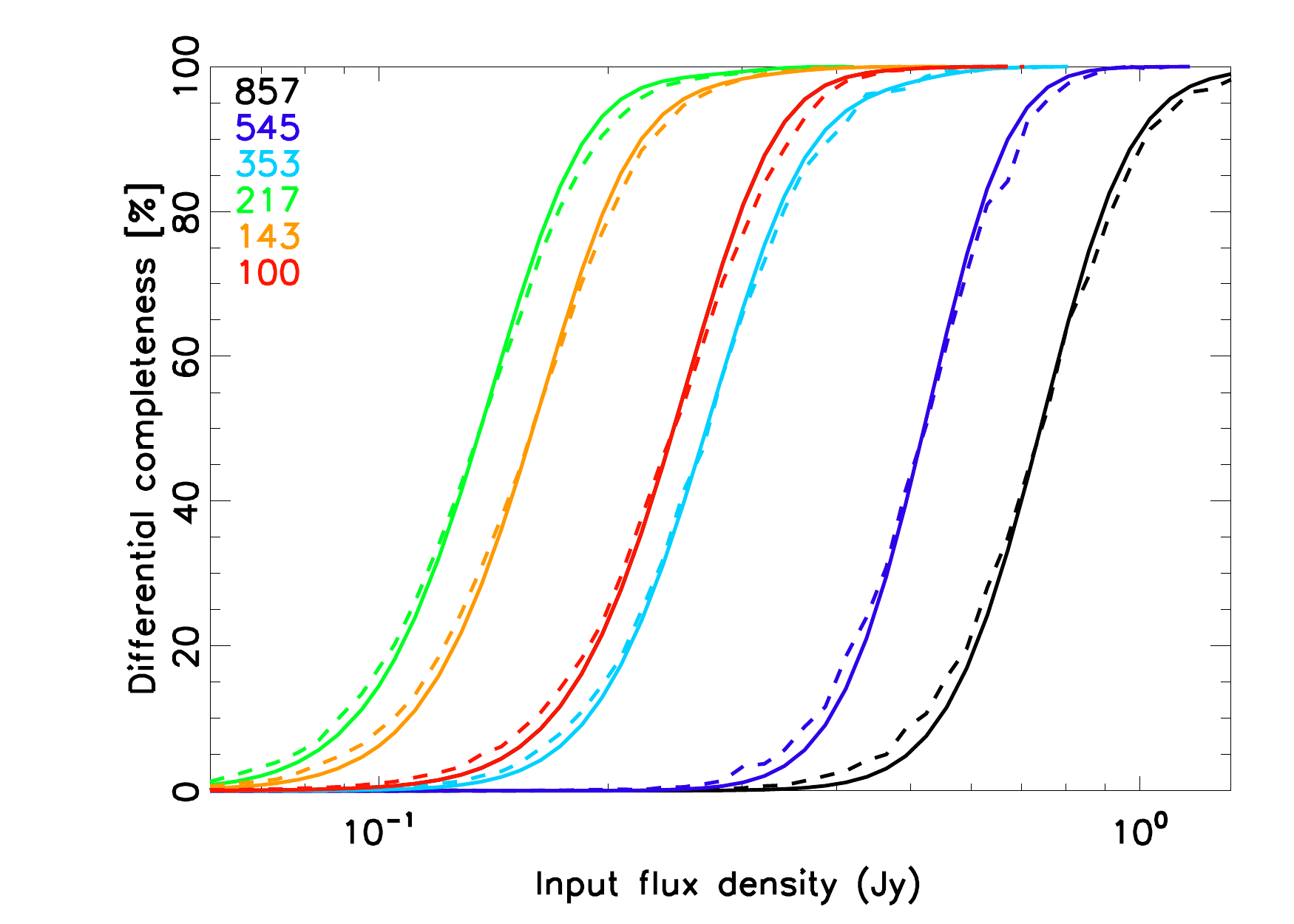}
\includegraphics[width=0.5\textwidth]{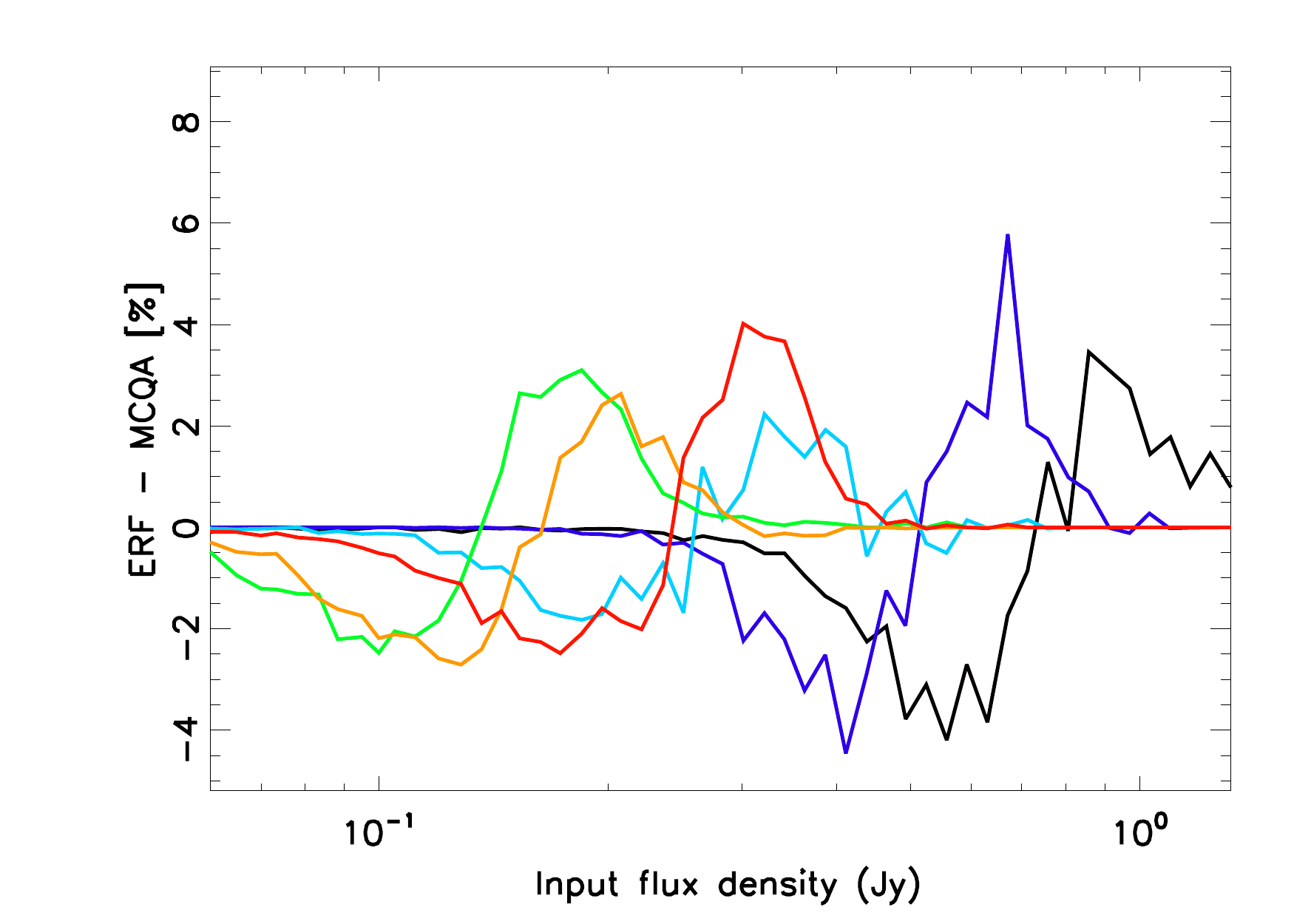}
\caption{Comparison of the error-function  semi-analytic completeness derived from the $\sigma_{s}$ threshold maps (Eq.~\ref{eq:erf}) and the Monte Carlo completeness estimates.  \textit{Top}: solid lines denote the error-function completeness and the dashed lines the Monte Carlo completeness. \textit{Bottom}: difference between the two completeness estimates.
\label{fig:HFI_compl_vs_erf}}
\end{center}
\end{figure}
There are no external full-sky catalogues in the HFI frequency range, so we rely on the injection of artificial sources into the \Planck\ maps to establish the completeness, following the power-law models fitted to the data described in Sect.~\ref{sec:hfi_reliability} and Table~\ref{tab:pow_law_mod}.

The completeness is determined from the injection of unresolved point sources into the real maps. Bias due to the superimposition of sources is avoided by preventing injection within an exclusion radius of $\sigma_{b}$, evaluated from the fitted FWHM in Table~\ref{tab:beam_data},  around both existing detections in the real map and previously injected sources.  We note that while superimposition of real sources will occur in the \Planck\ maps, it is beyond the scope of this paper to disentangle such effects.  Our definition of a point source here is a beam-shaped spike of emission, regardless of the make-up of astrophysical objects that produce the emission.

The flux from real and injected point sources contributes to the noise estimation for each patch, reducing the \snr\ of all detections and biasing the completeness.  We prevent this effect by determining the noise properties on the maps before injecting sources, and have verified that any remaining bias on detection and parameter estimates due to injected sources is negligible.  The injected sources are convolved with the effective beam computed using the {\tt FEBeCoP} algorithm \citep{mitra2010,planck2014-a05, planck2014-a08}.

We show the Monte Carlo completeness functions computed per channel in Fig.~\ref{fig:HFI_compl_MCQA}.  We have calculated them using several cuts based on the ``highest reliability catalogue'' column that indicates the highest reliability catalogue to which each source belongs (see Sect.~\ref{sec:hfi_reliability}); as the required catalogue reliability increases, the completeness decreases.  We also show the completeness for the unvalidated area of the sky containing the \ecat.  This area is significantly less complete than the main \plccs\ regions due to Galactic emission, where the completeness deteriorates markedly with frequency.
The \plccs\ product includes noise maps at each channel, which are also shown in the middle panels of Fig.~\ref{fig:maps}.  These maps contain the MHW2 detection noise, defined as the standard deviation of the MHW2 filtered patches, $\sigma_{s}(\theta,\phi)$.
A good approximation to the Monte Carlo completeness can be calculated for any subsection of the sky using the noise maps and the reliability threshold maps.  Assuming Gaussian noise in the filtered patches, the completeness is given by Eq. (\ref{eq:erf}).
We compare the error function and Monte Carlo completeness estimates for the 80\,\% reliability catalogues at each channel in Fig.~\ref{fig:HFI_compl_vs_erf}.  The effective $\sigma_s$ that we used, and included in the data release, has been normalized across the \plccs\ region to match the effective noise from the Monte Carlo tests.  The Monte Carlo completeness drop-off is slightly wider than the error-function completeness because it includes the effects of non-Gaussian noise from the background and varying asymmetric beams.  However, the discrepancy is less than $5$\,\%.

\subsection{Reliability}
\label{sec:qa_reliability}

The underlying philosophy of the \plccs\ was to provide a catalogue of sources with a reliability of at least 80\,\%, which could be cut if desired by a user to generate higher reliability subsets. This has required, as explained in Sect.~\ref{sec:construction}, the division of the HFI data into two subcatalogues, the \plccs\ and  \ecat.
The assignment to the \ecat\ is based on the position of the source on the sky. These locations are determined by the union of the Galactic region and the filament mask. Note that at 857\,GHz this union represents more than half of the sky. The \ecat\ therefore contains substantially more sources than the \plccs\ in the higher frequency channels since it includes all the Galactic plane sources.
\subsubsection{LFI: reliability assessment}
\label{sec:lfi_reliability}

The band-merged catalogue compared with ancillary data, as described in Sect.~\ref{sec:lfi_completeness}, has also been used to assess the reliability of the LFI catalogues.
All but 426 PCCS2 sources (out of 2039) were identified with known sources in the external radio source catalogues of CRATES, NEWPS, and AT20G. The initial estimate for the reliability, using just these external catalogues, is thus greater than $79$\,\%. The percentage of unidentified LFI sources as a function of their flux density is shown in the lower panel of Fig.~\ref{fig:LFI_comp_reliab}.
Associations for some of these initially unidentified sources were later found by performing a search in NED. This procedure was carried out on a source-by-source basis and some subjective judgement was required. Hence, we report a range in the possible number of sources thus identified. The sources that are positively identified by this approach are flagged as such and the associated sources are named in the \plccs\ catalogue.

Reliability at high Galactic latitudes using external catalogues: out of the 426 initially unidentified sources, 180 were at high Galactic latitudes,  $|b| \geq 20^{\circ}$. More than a quarter of these were positively identified using NED, leaving 132.  Given the 1161 sources detected over this latitude range, this implies a reliability of 92\,\% for high Galactic latitude sources. It should be noted that  7--8\,\% of the remaining 132 unidentified sources appear in multiple \Planck\ bands, and are therefore likely to be real. Hence, this reliability estimate is a lower bound.

Reliability using external catalogues and taking into account multiple band detections: we also searched for identifications for all sources that appeared in two or more \Planck\ bands; here we included the 100\,GHz band in the analysis.  Of the 426 initially unidentified sources, 133 appeared at two or more frequencies. Of these, the number that remained unidentified was 71. Given that more than half of the LFI sources (1149) appear at two or more frequencies, this implies a reliability $> 94\,\%$ for sources detected in two or more \Planck\ bands.

Reliability using external catalogues and sources detected in a single band and with SNR $>5$: we searched NED for identifications for sources detected in only a single \Planck\ band with \snr\ $>5$.  Not surprisingly, the rate of identification was lower, but a few positive identifications were added.  In the end we were left with 335 out of the original 426 that remained either unexamined or unidentified. We may therefore conclude that  the overall reliability of the LFI catalogue is at least 84\,\%.

The procedure described above was used to construct the \plccs\ for the three lowest \Planck\ bands, and additionally to populate the EXT\_VAL column for them. 
This column summarizes the comparison with external catalogues, and is described in Sect.~\ref{sec:content}. Note that the higher-frequency bands also provide information for each source in this column, but they are not used in the reliability assessment.

\subsubsection{HFI: Filament masks}
\label{sec:filament_masks}

There are regions on the sky even at high Galactic latitudes in which the detection of sources cannot be trusted to high levels of reliability.
The reason is that the Mexican hat wavelet algorithm used for the detection of compact sources is also efficient at edge detection. While this is  ideal for point sources embedded in a Gaussian noise background, it is not optimal for a map with non-Gaussian structures where there are other edges to detect, such as dusty filaments.
The impact of false detections due to these structures may in part be ameliorated by rejection criteria based on the number of connected pixels of the detection, as described in \cite{planck2013-p05}; however, this is not a complete solution.
We wish to place source detections occurring in these filamentary structures into the \ecat\ rather than the main \plccs. In order to do this we need to create filament masks, which describe the regions of the sky containing filamentary structures that pass into the wavelet filtered patches.

The obvious way to construct these masks would be to use the filtered patches themselves, after median filtering to remove the point-like objects from the patch. The difficulty with this approach is setting an appropriate threshold above which to mask.
Given that these dusty structures are far from uniformly distributed across the sky, local evaluations of this threshold will not be as successful as a global one. Any practical filament mask should also be continuous;  for both these reasons they cannot be created directly from the filtered patches.

A Mexican hat wavelet filtered map is very close to a difference-of-Gaussians (DOG) map if the ratio of the two FWHMs used is 2:1.  The DOG map is created by
smoothing the original map with two different Gaussians, creating two new maps that are then differenced.\footnote{This is the default approach used for finding compact sources in data from the SCUBA-2 instrument \citep[e.g.][]{MacKenzie11}.}
Here the smaller of the two FWHMs used is the fitted FWHM from Table~\ref{tab:beam_data}.
A full-sky DOG map may be trivially created, and while it is not identical to the filtered patches it traces the same structures in the original map that pass into the filtered patches.
 In order to remove point sources, the DOG map is median-filtered using a filter radius of twice the fitted FWHM (from Table~\ref{tab:beam_data}).
By thresholding this median-filtered DOG (MF-DOG) map, we can create the desired filament mask for each channel.
In order to find an appropriate threshold, we select the cleanest (faintest) 25\,\% of the sky based on the smoothed sky brightness of the 857\,GHz channel map, and create a histogram of the MF-DOG map from the pixels in this region.
We then fit a Gaussian to this histogram, and the filament mask is given by all the pixels in the MF-DOG map with values greater than 3 times the $\sigma$ of this fitted Gaussian. Negative fluctuations are not masked, since they cannot lead to spurious detections.
In the case of point sources in a Gaussian background, this procedure would result in an MF-DOG map that would contain solely Gaussian noise, and a mask created as above would be expected to mask 0.15\,\% of the sky.
Table~\ref{tab:mask_stats} shows the percentage area of the filament mask for the frequency channels at which it is used;  we see that the percentage area masked is well in excess of the expectation from Gaussian statistics. Additionally, we see that the area of the filament mask increases with frequency as the maps contain more and more emission from dust. Indeed, we do not need to use a filament mask for the lower two HFI frequency channels, because the dusty filamentary structures are not a problem at these frequencies.

Note that the risk of high \snr\ point sources located inside filamentary structures being placed inside the filament mask is limited by the fact that high \snr\ point sources in the DOG map are positive peaks surrounded by negative troughs. This means that  once the DOG map is median-filtered the resultant level in the MF-DOG map at the location of these point sources is likely to be below the level at which thresholding occurs.

\begin{table}
\begingroup
\newdimen\tblskip \tblskip=5pt
\caption{Fractions of sky covered by the \plccs\ catalogue, the filament mask, and the Galactic region for the HFI channels. No masks are required for the LFI channels, where the \plccs\ covers the entire sky.
The filament mask is not required for the lowest two HFI frequency channels. Some parts of the sky are covered by \textit{both} the filament mask and the Galactic region.}
\label{tab:mask_stats}
\nointerlineskip
\vskip -3mm
\footnotesize
\setbox\tablebox=\vbox{
   \newdimen\digitwidth
   \setbox0=\hbox{\rm 0}
   \digitwidth=\wd0
   \catcode`*=\active
   \def*{\kern\digitwidth}
\def\leaderfill{\leaders \hbox to 5pt{\hss.\hss}\hfil}
   \newdimen\signwidth
   \setbox0=\hbox{+}
   \signwidth=\wd0
   \catcode`!=\active
   \def!{\kern\signwidth}
\halign{\hbox to 0.5in{#\leaderfill}\tabskip=1em&\hfil#\hfil&\hfil#\hfil &\hfil#\hfil\tabskip=0pt\cr
\noalign{\doubleline}
\omit\hfil Channel\hfil&\plccs\ area& Filament mask&Galactic region\cr
\omit&\% sky&\% sky& \% sky\cr
\noalign{\vskip 3pt\hrule\vskip 5pt}
100&85.0&\ldots&15.0\cr
143&85.0&\ldots&15.0\cr
217&64.9&*2.2 &35.0\cr
353&47.6&*7.5 &52.0\cr
545&47.0&10.9 &52.0\cr
857&46.3&13.9 &52.0\cr
\noalign{\vskip 3pt\hrule\vskip 3pt}}}
\endPlancktable
\endgroup
\end{table}
\subsubsection{HFI: reliability assessment}
\label{sec:hfi_reliability}
\begin{table}
\begingroup
\newdimen\tblskip \tblskip=5pt
\caption{Power law model fit parameters obtained from the simulation of the source number counts in the HFI channels used in the reliability assessment.}
\label{tab:pow_law_mod}
\nointerlineskip
\vskip -3mm
\footnotesize
\setbox\tablebox=\vbox{
   \newdimen\digitwidth
   \setbox0=\hbox{\rm 0}
   \digitwidth=\wd0
   \catcode`*=\active
   \def*{\kern\digitwidth}
\def\leaderfill{\leaders \hbox to 5pt{\hss.\hss}\hfil}
   \newdimen\signwidth
   \setbox0=\hbox{+}
   \signwidth=\wd0
   \catcode`!=\active
   \def!{\kern\signwidth}
\halign{\hbox to 0.5in{#\leaderfill}\tabskip=1em & \hfil#\hfil & \hfil#\hfil &\hfil#\hfil\tabskip=0pt\cr
\noalign{\doubleline}
\omit\hfil Channel\hfil& $S_{\rm min}$ & $S_{\rm taper}$ & $\alpha$\cr
\omit\hfil & [Jy]& [Jy]& \cr
\noalign{\vskip 3pt\hrule\vskip 5pt}
100& 0.3 & 0.1 & 2.54\cr
143& 0.3 & 0.2 & 2.51\cr
217& 0.2 & 0.1 & 2.63\cr
353& 0.4 & 0.1 & 2.69\cr
545& 0.7 & 0.1 & 2.59\cr
857& 1.5 & 0.3 & 2.34\cr
\noalign{\vskip 3pt\hrule\vskip 3pt}}}
\endPlancktable
\endgroup
\end{table}

As described in Sect.~\ref{sec:selection}, we use two methods to assess the HFI reliability, the \emph{simulation reliability} and the \emph{injection reliability}.
The simulation reliability is assessed by injecting sources into simulated maps; this simulates both the real and spurious detection components of the total counts.
The injection reliability, however, involves injecting sources into the real maps; hence only the real detection component of the total counts is simulated.
The real component to inject,  in both cases, is evaluated using the \plccs\  above a given flux density threshold (above which we are complete) fitted to a single power-law model defined as $N(S) \propto S^{-\alpha}$. 
For each channel we provide three numbers in Table \ref{tab:pow_law_mod}: $\alpha$ is the power-law index (estimated per frequency with error in the range 0.05--0.1); $S_ {\rm min}$ is the minimum flux density considered when fitting the model (i.e. lower flux densities are excluded to allow for incompleteness); and $S_{\rm taper}$ is the flux density at which the power law  was truncated for the completeness simulations to avoid dominating the injected population with unobservable faint sources. Note that there was no truncation for the reliability work (simulation or injection). 
Simulation reliability is preferable to injection reliability because it provides a more complete understanding of the detection properties of the catalogue, including information on how the reliability varies as a function position on the sky, as well as on \snr.  

Previously, for the \lastcat\ (as described in \citealt{planck2013-p05}),  the simulation reliability was used for 100--217\,GHz, while injection reliability was used for 353--857\,GHz.
This was because the simulated maps at 353--857\,GHz could not be used to produce a simulation reliability estimate. The simulated number counts and completeness were not consistent with the real data, and the discrepancies arise from deficiencies in the simulation of diffuse dust emission near the beam scale.
Since then, improvements have been made in the simulations; using the FFP8 simulations \citep{planck2014-a14},  it is now possible to extend the use of simulation reliability to 353\,GHz.
These maps, however, include a leaked compact-source component from the \Planck\ maps from which they were derived.  As this can produce artificial, high \snr, spurious sources,  we screen these from the reliability estimates by considering any detection at \snr\ $>10$ to be real.
At 545 and 857\,GHz, we continue to use the injection reliability estimate. 

Table~\ref{tab:mask_stats} shows the percentage area of the sky occupied by the \plccs, for each HFI channel. 
This corresponds to the area in which the reliability assessment is performed.
The \plccs\ covers the region of the sky not excised by the filament mask or the Galactic region. 
The Galactic region is determined by the area, from each channel, that must be excluded from the reliability assessment  in order to achieve consistency between simulated and real catalogues.
The percentage areas of the Galactic regions and filament masks are also shown in Table~\ref{tab:mask_stats}.
Note that the Galactic region and the filament mask can and do overlap; their union forms the area of sky in which sources are assigned to the \ecat. Hence, the complement of the \plccs\ is the \ecat, and consequently no source can appear in both subcatalogues.

Each entry in the \plccs\ and \ecat\ catalogues contains a field, HIGHEST\_RELIABILITY\_CAT, which specifies the highest reliability subsample in which that source may be included. Hence, this field can be used to cut the catalogues to produce subsets with a reliability higher than the survey target of $80 \%$.

For the 100--353\,GHz channels we can perform the reliability assessment using simulation reliability. This allows us to define a local \snr\ threshold, $q(\theta,\phi,\mathcal{P})$, as a function of both sky position $(\theta,\phi)$ and target reliability, $\mathcal{P}$. This threshold gives a local reliability $\mathcal{P}$ within an 8\degr\ radius of $(\theta,\phi)$. This information allows us to populate the HIGHEST\_RELIABILITY\_CAT field for all sources in the \plccs\ catalogues in these channels, and a resolution of 1\,\% reliability.
Note that, for these four channels, the option to create a higher reliability subset will also apply to spatial subsets of the original catalogue. For example, one could create a catalogue of the north ecliptic pole region to a reliability of 97\,\%, if desired.

For the 545 and 857\,GHz channels the limitations of the injection reliability mean that we are unable to define a local \snr\ threshold for a  given reliability. Hence, we provide a global \snr\ threshold that will deliver the target reliability for the full catalogue. We use this approach to populate the HIGHEST\_RELIABILITY\_CAT field for these channels  in steps of 5\,\% in reliability. Since there is no local assessment, the option to create higher-reliability spatial subsets for these channels is not available, because the desired reliability will only apply to the full area covered by the catalogue.
We note that the \plccs\ survey \snr\ threshold at 857\,GHz is substantially higher than the threshold applied to build the \lastcat.  We have improved the modelling of the real extragalactic sources for the injection reliability estimate for the \plccs, which results in a shallower spectral index for the input source model and fewer injected sources at low \snr.  This produces a more realistic, and lower, reliability estimate at a given \snr\ relative to the \lastcat.  As the threshold has moved to higher \snr, the flux density at 90 \% completeness in Table~\ref{tab:all_pccs_stats} is now higher than for the \lastcat.

The \snr\ threshold maps which produce the survey target of 80\,\% integral reliability are shown in the right-hand panels of  Fig.~\ref{fig:maps}, for all the HFI channels. Note the flat \snr\ cut applied to the 545 and 857\,GHz channels in the \plccs\ regions, as well as the flat \snr\ = 5 cut applied to the \ecat\ regions. The \snr\ threshold HFI maps for $\mathcal{P}= 80\,\%$, 85\,\%, 90\,\%, and 95\,\% are included in the data release.

\subsubsection{HFI: Comparison with H-ATLAS}

As an external check on the reliability  of the \plccs\ we have exploited the catalogue of submillimetre sources extracted from the full \textit{Herschel} Astrophysical Terahertz Large Area Survey \citep[H-ATLAS;][]{eales10}. The survey covers an area of about $550\,\hbox{deg}^2$ at 250, 350, and 500 $\mu m$ (the two longer wavelength channels corresponding fairly closely with HFI's 857 GHz and 545 GHz). A public catalogue is available only for the H-ATLAS Science Demonstration Phase field (covering $\simeq 16\,\hbox{deg}^2$) so we have used, with permission, one that is unpublished and was released for internal use of the H-ATLAS consortium \citep{maddox15,valiante15}. Within the H-ATLAS fields, the \plccs\ catalogue contains 39, 44, and 121 sources at 353, 545, 857\,GHz, respectively, while  the \ecat\  contains two sources, both detected only at 857\,GHz.
We have identified, at each frequency, the \plccs\ with the H-ATLAS catalogue using a search radius equal to half the \Planck\ FWHM. Increasing the search radius to one FWHM does not add any additional reliable counterparts. As expected, given the large surface density of H-ATLAS sources, at least one source is always present within the search radius. We have taken as reliable counterparts to \Planck\ sources those with H-ATLAS flux densities within a factor of three of the \Planck\ APERFLUX ones. Since the \Herschel\ photometry does not extend to 353\,GHz, to look for reliable counterparts at this frequency we have extrapolated the H-ATLAS flux densities using the spectral properties measured at higher frequencies. In the present, preliminary, version of the H-ATLAS catalogue, aperture corrections of flux densities for extended sources have been applied only for a fraction of the area. It  is thus possible that a few H-ATLAS flux densities of large galaxies have been strongly underestimated and are therefore missed as reliable counterparts to \Planck\ sources by the flux-density criterion. To recover them we have looked for associations of \Planck\ sources with large, nearby optical galaxies. In this way we have recovered five H-ATLAS counterparts at each of the \Planck\ frequencies.
We find 26, 38, and 112 reliable counterparts to \plccs\ sources at 353, 545, and 857\,GHz, respectively, which translates into a reliability of  26/39\,~(67\,\%), 38/44\,~(86\,\%),  and 112/121\,~(93\,\%). Neither of the two sources in the \ecat\ has a reliable H-ATLAS counterpart.

The 353\,GHz channel is five identifications short of the target 80\,\% integral reliability. Since we require consistent flux densities between \Planck\ and the extrapolated values from \Herschel\ for an identification, this slight deficit should not raise too much concern about the reliability of the 353\,GHz catalogue.

An assessment of the expected number of matches due to chance between H-ATLAS and \Planck\ sources was made using the number of \Planck\, detections within the H-ATLAS area,  the size of the \Planck\ beams, and the number density of H-ATLAS sources with flux densities above the 90\,\% completeness limits of the \plccs\ (scaling these limits by a factor of three, because we have taken as reliable counterparts to \Planck\ sources those with H-ATLAS flux densities within a factor of three of the \Planck\ APERFLUX ones).
It was found that  the number of random associations is $<10^{-4}$ for all frequencies studied ( 353, 545, and 857 GHz).

\subsection{Astrometry}
\label{sec:astrometry}
 \begin{table*}
\begingroup
\newdimen\tblskip \tblskip=5pt
\caption{Positional offset between the \plccs\ and the VLA (in the sense \plccs\ $-$ VLA) in equatorial, Galactic, and ecliptic coordinates. The mean offsets in longitude and latitude (or RA and Dec.) are given, together with the standard error on the mean. 
These offsets place strong limits on any global positional offset between \Planck\ and the VLA, and hence the international celestial reference frame (ICRF).
The radial positional uncertainty for an individual source, given in the positional error column,  is evaluated from the standard deviations of the offset in the latitude and longitude positions, assuming no correlations between these offsets. 
}
\label{tab:pos_offsets}
\nointerlineskip
\vskip -3mm
\footnotesize
\setbox\tablebox=\vbox{
   \newdimen\digitwidth
   \setbox0=\hbox{\rm 0}
   \digitwidth=\wd0
   \catcode`*=\active
   \def*{\kern\digitwidth}
\def\leaderfill{\leaders \hbox to 5pt{\hss.\hss}\hfil}
   \newdimen\signwidth
   \setbox0=\hbox{+}
   \signwidth=\wd0
   \catcode`!=\active
   \def!{\kern\signwidth}
\halign{\hbox to 0.5in{#\leaderfill}\tabskip=1em&\hfil#\hfil&\hfil#\hfil&\hfil#\hfil&\hfil#\hfil&\hfil#\hfil&\hfil#\hfil&\hfil#\hfil&\hfil#\hfil \tabskip=0pt\cr
\noalign{\doubleline}
\omit&\multispan{2}{\hfil Equatorial offsets\hfil}&\multispan{2}{\hfil Galactic offsets\hfil}&\multispan{2}{\hfil Ecliptic offsets\hfil}\cr
\noalign{\vskip-6pt}
\omit&\multispan{2}\hrulefill&\multispan{2}\hrulefill&\multispan{2}\hrulefill&Positional&Number\cr
\omit\hfil Channel\hfil&RA&Dec&longitude&latitude&longitude&latitude&error&of sources\cr
\omit&[arcsec]&[arcsec]&[arcsec]&[arcsec]&[arcsec]&[arcsec]&[arcsec]&\cr
\noalign{\vskip3pt\hrule\vskip5pt}
*30&$*-9\pm6$&$-7\pm7$&$!16\pm7$&$!0\pm6$&$-1\pm7$&$*-3\pm8$&78.0&66\cr
*44&$-11\pm8$&$!8\pm9$&$!18\pm9$&$-7\pm8$&\ldots&\ldots&97.6&67\cr
*70&$*-3\pm5$&$-1\pm6$&$!*4\pm5$&$-7\pm6$&$-2\pm5$&$!*6\pm6$&58.9&70\cr
100&$*-1\pm5$&$-4\pm5$&$!*7\pm5$&$!7\pm5$&$!9\pm4$&$-14\pm5$&53.7&70\cr
143&$*-1\pm5$&$-9\pm4$&$*-6\pm4$&$!5\pm5$&$-1\pm5$&$-13\pm4$&48.8&66\cr
217&$*!5\pm4$&$-1\pm4$&$*-2\pm4$&$!3\pm4$&$!3\pm4$&$*-1\pm5$&37.5&48\cr
\noalign{\vskip 3pt\hrule\vskip 3pt}}}
\endPlancktable
\endgroup
\end{table*}
The astrometric accuracy and positional uncertainties of the \plccs\ and \ecat\ were determined using both internal and external tests.  The external validation was based on a comparison of \plccs\ source positions with those measured with the Karl~G.\ Jansky Very Large Array (VLA), which in turn are tied to the ICRF reference frame to an accuracy well below 1\arcsec.  This comparison both validates the astrometric accuracy of the \Planck\ catalogues and provides estimates of positional uncertainties, channel by channel.  We also used simulations, in which we  injected sources to test the accuracy of the positions determined by the detection algorithms.
\subsubsection{External tests}
\label{sec:external-astrometry}
For the six lowest frequency channels, a direct comparison was made between the \plccs\ coordinates of bright, non-thermal radio sources and the subarcsecond precision positions determined by the VLA.   Table~\ref{tab:pos_offsets}  summarizes the observed offsets in position between \Planck\ and VLA observations for up to 70 (depending on the frequency) compact, unconfused synchrotron sources.
The positional offsets recorded in Table~\ref{tab:pos_offsets} are the averages (together with the error on the mean) of all the sources in this study; hence these results limit any \textit{global} positional offset between the VLA and \Planck, and do not represent the positional offset for a single source. 
The positional errors given in Table~\ref{tab:pos_offsets} are evaluated from the standard deviations of the offsets in latitude and longitude, and hence provide the radial error in the position of an individual source in each channel.
The numbers of sources used at each channel are also given.
We also examined positional offsets and uncertainties of a subset of the brightest \Planck-VLA sources (38 sources with 70\,GHz flux density $>1.5$\,Jy, 35 sources with 30\,GHz flux density $>1.9$\,Jy).  As expected, the positional uncertainty was reduced slightly (e.g. from 59\arcsec\ to 42\arcsec\ at 70\,GHz). The offsets in equatorial and ecliptic coordinates changed by 0.5 to 1.0 $\sigma$ for these bright sources.  In particular the large offset in Galactic longitude at 30\,GHz is reduced from 16\arcsec\ to 7\arcsec\ and the large offset in ecliptic latitude at 100\,GHz is reduced from $-14$\arcsec\ to $-7$\arcsec.
\subsubsection{Internal tests}
\label{sec:internal-astrometry}
The positional accuracy  of individual sources in the \plccs\ depends on the accuracy of the positions in the detection pipelines. This may be evaluated by the injection of sources into the real maps. The recovered positions are compared against the known positions of the injected sources.  Table~\ref{tab:qa} shows the resulting estimates of positional accuracy.  Table~\ref{tab:qa2} shows the same thing, but here the population of injected sources was limited to those with \snr\ $>20$. Note that all of the positional errors are less than the width of one HFI map pixel, which is half the width of an LFI map pixel. 
Also note that there is good agreement between the positional errors found from the VLA study and those found from these simulations. For the lower frequencies the position errors for sources below the 100\,\% completeness  limit must be included in the average to bring the position errors for these channels into agreement with those found from the VLA study.

Additionally, we fitted the following functional form relating the position error, $\sigma_r$ to the \snr\ of the detection:
\begin{equation}
\label{eqn:pos_err}
\sigma_r^2 = \left ( \frac{{\rm FWHM}}{ d \times {\rm \snr}} \right)^2 + \sigma_0^2 \ ,
\end{equation}
where the value used for the FWHM was the fitted FWHM from Table~\ref{tab:beam_data}, while the values of the parameters $\sigma_0$ and $d$ are shown in Table~\ref{tab:astro_error_snr}.

\subsection{Photometry}
\label{sec:qa_photometry}

The photometric accuracy of \Planck\ is very high \citep[see][]{planck2014-a01,planck2014-a03,planck2014-a08}.  The consistency of  the \Planck\ calibration is shown in \cite{planck2014-a01} to be approximately 0.2\,\% in most \Planck\ bands. The  calibration of \Planck, however, is based on the measurements of a dipole signal, and it is appropriate to ask if that accuracy extends to much smaller angular scales.
The calibration depends on our knowledge of the instruments and the window functions, which in turn depends on our understanding of the beam properties in each \Planck\ band.  Both beam properties and calibration can also be tested by comparing, on a statistical basis, the flux densities of compact sources at different \Planck\ frequencies or using different photometric methods.  We refer to these tests, among others, as ``internal.''  We have also undertaken a direct comparison of \plccs\ flux densities with ground-based or other observations of bright sources.  We refer to such comparisons as ``external'' tests.

\subsubsection{Internal Consistency}
\label{sec:internal_consistency}

\begin{table}
\begingroup
\newdimen\tblskip \tblskip=5pt
\caption{Native photometry (DETFLUX) bias, $\langle \Delta_{S} \rangle$, photometric recovery uncertainty, and radial position uncertainty.  The radial position uncertainty is the 63\,\% error radius.  All were determined from source injection into the maps, using only those injected sources with input flux above the 100\,\% completeness threshold. In the three lowest channels, the DETFLUX photometry provided in the catalogues has been corrected for this bias.}
\label{tab:qa}
\nointerlineskip
\vskip -3mm
\footnotesize
\setbox\tablebox=\vbox{
   \newdimen\digitwidth
   \setbox0=\hbox{\rm 0}
   \digitwidth=\wd0
   \catcode`*=\active
   \def*{\kern\digitwidth}
\def\leaderfill{\leaders \hbox to 5pt{\hss.\hss}\hfil}
   \newdimen\signwidth
   \setbox0=\hbox{+}
   \signwidth=\wd0
   \catcode`!=\active
   \def!{\kern\signwidth}
\halign{\hbox to 0.5in{#\leaderfill}\tabskip=1em &\hfil#\hfil&\hfil#\hfil &\hfil#\hfil\tabskip=0pt\cr
\noalign{\doubleline}
\omit\hfil Channel\hfil&DETFLUX bias& stdev($\Delta_S/\sigma_S$)& Position error\cr
\omit&[\%]&\omit&[arcsec]\cr
\noalign{\vskip 3pt\hrule\vskip 5pt}
*30&*$-2.34$&0.33&50.49\cr
*44&*$-4.12$&1.67&59.57\cr
*70&$-12.05$&3.69&44.07\cr
100&!*$1.10$&1.22&51.96\cr
143&*$-0.91$&1.44&43.68\cr
217&*$-2.36$&1.82&39.94\cr
353&*$-3.72$&1.85&39.59\cr
545&*$-1.59$&2.13&39.58\cr
857&*$-3.51$&2.51&39.41\cr
\noalign{\vskip 3pt\hrule\vskip 3pt}}}
\endPlancktable
\endgroup
\end{table}

\begin{table}
\begingroup
\newdimen\tblskip \tblskip=5pt
\caption{As Table~\ref{tab:qa}, but for all detections with $S/N>20$.}
\label{tab:qa2}
\nointerlineskip
\vskip -3mm
\footnotesize
\setbox\tablebox=\vbox{
   \newdimen\digitwidth
   \setbox0=\hbox{\rm 0}
   \digitwidth=\wd0
   \catcode`*=\active
   \def*{\kern\digitwidth}
\def\leaderfill{\leaders \hbox to 5pt{\hss.\hss}\hfil}
   \newdimen\signwidth
   \setbox0=\hbox{+}
   \signwidth=\wd0
   \catcode`!=\active
   \def!{\kern\signwidth}
\halign{\hbox to 0.5in{#\leaderfill}\tabskip=1em &\hfil#\hfil&\hfil#\hfil&\hfil#\hfil\tabskip=0pt\cr
\noalign{\doubleline}
\omit\hfil Channel\hfil&DETFLUX bias&stdev($\Delta_S/\sigma_S$)&Position error\cr
\omit&[\%]&\omit&[arcsec]\cr
\noalign{\vskip 3pt\hrule\vskip 5pt}
*30&*$-2.35$&*1.16&37.88\cr
*44&*$-3.15$&*1.98&44.35\cr
*70&$-13.75$&11.39&39.69\cr
100&*!$0.58$&*1.45&45.80\cr
143&*$-1.18$&*1.76&39.53\cr
217&*$-2.06$&*2.15&38.33\cr
353&*$-3.24$&*2.14&38.57\cr
545&*$-0.81$&*2.54&37.85\cr
857&*$-2.27$&*2.79&37.99\cr
\noalign{\vskip 3pt\hrule\vskip 3pt}}}
\endPlancktable
\endgroup
\end{table}

 \begin{table}
\begingroup
\newdimen\tblskip \tblskip=5pt
\caption{Parameters  $\sigma_0$  and $d$ determined by fitting Eq.~(\ref{eqn:pos_err}), relating the position error, $\sigma_r$, from the simulations to the \snr\ of the detection.}
\label{tab:astro_error_snr}
\nointerlineskip
\vskip -3mm
\footnotesize
\setbox\tablebox=\vbox{
   \newdimen\digitwidth
   \setbox0=\hbox{\rm 0}
   \digitwidth=\wd0
   \catcode`*=\active
   \def*{\kern\digitwidth}
\def\leaderfill{\leaders \hbox to 5pt{\hss.\hss}\hfil}
   \newdimen\signwidth
   \setbox0=\hbox{+}
   \signwidth=\wd0
   \catcode`!=\active
   \def!{\kern\signwidth}
\halign{\hbox to 0.5in{#\leaderfill}\tabskip=2em &\hfil#\hfil&\hfil#\hfil\tabskip=0pt\cr
\noalign{\doubleline}
\omit\hfil Channel\hfil&$\sigma_{0}$&$d$\cr
\omit&[arcmin]&\cr
\noalign{\vskip 3pt\hrule\vskip 5pt}
*30&$0.267\pm0.001$*&$2.14\pm0.44$\cr
*44&$0.217\pm0.001$*&$1.74\pm0.59$\cr
*70&$0.538\pm0.001$*&$1.68\pm0.23$\cr
100&$0.685\pm0.001$*&$1.61\pm0.01$\cr
143&$0.615\pm0.001$*&$1.60\pm0.01$\cr
217&$0.580\pm0.002$*&$1.38\pm0.02$\cr
353&$0.578\pm0.002$*&$1.43\pm0.03$\cr
545&$0.539\pm0.002$*&$1.48\pm0.04$\cr
857&$0.546\pm0.0004$&$1.46\pm0.02$\cr
\noalign{\vskip 3pt\hrule\vskip 3pt}}}
\endPlancktable
\endgroup
\end{table}

\paragraph{Simulations} For the HFI channels we characterize the accuracy of source photometry by comparing the native flux-density estimates (DETFLUX) of matched sources to the known flux densities of sources injected into the real maps.
The photometric accuracy is a function of \snr, and the faint detections are affected by upward bias due to noise fluctuations. In the previous \lastcat, at the higher HFI frequencies, the DETFLUX estimates were found to be biased low \citep{planck2013-p05}. This has been corrected in the construction of the \plccs.
Tables \ref{tab:qa} and \ref{tab:qa2} show the DETFLUX bias per channel as well as the standard deviation of $\Delta_S / \sigma_S$, the difference between the input and recovered flux densities normalized by the uncertainty on the flux density, which would be unity for Gaussian noise.

\begin{figure}
\begin{center}
\includegraphics[width=0.5\textwidth]{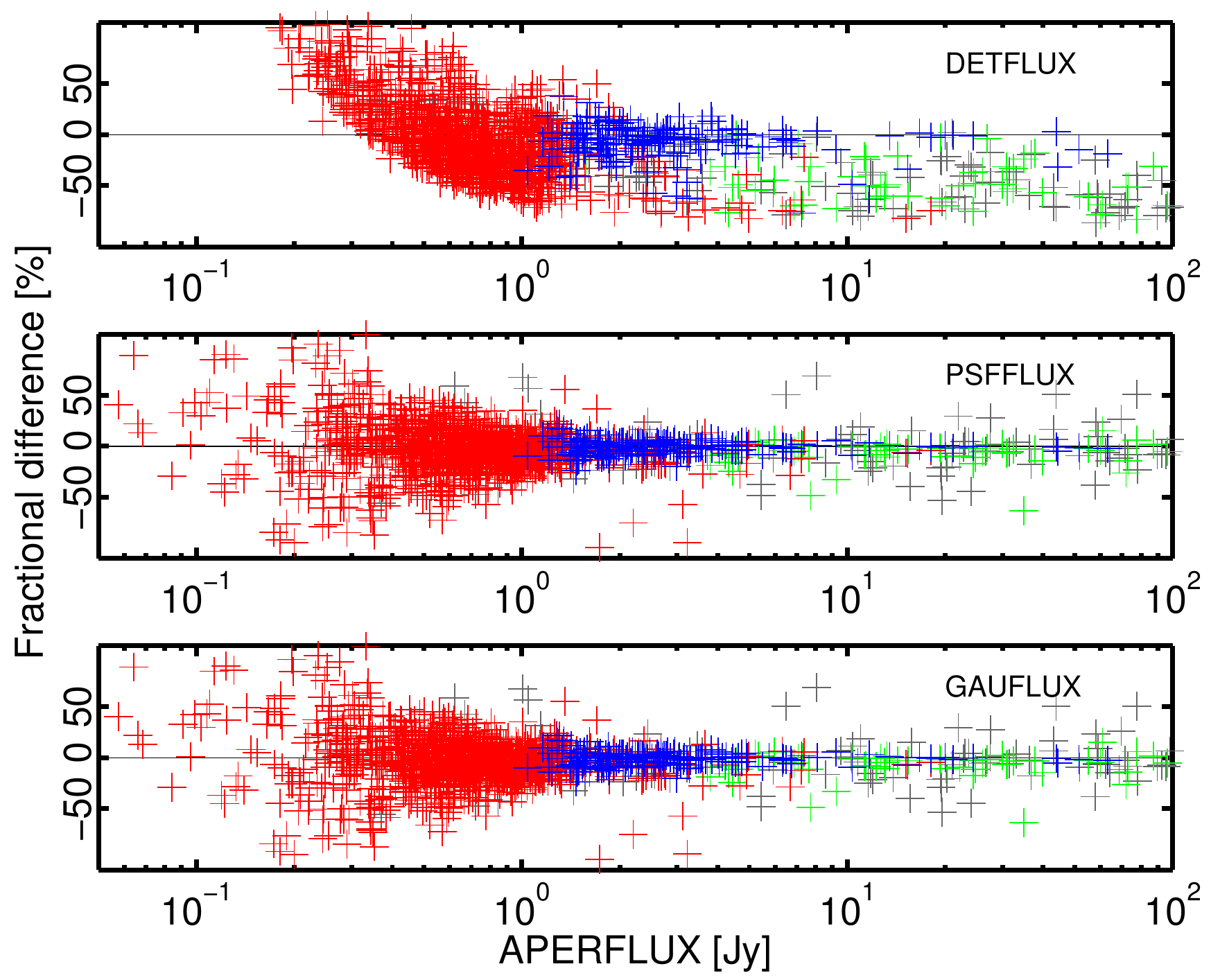}
\caption{Comparison of the DETFLUX, PSFFLUX, and GAUFLUX flux-density estimates with APERFLUX for the \plccs\ 30\,GHz catalogue. The fractional difference is defined as $(S-S_{\rm APERFLUX})/S_{\rm APERFLUX}$. The green and blue points correspond to sources where $S_{\rm APERFLUX}/S_{\rm APERFLUX\_ERR} > 5$. Grey and green points correspond to sources with $|b| < 5\degr$, while the red and blue have $|b| > 5\degr$. 
\label{fig:PCCS2_comp_phots30}}
\end{center}
\end{figure}

\begin{figure}
\begin{center}
\includegraphics[width=0.5\textwidth]{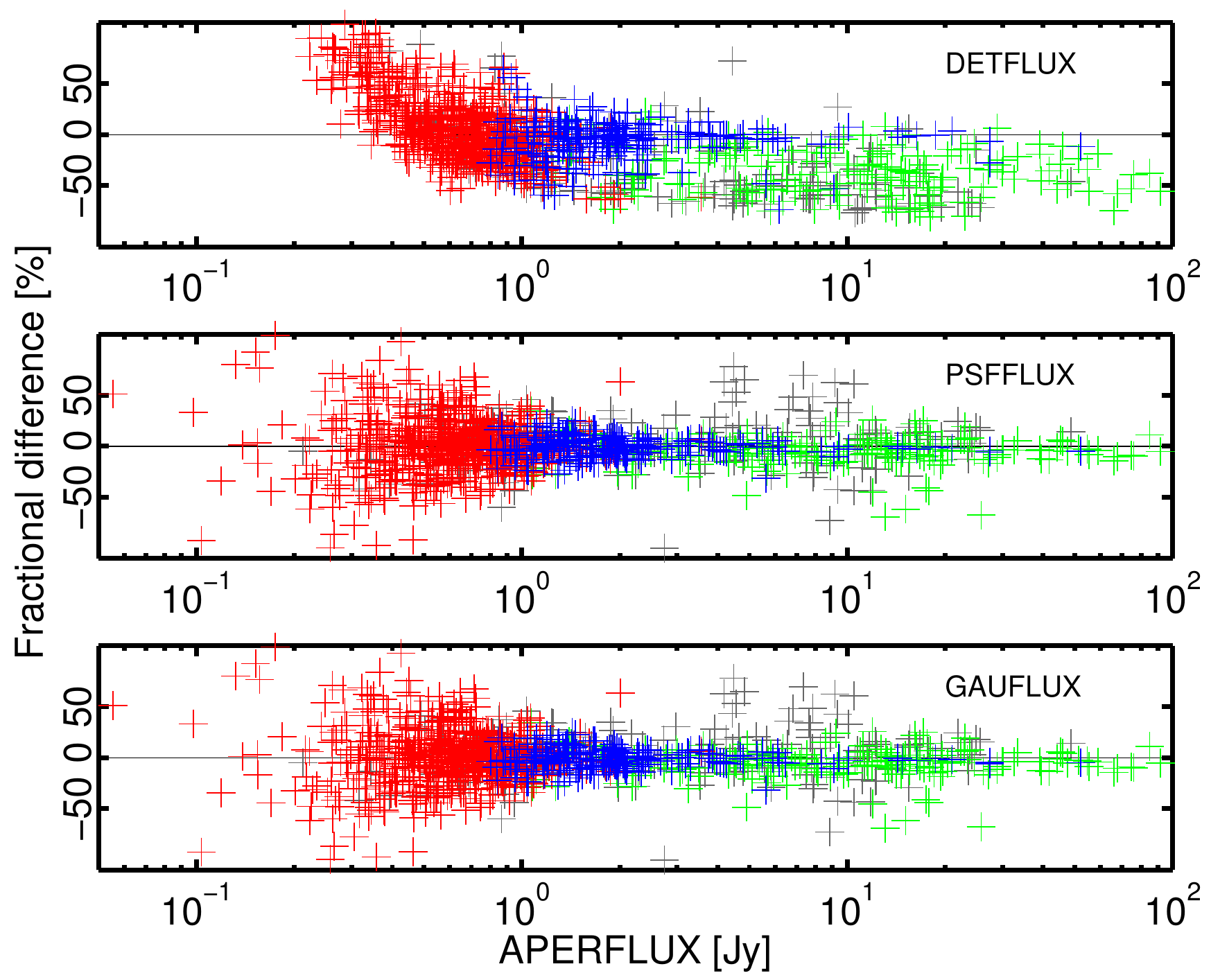}
\caption{Comparison of the DETFLUX, PSFFLUX, and GAUFLUX flux-density estimates with APERFLUX for the \plccs\ 70\,GHz catalogue. The fractional difference is defined as $(S-S_{\rm APERFLUX})/S_{\rm APERFLUX}$. The green and blue points correspond to sources where $S_{\rm APERFLUX}/S_{\rm APERFLUX\_ERR} > 5$. Grey and green points correspond to sources with $|b| < 5\degr$, while the red and blue have $|b| > 5\degr$. 
\label{fig:PCCS2_comp_phots70}}
\end{center}
\end{figure}

\begin{figure}
\begin{center}
\includegraphics[width=0.5\textwidth]{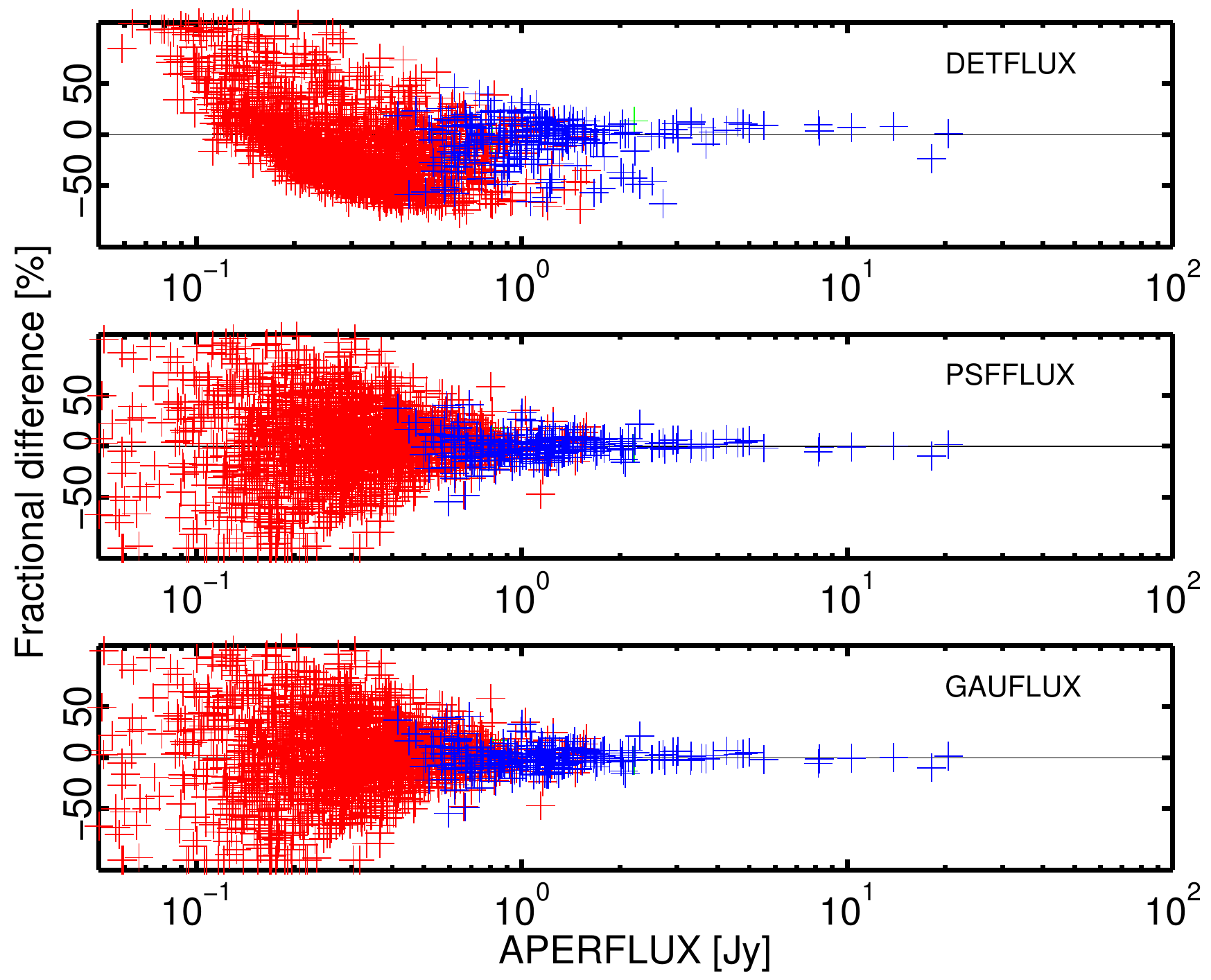}
\caption{Comparison of the DETFLUX, PSFFLUX, and GAUFLUX flux-density estimates with APERFLUX for the \plccs\ 143\,GHz catalogue. The fractional difference is defined as $(S-S_{\rm APERFLUX})/S_{\rm APERFLUX}$. The blue points correspond to sources where $S_{\rm APERFLUX}/S_{\rm APERFLUX\_ERR} > 5$. 
\label{fig:PCCS2_comp_phots143}}
\end{center}
\end{figure}

\begin{figure}
\begin{center}
\includegraphics[width=0.5\textwidth]{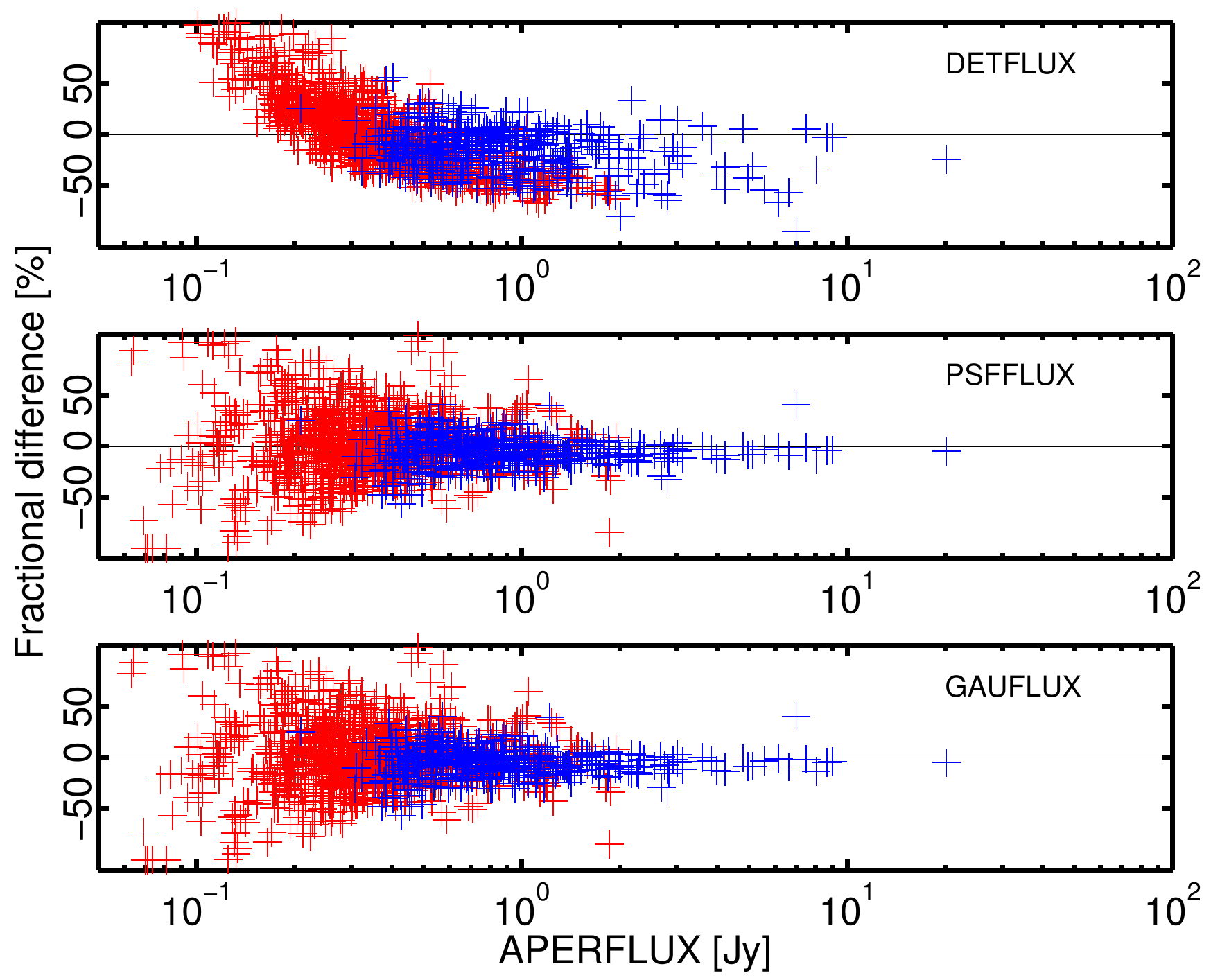}
\caption{Comparison of the DETFLUX, PSFFLUX, and GAUFLUX flux-density estimates with APERFLUX  for the \plccs\ 353\,GHz catalogue. The fractional difference is defined as $(S-S_{\rm APERFLUX})/S_{\rm APERFLUX}$. The blue points correspond to sources where $S_{\rm APERFLUX}/S_{\rm APERFLUX\_ERR} > 5$. 
\label{fig:PCCS2_comp_phots353}}
\end{center}
\end{figure}

\begin{figure}
\begin{center}
\includegraphics[width=0.5\textwidth]{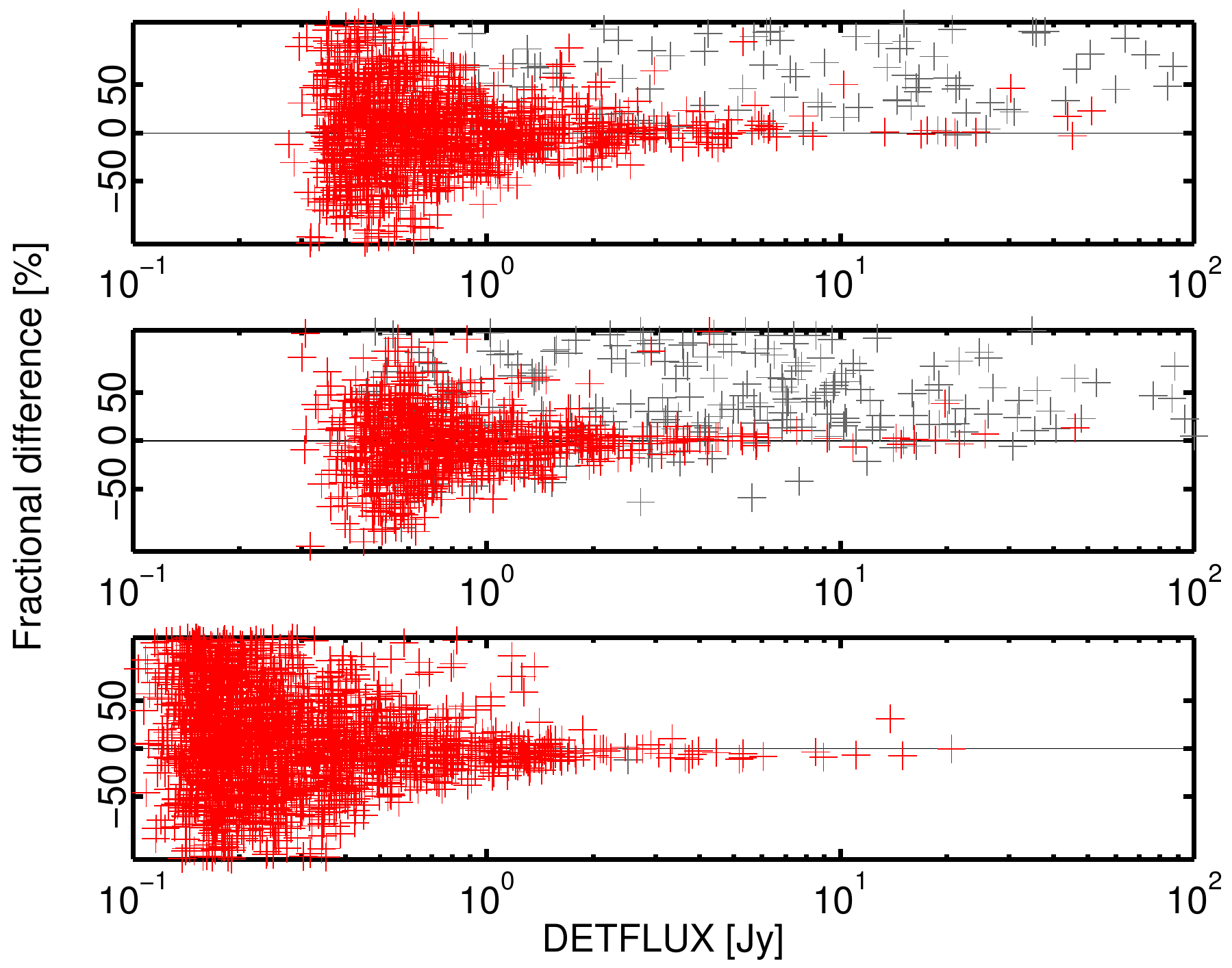}
\caption{Comparison of the APERFLUX flux-density estimates with DETFLUX for the \plccs\ 30\,GHz (top panel), 70\,GHz (middle panel) and 143\,GHz (bottom panel) catalogues. The fractional difference is defined as $(S_{\rm APERFLUX}-S_{\rm DETFLUX})/S_{\rm DETFLUX}$. Grey point are sources at $|b| < 5\degr$; while red points have $|b| > 5\degr$.  The greater depth of the 143\,GHz catalogue is clear.
\label{fig:PCCS2_comp_phots_detflux}}
\end{center}
\end{figure}

\paragraph{Comparisons of the four different flux-density estimates} We next compare values derived from the four different methods of assessing flux densities.   Figures~\ref{fig:PCCS2_comp_phots30},~\ref{fig:PCCS2_comp_phots70}, \ref{fig:PCCS2_comp_phots143}, and~\ref{fig:PCCS2_comp_phots353} show the results for four \Planck\ channels.
These comparisons are made against APERFLUX. This flux estimation method is the simplest and makes the smallest number of assumptions about the data. However, DETFLUX has smaller uncertainties than APERFLUX, which may be seen in these figures by the upward curve towards lower APERFLUX values in the comparison with DETFLUX.  This may be understood as there being a clear signal present in DETFLUX when the APERFLUX is compatible with noise.  Figure~\ref{fig:PCCS2_comp_phots_detflux} shows  the comparison against DETFLUX for APERFLUX at 30, 70, and 143\,GHz. In these plots, as expected,  the curvature disappears, and we see good agreement between the methods that becomes progressively noisier towards lower values of DETFLUX.
For unresolved sources in regions where there is little non-Gaussianity present in the background, DETFLUX is the flux estimation method of choice, given its greater sensitivity.
However, in regions of high non-Gaussian background emission, DETFLUX is less robust. This may been seen by the lack of consistency between DETFLUX and APERFLUX, in these figures, at least for the green and grey points representing sources that lie within $5\degr$ of the Galactic plane. 
As the frequency increases so do the levels of non-Gaussian emission. In the comparison between DETFLUX and APERFLUX at 353\,GHz, which only contains sources at the higher Galactic latitudes, there is a large degree of scatter. Indeed, at 353\,GHz and above it is advisable to favour APERFLUX over DETFLUX.

\paragraph{Interband Comparisons}
As an additional internal test, we compared \plccs\ flux densities at one band with those in neighbouring bands.  We performed this comparison for the six lowest \Planck\ channels. We began by selecting all \plccs\ sources at 70\,GHz with the following restrictions: flux density $S(70)\ge0.9$\,Jy (virtually all such sources had \snr\,$>$7); Galactic latitude $|b| \ge 10^{\circ}$; and no evidence of extension.  
We then matched these with sources in the \plccs\ at 30, 44, 100, 143, and 217\,GHz.  Of the 203 sources, more than 99\,\% were detected at 30, 44, and 100\,GHz as well as 70\,GHz, and 97\,\% at 143\,GHz.  Since virtually all the sources had synchrotron spectra, and, in general, negative spectral indices, it is not surprising that only 90\,\% of the sources could be identified at 217\,GHz.  We note that some of the sources not found in the \plccs\ at 217\,GHz did appear in the \ecat; these few sources, however, were not used in this internal test.

We used this merged catalogue to calculate spectral indices for each of the 203 sources.  The spectral indices were then used to make the small colour-corrections to the flux density in each band for each source.  For the colour-corrections at 30\,GHz we used the 30--44 spectral index, and for 353\,GHz the 217--353 spectral index.  For the other five bands, for band $N$, we used the spectral index found between bands $N-1$ and $N+1$ (e.g. for 44\,GHz, we used the 30--70 spectral index).  The colour-corrections we used are tabulated in \cite{planck2014-a03} and \cite{planck2014-a08}.The amplitude of these corrections in a given \Planck\ band ranges from around 2.5\,\% at 70\,GHz to less than 1\,\% at 30\,GHz.  For a given source, the precision of the colour-corrections was typically 0.2\,\%.

Next, we used the colour-corrected flux densities to recompute spectral indices for each source.  These spectral indices were then used to predict a flux density for each source in frequency band $N$ by assuming a constant spectral index between bands $N-1$ and $N+1$.  For instance, we interpolated between the 28.4 and 70.4\,GHz flux densities to predict a 44.1\,GHz flux density for each source, using the calculated spectral index for that source. These predicted values were then plotted against the actual (colour-corrected) measurements at 44.1\,GHz.  This operation was repeated for 70, 100, 143, and 217\,GHz.
If the flux-density scales of \Planck\ are consistent across bands and if the spectral index is constant as assumed, we expect to see lines of unit slope.  In fact, the slopes were close to unity for all five bands tested, as shown in Table~\ref{tab:interband_comp}.
\begin{table}
\begingroup
\newdimen\tblskip \tblskip=5pt
\caption{Slope found in the plots of predicted versus measured flux densities in five of the \Planck\ bands.  A slope $>1$ implies that the measured flux density exceeds the predicted value.  Departures from unit slope can be explained by slight {\it curvature} of the spectra; we show the required change in spectral index,  $\Delta\alpha$, to explain the departure from unity.}
\label{tab:interband_comp}
\nointerlineskip
\vskip -3mm
\footnotesize
\setbox\tablebox=\vbox{
   \newdimen\digitwidth
   \setbox0=\hbox{\rm 0}
   \digitwidth=\wd0
   \catcode`*=\active
   \def*{\kern\digitwidth}
\def\leaderfill{\leaders \hbox to 5pt{\hss.\hss}\hfil}
   \newdimen\signwidth
   \setbox0=\hbox{+}
   \signwidth=\wd0
   \catcode`!=\active
   \def!{\kern\signwidth}
\halign{\hbox to 0.5in{#\leaderfill}\tabskip=1em &\hfil#\hfil&\hfil#\hfil\tabskip=0pt\cr
\noalign{\doubleline}
\omit\hfil Channel\hfil&Slope&Required $\Delta\alpha$\cr
\noalign{\vskip 3pt\hrule\vskip 5pt}
*44&$0.985 \pm 0.005$&$+0.066$\cr
*70&$1.025 \pm 0.005$&$-0.125$\cr
100&$1.038 \pm 0.005$&$-0.210$\cr
143&$0.997 \pm 0.005$&$+0.015$\cr
217&$1.017 \pm 0.005$&$-0.076$\cr
\noalign{\vskip 3pt\hrule\vskip 3pt}}}
\endPlancktable
\endgroup
\end{table}

We examined several different possibilities for the slight departures from unit slopes.  First, we explored the possibility that CO line emission could influence the results by perturbing the flux densities. The \plccs\ flux densities used were the MHW2 estimates. The filtering removes all foreground emissions on large scales; only scales approaching the size of the beam could affect the MHW2 flux density estimates. Thus, Galactic CO emission could introduce scatter in the 100\,GHz values, but should not give significant coherent offsets. There is, however,  the question of redshifted CO line emission from the sources themselves. 
Since the sources meeting our selection criteria are mostly bright blazars \citep{planck2011-6.2}, the ratio of CO line flux to continuum emission is expected to be very small. 

Next, we investigated whether the statistically significant departures from unit slope in the plots of predicted versus measured flux could reasonably be explained by a breakdown in the assumption that the spectral index of the sources stays constant from band $N-1$ to band $N+1$. For instance, the measured slope at 70\,GHz is 1.0125; that is, the measured 70\,GHz flux densities are about 2.55\,\% higher than we would find by interpolating between 44 and 100\,GHz, assuming no change in the spectral index between 44 and 100\,GHz.  If instead we allow for spectral {\it curvature}, the small discrepancy can be reduced or eliminated.  With the simplest assumption, a sharp change of spectral index, $\Delta\alpha$, at 70\,GHz, we find that the small excess of measured over predicted flux can be explained by $\Delta\alpha=-0.125$ at 70\,GHz.  We performed similar calculations for this simple model in the other \Planck\ bands (see Table~\ref{tab:interband_comp}).  Spectral index changes of this magnitude are reasonable (electron ageing can account for a change in spectral index of around $-0.5$); see Sect.~\ref{sec:characteristics} for plots of spectral index distributions for all sources in the \plccs, not just those used in this analysis.

\subsubsection{External consistency}
\label{sec:external_consistency}
The calibration of \Planck\ is precise, and we have demonstrated the internal consistency of the flux densities in Sect.~\ref{sec:internal_consistency}.  The calibration of \Planck\ is also \textit{absolute} in the sense that it depends only on the motion of the satellite and the 0.02\,\% accurate measurement of the CMB temperature \citep{Fixsen09}. Consequently, comparing \Planck\ flux densities to those measured by other instruments is actually a check on the accuracy of the latter.  Indeed, \cite{Butler15} have employed \Planck\ measurements to refine the centimetre-wavelength flux-density scales used at the VLA and the ATCA.  Here, we summarize the results of that study and of comparisons with other CMB and submillimetre instruments and missions.
%
\begin{figure}
\begin{center}
\includegraphics[width=0.53\textwidth]{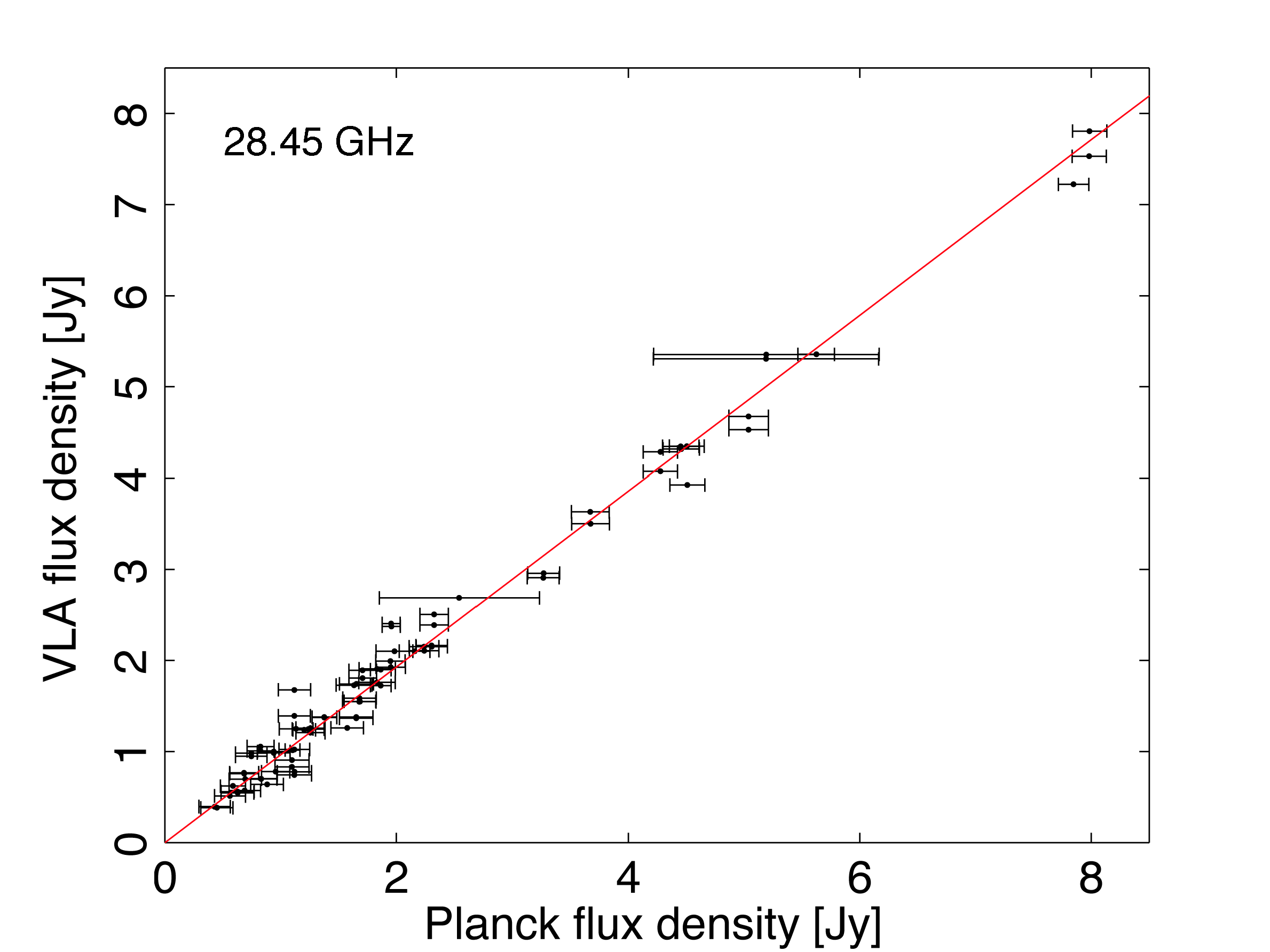}
\includegraphics[width=0.53\textwidth]{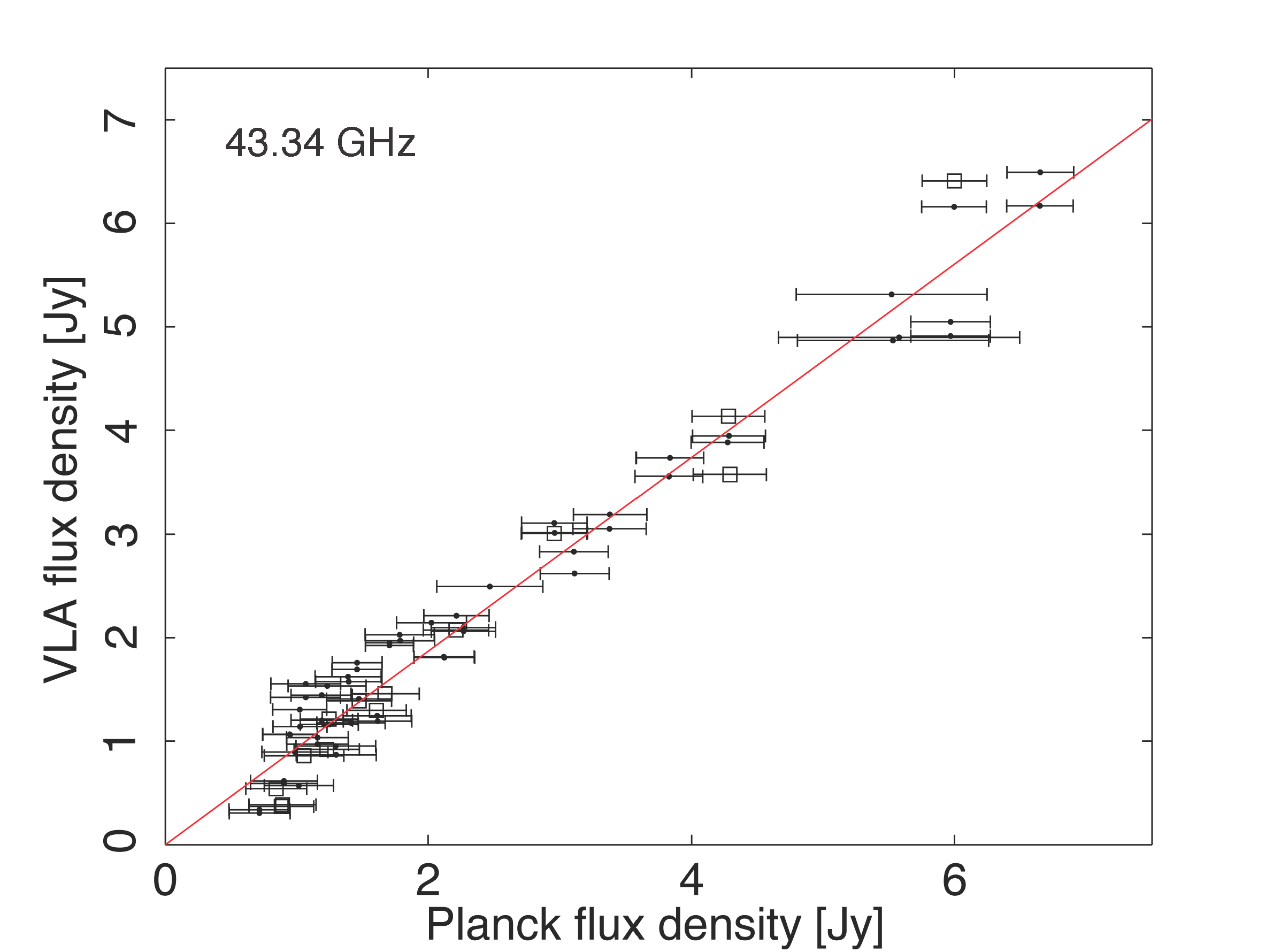}
\caption{Comparison of simultaneous colour-corrected flux density measurements by \Planck\ and the VLA at 28.45\,GHz (top) and \Planck\ and the VLA (dots) or ATCA (open squares) 43.34\,GHz (bottom). The best fit line is shown in red. The \Planck\ 44\,GHz channel is noisier than the 30\,GHz channel and shows Eddington bias at low flux densities. Some sources have larger error bars than others because they lie close to the Galactic plane where the uncertainties in the measured flux densities can be large. The tiny VLA error bars have not been plotted. }
\label{fig:LFI_VAL_VLA}
\end{center}
\end{figure}
\begin{table}
\begingroup
\newdimen\tblskip \tblskip=5pt
\caption{Comparison of  the corrected \Planck\ flux-density measurements with the ATCA, VLA, and ACT values.  
The error in the flux-density ratios is purely due to statistical uncertainties; the error in the \% increment includes the uncertainties due to the \Planck\ beams and calibration. }
\label{tab:planck_v_atca_and_vla}
\nointerlineskip
\vskip -3mm
\footnotesize
\setbox\tablebox=\vbox{
   \newdimen\digitwidth
   \setbox0=\hbox{\rm 0}
   \digitwidth=\wd0
   \catcode`*=\active
   \def*{\kern\digitwidth}
\def\leaderfill{\leaders \hbox to 5pt{\hss.\hss}\hfil}
   \newdimen\signwidth
   \setbox0=\hbox{.}
   \signwidth=\wd0
   \catcode`!=\active
   \def!{\kern\signwidth}
\halign{\hbox to 1.0in{#\leaderfill}\tabskip=1em &\hfil#\hfil&\hfil#\hfil\tabskip=0pt\cr
\noalign{\doubleline}
\omit\hfil Frequency\hfil&Flux density ratio&\% increment of \Planck\ over \cr
\omit\hfil[GHz]\hfil&(ground/\Planck)&ground-based measures\cr
\noalign{\vskip 3pt\hrule\vskip 5pt}
*22.45 (ATCA)&$0.99 \pm 0.017$&$1 \pm 1.7$\cr
*28.45 (VLA)&$0.97 \pm 0.008$&$3 \pm 1.2$\cr
*43.34 (VLA, ATCA)&$0.94 \pm 0.013$&$6 \pm 1.7$\cr
147.70 (ACT)&$0.97 \pm 0.03$*&$3 \pm 3$!*\cr
217.60 (ACT)&$0.96 \pm 0.03$*&$4 \pm 3$!*\cr
\noalign{\vskip 3pt\hrule\vskip 3pt}}}
\endPlancktable
\endgroup
\end{table}
\paragraph{The 30 and 44\,GHz channels}
The comparison between \Planck\ flux densities at 28.4 and 44.1\,GHz and ground based observations at 22.45, 28.45, and 43.34\,GHz is based on observations carried out at the ATCA and the VLA in April and May 2013.  Both instruments observed a set of strong, unresolved, unconfused radio sources also scanned by \Planck\ in this time interval.  These observations are part of a wider effort by Perley and Stevens (2015, in preparation) to compare the flux-density scales of the two interferometers, the VLA in the north and ATCA in the south.  The \Planck\ DETFLUX measurements were derived not from the \plccs, which averages over the four years of LFI observations, but from a special map constructed using only data from 1 April to 30 June 2013.  This was necessary in order to minimize the effects of source variability.
Since the central frequencies of the \Planck\ bands did not exactly match the frequencies employed by the ground-based instruments, we interpolated and  colour-corrected the \Planck\ measurements to 22.45, 28.45, and 43.34\,GHz. For both purposes, we used spectral indices derived from the far more precise interferometric measurements (VLA or ATCA).  A comparison of the corrected \Planck\ measurements with the ATCA and VLA values is shown in Table~\ref{tab:planck_v_atca_and_vla} and Fig.~\ref{fig:LFI_VAL_VLA}. The \Planck\ measurements are consistently slightly higher across all frequencies.
These results are summarized in \cite{Butler15}, and described in greater detail in \citet{2015arXiv150602892P} where various tests of the validity of the results are presented.  In particular, the effect of dropping sources found to have varied over the three-month period of the \Planck\ observations is examined.  The results in Table~\ref{tab:planck_v_atca_and_vla} are estimates derived from \cite{Butler15}.  The discrepancies between the satellite and ground-based values lie close to or within the estimated error in the latter.  For the flux-density scale of \cite{Perley13}, employed at the VLA, this uncertainty is estimated to be 5\,\%, and roughly the same level of precision may be assigned to the flux-density scale employed at ATCA.  
We also compared the ATCA and VLA measurements to \plccs\ 30 and 44\,GHz flux densities (averaged over 4 years) and found consistent results, for the same sources.  As expected, source variability substantially increased the scatter.  Excluding three manifestly variable sources, we find VLA/\Planck\ = $0.96\pm0.02$ at 28.45\,GHz and (VLA\&ATCA)/\Planck\ = $0.93\pm0.03$  at 43.34\,GHz.  These are consistent with the more precise comparison described above. The $3\%$ and $6\%$ differences may be compared to the quoted $5\%$ uncertainty in the flux density scales as given by \cite{Perley13}.

\begin{figure}
\begin{center}
\includegraphics[width=0.53\textwidth]{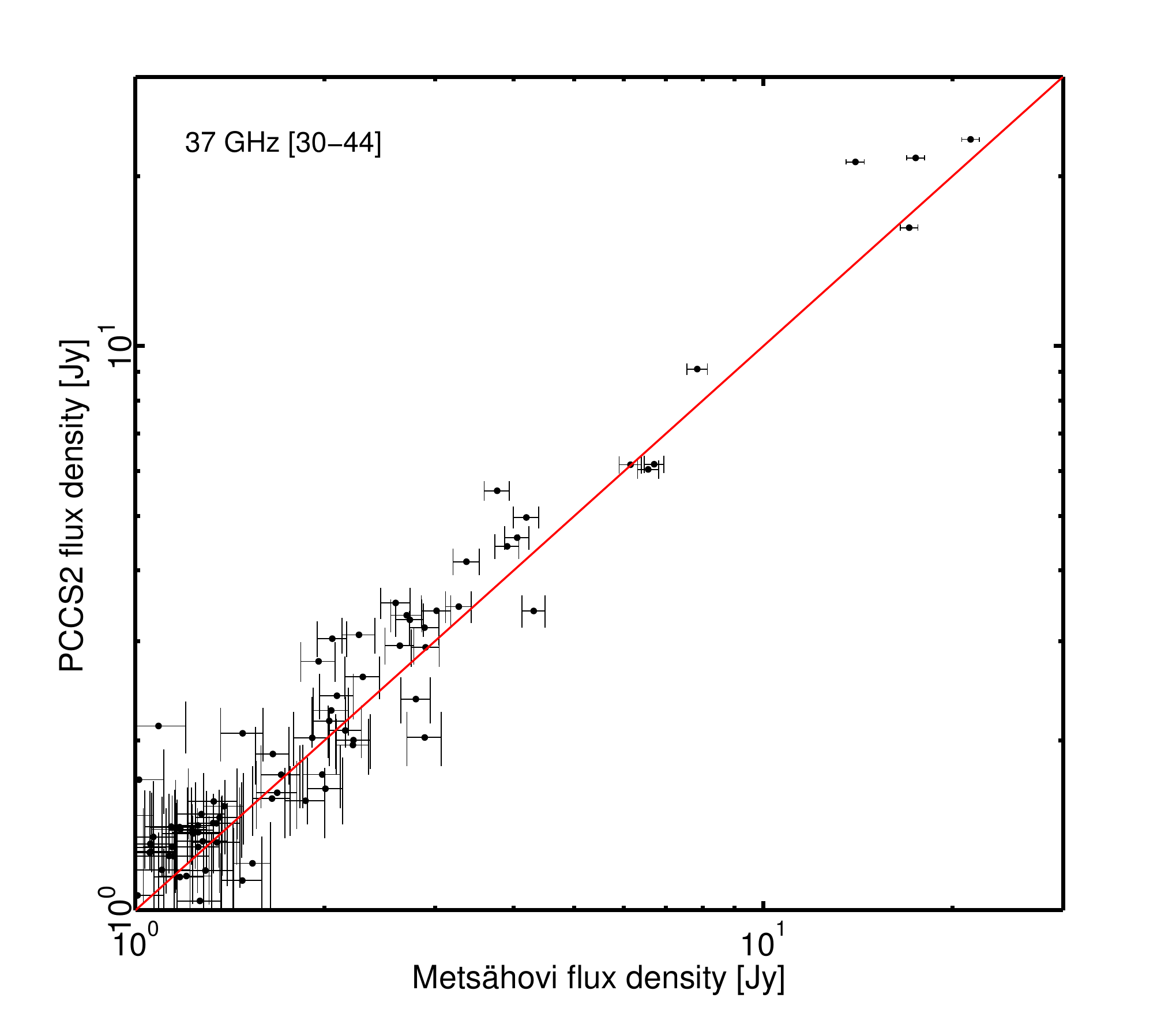}
\includegraphics[width=0.53\textwidth]{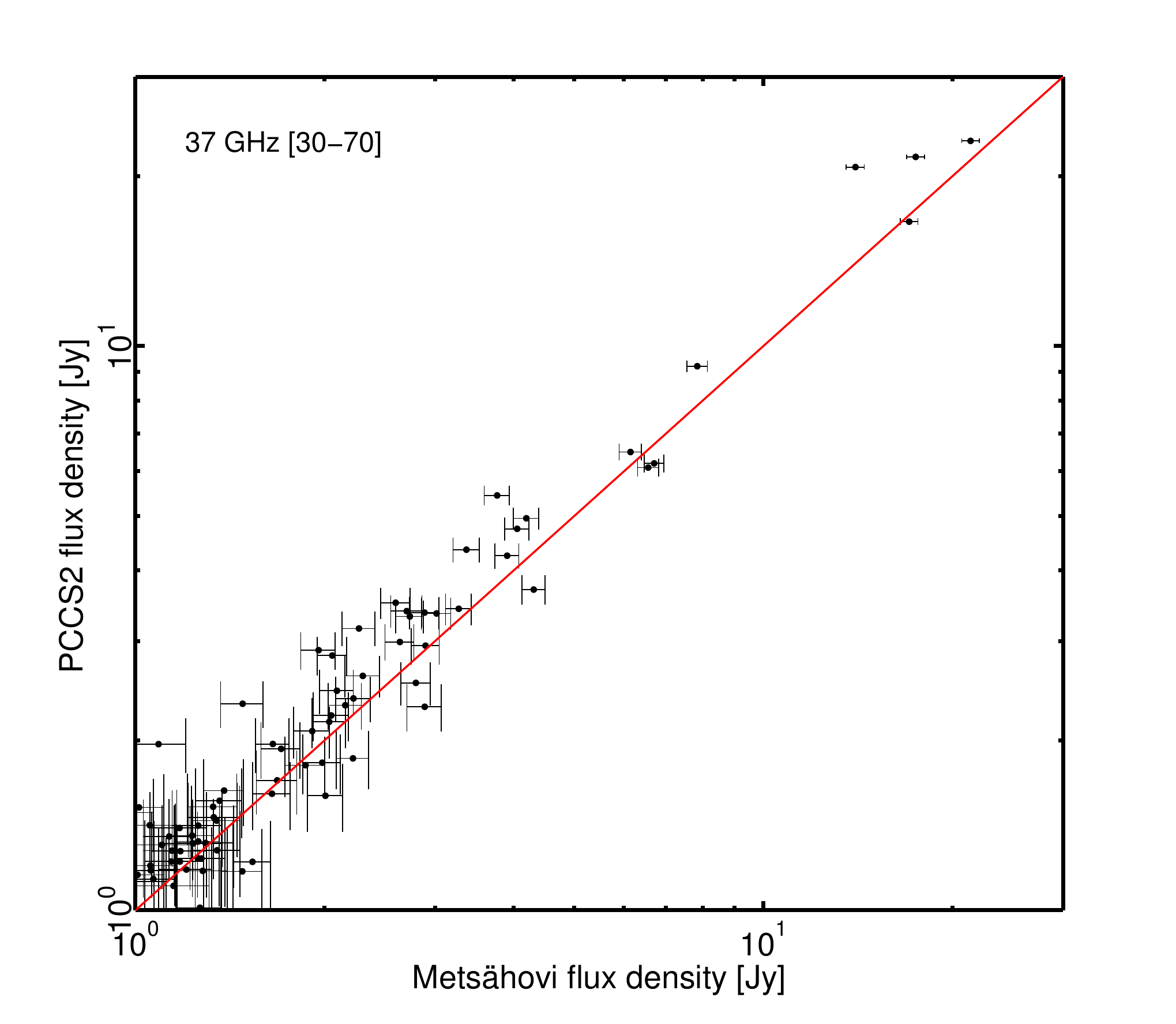}
\caption{Comparison between the Mets{\"a}hovi and the colour-corrected \plccs\ flux densities (DETFLUX) interpolated to 37\,GHz using 30 and 44\,GHz (\textit{top}) and 30 and 70\,GHz (\textit{bottom}). The multiple observations of each source have been averaged to a single flux density; the averaging was performed over the period of the full \Planck\ mission, not just the epochs at which each source was observed by \Planck. The uncertainties, therefore,  reflect the variability of the sources instead of the flux-density accuracy of the measurements, which is of the order of a few mJy.}
\label{fig:LFI_MET}
\end{center}
\end{figure}

The Mets{\"a}hovi Observatory is continuously monitoring bright radio sources in the northern sky at 37\,GHz \citep{Terasranta04}. From their sample, sources brighter than 1\,Jy were selected and their flux densities averaged over the period of \Planck\ observations used for the \plccs\  (Planck Intermediate results, in preparation); note that this period corresponds to the full duration of the \Planck\ mission.  Hence, the uncertainties in Fig.~\ref{fig:LFI_MET} reflect the variability of the sources during the \Planck\ mission. The \Planck\ measurements were colour-corrected and extrapolated to the Mets{\"a}hovi frequency before making the comparison. The \Planck\ and Mets{\"a}hovi flux densities agree at the $0.3$\,\% and $0.1$\,\% level, and an uncertainty of $\pm$4\,\%, at 30--44\,GHz and 30--70\,GHz, respectively.
\paragraph{The 143 and 217\,GHz channels}
Flux densities at 143 and 217\,GHz from the earlier \lastcat\ were compared to measurements made at the Atacama Cosmology Telescope (ACT) by \cite{Louis14}.  We repeated those comparisons using the new \plccs\ values, and extended the comparison to South Pole Telescope (SPT) flux densities from \cite{Mocanu13}.  We employed the DETFLUX values from the \plccs.  As in \cite{Louis14}, the \Planck\ flux densities were colour-corrected and extrapolated to the central frequencies of both ACT (147.6 and 217.6\,GHz) and SPT (152.9 and 218.1\,GHz), using the spectral index appropriate for each source.  It is important to bear in mind that the ground-based measurements were in almost all cases far from simultaneous with those of  \Planck. Thus the variability of sources, virtually all AGN, caused significant scatter.  In the case of the ACT equatorial sources, we had measurements from both the 2009 and the 2010 seasons.  That allowed us to find and drop two manifestly variable sources.  We also dropped two sources with thermal spectra. 
For SPT, \cite{Mocanu13} present just a single flux density for each source, so in most cases we had no means of discovering and removing sources that were variable.  Four sources, however, were detected by all three experiments, ACT and SPT in 2008 and \Planck\ integrated over the mission.  Three of these were evidently variable (and are discussed further below).
For the comparison between corrected \Planck\ flux densities and those of ACT at 147.6\,GHz, we were left with 58 sources in common.  ACT flux densities were on average 0.97$\pm$0.03 times \Planck's.  At 217.6\,GHz, fewer ACT sources (50) were detected by \Planck, and we find ACT = 0.89$\pm$0.03 times \Planck.

We now consider the effect on these results of sources known to be variable. First, comparison of ACT measurements made in 2009 and 2010 showed that two sources varied strongly.  Dropping them changed the 147.6\,GHz slope to 0.95.  On the other hand, if we drop the three variable sources detected by all three experiments, the slope becomes 0.98. If we drop all five sources for which we have direct evidence of variability, the slope at 147.6\,GHz settles to 0.97$\pm$0.02. 
Since the exclusion of variable sources  moves the slope both up and down in amplitude, we adopt 0.97$\pm$0.03 for the relation between ACT and \Planck\ flux densities; ACT flux densities are about $3$\,\% lower than \Planck's.  At 217.6\,GHz, if we remove the few sources for which we have direct evidence of variability, the slope changes to 0.96$\pm$0.03.

\Planck\ found fewer SPT sources than ACT sources (25 at 152.9\,GHz and 30 at 218.1\,GHz), and the scatter due to source variability was larger.  If we include all sources, we again find the ground-based flux densities are lower than \Planck's: 0.95$\pm$0.05 at 152.9\,GHz and 0.88$\pm$0.05 at 218.1\,GHz. If we now exclude the three sources that were seen to vary between 2008 and the later \Planck\ mission, the SPT results become 0.99$\pm$0.05 at 152.9\,GHz and 0.97$\pm$0.05 at 218.1\,GHz.  We adopt these values, all lying between 0.96 and 0.99, as evidence of good agreement between the ground-based and \Planck\ flux density scales at 143 and 217\,GHz. The small differences between the \Planck\ flux density scales at these frequencies and those measured from the ground can be compared to the  following uncertainties quoted for the ground based experiments: SPT, $1.6\%$ and $2.4\%$ calibration uncertainty in the maps at 150 and 220\,GHz, respectively \citep{Mocanu13}; and ACT, $6\%$ uncertainty in the flux density scale at 148\,GHz \citep{Louis14}.

\paragraph{The 857\,GHz channel}
\begin{figure}
\begin{center}
\includegraphics[width=0.55\textwidth]{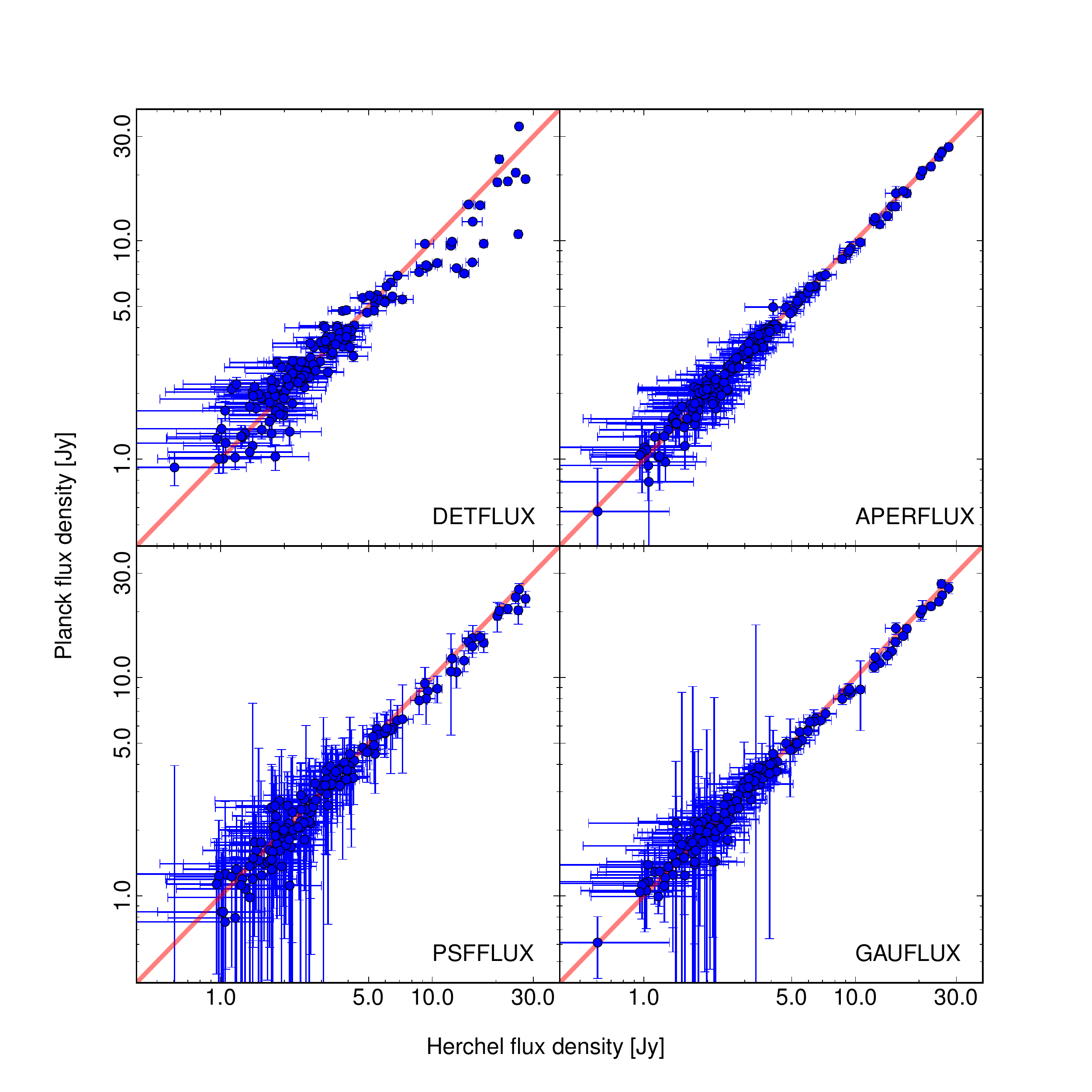}
\caption{Comparison of flux densities from \Planck\ and \Herschel\ at 350\,$\mu$m (857\,GHz) using the \Herschel\ Reference Survey. The one-to-one lines are shown in red.}
\label{fig:HFI_Herchel_comp}
\end{center}
\end{figure}
At this frequency, there are no other all-sky surveys available matching in frequency against which the \plccs\ fluxes can be compared. However, the 350\,$\mu$m channel of the SPIRE instrument \citep{griffin10} on \Herschel\ \citep{pilbratt10} is a close match to the 857\,GHz passband of \Planck. Pointed observations of compact extragalactic targets obtained with SPIRE can thus be used to validate the flux density measurements made in the \plccs.
The most useful set of observations from SPIRE for our purposes are those from the \Herschel\ Reference Survey (HRS) \citep{boselli10}, a survey of the far-infared and submillimetre properties of local bright galaxies. The published SPIRE photometry for this sample \citep{ciesla12} uses either PSF fitting for sources unresolved by \Herschel\ or apertures matched to the observed sizes of the sources in the \Herschel\ maps. In both these cases the apertures used will be much smaller than the \Planck\ beam. Since several of the HRS galaxies have nearby bright companions, and since other sources might be included in the large \Planck\ beam, new flux density values were extracted for the HRS sources using apertures matched to the size of the \Planck\ beams (Eales et al., private communication).  This allows a direct comparison of the SPIRE 350\,$\mu$m flux densities to the \Planck\ 857\,GHz fluxes for the same 141 objects.
It should be noted that only three of these objects have not been flagged as EXTENDED by \Planck. We should, therefore, not be surprised if the flux extraction methods (that assume a single point-like source) will be biased low.
In Fig.~\ref{fig:HFI_Herchel_comp} and Table~\ref{tab:planck_v_hrs} we show the results of this comparison for the four different flux extraction methods used in the \plccs; namely DETFLUX, APERFLUX, PSFFLUX, and GAUFLUX. 
The best performing \Planck\ flux-density extraction method is, perhaps not surprisingly, the method that most closely resembles the flux-density extraction method applied to the \Herschel\ maps, namely APERFLUX, which shows good agreement between \Herschel\ and \Planck\ fluxes over the full range of source brightness.
The worst performing method, in contrast, is DETFLUX, which shows an increased scatter, and an overall bias to lower \Planck\ values for the brighter sources. This bias is expected if the sources are not truly point-like. For a population of slightly extended sources we expect a noticeable downward bias for the brighter sources, which disappears into the noise for fainter objects.  This pattern of bias for the brighter sources which disappears once these sources are excluded is also seen for PSFFLUX.
The increased scatter seen for the brighter sources with DETFLUX probably arises from errors in the recovered position. This is due to the relationship between the scale of the wavelet used and the pixel size in the maps. For this channel (857\,GHz) the small beam size and the increased level of foregrounds in the map mean that the optimum scale for the wavelet is very narrow with respect to the map pixels. This makes the flux-density estimate extremely sensitive to errors in the recovered position of the source. This effect is most pronounced at this channel and for the brighter sources.
A similar bias seen in GAUFLUX is a little more difficult to explain, but again may result from the assumption that sources are single and point-like.  A double source, for instance, may pose difficulties in the flux-density estimation resulting in a bias low.
We conclude that there is a good match between the \plccs\ flux densities at 857\,GHz and \Herschel\ flux densities in the matching 350\,$\mu$m SPIRE band. We also conclude that, for most purposes concerned with the flux-density measurement of compact sources like the galaxies discussed here, APERFLUX is the most appropriate flux-density measure to use for the higher frequency channels.
 \begin{table}
\begingroup
\newdimen\tblskip \tblskip=5pt
\caption{Comparison of flux-density measurements from \Planck\ and \Herschel\ at 857\,GHz (350\,$\mu$m). The \Planck\ measurements for all flux-extraction methods are lower than the \Herschel\ results. If, however, the 17 sources in the HRS sample with flux densities greater than 10\,Jy are excluded from the analysis, then the flux-density ratios for all flux-extraction methods are within 1\,$\sigma$ of unity.} 
\label{tab:planck_v_hrs}
\nointerlineskip
\vskip -3mm
\footnotesize
\setbox\tablebox=\vbox{
   \newdimen\digitwidth
   \setbox0=\hbox{\rm 0}
   \digitwidth=\wd0
   \catcode`*=\active
   \def*{\kern\digitwidth}
\def\leaderfill{\leaders \hbox to 5pt{\hss.\hss}\hfil}
   \newdimen\signwidth
   \setbox0=\hbox{+}
   \signwidth=\wd0
   \catcode`!=\active
   \def!{\kern\signwidth}
\halign{\hbox to 1.0in{#\leaderfill}\tabskip=1em &\hfil#\hfil&\hfil#\hfil\tabskip=0pt\cr
\noalign{\doubleline}
\omit &\multispan{2}{\hfil Flux density ratio (\Herschel/\Planck)\hfil}\cr
\noalign{\vskip-6pt}
\omit \hfil Extraction\hfil&\multispan{2}{\hrulefill}\cr
\omit \hfil method\hfil&All 141 HRS sources&Discarding $S_{\rm HRS} >$10\,Jy\cr
\noalign{\vskip 3pt\hrule\vskip 5pt}
DETFLUX&$1.045\pm0.010$&$0.982\pm0.019$\cr
APERFLUX&$1.019\pm0.011$&$1.009\pm0.021$\cr
PSFFLUX&$1.065\pm0.022$&$1.016\pm0.038$\cr
GAUFLUX&$1.056\pm0.013$&$1.024\pm0.024$\cr
\noalign{\vskip 3pt\hrule\vskip 3pt}}}
\endPlancktable
\endgroup
\end{table}
 
\begin{figure}
\begin{center}
\includegraphics[width=0.52\textwidth]{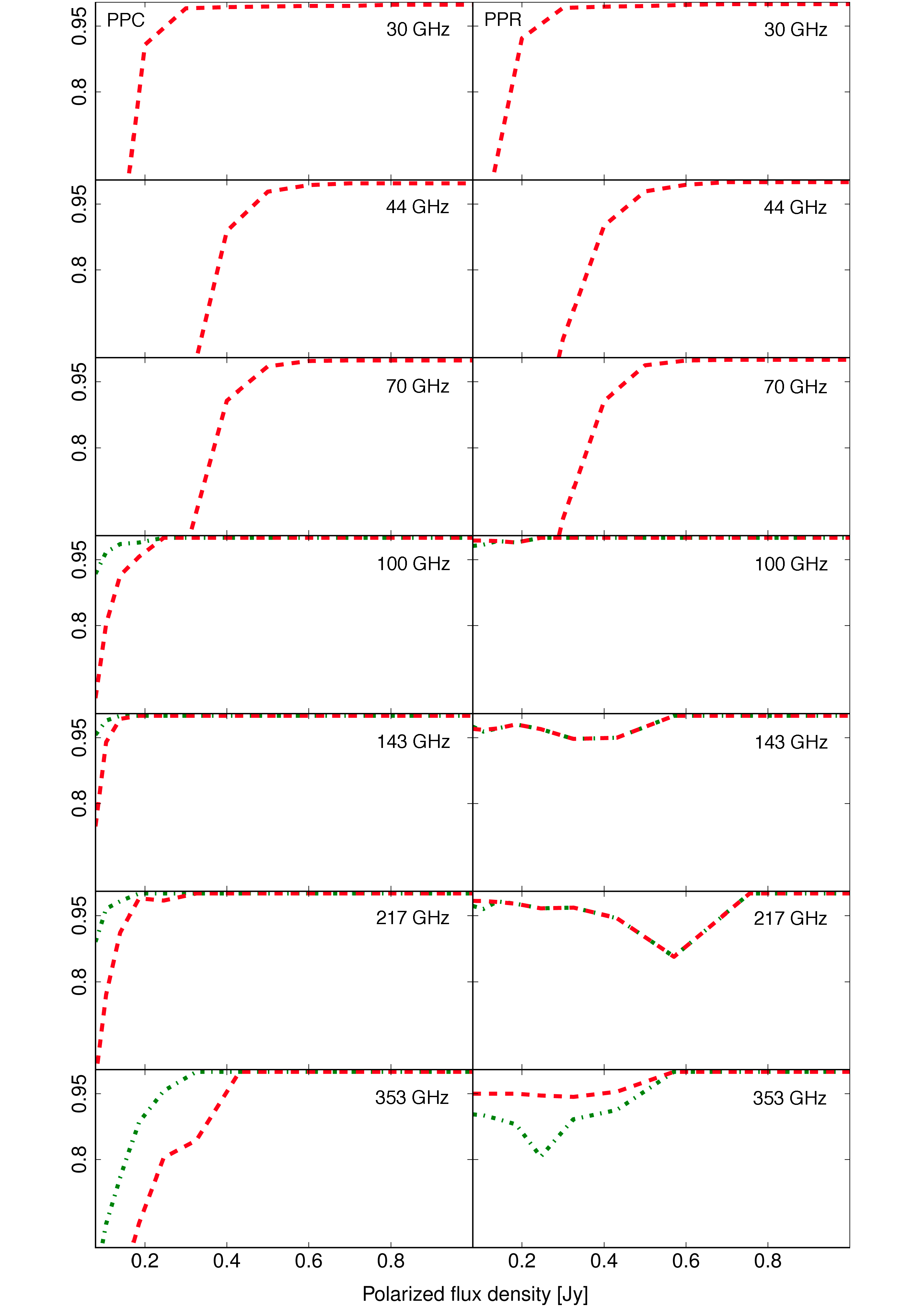}
\caption{
\textit{Left}: polarized photometric ompleteness (PPC). \textit{Right}: polarized photometric reliability (PPR).
Red dashed lines: derived with the common method applied to both LFI and HFI. 
Green dot-dashed lines: including the marginal polarization data (HFI only). From top to bottom, 30--353\,GHz.
The plots were constructed using the \plccs catalogue only. 
\label{fig:ppc_and_ppr}}
\end{center}
\end{figure}
\begin{figure}
\begin{center}
\includegraphics[width=0.49\textwidth]{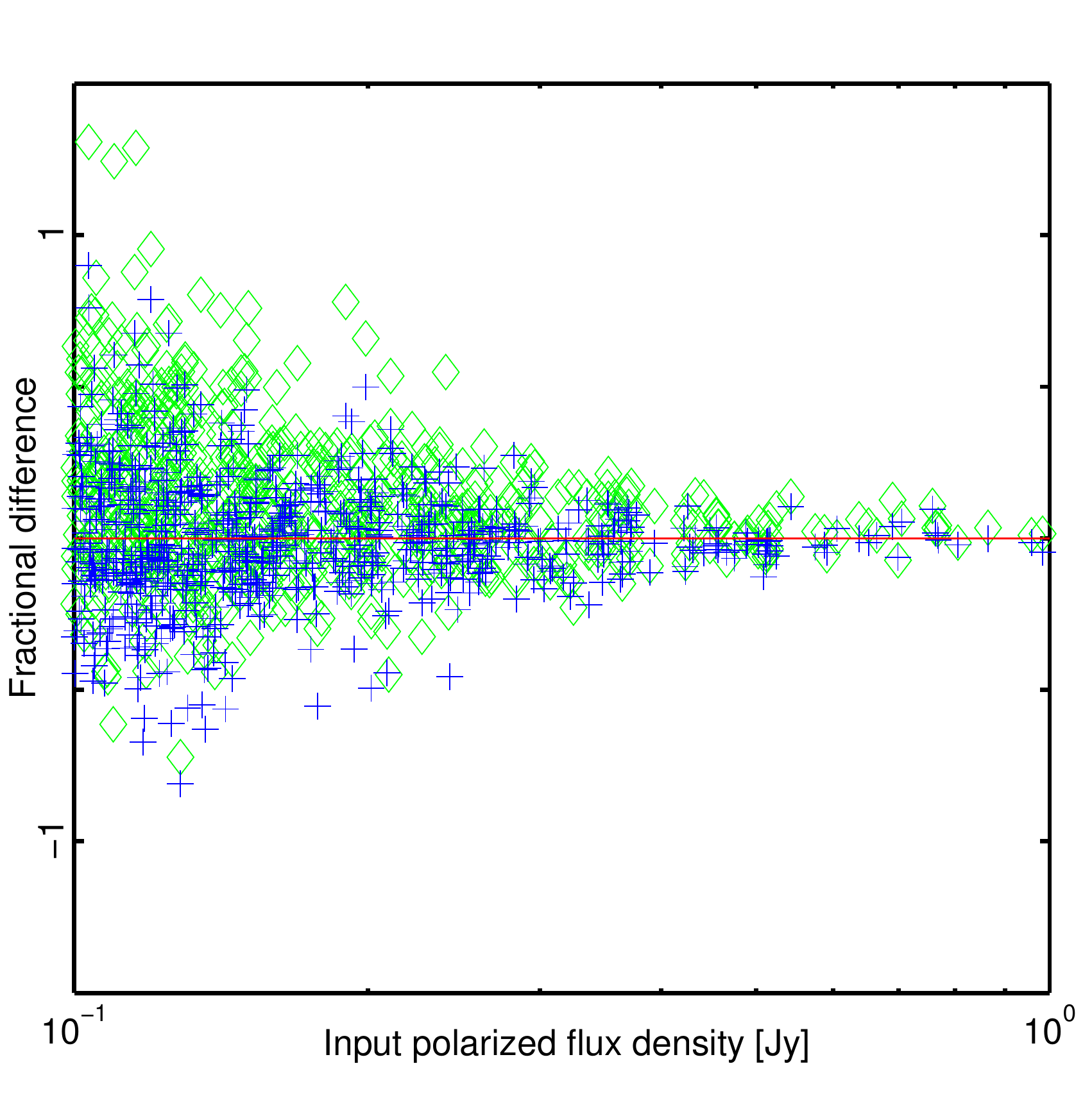}
\caption{Comparison between the LFI and HFI implementations of the maximum-likelihood estimator code to measure polarized flux densities. 
 The comparison was performed using the 100\,GHz FFP8 maps, and the locations of the point sources included in the simulation, because the polarization pipelines perform non-blind extractions. The recovered fractional difference of the polarized flux densities of the two methods, defined as (recovered $-$ true)/true, is plotted against the true polarized flux density, as simulated in the FFP8 maps. The results from the LFI pipeline, {\tt IFCAPOL}, are shown by the green diamonds, and the results from the HFI pipeline, {\tt PwSPOL}, are shown by the blue crosses.}
\label{fig:LFI_HFI_P_comparison}
\end{center}
\end{figure}
\begin{figure}
\begin{center}
\includegraphics[width=0.52\textwidth]{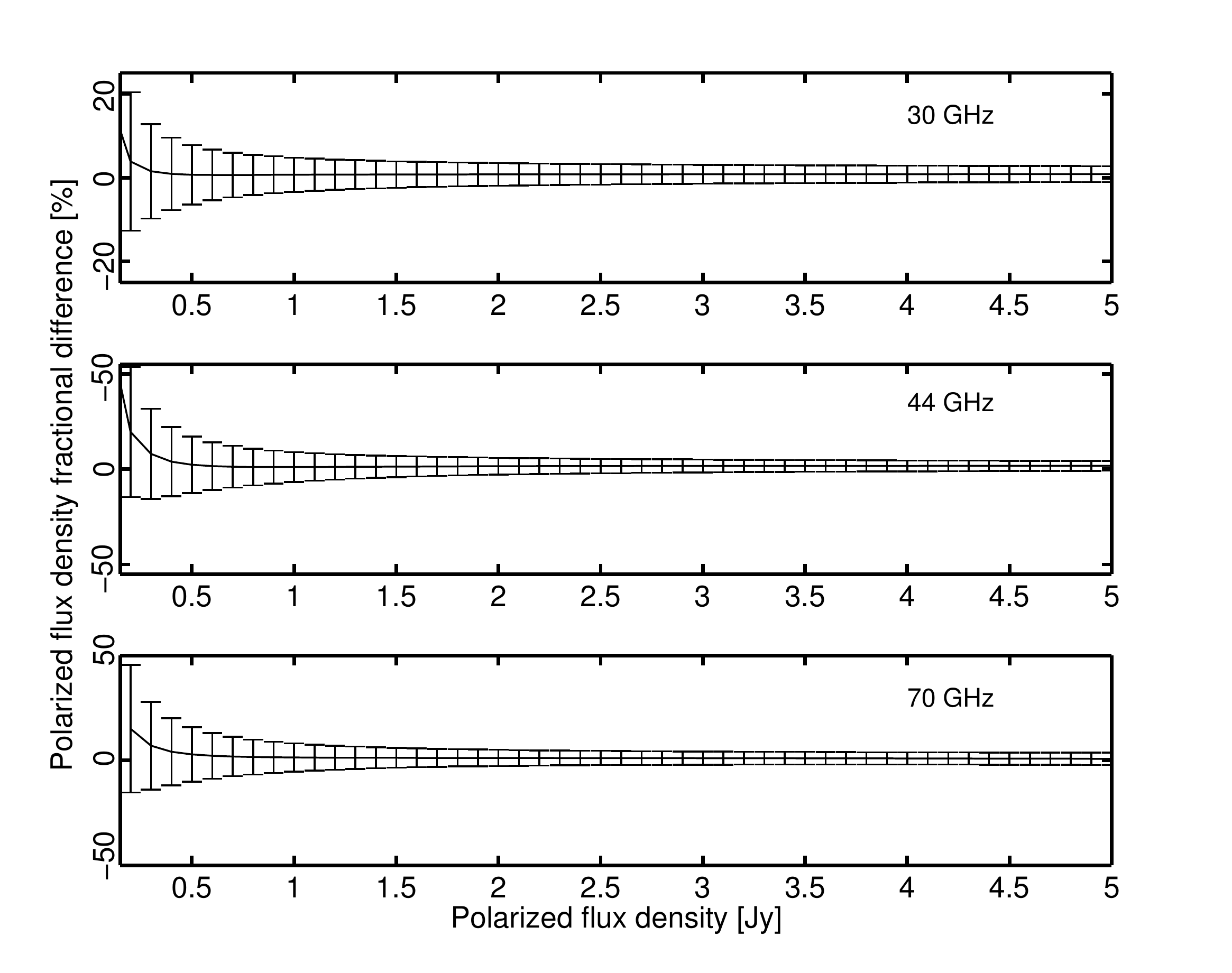}
\caption{Internal validation using Monte Carlo simulations to assess the recovery of the polarized flux density for the 30, 44, and 70\,GHz channels. In these simulations, point sources were injected into the $Q$ and $U$ maps with 50 different polarized flux density values, starting at 0.2\,Jy and increasing with a step size of 0.1\,Jy.  The fractional difference, defined as (recovered $-$ true)/true, is plotted against the true polarized flux density.  The recovered polarized flux densities are unbiased, except for the faintest sources, were the effect of Eddington-type bias is seen.
\label{fig:LFI_P_MC}}
\end{center}
\end{figure}
\begin{figure}
\begin{center}
\includegraphics[width=0.52\textwidth]{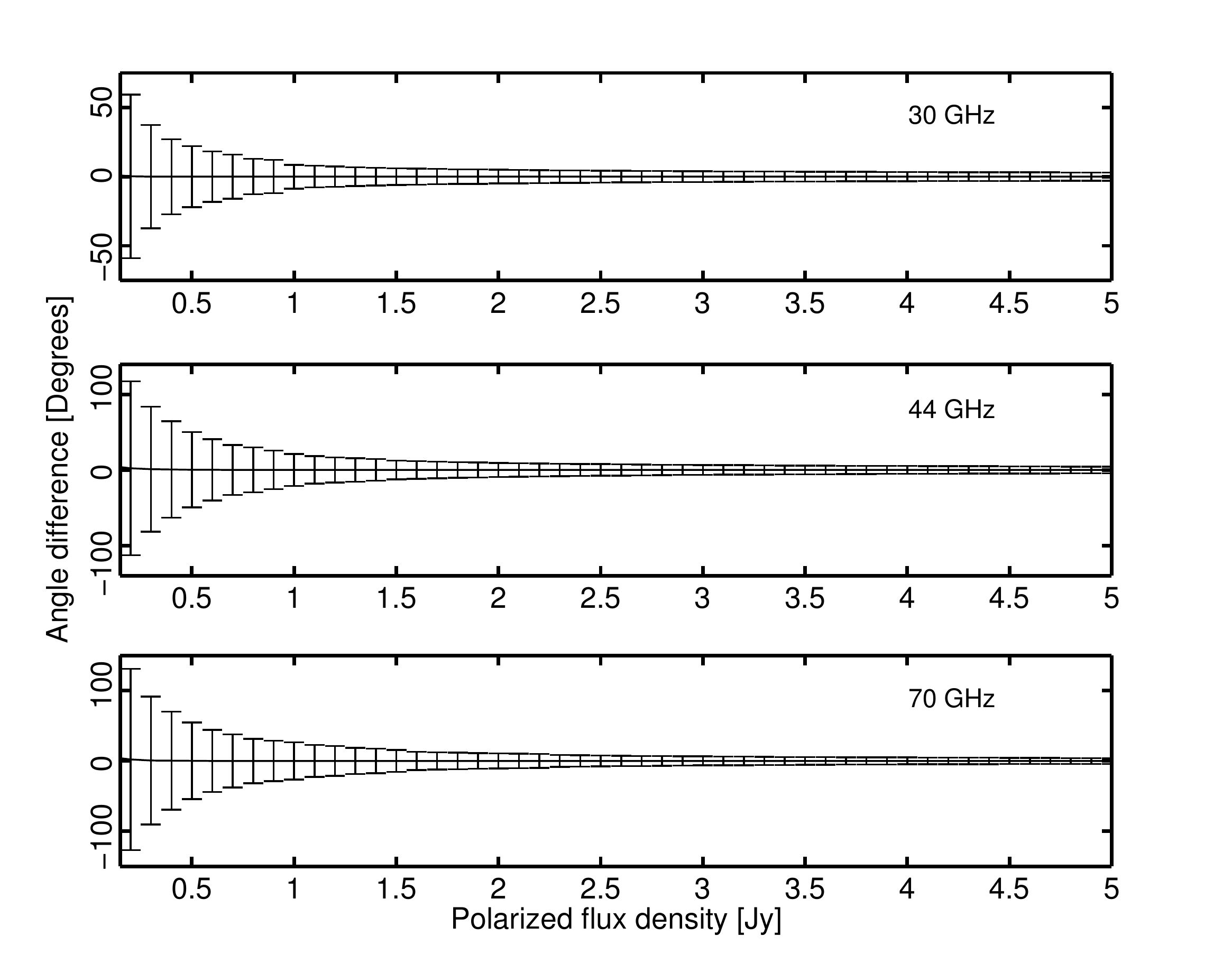}
\caption{Internal validation using Monte Carlo simulations to assess the recovery of the polarization angle of the  30, 44, and 70\,GHz channels. In these simulations, point sources spanning the full range of polarization angles were injected into the $Q$ and $U$ maps with 50 different polarized flux density values, starting at 0.2\,Jy and increasing with a step size of 0.1\,Jy.
The difference between the recovered and true angles is plotted against the true polarized flux density. Here we see there are no biases in the recovery of the polarization angle, although obviously the uncertainty increases for fainter sources.
\label{fig:LFI_Pang_MC}}
\end{center}
\end{figure}
\begin{figure}
\begin{center}
\includegraphics[width=0.49\textwidth]{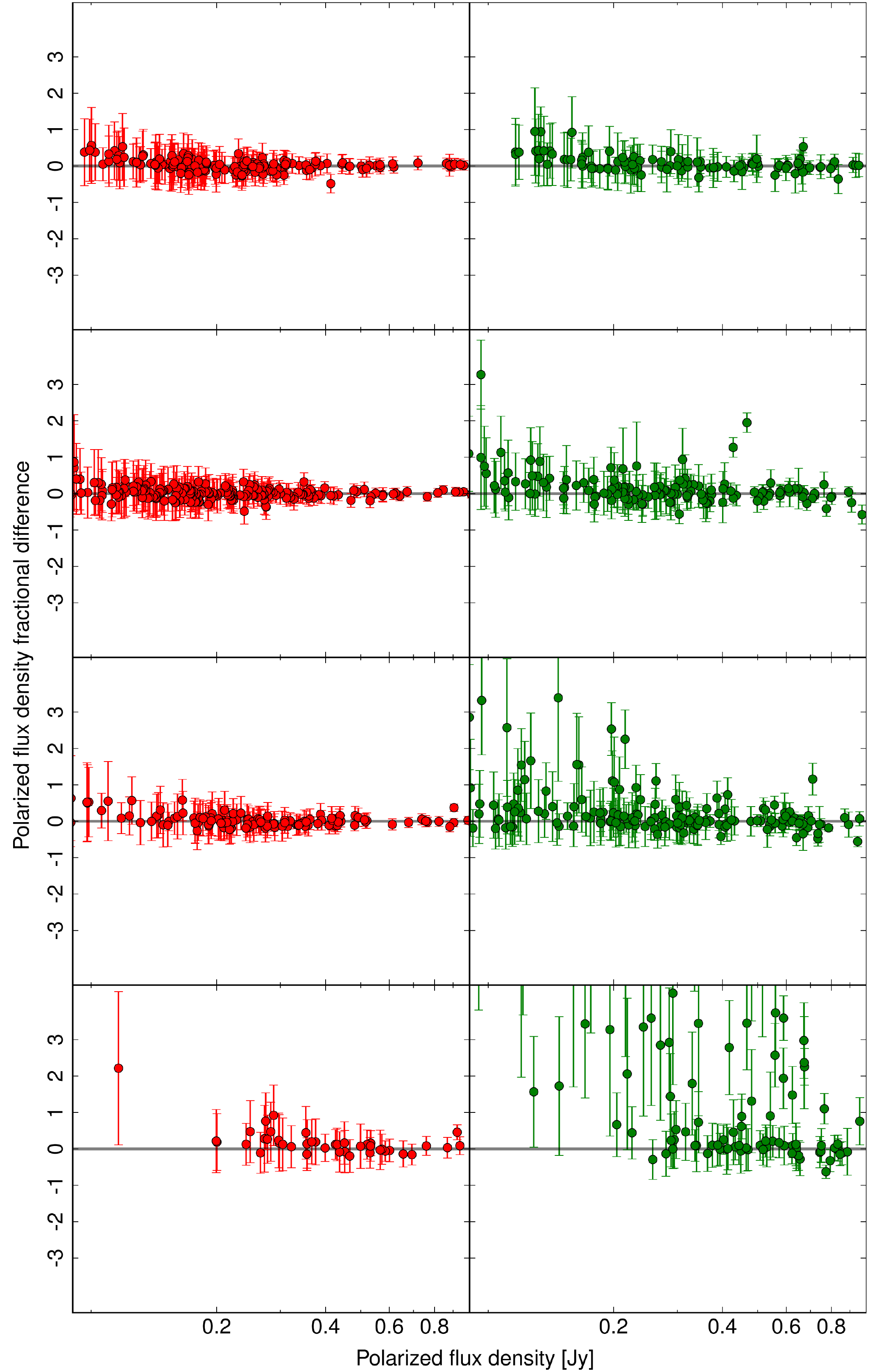}
\caption{Fractional difference between the true and recovered polarized flux densities, from top to bottom, for the 100--353\,GHz FFP8 simulations. 
\textit{Left} (red):  \plccs.
\textit{Right} (green): \ecat.  Here the fractional difference is defined as recovered minus true divided by the true value.
These are the significantly polarized sources, as found by the common method, and the uncertainties associated with the best-fit estimates are $\pm 3\,\sigma$ error bars.
Eddington-type bias is seen at lower polarized flux densities.
\label{fig:HFI_FFP8_FLUX}}
\end{center}
\end{figure}
%
\begin{figure}
\begin{center}
\includegraphics[width=0.49\textwidth]{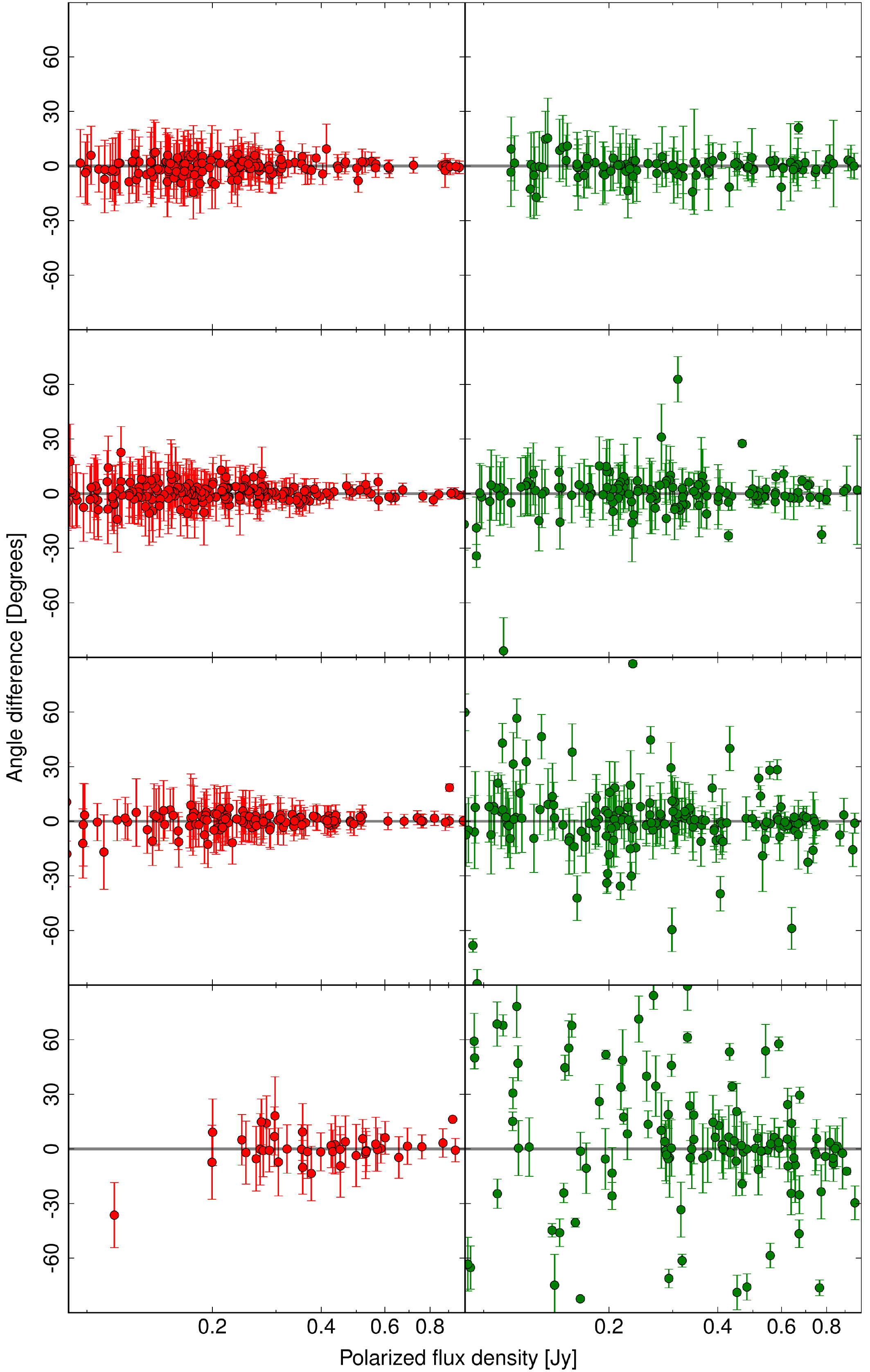}
\caption{Difference between the recovered and true polarization angles, from top to bottom, for the 100--353\,GHz FFP8 simulations.
\textit{Left} (red):  \plccs.
\textit{Right} (green): \ecat.  
These are the significantly polarized sources, as found by the common method, and the uncertainties associated with the best-fit estimates are $\pm 3\,\sigma$ error bars.
\label{fig:HFI_FFP8_ANGLE}}
\end{center}
\end{figure}
%

\subsection{Polarization measures}
\label{sec:qa_polarization}

To validate the polarization measurements in the \plccs\ and \ecat\ catalogues, we rely on simulations based on the injection of sources with known properties into the \Planck\ polarization maps between 30 and\,70\,GHz and the FFP8 simulated maps \citep{planck2014-a14} between 100 and 353\,GHz.  
It should be noted that the injection of sources into the real maps has the advantage that many thousands of sources can be used to test the analysis; however this procedure does not simulate the leakage due to the bandpass mismatch. The FFP8 maps do contain this effect, although there are only hundreds rather than thousands of polarized sources.
Both approaches, however, allow us to test the fidelity of the polarized flux densities and polarization angles produced by our analysis pipelines.  They also allow us to compute measures for the completeness and reliability of recovered polarization measurements.
Since the extraction of these measurements is non-blind, based on the positions provided by the analysis of the temperature maps, these terms for the polarization pipelines only have any meaning given that the source is real.

We define the ``polarization photometric completeness'' (PPC) as the percentage of polarized sources correctly identified as polarized above a given true polarized flux density, and ``polarization photometric reliability'' (PPR) as the  percentage of polarized sources whose polarized flux density is contained in the interval defined by the best-fit value and $\pm$3\,$\sigma$ errors.
Figure~\ref{fig:ppc_and_ppr} shows the PPC and PPR as a function of the true  and estimated polarized flux density, respectively, for the \plccs\ for all the polarized channels. The red dashed lines show the results for the subsets of significantly polarized sources, whose measurements are provided for both LFI and HFI,
while the green dot-dashed lines show the results when marginally polarized sources are also included in the analysis. Here we see that the inclusion of the marginal data increases the completeness with only a minimal decrease in the 353\,GHz channel reliability. The negative ``kinks'' in the reliability curves of the 217 and 353\,GHz channels are caused by the same single source. This source was detected in the intensity maps with \snr\ = 151.5 (217\,GHz) and
the recovered flux density was underestimated by around $23$\,$\sigma$, where $\sigma$ is the estimated error on the flux density. The dramatic underestimation of the flux density was caused by the recovered position of the source being offset from the true position by 1.27\,arcmin. In this \snr\ regime such an offset is sufficient to explain the failure of the PPR criterion, as the polarized flux density will be underestimated by more than 3\,$\sigma$.

A comparison between the LFI and HFI  implementations of the common procedure for extracting the polarization measurements for the significantly polarized sources was performed. Figure~\ref{fig:LFI_HFI_P_comparison} shows a comparison of the recovered polarized flux densities (for each implementation), and the true polarized flux density. 
We have assessed the performance of each method, and find that above 250\,mJy, where we are complete, the average recovered polarized flux density is within 1\,\% of the true value for {\tt IFCAPOL} and within 0.8\,\% for {\tt PwSPOL}, and that they are within 0.2\,\% of each other.
\subsubsection{Internal consistency}
\paragraph{30--70\,GHz}
In order to assess the performance of the maximum likelihood estimator used to provide the polarization measurements in the three lowest \Planck\ channels ({\tt IFCAPOL}), we used Monte Carlo simulations. Point sources were simulated and convolved with the appropriate Gaussian effective beam from Table~\ref{tab:beam_data} for each channel, and  were injected into \healpix\ $N_{\rm side}=4096$ maps. These maps were then downgraded to $N_{\rm side}=1024$ to match the pixelization of the LFI maps, and the source maps were added to the \textit{Planck} 2015 (PR2) $Q$ and $U$ maps.  The sources were injected away from known bright radio sources. In all, 37\,000 sources, in 50 flux-density bins, were injected into the $Q$ and $U$ maps. The pipelines were run given the positions of these simulated sources, producing polarized flux density and polarization angle measurements.
The results of the Monte Carlo simulations on the recovery of the polarized flux densities are shown in Fig.~\ref{fig:LFI_P_MC}. We can see that the polarized flux densities are recovered in an unbiased way for strongly polarized sources, and that the faintest ones suffer from Eddington-type bias. Figure~\ref{fig:LFI_Pang_MC} shows the equivalent plot for the polarization angle and here we can see that the angle is recovered successfully for all channels, where the uncertainties increase towards fainter polarized flux densities.
Based on these simulations, we computed the polarization photometric completeness and reliability, shown in Fig.~\ref{fig:ppc_and_ppr}.  In the simulations we injected sources at all Galactic latitudes, and we did not apply any Galactic cut. We find that our catalogues are complete at the 90\,\% level at 200\,mJy (polarized flux density) at 30\,GHz, and at 400\,mJy at 44 and 70\,GHz. At the 600--700\,mJy level the three catalogues are complete. A summary of these results can be found in Table~\ref{tab:pccs_pol} (Sect.~\ref{sec:characteristics}).
\paragraph{100--353\,GHz}
The recovery of the polarized flux densities from the FFP8 simulations, using the method common to both LFI and HFI, is shown in Fig.~\ref{fig:HFI_FFP8_FLUX}  for the \plccs\ and the \ecat. 
The \plccs\ polarized flux density estimates are reliable across the full range of flux densities and considerably more reliable than the \ecat. Given that the \ecat\ catalogue covers the Galactic plane region this is hardly unexpected.
The polarized flux densities are unbiased over the range of values where the survey is complete, but for the fainter sources a positive bias is present, as expected from Eddington-type bias.
Figure~\ref{fig:HFI_FFP8_ANGLE} shows the recovery of the polarization angle,  which is unbiased over the full polarized flux-density range, although the errors in its recovery increase as the polarized flux density of the source decreases.
This behaviour is exactly what is expected if our assumptions made in Sect.~\ref{subsec:polHFI} hold; namely that $\sigma_Q \approx \sigma_U$, and that they are uncorrelated.
Consider sources with a given polarization angle in the range of polarized flux densities which suffer from Eddington bias: the required upward noise fluctuations could come from $Q$ or $U$ or both. Once the polarization angle is evaluated, if our assumptions hold, then the average value found for the angle will be unbiased but its dispersion will be much larger than for the brighter sources.
In Fig.~\ref{fig:HFI_FFP8_ANGLE} we also see that the measurements in the \ecat\ are again less reliable than the \plccs.

\subsubsection{External consistency}
The limited number of polarimetric millimetre surveys, the small number of bright \Planck\ sources with a high significance in polarization, and the fact that the majority of polarized sources are variable, makes it difficult to validate the \Planck\ polarized flux densities with external datasets.

One of the objects from the catalogue that we have studied in detail is Tau\,A, also known as M1 or the Crab Nebula.  This object is resolved in the higher frequency \Planck\ channels and may not be the best source for validation at these frequencies; nevertheless it is the brightest compact source in polarization in \Planck\ and has been thoroughly studied in other experiments. In Table~\ref{tab:planck_wmap_pol} we compare the total intensity and polarized flux densities, polarization fraction, and polarization position angle for Tau\,A with measurements from WMAP \citep{Weiland11}  and the IRAM 30m telescope at 89\,GHz \citep{Aumont10}. In general, it is assumed that the polarization position angle of Tau\,A is constant across the frequency range of interest, up to at least 353\,GHz. In Table \ref{tab:planck_wmap_pol} one can see that the \Planck\ polarization position angles (as measured by the maximum likelihood filtering method and by aperture photometry) are significantly different from those of WMAP and IRAM at some frequencies.  We have investigated these discrepancies and found that there are multiple factors in LFI and HFI affecting our measurements of the position angles of Tau\,A that deserve further attention.

First,  at the position of Tau\,A in the Stokes $U$ maps one can see a small spurious signal that is affecting the angle measurements. For this object, most of the polarized signal is in the Stokes $Q$ map, and $U$ makes very little contribution to the total polarized flux density (quadrature sum of $Q$ and $U$). However, as shown in Table~\ref{tab:planck_wmap_pol}, when we calculate the polarization position angle with either method, the contribution from the spurious signal has the effect of changing the position angles by up to $5\degr$ from the $-88\fdg2$ measured at IRAM \citep{Aumont10}. This spurious signal has been introduced by the complex cross-terms in the polarized beams, which are normally expected to be very small and below the map noise  level, but show up here because Tau A is so bright. In an attempt  to remove this effect, we have produced maps where we have deconvolved the beam and fitted for spurious signals. Results for LFI frequencies, from the {\tt ArtDeco} map-making pipeline at the LFI DPC, are shown in column~7 of Table~\ref{tab:planck_wmap_pol}. These new measures of position angle agree with the expected value from IRAM and WMAP. However, we caution that these new maps are under development and the measurements should be used with caution. There is an ongoing effort, not yet completed, to generate a similar set of maps for the HFI frequencies, in order to understand whether similar spurious signals affect the HFI polarization angle measurements. The tests that we have carried out with the new LFI maps indicate that the polarized flux densities and angles of the other sources in the catalogue are mostly unchanged, but some sources with high polarized flux density may be marginally affected.

Second, the angular size of Tau\,A measured with IRAM at 89\,GHz  \citep{Aumont10} shows that it will be slightly resolved by \Planck\ in the HFI channels. This could have an impact on the flux densities and polarization position angles measured with both the filtering method and with aperture photometry. In particular, the filtering method assumes that the sources are unresolved, so the flux densities derived with this method should be regarded as lower limits if the source is extended. In the case of aperture photometry, the integration radius that we use in HFI is too small for a source as large as Tau\,A. We have therefore increased the aperture radii to $2\times({\rm FWHM}^2 + \theta_{\rm Tau\,A}^2)^{1/2}$, where $\theta_{\rm Tau\,A}=4\arcmin$ as measured by IRAM, and recomputed the angles for the four HFI channels with polarization capabilities. This improves the agreement between the HFI position-angle measurements and those from WMAP and IRAM,  except at 353\,GHz. The results are shown in column~7 of Table~\ref{tab:planck_wmap_pol}. 

\begin{figure}
\begin{center}
\includegraphics[width=0.49\textwidth]{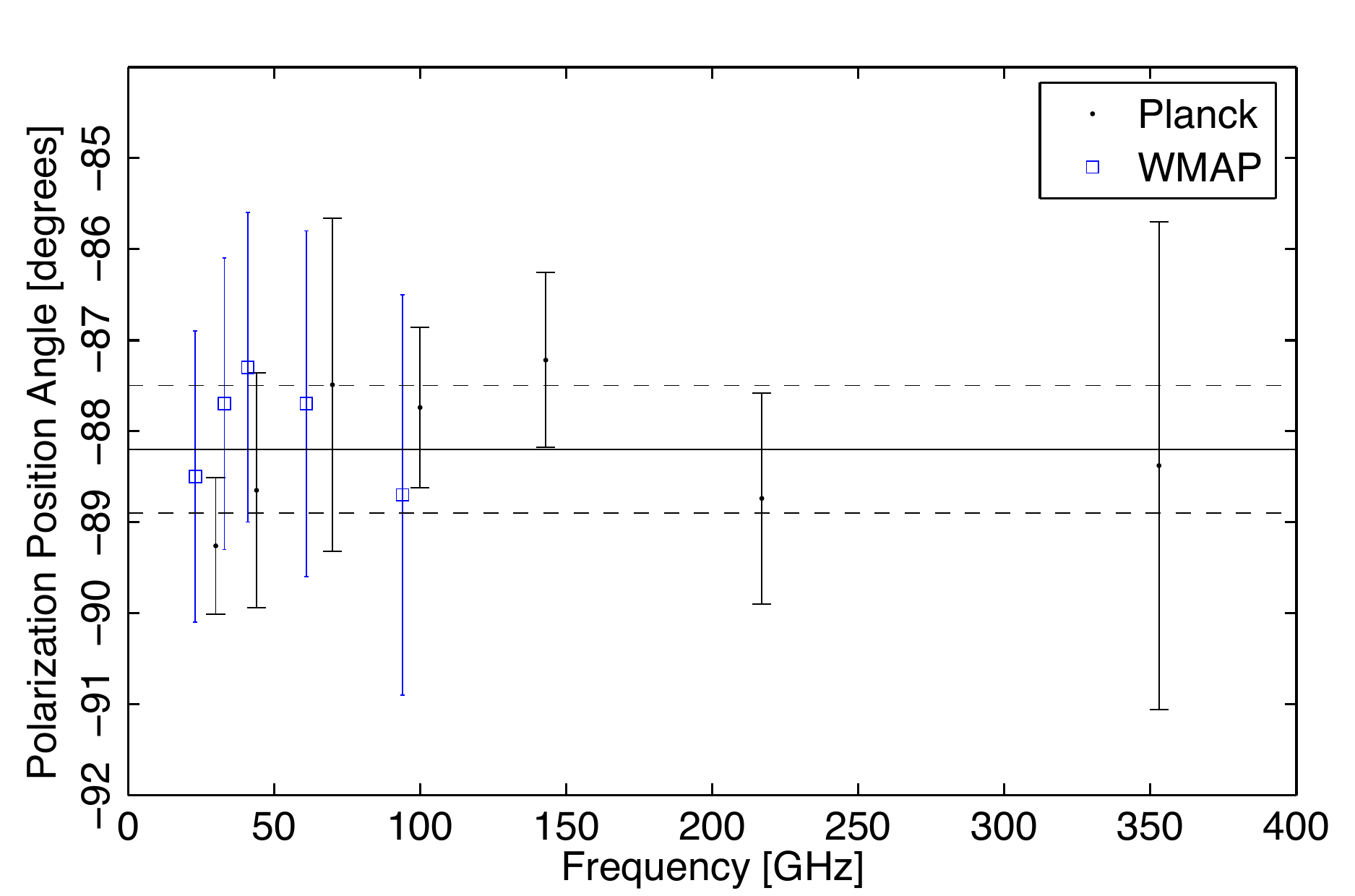}
\caption{Polarization position angles for Tau\,A from WMAP (blue squares, \citealt{Weiland11}) and \Planck\ (black dots). 
The IRAM measurement from \citet{Aumont10} is also shown as a solid line, where a $\pm1\sigma$ uncertainty is shown with dashed lines.}
\label{fig:TauA_ANGLE}
\end{center}
\end{figure}

Third, as shown in \cite{planck2014-a07} and \cite{planck2014-a09}, different methods have been used in LFI and HFI for correcting the  frequency maps for a global leakage signal from total intensity into polarization due to the bandpass mismatch. In practice, this means that the LFI frequency maps have not been corrected, whereas the HFI ones have. The global bandpass corrections applied to the maps are not the same as  the local corrections that we apply in this paper for point sources, which depend on the spectrum of each source (see Appendix \ref{sec:mismatch}). However, even though our techniques to extract the flux densities of the sources in the $Q$ and $U$ maps will remove most (if not all) of the global correction if the sources are point-like, this may not be true for extended sources. To test this, we have extracted the polarization position angles from HFI maps where the global bandpass correction had not been applied. The results are shown in Table~\ref{tab:planck_wmap_pol} and in Fig.~\ref{fig:TauA_ANGLE}, where we compare the polarization position angles from WMAP and IRAM  with the new measurements from \Planck. The recovered angles between 100 and 217\,GHz are very similar to the case where the global bandpass mismatch had been applied, except at 353\,GHz, where the new angle is in much better agreement with the other \Planck, WMAP, and IRAM measurements. 

These analyses show the complexity of the \Planck\ maps in polarization, particularly for bright extended objects like Tau\,A, where we had to fit and remove the spurious signal in the LFI, increase the integration aperture radii in the HFI, and remove the global bandpass correction at 353\,GHz, in order to achieve a consistency within $2\degr$ in polarization position angle between \Planck\ and WMAP or IRAM. Tau\,A is the only very bright polarized source in our maps, and the limited amount of polarization information available at \Planck\ frequencies for other sources limits our ability to conclude that the polarization angles of the rest of the sources in the catalogue are not affected by these issues. Therefore, as in the case of Tau\,A, the polarization position angles should be used with caution. In future releases of  the \Planck\ products we will revisit this issue.

%
%

\begin{table*}
\begingroup
\newdimen\tblskip \tblskip=5pt
\caption{Total intensity and polarized flux densities, polarization fraction, and polarization position angles for Tau\,A measured from the \Planck\ full-mission maps at 30 to 353\,GHz,  WMAP at 23 to 94\,GHz \citep{Weiland11}, and with the IRAM 30\,m telescope at 89\,GHz \citep{Aumont10}.}
\label{tab:planck_wmap_pol}
\nointerlineskip
\vskip -3mm
\footnotesize
\setbox\tablebox=\vbox{
   \newdimen\digitwidth
   \setbox0=\hbox{\rm 0}
   \digitwidth=\wd0
   \catcode`*=\active
   \def*{\kern\digitwidth}
   \newdimen\signwidth
   \setbox0=\hbox{+}
   \signwidth=\wd0
   \catcode`!=\active
   \def!{\kern\signwidth}
\halign{#\hfil\tabskip=0.5em &\hfil#\hfil&\hfil#\hfil&\hfil#\hfil&\hfil#\hfil&\hfil#\hfil&\hfil#\hfil&\hfil#\hfil\tabskip=0pt \cr
\noalign{\doubleline}
\omit\hfil Freq\hfil&$I$&$P$&$P/I$&\omit\hfil Pol. Angle$^{a}$ \hfil &\omit\hfil Pol. Angle$^{b}$\hfil &\omit\hfil Pol. Angle$^{c,d}$\hfil &\omit\hfil Pol. Angle$^{e}$\hfil \cr
\omit\hfil[GHz]\hfil&[Jy] &[Jy] &[\%]&[deg]&[deg] &[deg]\cr
\noalign{\vskip 3pt\hrule\vskip 5pt}
*30&$344.23\pm  0.27$&$24.44  \pm  1.05$&$*7.10 \pm 0.33$ &$-84.54\pm  0.54\pm 0.50$& $-83.71\pm 1.40\pm 0.50$&$-89.26\pm 0.25\pm 0.50$&\ldots\cr
*44&$292.68\pm  0.23$&$19.07  \pm  1.10$&$*6.51 \pm 0.51$ &$-88.34\pm 0.32\pm 0.50$&$-86.93 \pm 0.47\pm 0.50$&$-88.65\pm 0.79\pm 0.50$&\ldots\cr
*70&$259.99\pm  0.11$&$20.55   \pm 0.61$&$*7.90 \pm 0.32$ &$-84.24\pm 0.23\pm 0.50$&$-85.03 \pm  1.32\pm 0.50$&$-87.49\pm 1.33\pm 0.50$&\ldots\cr
100&$215.16\pm 0.06$&$15.54  \pm  0.14$&$*7.22 \pm 0.06$ &$-88.53\pm  0.11\pm 0.62$&$-87.52 \pm 0.13\pm 0.62$&$-87.59\pm 0.26\pm 0.62$& $-87.74\pm 0.26\pm 0.62$\cr
143&$167.10\pm 0.04$&$12.02  \pm  0.08$&$*7.19 \pm 0.05$ &$-84.85\pm  0.13\pm 0.62$&$-85.72 \pm 0.15\pm 0.62$&$-87.03\pm  0.35\pm 0.62$& $-87.22\pm  0.34\pm 0.62$\cr
217&$124.21\pm  0.04$&$10.09  \pm  0.08$&$*8.12 \pm 0.06$ &$-87.33\pm  0.12\pm 0.62$&$-88.73 \pm 0.18\pm 0.62$&$-88.84\pm 0.55\pm 0.62$& $-88.74\pm 0.55\pm 0.62$\cr
353&$*82.17\pm 0.67$&$*9.88  \pm   0.17$&$12.02 \pm  0.23$ &$-86.11\pm 0.37\pm 0.62$&$-85.15 \pm 0.37\pm 0.62$&$-85.16\pm 1.93\pm 0.62$& $-88.38\pm 2.06\pm 0.62$\cr
\noalign{\vskip 3pt\hrule\vskip 5pt}
*23&$383.80 \pm 9.60$&$27.17 \pm 0.68$&$*7.08 \pm 0.25$ &$-88.50 \pm 0.10\pm 1.50$&\ldots&\ldots&\ldots\cr
*33&$342.80 \pm 6.40$&$23.80 \pm 0.44$&$*6.94 \pm 0.18$ &$-87.70 \pm 0.10\pm 1.50$&\ldots&\ldots&\ldots\cr
*41&$317.70 \pm 8.60$&$22.12 \pm 0.60$&$*6.97 \pm 0.27$ &$-87.30 \pm 0.20\pm 1.50$&\ldots&\ldots&\ldots\cr
*61&$276.00 \pm 5.20$&$19.31 \pm 0.36$&$*7.00 \pm 0.19$ &$-87.70 \pm 0.40\pm 1.50$&\ldots&\ldots&\ldots\cr
*94&$232.80 \pm 9.70$&$16.60 \pm 0.73$&$*7.13 \pm 0.43$ &$-88.70 \pm 0.70\pm 1.50$&\ldots&\ldots&\ldots\cr
\noalign{\vskip 3pt\hrule\vskip 5pt}
*89&$195.00\pm11.0$&$14.50 \pm 3.20$&$*8.80 \pm 0.02$ &$-88.20 \pm 0.20 \pm 0.50$&\ldots&\ldots&\ldots\cr
\noalign{\vskip 3pt\hrule\vskip 3pt}}}
\endPlancktable
 
$^{a}$ Position angle for \Planck\ calculated using the maximum likelihood filtering method. \par
$^{b}$ Position angle for \Planck\ calculated using aperture photometry. \par
$^{c}$ Position angle for \Planck\ LFI channels (30, 44, and 70\,GHz) calculated using the maximum likelihood filtering method on the special beam deconvolved maps. \par
$^{d}$ Position angle for \Planck\ HFI channels (100, 143, 217, and 353\,GHz) calculated using wider aperture photometry to allow for the angular extent of Tau\,A.\par
$^{e}$ Position angle for \Planck\ HFI channels (100, 143, 217, and 353\,GHz) calculated using wider aperture photometry, and  using HFI maps for which the diffuse bandpass correction has not been applied. \par
Note: In the \Planck\ channels, the statistical error bars in the polarization position angle of Tau\,A do not reflect the true uncertainties of these measurements. In addition to the statistical error, a 0\fdg5  systematic error has to be added to the LFI measurements \citep{planck2014-a04} and 0\fdg62 to the HFI measurements \citep{planck2014-a09} as indicated by the second $\pm$ symbol. Similarly, the WMAP errors in the polarized position angle are statistical and a systematic error of 1\fdg5 has been added \citep{Weiland11}.  
\endgroup
\end{table*}
 
%
In addition, we have cross-matched the Plateau de Bure interferometer (PdBI) polarimetric survey of 86 AGN at 100\,GHz  \citep{trippe10} with the \Planck\ 70 and 100\,GHz \plccs\ catalogues, finding two sources in common (PKS\,0851+202 and 3C\,273). For these sources we see good agreement: the \Planck\ polarized flux densities are $509\pm106$\,mJy and $515\pm102$\,mJy at 70\,GHz, and $566\pm38$\,mJy and $503\pm36$\,mJy at 100\,GHz,  as compared with $561\pm156$\,mJy and $418\pm118$\,mJy measured by PdBI, for PKS 0851+202 and 3C273, respectively. 

We have also cross-matched the recent IRAM polarimetric survey from \cite{Agudo14} at 86\,GHz, again with the \Planck\ 70 and 100\,GHz \plccs\ catalogues. Although there are 130 and 133 sources common to both samples, if we restrict the comparison to those with measured \Planck\ polarized flux densities (rather than upper limits), we are left with five and seven sources at 70 and 100\,GHz, respectively. At 70\,GHz, two of the five sources (3C\,273 and 3C\,279) have similar polarized flux densities in both data sets, $559$ and $522$\,mJy in IRAM, as compared with $519\pm93$ and $368\pm85$ in \Planck. At 100\,GHz, three of the seven sources (3C\,273, 3C\,279, and PKS 1055+01) have similar polarized flux densities, $522$, $559$ and $305$\,mJy in IRAM as compared with $566\pm38$, $430\pm37$, and $349\pm31$\,mJy in \Planck, respectively. 

Additionally, we have compared the polarized flux densities found for 3C\,273, the only bright source in polarization in the sample of sources observed simultaneously with the VLA and \Planck\ in the spring of 2013.
The \Planck\ 30\,GHz polarized flux density (colour-corrected for comparison with the VLA) was $854\pm82$\,mJy, while the VLA observed $843\pm50$\,mJy.
At 44\,GHz the \Planck\ polarized flux density was $567\pm131$\,mJy, as compared with $623\pm70$\,mJy seen by the VLA.
\subsection{Summary of validation}
\label{sec:summary_val}
The several internal and external validation tests described in Sect.~\ref{sec:validation} allow us to assess reliability and completeness as a function of flux density in each \Planck\ band.  These tests also allow us to assess the accuracy of positions and flux densities tabulated in the \plccs\ and \ecat\ catalogues.  With the possible exception of a $6\%$ or $1$--$1.5\,\sigma$ difference between \Planck\ flux densities measured in the noisy 44\,GHz map and those measured from the ground, no clearly significant discrepancies in any of these quantities was found.  We thus conclude that both positions and flux densities in the \plccs\ are valid within the tabulated statistical errors in total intensity.  Note, however, that we have no direct test of the flux densities scales at 353 and 545\,GHz. Regarding the measurements in polarization, the number of external surveys available at these frequencies in polarization is very limited. In particular, we have compared the flux densities in polarization of Tau\,A, the Crab nebula with recent measurements of WMAP and other high resolution instruments on the ground and we have not found significant discrepancies in the polarized flux densities. However, we find a small discrepancy in the measurement of the polarization angle of Tau\,A in some of the \Planck\ channels. For this object, one of the brightest compact sources in polarization, there is a small amount of signal in the $U$ maps at the position of the source, where little or no signal is expected for this object. This signal, much smaller than the signal in the $Q$ map, does not have an effect in the measurement of the total polarized flux density since the $Q$ and $U$ flux densities are added in quadrature (as shown in Eq.~\ref{eqn:P_def}), but it can explain the discrepancy in the polarization angle with respect to external measurements. Since our statistical errors do not account for this small systematic effect, when we propagate the errors in the measurements of the polarized flux density into the errors in the  measurement of the polarization angle, the errors that we obtain can be underestimated. 
\section{Characteristics of the \plccs}
\label{sec:characteristics}
\begin{figure}
\begin{center}
\includegraphics[width=0.5\textwidth]{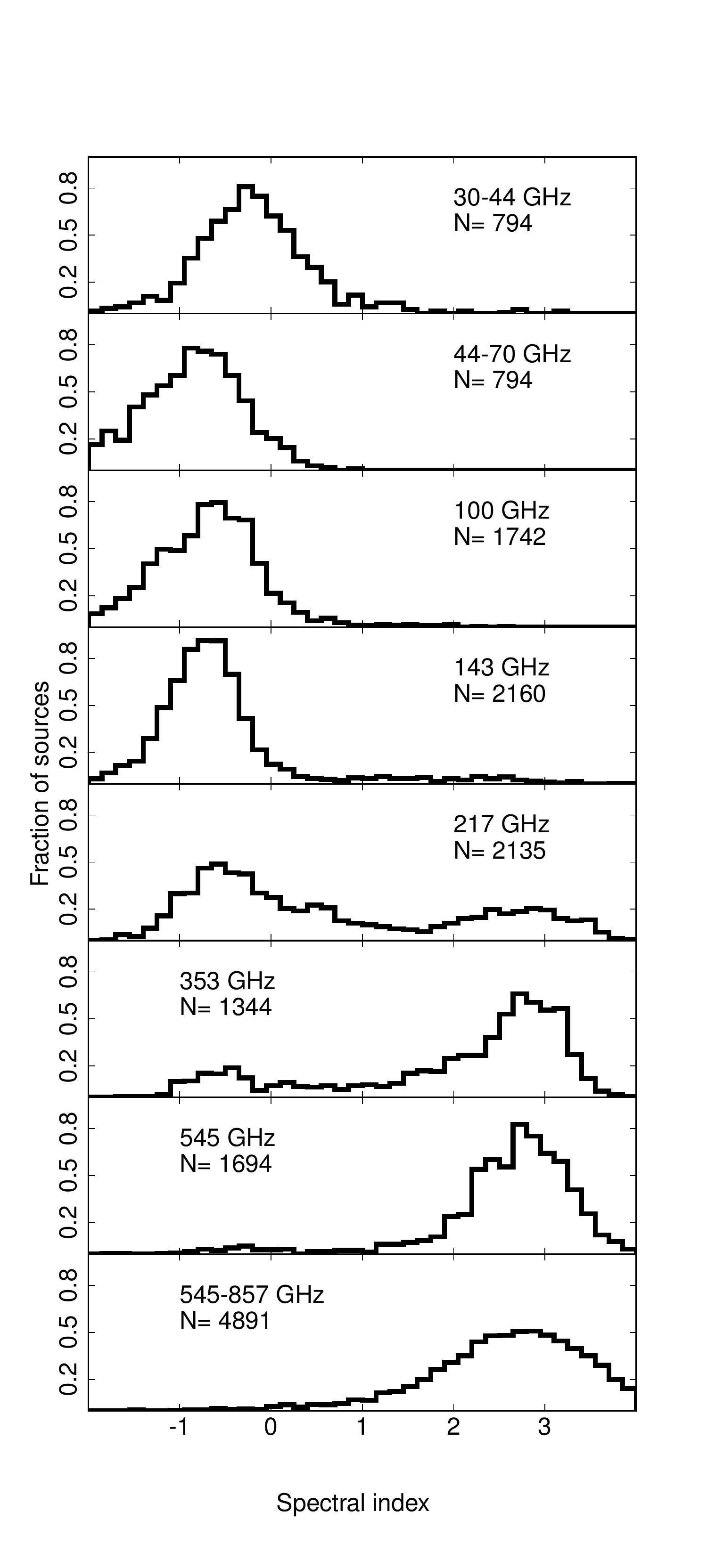}
\caption{Histograms of spectral index for sources in \plccs. The changes in the source populations with frequency are clearly visible, the lower frequencies are dominated by synchrotron sources and the higher frequencies by dusty ones. At the intermediate frequencies both source populations are discernible.
Between the top two panels, there is a visible shift in the peak of the histogram. This is due to a steepening of the spectral indices of the radio sources. 
In each panel we give $N$, the number of sources in the histogram.
}
\label{fig:spectral_indices}
\end{center}
\end{figure}
\begin{figure}
\begin{center}
\includegraphics[width=0.465\textwidth]{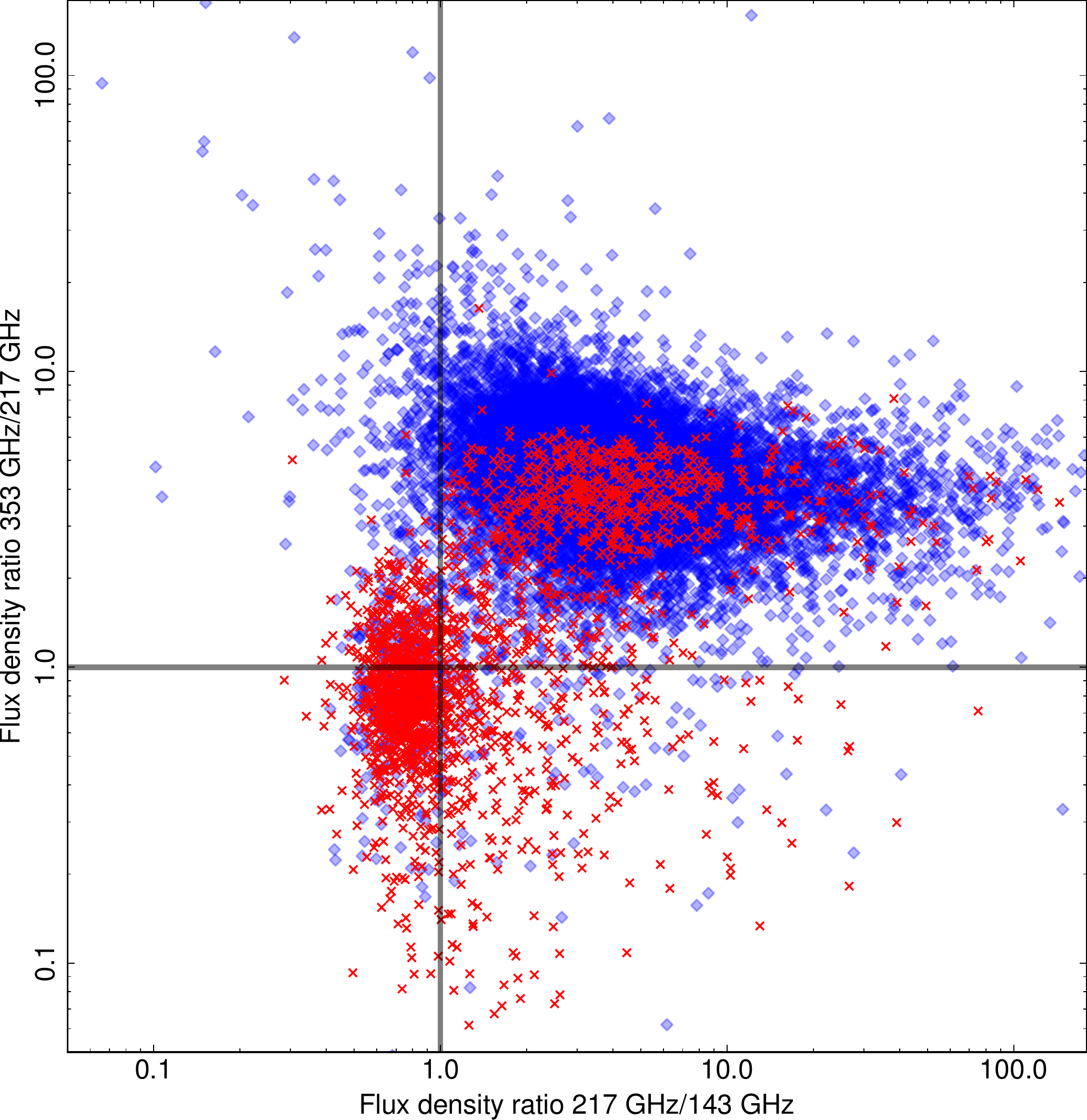}
\includegraphics[width=0.465\textwidth]{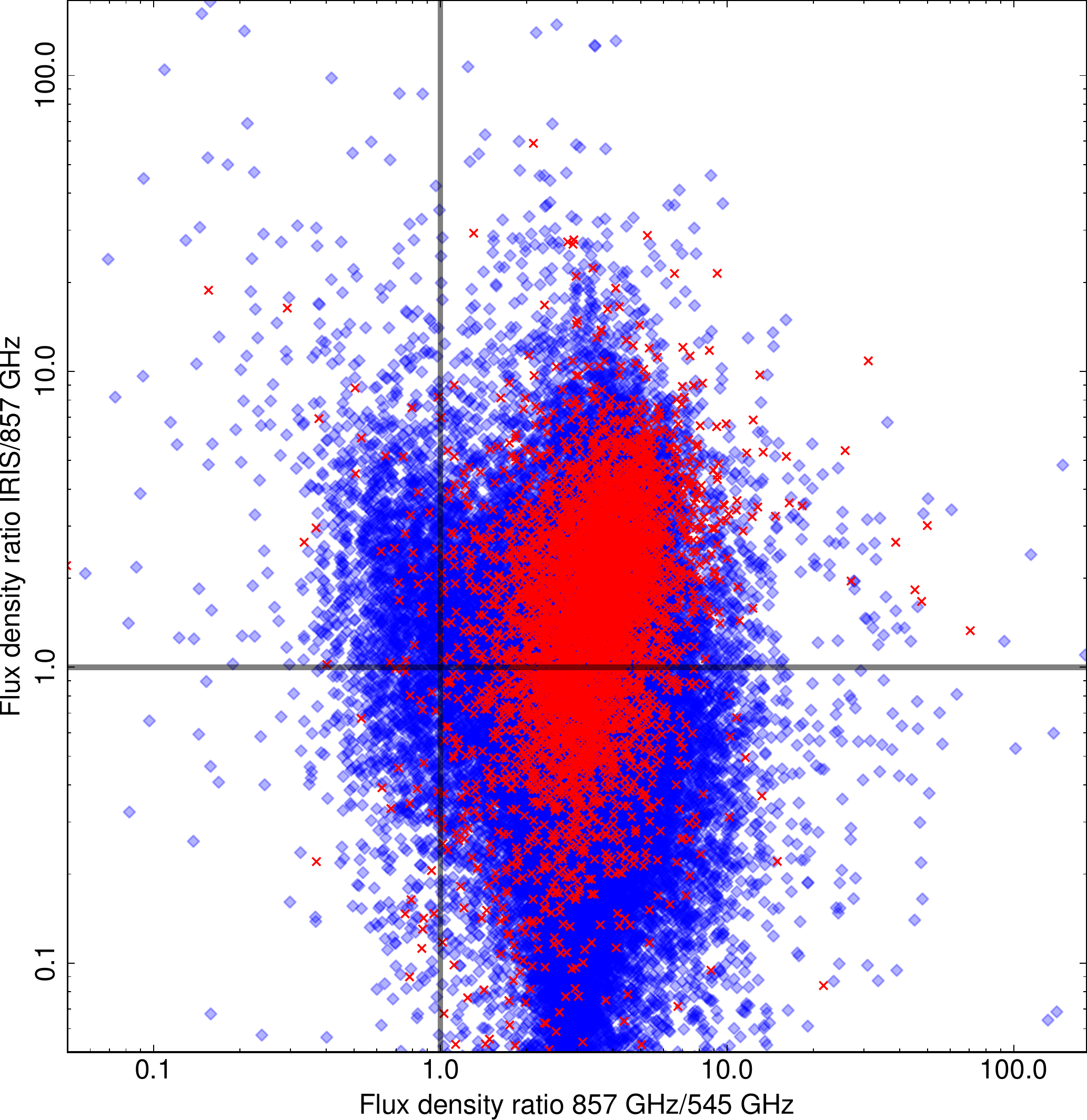}
\caption{Colour-colour plots. Red crosses represent sources from the \plccs\ and blue diamonds sources from the \ecat. 
\textit{Top}: common frequency 217\,GHz. We can see the non-thermal and thermal source populations of the \plccs; the \ecat\ contains significantly more thermal sources than the \plccs, as  expected given the that \ecat\ contains the Galactic plane region. 
\textit{Bottom}: common frequency 857\,GHz. The \plccs\ is consistent with a population of cold sources spanning a narrow range in temperature, whereas the \ecat\ shows a wider distribution of source properties.
\label{fig:colour-colour}}
\end{center}
\end{figure}
In Fig.~\ref{fig:sensitivity}, we displayed the sensitivity of the \plccs\ compared with the \lastcat, the ERCSC, and several other CMB projects. 
For the \plccs\ we define the sensitivity to be the flux density at the 90\,\% completeness limit for each \Planck\ channel.
The improvements between the \plccs\ and the \lastcat\ are most apparent for the LFI channels. This is to be expected given the larger increase of data for LFI than for HFI in the full mission. Additionally, for the higher frequency channels the foregrounds are a significant noise source for the detection of compact sources; the reduction of the instrumental noise resulting from longer integration may therefore not increase the depth of the catalogue as much as might be expected. Finally, the estimated sensitivity of the catalogue is worse at 857\,GHz (Sect.~\ref{sec:hfi_reliability}), owing to the improved understanding of the reliability.
%

Table~\ref{tab:all_pccs_stats} compares the characteristics of the \plccs, the \ecat,  and the \lastcat.  The total number of sources in each catalogue is given as well as the number outside the Galactic region. The numbers of sources in the extragalactic zone in general increase between the \lastcat\ and the union of the \plccs\ and \ecat.
For the highest three frequency channels,  the union of the \plccs\ and \ecat\ catalogues contains many more sources than the \lastcat\, due to the lower \snr\ threshold applied in the \ecat\ than in the Galactic region of the \lastcat.
In the extragalactic zone, we compare the average uncertainty on the flux density and the 90\,\% completeness values. We can see that the \plccs\ has lower uncertainties and is consequently more complete than the \lastcat, except at 857\,GHz, where the completeness has dropped (see Sect.~\ref{sec:hfi_reliability}).
%

Table~\ref{tab:pccs_pol} shows the characteristics of the subset of sources with significant polarized emission, for both the \plccs\ and \ecat\ catalogues. The majority of the significantly polarized sources are in the Galactic plane region; hence for the HFI channels the majority of these sources are in the \ecat\ catalogue.

Table~\ref{tab:pccs2_stats} shows the numbers of sources internally matched in adjacent frequency channels, within the union of the \plccs\ and \ecat. It shows the number of sources matched \textit{both} above and below in frequency (e.g. sources at 100\,GHz found in both the 70 and 143\,GHz catalogues), those matched \textit{either} above or below in frequency (a less stringent criterion), and the percentage of sources so matched.
Note that sources matched by the ``above or below criterion'' will include as a subset those sources meeting the more stringent ``above \textit{and} below'' criterion.
A source is considered to be matched if the positions are closer than the larger FWHM of the two channels. A catalogue was extracted from the IRIS 100 \,\micron\ map~\citep{mdeschenes05} using the MHW2 pipeline, and used as the neighbouring channel above 857\,GHz. The IRIS mask, which removes around 2.1\,\% of the sky, was applied to the 857\,GHz catalogue before performing this comparison, and this reduces the number of sources in the union to 47\,156, a decrease of about 2.1\,\%. The number of matches given for the 857\,GHz channel only includes sources outside the IRIS mask. For the 30\,GHz channel, the matches were evaluated using only the channel above, 44\,GHz. The low percentage of internal matches of the 30\,GHz channel  results from two factors: the generally negative spectral index of the sources at these frequencies; and the relatively low sensitivity of the 44\,GHz receivers.

Figure~\ref{fig:spectral_indices} shows histograms of the spectral indices obtained via the non-blind flux density extraction from the neighbouring channels. As expected, the high-frequency channels (545 and 857\,GHz) are dominated by dusty galaxies and the low-frequency ones are dominated by synchrotron sources, where the change in the dominant source population occurs between 217 and 353\,GHz. It can also be seen that there is a shift in the peak of the histogram between the top two panels, between 30--44\,GHz and 44--70\,GHz and above. The reason for this is a steepening of the spectral indices of radio sources, which  has been seen previously \citep{massardi09,planck2011-6.1,planck2013-p05}.

Figure~\ref{fig:colour-colour} shows colour-colour plots for sources from the \plccs\ and the \ecat. The positions in the catalogue at the common frequency in the colour-colour plot are used to perform a non-blind extraction of the flux densities from the maps at the neighbouring frequencies. These are used together with the common frequency-channel flux densities to construct the colour-colour plot.
The common frequency in the top panel is 217\,GHz, and here we can see the two populations, which are also seen in the spectral-index histogram (Fig.~\ref{fig:spectral_indices}),  for the \plccs. Also apparent is the domination of the \ecat\ at this frequency by thermal sources, since the \ecat\ includes sources in the Galactic plane.
The bottom panel uses the IRIS map and the \Planck\ 545\,GHz maps for the non-blind extraction of flux densities at the locations of the 857\,GHz catalogues. The \plccs\ colour-colour distribution is consistent with a population of cold sources with a narrow range of temperatures (around 10--40\,K) in the range of spectral indices is shown in Fig.~\ref{fig:spectral_indices}.  The larger dispersion of values in the \ecat\, by comparison, is indicative of the greater noise levels in the region of sky corresponding to the \ecat\, as well as a broader distribution of temperatures for the sources.
%
 \begin{table*}
\begingroup
\newdimen\tblskip \tblskip=5pt
\caption{\plccs\ and \ecat\ characteristics compared with those of  the \lastcat.}
\label{tab:all_pccs_stats}
\nointerlineskip \vskip -3mm \footnotesize
\setbox\tablebox=\vbox{
\newdimen\digitwidth
\setbox0=\hbox{\rm 0}
\digitwidth=\wd0
\catcode`*=\active
\def*{\kern\digitwidth}
\newdimen\signwidth
\setbox0=\hbox{+}
\signwidth=\wd0
\catcode`!=\active
\def!{\kern\signwidth}
\halign to \hsize{\hbox to 1.7in{#\leaderfil}\tabskip 2.2em plus 1em&\hfil#\hfil&\hfil#\hfil&\hfil#\hfil&\hfil#\hfil&\hfil#\hfil&\hfil#\hfil&\hfil#\hfil&\hfil#\hfil&\hfil#\hfil\tabskip=0pt\cr
\noalign{\doubleline}
\omit Channel&30&44&70&100&143&217&353&545&857\cr
\noalign{\vskip 3pt\hrule\vskip 5pt}
Freq [GHz]&28.4&44.1&70.4&100.0&143.0&217.0&353.0&545.0&857.0\cr
$\lambda$ [$\mu$m]&10561*&6807\,&4260\,&\,3000&\,2098&\,1382&*\,850&*\,550&*\,350\cr
\noalign{\vskip 10pt}
\noalign{\textit{Number of sources}}
\noalign{\vskip 5pt}
\plccs\ &1560&934&1296&1742&2160&*2135&*1344&*1694&*4891\cr
\ecat\ &\ldots&\ldots&\ldots&2487&4139&16842&22665&31068&43290\cr
Union \plccs+\ecat &\ldots&\ldots&\ldots&4229&6299&18977&24009&32762&48181\cr
\lastcat$^a$&1256&731&*939&3850&5675&16070&13613&16933&24381\cr
\noalign{\vskip 10pt}
\noalign{\textit{Number of sources}}
\noalign{\textit{ - in extragalactic zone$^b$}}
\noalign{\vskip 5pt}
\plccs\,&745&367&504&1742&2160&*2135&*1344&*1694&*4891\cr
\ecat\,&\ldots&\ldots&\ldots&***0&***0&***26&**289&**839&*2097\cr
Union \plccs+\ecat&\ldots&\ldots&\ldots&1742&2160&*2161&*1633&*2533&*6988\cr
\lastcat$^a$&572&258&332&1483&1779&*1745&*1424&*3566&*7270\cr
\noalign{\vskip 10pt}
\noalign{\textit{Flux densities {\rm [mJy]}}}
\noalign{\textit{ - in extragalactic zone$^b$}}
\noalign{\vskip 5pt}
\plccs* : minimum$^{c}$&*376&*603&*444&*232&*147&**127&**242&**535&**720\cr
******* : 90\% completeness&*426&*676&*489&*269&*177&**152&**304&**555&**791\cr
******* : uncertainty&**87&*134&*101&**55&**35&***29&***55&**105&**168\cr
\noalign{\vskip 3pt}
\ecat\,: minimum$^{c}$&**\ldots&**\ldots&**\ldots&**\ldots&**\ldots&**189&**350&**597&**939\cr
******* : 90\% completeness&**\ldots&**\ldots&**\ldots&**\ldots&**\ldots&**144&**311&**557&**927\cr
******* : uncertainty&**\ldots&**\ldots&**\ldots&**\ldots&**\ldots&***35&***73&**144&**278\cr
\noalign{\vskip 3pt}
\lastcat$^a$\,* : minimum$^{c}$&*461&*825&*566&*267&*169&**140&**273&**445&**668\cr
******* : 90\% completeness&*575&1047&*776&*300&*190&**180&**330&**570&**680\cr
******* : uncertainty&*109&*198&*149&**62&**39&***33&***65&**119&**188\cr
\noalign{\vskip 5pt\hrule\vskip 3pt}}}
\endPlancktablewide
\tablenote a \cite{planck2013-p05}\par
\tablenote b  30--70\,GHz: as in \lastcat, the extragalactic zone is given by $|b|>30^{\circ}$. 100--857\,GHz: outside of Galactic region where the reliability cannot be accurately assessed. Note that for the \ecat\ the only sources that occur in this region lie in the filament mask.\par
\tablenote c Minimum flux density of the catalogue in the extragalactic zone after excluding the faintest 10\,\% of sources.\par
\endgroup
\end{table*}
\begin{table*}
\begingroup
\newdimen\tblskip \tblskip=5pt
\caption{\plccs\ and \ecat\ polarization characteristics for sources with polarized emission with significance  $>99.99\%$.}
\label{tab:pccs_pol}
\nointerlineskip \vskip -3mm \footnotesize
\setbox\tablebox=\vbox{
\newdimen\digitwidth
\setbox0=\hbox{\rm 0}
\digitwidth=\wd0
\catcode`*=\active
\def*{\kern\digitwidth}
\newdimen\signwidth
\setbox0=\hbox{+}
\signwidth=\wd0
\catcode`!=\active
\def!{\kern\signwidth}
\halign to \hsize{\hbox to 3.25in{#\leaderfil}\tabskip 2.2emplus 1em&\hfil#\hfil&\hfil#\hfil&\hfil#\hfil&\hfil#\hfil&\hfil#\hfil&\hfil#\hfil&\hfil#\hfil\tabskip=0pt\cr
\noalign{\doubleline}
\omit Channel&*30&*44&*70&*100&*143&*217&*353\cr
\noalign{\vskip 3pt\hrule\vskip 5pt}
\noalign{\vskip 5pt}
Number of significantly polarized sources in \plccs&122&**30&*34&*20&*25&*11&**1\cr
Minimum polarized flux density$^{a}$ [mJy]&117&*181&284&138&148&166&453\cr
Polarized flux density uncertainty [mJy]&*46&**88&*91&*30&*26&*30&*81\cr
Minimum polarized flux density completeness 90\% [mJy]&199&*412&397&135&100&136&347\cr
Minimum polarized flux density completeness 95\% [mJy]&251&*468&454&160&111&153&399\cr
Minimum polarized flux density completeness 100\% [mJy]&600&*700&700&250&147&257&426\cr
\noalign{\vskip 5pt}
Number of significantly polarized sources in \ecat&*\ldots&*\ldots&*\ldots&*43&111&325&666\cr
Minimum polarized flux density$^{a}$ [mJy]&*\ldots&*\ldots&*\ldots&121&*87&114&348\cr
Polarized flux density uncertainty [mJy]&*\ldots&*\ldots&*\ldots&*52&*44&*55&178\cr
Minimum polarized flux density completeness 90\% [mJy]&*\ldots&*\ldots&*\ldots&410&613&270&567\cr
Minimum polarized flux density completeness 95\% [mJy]&*\ldots&*\ldots&*\ldots&599&893&464&590\cr
Minimum polarized flux density completeness 100\% [mJy]&*\ldots&*\ldots&*\ldots&835&893&786&958\cr
\noalign{\vskip 5pt\hrule\vskip 3pt}}}
\endPlancktablewide
\tablenote a Minimum polarized flux density of the catalogue of significantly polarized sources after excluding the faintest 10\,\% of sources. For the LFI channels we have not considered the sources that have been flagged as unidentified. There are nine, one, and one of these unidentified sources at 30, 44, and 70\,GHz, respectively. \par
\endgroup
\end{table*}
\begin{table}
\begingroup
\newdimen\tblskip \tblskip=5pt
\caption{Sources matched between neighbouring channels.
Given that, for the HFI channels, the sky area corresponding to the \plccs\ and \ecat\ is different for every channel, the comparison between frequency channels must necessarily be performed on the union of the \plccs\ and \ecat.
The fraction matched from the same analysis applied to the previous \lastcat\ is shown in brackets in the last column, for comparison purposes.
Note that 217\,GHz is different from other bands in the ratio of the number of above-and-below matches compared to the number of above-or-below. This is of course because of the change of spectral index at that point, as discussed in Sect.~\ref{sec:characteristics} and Figs.~\ref{fig:spectral_indices} and~\ref{fig:colour-colour}. 
}
\label{tab:pccs2_stats}
\nointerlineskip \vskip -3mm \footnotesize
\setbox\tablebox=\vbox{
\newdimen\digitwidth
\setbox0=\hbox{\rm 0}
\digitwidth=\wd0
\catcode`*=\active
\def*{\kern\digitwidth}
\newdimen\signwidth
\setbox0=\hbox{+}
\signwidth=\wd0
\catcode`!=\active
\def!{\kern\signwidth}
\halign{\hbox to 0.5in{#\leaderfil}\tabskip 1.5em &\hfil#\hfil&\hfil#\hfil&\hfil#\hfil&\hfil#\hfil&\hfil#\hfil \tabskip=0pt\cr
\noalign{\doubleline}
\omit\hfil&\omit&\multispan{2}{\hfil No. matched\hfil}&\multispan{2}{\hfil Fraction matched\hfil}\cr
\noalign{\vskip-6pt}
\omit&\omit&\multispan{2}{\hrulefill}&\multispan{2}{\hrulefill}\cr
\omit &No.&Above&Above&\plccs&*\cr
\omit \hfil Channel&sources&and&or&and&\cr
\omit &\omit&below&below&\ecat&(\lastcat)\cr
\noalign{\vskip 5pt\hrule\vskip 3pt}
*30$^{a}$&*1560&**\ldots&**799&**51.2\%&(50.1\%)\cr
*44&**934&**700&**851&**91.1\%&(90.8\%)\cr
*70&*1296&**735&*1113&**85.9\%&(86.8\%)\cr
100&*4229&*1047&*3049&**72.1\%&(71.6\%)\cr
143&*6299&*2734&*5163&**81.9\%&(81.9\%)\cr
217&18977&*3837&14928&**78.7\%&(66.1\%)\cr
353&24009&12171&20867&**86.9\%&(88.7\%)\cr
545&32762&17003&28423&**86.8\%&(85.8\%)\cr
857$^{b}$&47156&14578&35390&**75.0\%&(74.9\%)\cr
\noalign{\vskip 5pt\hrule\vskip 3pt}}}
\endPlancktable
\tablenote a The $30$\,GHz channel is only matched with the 44\,GHz channel above.\par
\tablenote b The $857$\,GHz channel is matched above with a catalogue extracted from the IRIS maps using the HFI--MHW. Both catalogues are cut with the IRIS mask prior to matching.\par
\endgroup
\end{table}
\section{The \plccs: access, content and usage}
\label{sec:content}

The \plccs\, is available from the Planck Legacy Archive.\footnote{\url{http://www.cosmos.esa.int/web/planck/pla}} It is composed of 15 single frequency catalogue FITS files, one per LFI channel and two per HFI channel. In addition there are associated maps, again provided as FITS files, which are described further in Sect.~\ref{sec:associated_maps}.
Additional information about the catalogue content and format can be found in the Explanatory Supplement,\footnote{ \url{http://wiki.cosmos.esa.int/planckpla2015}} in the FITS file headers, and in the first PCCS paper \citep{planck2013-p05}.  Here we summarize the catalogue contents, focusing on the additional features of the \plccs\ catalogues.

\begin{itemize}
\item{Source identification: NAME (e.g. PCCS2 030\,G184.54$-$05.78).}
\item{Position: GLON and GLAT contain the Galactic coordinates, and RA and DEC give the same information in equatorial coordinates (J2000).}
\item{Flux density: the four estimates of flux density (DETFLUX, APERFLUX, PSFFLUX, and GAUFLUX) in mJy, and their associated uncertainties.}
\item{Source shape: the elliptical Gaussian fit to the source; i.e. the semi-axes, and orientation}.
\item{Polarization measurements: the polarized flux density and polarization angle and their associated errors for significantly polarized sources; provided for all seven of the nine \Planck\ channels that have polarization data (30--353\,GHz).
The polarization angles are defined as increasing anticlockwise (north through east) following the IAU convention; the position angle zero is the direction of the north Galactic pole.
}
\item{Marginal polarization measurements: measurements for less significantly polarized sources, as described in Sect.~\ref{subsec:polHFI}; these are provided for the 100--353\,GHz channels only.}
\item{Source extension: the EXTENDED flag is set to 1 if a source is extended. A source is extended if the geometric mean of the elliptic Gaussian fit to the source is greater than one-and-a-half times the fitted FWHM from Table~\ref{tab:beam_data}}.
\item{External validation: EXT\_VAL contains a summary of the inter-channel and external validation. See the definition below.}
\item{Positional coincidence identification with a previous \Planck\ catalogue: the ERCSC and \lastcat\ columns indicate the names of the ERCSC and \lastcat\ counterparts, if they exist, at that channel.}
\item{Degree of reliability: in the \plccs\ catalogue the HIGHEST\_RELIABILITY\_CAT column contains the highest reliability catalogue to which the source belongs.}
\item{Reason for inclusion in \ecat: the WHICH\_ZONE flag encodes why the source has been placed in the \ecat.}
\item{Cirrus indicators: a fraction of the sources detected in the upper HFI bands could be associated with Galactic interstellar medium features or cirrus. For the 217--857\,GHz channels, the CIRRUS\_N and the new column, SKY\_BRIGHTNESS, may be used as cirrus indicators. The CIRRUS\_N column is defined as in \lastcat, the number of sources detected at 857\,GHz (using a uniform S/N threshold of 5) within a $1\deg$ radius of the source. The SKY\_BRIGHTNESS is defined as the mean 857\,GHz brightness within a $2^{\circ}$ radius of the source. See Sect. \ref{sec:cautionary} for details}.
\end{itemize}

Note that all flags are evaluated per frequency channel. Two flags require information from the 857\,GHz channel for their evaluation, CIRRUS\_N  and SKY\_BRIGHTNESS.
These are, however, evaluated independently for each frequency channel for which they are provided. This means, for example,  a source that as been observed at 545 and 857\,GHz at slightly different positions in each channel (but close enough to be considered to be the same source), may have different values for the CIRRUS\_N  and SKY\_BRIGHTNESS flags in each channel.

The EXT\_VAL column summarizes the cross-identification with external catalogues. Its definition has been modified slightly with respect to the \lastcat. The EXT\_VAL flag now has a value of 0, 1, 2,  or 3 as described below:
\begin{enumerate}
\item[0:] The source has no clear counterpart in any of the external catalogues and it has not been detected in a neighbouring \Planck\ channel.
\item[1:] The source has no clear counterpart in any of the external catalogues, has not been detected in a neighbouring \Planck\ channel, but was detected in the same channel in the \lastcat.
\item[2:] The source has no clear counterpart in any of the external catalogues, but it has been detected in a neighbouring \Planck\ channel in this release.  For the HFI channels,  we consider the catalogues extracted from the IRIS maps as a neighbouring \Planck\ channel, given the common detection algorithm applied to both data sets.
\item[3:] The source has a clear counterpart in one of the radio catalogues (CRATES, \citealt{healey07}; NEWPS, \citealt{caniego07}; AT20G, \citealt{murphy10}),  the  Revised IRAS-FSC Redshift Catalogue \citep[RIFSCz;][]{wang14}, or the submillimetre catalogue of H-ATLAS \citep{eales10}.
\end{enumerate}
This flag provides extra information about the reliability of individual sources: those flagged as EXT\_VAL = 3 are already known; those with EXT\_VAL = 2  (or 1) have been detected in other \Planck\ channels (or maps) and are therefore potentially new sources, and those with EXT\_VAL = 0 appear in only a single channel and only in this release, and hence are more likely to be spurious. 
%

The EXT\_VAL flag requires data from the neighbouring channels for its evaluation. It is, however, evaluated per frequency channel. If, for instance, a source is identified with an external catalogue and a source in a neighbouring channel, the source in the neighbouring channel will only be identified with the external catalogue if it also satisfies the identification criteria with the external catalogue.

It should be noted that there is no column that contains the coordinate uncertainties for each source. The errors in position are purely statistical and may be determined for each source using Eq.~(\ref{eqn:pos_err}) with the parameters given in  Table~\ref{tab:astro_error_snr}.
There is also no column that contains the \snr\ of the detection, since this is given by DETFLUX/DETFLUX\_ERR.

\subsection{Maps associated with the catalogues}
\label{sec:associated_maps}
Along with the source catalogues we provide associated maps for the HFI channels. 
There are three types of maps provided, which are shown in Fig~\ref{fig:maps}. These are a zone mask, a noise level map, and a \snr\ threshold map.
%

The zone map shows the areas of the sky covered by the \plccs\ and \ecat\ catalogues. The zone map takes the value of zero in the areas corresponding to the \plccs\ and is non-zero in areas corresponding to the \ecat.
It is related to the WHICH\_ZONE flag, in that a non-zero value encodes the reason why sources in that patch of sky are placed in the \ecat. 
The value 1 corresponds to the filament mask outside of the Galactic region, 2 corresponds to the Galactic region, and 3 corresponds to the filament mask inside the Galactic region.

The noise level map corresponds to the detection noise for compact sources.  This is not the same as the instrumental noise, because it includes ``noise'' from all signals other than compact sources.

The \snr\ threshold map contains the thresholds required at each location on the sky to produce a catalogue of the stated reliability.  Threshold maps for the 80 \%, 85 \%, 90 \%, and 95 \% reliability catalogues are provided. The  \snr\ threshold for the 80\% reliability map contains the \snr\ cut applied to the \ecat\ in the area of the sky where the zone map is non-zero. The \snr\ threshold maps for higher reliabilities contain null values in this region.

In total there are, for each HFI channel, six associated maps: one zone map, one noise level map, and four \snr\ threshold maps.
As described in Sect.~\ref{sec:hfi_completeness}, the completeness of the \plccs\ and \ecat\ may be evaluated using these maps. By providing the \snr\ threshold maps for the higher reliabilities the completenesses for the higher reliability subsets of the \plccs\ may also be evaluated.
Indeed the completeness for a given reliability and region of the sky may be assessed using these data.

\subsection{Cautionary notes on the use of catalogues}
\label{sec:cautionary}
The \plccs\ supersedes previous \Planck\ compact source catalogues (ERCSC and \lastcat), because it has been produced using the full mission data and the latest processing and calibration pipelines. Since the three sets of catalogues have been produced from the analysis of maps that average different amounts of data, some idea of the variability of the sources could be obtained from the comparison of the three, although for this purpose it would be better to analyse the single survey maps, the time-ordered data or the user specified time interval maps that can be obtained from the Planck Legacy Archive. 

As noted earlier, the aim of the \plccs\ is to provide as complete a list as possible of \Planck\ sources with a reasonable, and user-adjustable degree of reliability. The criteria used to include or exclude candidate sources differ from channel to channel and in different parts of the sky; they also are based on different \snr\ levels between channels and as a function of position on the sky.  These differences are consequences of our desire to make the catalogue as complete as possible, yet maintain $> 80$\,\% reliability; the differences have to be taken into account when using the \plccs\ for statistical studies.  

We now turn to several specific cautions and comments for users of the \plccs\ and \ecat.

\paragraph{Bandpass corrections to polarization:} For many sources in the three lowest \Planck\ frequency channels, the bandpass correction of the $Q$ and $U$ flux densities is not negligible. Even though we have attempted to correct for this effect on a source by source basis and have propagated this uncertainty into the error bars on the polarized flux densities and polarization angles, there is still room for improvement. This can be seen in the residual leakage present at the position of Tau\,A in the Stokes $U$ maps.  It is anticipated that there will be future updates to the LFI catalogues once the bandpass corrections and errors have been improved.

\paragraph{Variability:} At radio frequencies, up to and including 217\,GHz, many of the extragalactic sources are variable. The measurements of their flux densities provided in the catalogues are, however, averages over the full \Planck\ mission. It should be noted therefore that follow-up observations of these sources  may show significant differences from those provided. 

\paragraph{Contamination from CO:} At infrared/submillimetre frequencies (100\,GHz and above), the \Planck\ bandpasses straddle energetically significant CO lines. The effect is the most significant at 100\,GHz, where the line might contribute more than 50\,\% of the measured flux density for some Galactic sources. 

\paragraph{Photometry:} Each source has multiple estimates of flux density: DETFLUX, APERFLUX, GAUFLUX, and PSFFLUX, as defined above. 
The evaluation of APERFLUX makes the smallest number of assumptions about the data and hence is the most robust, especially in regions of high non-Gaussian background emission, but it may have larger uncertainties than the other methods.  
Hence, a general recommendation for which estimate to use for unresolved sources would be DETFLUX for 30 to 217\,GHz and APERFLUX for 353 to 857\,GHz.  Note that for a specific source the nature of the local background will influence the best choice for the flux estimator.
For bright \textit{resolved} sources, GAUFLUX is recommended, taking int account that it may not be robust for sources close to the Galactic plane due to the strong backgrounds.
In the \plccs\ and \ecat\ we provide polarized flux densities for two methods, one obtained from the measured flux densities in the filtered maps of $Q$ and $U$ and the other obtained from the measured flux densities on the unfiltered maps of $Q$ and $U$ using aperture photometry. Both methods agree for the brightest sources but, because the  noise level is higher in the unfiltered maps, for weaker sources the polarized flux densities obtained with the filtering method are more robust. In addition, we have found that at the position of very bright sources in polarization one can see a spurious signal introduced by the complex beams in polarization. This spurious signal is small compared with the flux density of the sources, but in cases like Tau\,A, where most of the signal is in the $Q$ map, this signal can have an impact on the flux density measured in the $U$ map, which is particularly important when calculating the polarization position angle. For this reason position angles should be used with caution.

\paragraph{Calibration:} The absolute calibration uncertainties of \Planck\ are well below 1\,\% for 30--353\,GHz \citep{planck2014-a06,planck2014-a09}, while for 545 and 857\,GHz the absolute calibration uncertainty is $<7$\,\%, which is primarily due to a 5\,\% systematic uncertainty arising from the planet models \citep{planck2014-a09}. This systematic uncertainty is not included in the internal validation (not simulated) or in the comparison with \Herschel\ data (which use the same planet models; \citealt{griffin10}). In addition, there is a $0\pdeg5$ and $0\pdeg3$ systematic uncertainty in the polarization position angles in the LFI and HFI, respectively.

\paragraph{Colour-correction:} The flux-density estimates have not been colour-corrected because this implies fixing the spectral index that describes each source, and the user may want to apply a colour-correction based on a specific spectral index determined with a higher-resolution experiment. In the Planck Legacy Archive there is a tool to apply the colour-correction to the  source flux density for a user supplied spectral index.  Colour-corrections are described in \cite{planck2014-a03} and \cite{planck2014-a08}. Note that the term bandpass correction in this paper refers to the correction required due to the mismatch in the bandpass between orthogonally polarized detectors and not to the colour-correction of the flux density in Stokes $I$.

\paragraph{Cirrus/ISM:}
The upper bands of HFI could be contaminated by apparent sources associated with Galactic interstellar medium features (ISM) or cirrus. The values of the parameters CIRRUS\_N  and SKY\_BRIGHTNESS can be used as indicators of contamination.
CIRRUS\_N  can be used to flag sources that might be clustered together and thereby associated with ISM structure. 
In order to provide some indications of the range of values of these parameters that could indicate contamination, we compared the properties of the IRAS-identified and non-IRAS-identified sources for both the \plccs\ and the \ecat,  outside  the Galactic plane.  At Galactic latitudes $|b| > 20^{\circ}$, we can use the RIFSCz \citep{wang14} to provide a guide to the likely nature of sources.
 
We compare the \plccs\ 857\,GHz catalogue and the \ecat\ 857\,GHz catalogue with the IRAS sources in the RIFSCz using a 3\,arcmin matching radius. Of the 4891 sources in the \plccs\ 857\,GHz catalogue, 3094 have plausible IRAS counterparts, while 1797 do not. Examination of histograms of the CIRRUS\_N and SKY\_BRIGHTNESS parameters in the \plccs\ show that these two classes of objects behave rather differently. The IRAS-identified sources have a peak sky brightness at about 1\,MJy\,${\rm{sr}^{-1}}$. 
The non-IRAS-identified sources have a bimodal distribution with a slight peak at 1\,MJy\,${\rm{sr}^{-1}}$ and a second peak at about 2.6\,MJy\,${\rm{sr}^{-1}}$. Both distributions have a long tail, but the non-IRAS-identified tail is much longer.  On this basis, sources with SKY\_BRIGHTNESS $>$ 4\,MJy\,${\rm{sr}^{-1}}$ should be treated with caution. 
In contrast non-IRAS-identified sources with SKY\_BRIGHTNESS $<$ 1.4\,MJy\,${\rm{sr}^{-1}}$ are likely to be reliable. Examination of their sky distribution, for example, shows that many such sources lie in the IRAS coverage gaps.
The CIRRUS\_N flag tells a similar story. Both IRAS-identified and IRAS non-identified sources have a peak CIRRUS\_N value of 2, but the non-identified sources have a far longer tail. Very few IRAS-identified sources have a value $>8$ but many unidentified sources do. These should be treated with caution.

The \ecat\ 857\,GHz catalogue contains many more sources with $|b|>20^{\circ}$ of which 1235 are identified with IRAS sources in the RIFSCz and 9235 are not. As with the \plccs\ catalogue the distributions of CIRRUS\_N and SKY\_BRIGHTNESS are different, where the differences are even more pronounced for these \ecat\ sources. Once again, few IRAS-identified sources have SKY\_BRIGHTNESS $> 4$\,MJy\,${\rm{sr}^{-1}}$, but the unidentified sources have brightnesses extending to $>$55\,MJy\,${\rm{sr}^{-1}}$. Similarly, hardly any of the IRAS-identified sources have CIRRUS\_N $>$ 8, but nearly half the unmatched sources do.
Of the 9235 \ecat\ 857\,GHz sources that are not identified with an IRAS source and that lie in the region $|b|>20^{\circ}$, 1850 (20\,\%) have WHICH\_ZONE  = 1, 2637 (29\,\%) have WHICH\_ZONE = 2, and 4748 (51\,\%) have WHICH\_ZONE = 3.
The \ecat\ covers 30.36\,\% of the region $|b|>20^{\circ}$, where 2.47\,\% is in the filament mask, 23.15\,\% in the Galactic region and 4.74\,\% in both.
If the 9235 unidentified detections were distributed uniformly over the region $|b|>20^{\circ}$, we can predict the number of unidentified sources in each zone and compare this to the values we have. 
We find that there are 2.5 and 3.3 times more sources than expected in zones 1 and 3, showing that the filament mask is indeed a useful criterion for regarding sources detected within it as suspicious.

It should be noted that the EXTENDED flag could also be used to identify ISM features, but nearby Galactic and extragalactic sources that are extended at \Planck\ spatial resolution will also meet this criterion. 

\section{Conclusions}
\label{sec:conclusions}

The Second Planck Catalogue of Compact Sources has been produced using the \Planck\ full mission data. The catalogue lists sources detected in total intensity in each of its nine frequency bands between 30 and 857\,GHz and polarization measurements at the positions of these sources for seven of the frequencies between 30 and 353\,GHz. Its format has changed with respect to the ERCSC and the \lastcat. We have divided the catalogue into two parts, the \plccs\ and \ecat, based on our ability to provide a measure of the reliability of each source detected at 100\,GHz and above, where the available external catalogues of compact sources are not able to fully assess the reliability of the detections. Sources located inside the defined Galactic plane masks or situated along dusty filamentary structures (as defined in the cirrus masks) are in the \ecat because the uncertainties in the number counts of the Galactic sources and the difficulty  of simulating the diffuse dust emission near the beam scale in the higher frequency channels do not allow us to  achieve the necessary consistency between the catalogues of input and detected sources in our reliability assessments.

Given the increase in the volume of data between the nominal mission and the full mission, and the improvements in the data processing and calibration of the frequency channel maps, the \plccs\ supersedes previous \Planck\ catalogues. The new catalogue is more complete than the \lastcat, in particular for the LFI channels, owing to the large increase in the data available, eight sky surveys compared with the two and a half sky surveys of \lastcat . In addition, improvements have been made in some of the techniques used to perform the photometry analysis, and in the reliability assessment of the catalogues. The completeness of the 857\,GHz channel, however, has not improved because improvements in the reliability assessment resulted in a higher \snr\ threshold being applied in the formation of this catalogue.  It should be noted, however, that the quality of the \plccs\ catalogue at this channel is better than that of the earlier \lastcat\ due to its greater reliability.

The division of the HFI catalogues into the \plccs\ and \ecat\ has permitted the addition of a parameter in the \plccs\ catalogue that will allow a user to define subsets of the catalogue with higher reliability levels than the target integral reliability of 80\,\%.
Associated maps are provided that will allow the user to evaluate the completeness of their chosen reliability subset, or indeed of the catalogue as a whole.
This added functionality gives the users of the \plccs\ the option of extracting high-reliability subcatalogues, and, in addition, provides a much more complete full catalogue, allowing studies of more sources and to fainter flux densities.
There are ongoing efforts in the \Planck\ Collaboration to produce multifrequency catalogues that will complement the \plccs; when completed, they will be available in the Planck Legacy Archive.
\begin{acknowledgements}
The Planck Collaboration acknowledges the support of: ESA; CNES and CNRS/INSU-IN2P3-INP (France); ASI, CNR, and INAF (Italy); NASA and DoE (USA); STFC and UKSA (UK); CSIC, MINECO, JA, and, RES (Spain); Tekes, AoF, and CSC (Finland); DLR and MPG (Germany); CSA (Canada); DTU Space (Denmark); SER/SSO (Switzerland); RCN (Norway); SFI (Ireland); FCT/MCTES (Portugal); ERC and PRACE (EU). A description of the Planck Collaboration and a list of its members, indicating which technical or scientific activities they have been involved in, can be found at \href{http://www.cosmos.esa.int/web/planck/planck-collaboration}{\texttt{http://www.cosmos.esa.int/web/planck/planck-collaboration}}. We are grateful to the H-ATLAS Executive Committee and primarily to the PIs, S. Eales and L. Dunne, for permission to use the unpublished H-ATLAS catalogue for the validation of the present catalogue. This research has made use of the ``Aladin sky atlas'' \citep{Bonnarel10}, developed at CDS, Strasbourg Observatory, France. Part of this work was performed using the Darwin Supercomputer of the University of Cambridge High Performance Computing Service (\url{http://www.hpc.cam.ac.uk/}), provided by Dell Inc. using Strategic Research Infrastructure Funding from the HEFCE and funding from the STFC. This research has made use of the NASA/IPAC Extragalactic Database (NED) which is operated by the Jet Propulsion Laboratory, California Institute of Technology, under contract with the National Aeronautics and Space Administration.
\end{acknowledgements}
\bibliographystyle{aat} 

\bibliography{Planck_bib,pccs}

\begin{thebibliography}{295}
\expandafter\ifx\csname natexlab\endcsname\relax\def\natexlab#1{#1}\fi

\bibitem[{{Ade} {et~al.}(2010){Ade}, {Savini}, {Sudiwala}, {Tucker},
  {Catalano}, {Church}, {Colgan}, {Desert}, {Gleeson}, {Jones}, {Lamarre},
  {Lange}, {Longval}, {Maffei}, {Murphy}, {Noviello}, {Pajot}, {Puget},
  {Ristorcelli}, {Woodcraft}, \& {Yurchenko}}]{ade2010}
{Ade}, P.~A.~R., {Savini}, G., {Sudiwala}, R., {et~al.} 2010, \aap, 520, A11.
\newblock {\textit{Planck} pre-launch status: The optical architecture of the
  HFI}

\bibitem[{Anderson \& Dahleh(1996)}]{anderson1996}
Anderson, C. \& Dahleh, M. 1996, SIAM J. Sci. Comput., 17, 913.
\newblock {Rapid Computation of the Discrete Fourier Transform}

\bibitem[{{Armitage-Caplan} \& {Wandelt}(2009)}]{armitage-caplan2009}
{Armitage-Caplan}, C. \& {Wandelt}, B.~D. 2009, \apjs, 181, 533.
\newblock {PReBeaM for Planck: A Polarized Regularized Beam Deconvolution
  Map-Making Method}

\bibitem[{{Artal} {et~al.}(2009){Artal}, {Aja}, {de la Fuente}, {Pascual},
  {Mediavilla}, {Martinez-Gonzalez}, {Pradell}, {de Paco}, {Bara}, {Blanco},
  {Garc{\'{\i}}a}, {Davis}, {Kettle}, {Roddis}, {Wilkinson}, {Bersanelli},
  {Mennella}, {Tomasi}, {Butler}, {Cuttaia}, {Mandolesi}, \&
  {Stringhetti}}]{artal2009}
{Artal}, E., {Aja}, B., {de la Fuente}, M.~L., {et~al.} 2009, Journal of
  Instrumentation, 4, 2003.
\newblock {LFI 30 and 44 GHz receivers Back-End Modules}

\bibitem[{{Ashdown} {et~al.}(2007{\natexlab{a}}){Ashdown}, {Baccigalupi},
  {Balbi}, {Bartlett}, {Borrill}, {Cantalupo}, {de Gasperis}, {G{\'o}rski},
  {Heikkil{\"a}}, {Hivon}, {Keih{\"a}nen}, {Kurki-Suonio}, {Lawrence},
  {Natoli}, {Poutanen}, {Prunet}, {Reinecke}, {Stompor}, \&
  {Wandelt}}]{ashdown2007a}
{Ashdown}, M.~A.~J., {Baccigalupi}, C., {Balbi}, A., {et~al.}
  2007{\natexlab{a}}, \aap, 471, 361.
\newblock {Making maps from Planck LFI 30 GHz data}

\bibitem[{{Ashdown} {et~al.}(2007{\natexlab{b}}){Ashdown}, {Baccigalupi},
  {Balbi}, {Bartlett}, {Borrill}, {Cantalupo}, {de Gasperis}, {G{\'o}rski},
  {Hivon}, {Keih{\"a}nen}, {Kurki-Suonio}, {Lawrence}, {Natoli}, {Poutanen},
  {Prunet}, {Reinecke}, {Stompor}, {Wandelt}, \& {The Planck CTP Working
  Group}}]{ashdown2007b}
{Ashdown}, M.~A.~J., {Baccigalupi}, C., {Balbi}, A., {et~al.}
  2007{\natexlab{b}}, \aap, 467, 761.
\newblock {Making sky maps from Planck data}

\bibitem[{{Ashdown} {et~al.}(2009){Ashdown}, {Baccigalupi}, {Bartlett},
  {Borrill}, {Cantalupo}, {de Gasperis}, {de Troia}, {G{\'o}rski}, {Hivon},
  {Huffenberger}, {Keih{\"a}nen}, {Keskitalo}, {Kisner}, {Kurki-Suonio},
  {Lawrence}, {Natoli}, {Poutanen}, {Pr{\'e}zeau}, {Reinecke}, {Rocha},
  {Sandri}, {Stompor}, {Villa}, {Wandelt}, \& {The Planck Ctp Working
  Group}}]{ashdown2009}
{Ashdown}, M.~A.~J., {Baccigalupi}, C., {Bartlett}, J.~G., {et~al.} 2009, \aap,
  493, 753.
\newblock {Making maps from Planck LFI 30 GHz data with asymmetric beams and
  cooler noise}

\bibitem[{{Aumont} {et~al.}(2010){Aumont}, {Conversi}, {Thum}, {Wiesemeyer},
  {Falgarone}, {Mac{\'{\i}}as-P{\'e}rez}, {Piacentini}, {Pointecouteau},
  {Ponthieu}, {Puget}, {Rosset}, {Tauber}, \& {Tristram}}]{aumont2010}
{Aumont}, J., {Conversi}, L., {Thum}, C., {et~al.} 2010, \aap, 514, A70.
\newblock {Measurement of the Crab nebula polarization at 90\,GHz as a
  calibrator for CMB experiments}

\bibitem[{{Baccigalupi}(1999)}]{baccigalupi1999}
{Baccigalupi}, C. 1999, \prd, 59, 123004.
\newblock {Cosmic microwave background: Polarization and temperature
  anisotropies from symmetric structures}

\bibitem[{{Barnes} {et~al.}(2003){Barnes}, {Hill}, {Hinshaw}, {Page},
  {Bennett}, {Halpern}, {Jarosik}, {Kogut}, {Limon}, {Meyer}, {Tucker},
  {Wollack}, \& {Wright}}]{barnes2003}
{Barnes}, C., {Hill}, R.~S., {Hinshaw}, G., {et~al.} 2003, \apjs, 148, 51.
\newblock {First-Year Wilkinson Microwave Anisotropy Probe (WMAP) Observations:
  Galactic Signal Contamination from Sidelobe Pickup}

\bibitem[{{Battaglia} {et~al.}(2009){Battaglia}, {Franceschet}, {Zonca},
  {Bersanelli}, {Butler}, {D'Arcangelo}, {Davis}, {Galeotta}, {Guzzi},
  {Hoyland}, {Hughes}, {Jukkala}, {Kettle}, {Laaninen}, {Leonardi}, {Maino},
  {Mandolesi}, {Meinhold}, {Mennella}, {Platania}, {Terenzi}, {Tuovinen},
  {Varis}, {Villa}, \& {Wilkinson}}]{battaglia2009}
{Battaglia}, P., {Franceschet}, C., {Zonca}, A., {et~al.} 2009, Journal of
  Instrumentation, 4, 2014.
\newblock {Advanced modelling of the Planck-LFI radiometers}

\bibitem[{{Bennett} {et~al.}(2003{\natexlab{a}}){Bennett}, {Halpern},
  {Hinshaw}, {Jarosik}, {Kogut}, {Limon}, {Meyer}, {Page}, {Spergel}, {Tucker},
  {Wollack}, {Wright}, {Barnes}, {Greason}, {Hill}, {Komatsu}, {Nolta},
  {Odegard}, {Peiris}, {Verde}, \& {Weiland}}]{bennett2003a}
{Bennett}, C.~L., {Halpern}, M., {Hinshaw}, G., {et~al.} 2003{\natexlab{a}},
  \apjs, 148, 1.
\newblock {First-Year Wilkinson Microwave Anisotropy Probe (WMAP) Observations:
  Preliminary Maps and Basic Results}

\bibitem[{{Bennett} {et~al.}(2011){Bennett}, {Hill}, {Hinshaw}, {Larson},
  {Smith}, {Dunkley}, {Gold}, {Halpern}, {Jarosik}, {Kogut}, {Komatsu},
  {Limon}, {Meyer}, {Nolta}, {Odegard}, {Page}, {Spergel}, {Tucker}, {Weiland},
  {Wollack}, \& {Wright}}]{bennett2010}
{Bennett}, C.~L., {Hill}, R.~S., {Hinshaw}, G., {et~al.} 2011, \apjs, 192, 17.
\newblock {Seven-year Wilkinson Microwave Anisotropy Probe (WMAP) Observations:
  Are There Cosmic Microwave Background Anomalies?}

\bibitem[{{Bennett} {et~al.}(2003{\natexlab{b}}){Bennett}, {Hill}, {Hinshaw},
  {Nolta}, {Odegard}, {Page}, {Spergel}, {Weiland}, {Wright}, {Halpern},
  {Jarosik}, {Kogut}, {Limon}, {Meyer}, {Tucker}, \& {Wollack}}]{bennett2003b}
{Bennett}, C.~L., {Hill}, R.~S., {Hinshaw}, G., {et~al.} 2003{\natexlab{b}},
  \apjs, 148, 97.
\newblock {First-Year Wilkinson Microwave Anisotropy Probe (WMAP) Observations:
  Foreground Emission}

\bibitem[{{Bennett} {et~al.}(2013){Bennett}, {Larson}, {Weiland}, {Jarosik},
  {Hinshaw}, {Odegard}, {Smith}, {Hill}, {Gold}, {Halpern}, {Komatsu}, {Nolta},
  {Page}, {Spergel}, {Wollack}, {Dunkley}, {Kogut}, {Limon}, {Meyer}, {Tucker},
  \& {Wright}}]{bennett2012}
{Bennett}, C.~L., {Larson}, D., {Weiland}, J.~L., {et~al.} 2013, \apjs, 208,
  20.
\newblock {Nine-year Wilkinson Microwave Anisotropy Probe (WMAP) Observations:
  Final Maps and Results}

\bibitem[{{Bersanelli} {et~al.}(2010){Bersanelli}, {Mandolesi}, {Butler},
  {Mennella}, {Villa}, {Aja}, {Artal}, {Artina}, {Baccigalupi}, {Balasini},
  {Baldan}, {Banday}, {Bastia}, {Battaglia}, {Bernardino}, {Blackhurst},
  {Boschini}, {Burigana}, {Cafagna}, {Cappellini}, {Cavaliere}, {Colombo},
  {Crone}, {Cuttaia}, {D'Arcangelo}, {Danese}, {Davies}, {Davis}, {de Angelis},
  {de Gasperis}, {de La Fuente}, {de Rosa}, {de Zotti}, {Falvella}, {Ferrari},
  {Ferretti}, {Figini}, {Fogliani}, {Franceschet}, {Franceschi}, {Gaier},
  {Garavaglia}, {Gomez}, {Gorski}, {Gregorio}, {Guzzi}, {Herreros},
  {Hildebrandt}, {Hoyland}, {Hughes}, {Janssen}, {Jukkala}, {Kettle},
  {Kilpi{\"a}}, {Laaninen}, {Lapolla}, {Lawrence}, {Lawson}, {Leahy},
  {Leonardi}, {Leutenegger}, {Levin}, {Lilje}, {Lowe}, {Lubin}, {Maino},
  {Malaspina}, {Maris}, {Marti-Canales}, {Martinez-Gonzalez}, {Mediavilla},
  {Meinhold}, {Miccolis}, {Morgante}, {Natoli}, {Nesti}, {Pagan}, {Paine},
  {Partridge}, {Pascual}, {Pasian}, {Pearson}, {Pecora}, {Perrotta},
  {Platania}, {Pospieszalski}, {Poutanen}, {Prina}, {Rebolo}, {Roddis},
  {Rubi{\~n}o-Martin}, {Salmon}, {Sandri}, {Seiffert}, {Silvestri},
  {Simonetto}, {Sjoman}, {Smoot}, {Sozzi}, {Stringhetti}, {Taddei}, {Tauber},
  {Terenzi}, {Tomasi}, {Tuovinen}, {Valenziano}, {Varis}, {Vittorio}, {Wade},
  {Wilkinson}, {Winder}, {Zacchei}, \& {Zonca}}]{bersanelli2010}
{Bersanelli}, M., {Mandolesi}, N., {Butler}, R.~C., {et~al.} 2010, \aap, 520,
  A4.
\newblock {\textit{Planck} pre-launch status: Design and description of the Low
  Frequency Instrument}

\bibitem[{{Bhandari} {et~al.}(2004){Bhandari}, {Prina}, {Bowman}, {Paine},
  {Pearson}, \& {Nash}}]{bhandari2004}
{Bhandari}, P., {Prina}, M., {Bowman}, R.~C., {et~al.} 2004, Cryogenics, 44,
  395.
\newblock {Sorption coolers using a continuous cycle to produce 20 K for the
  Planck flight mission}

\bibitem[{{BICEP2 Collaboration}(2014)}]{BicepDetection}
{BICEP2 Collaboration}. 2014, \prl, 112, 241101.
\newblock {Detection of B-Mode Polarization at Degree Angular Scales by BICEP2}

\bibitem[{{BICEP2/Keck Array and Planck Collaborations}(2015)}]{pb2015}
{BICEP2/Keck Array and Planck Collaborations}. 2015, \prl, 114, 101301.
\newblock {Joint Analysis of BICEP2/Keck Array and Planck Data}

\bibitem[{{Buehler} {et~al.}(1996){Buehler}, {Desorgher}, \&
  {Zehnder}}]{buehler1996}
{Buehler}, P., {Desorgher}, L., \& {Zehnder}, A. 1996, in ESA Special
  Publication, Vol. 392, Environment Modeling for Space-Based Applications, ed.
  {T.-D.~Guyenne \& A.~Hilgers}, 87

\bibitem[{{Cantalupo} {et~al.}(2010){Cantalupo}, {Borrill}, {Jaffe}, {Kisner},
  \& {Stompor}}]{cantalupo2010}
{Cantalupo}, C.~M., {Borrill}, J.~D., {Jaffe}, A.~H., {Kisner}, T.~S., \&
  {Stompor}, R. 2010, \apjs, 187, 212.
\newblock {MADmap: A Massively Parallel Maximum Likelihood Cosmic Microwave
  Background Map-maker}

\bibitem[{{Cappellini} {et~al.}(2003){Cappellini}, {Maino}, {Albetti},
  {Platania}, {Paladini}, {Mennella}, \& {Bersanelli}}]{cappellini2003}
{Cappellini}, B., {Maino}, D., {Albetti}, G., {et~al.} 2003, \aap, 409, 375.
\newblock {Optimized in-flight absolute calibration for extended CMB surveys}

\bibitem[{{Cardoso} {et~al.}(2008){Cardoso}, {Martin}, {Delabrouille},
  {Betoule}, \& {Patanchon}}]{cardoso2008}
{Cardoso}, J., {Martin}, M., {Delabrouille}, J., {Betoule}, M., \& {Patanchon},
  G. 2008, IEEE Journal of Selected Topics in Signal Processing, 2, 735,
  special issue on Signal Processing for Astronomical and Space Research
  Applications.
\newblock {Component separation with flexible models. Application to the
  separation of astrophysical emissions}

\bibitem[{{Carvalho} {et~al.}(2009){Carvalho}, {Rocha}, \&
  {Hobson}}]{carvalho2009}
{Carvalho}, P., {Rocha}, G., \& {Hobson}, M.~P. 2009, \mnras, 393, 681.
\newblock {A fast Bayesian approach to discrete object detection in
  astronomical data sets - PowellSnakes I}

\bibitem[{Catalano {et~al.}(2010)Catalano, Coulais, \& Lamarre}]{catalano2010}
Catalano, A., Coulais, A., \& Lamarre, J.-M. 2010, Appl. Opt., 49, 5938.
\newblock Analytical approach to optimizing alternating current biasing of
  bolometers

\bibitem[{{Colombi} {et~al.}(2009){Colombi}, {Jaffe}, {Novikov}, \&
  {Pichon}}]{colombi2009}
{Colombi}, S., {Jaffe}, A., {Novikov}, D., \& {Pichon}, C. 2009, \mnras, 393,
  511.
\newblock {Accurate estimators of power spectra in N-body simulations}

\bibitem[{{Colombo} {et~al.}(2009){Colombo}, {Pierpaoli}, \&
  {Pritchard}}]{colombo2009}
{Colombo}, L.~P.~L., {Pierpaoli}, E., \& {Pritchard}, J.~R. 2009, \mnras, 398,
  1621.
\newblock {Cosmological parameters after WMAP5: forecasts for Planck and future
  galaxy surveys}

\bibitem[{{Cuttaia} {et~al.}(2009){Cuttaia}, {Mennella}, {Stringhetti},
  {Maris}, {Terenzi}, {Tomasi}, {Villa}, {Bersanelli}, {Butler}, {Cappellini},
  {Cuevas}, {D'Arcangelo}, {Davis}, {Frailis}, {Franceschet}, {Franceschi},
  {Gregorio}, {Hoyland}, {Leonardi}, {Lowe}, {Mandolesi}, {Meinhold}, {Mendes},
  {Roddis}, {Sandri}, {Valenziano}, {Wilkinson}, {Zacchei}, {Zonca},
  {Battaglia}, {De Nardo}, {Grassi}, {Lapolla}, {Leutenegger}, {Miccolis}, \&
  {Silvestri}}]{cuttaia2009}
{Cuttaia}, F., {Mennella}, A., {Stringhetti}, L., {et~al.} 2009, Journal of
  Instrumentation, 4, 2013.
\newblock {Planck-LFI radiometers tuning}

\bibitem[{Dalhaus(1988)}]{dalhaus1988}
Dalhaus, R. 1988, The Annals of Statistics, 16, 808.
\newblock {Small Sample Effects in Time Series Analysis: A New Asymptotic
  Theory and a New Estimate}

\bibitem[{{Dame} {et~al.}(2001){Dame}, {Hartmann}, \& {Thaddeus}}]{dame2001}
{Dame}, T.~M., {Hartmann}, D., \& {Thaddeus}, P. 2001, \apj, 547, 792.
\newblock {The Milky Way in Molecular Clouds: A New Complete CO Survey}

\bibitem[{{D'Arcangelo} {et~al.}(2009{\natexlab{a}}){D'Arcangelo}, {Figini},
  {Simonetto}, {Villa}, {Pecora}, {Battaglia}, {Bersanelli}, {Butler},
  {Cuttaia}, {Garavaglia}, {Guzzi}, {Mandolesi}, {Mennella}, {Morgante},
  {Pagan}, \& {Valenziano}}]{darcangelo2009a}
{D'Arcangelo}, O., {Figini}, L., {Simonetto}, A., {et~al.} 2009{\natexlab{a}},
  Journal of Instrumentation, 4, 2007.
\newblock {The Planck-LFI flight model composite waveguides}

\bibitem[{{D'Arcangelo} {et~al.}(2009{\natexlab{b}}){D'Arcangelo}, {Simonetto},
  {Figini}, {Pagana}, {Villa}, {Pecora}, {Battaglia}, {Bersanelli}, {Butler},
  {Garavaglia}, {Guzzi}, {Mandolesi}, \& {Sozzi}}]{darcangelo2009b}
{D'Arcangelo}, O., {Simonetto}, A., {Figini}, L., {et~al.} 2009{\natexlab{b}},
  Journal of Instrumentation, 4, 2005.
\newblock {The Planck-LFI flight model ortho-mode transducers}

\bibitem[{{Davis} {et~al.}(2009){Davis}, {Wilkinson}, {Davies}, {Winder},
  {Roddis}, {Blackhurst}, {Lawson}, {Lowe}, {Baines}, {Butlin}, {Galtress},
  {Shepherd}, {Aja}, {Artal}, {Bersanelli}, {Butler}, {Castelli}, {Cuttaia},
  {D'Arcangelo}, {Gaier}, {Hoyland}, {Kettle}, {Leonardi}, {Mandolesi},
  {Mennella}, {Meinhold}, {Pospieszalski}, {Stringhetti}, {Tomasi},
  {Valenziano}, \& {Zonca}}]{davis2009}
{Davis}, R.~J., {Wilkinson}, A., {Davies}, R.~D., {et~al.} 2009, Journal of
  Instrumentation, 4, 2002.
\newblock {Design, development and verification of the 30 and 44 GHz front-end
  modules for the Planck Low Frequency Instrument}

\bibitem[{{de Gasperis} {et~al.}(2005){de Gasperis}, {Balbi}, {Cabella},
  {Natoli}, \& {Vittorio}}]{degasperis2005}
{de Gasperis}, G., {Balbi}, A., {Cabella}, P., {Natoli}, P., \& {Vittorio}, N.
  2005, \aap, 436, 1159.
\newblock {ROMA: A map-making algorithm for polarised CMB data sets}

\bibitem[{{Delabrouille}(1998)}]{delabrouille1998}
{Delabrouille}, J. 1998, \aaps, 127, 555.
\newblock {Analysis of the accuracy of a destriping method for future cosmic
  microwave background mapping with the PLANCK SURVEYOR satellite}

\bibitem[{{Delabrouille} {et~al.}(2013){Delabrouille}, {Betoule}, {Melin},
  {Miville-Desch{\^e}nes}, {Gonzalez-Nuevo}, {Le Jeune}, {Castex}, {de Zotti},
  {Basak}, {Ashdown}, {Aumont}, {Baccigalupi}, {Banday}, {Bernard}, {Bouchet},
  {Clements}, {da Silva}, {Dickinson}, {Dodu}, {Dolag}, {Elsner}, {Fauvet},
  {Fa{\"y}}, {Giardino}, {Leach}, {Lesgourgues}, {Liguori}, {Macias-Perez},
  {Massardi}, {Matarrese}, {Mazzotta}, {Montier}, {Mottet}, {Paladini},
  {Partridge}, {Piffaretti}, {Prezeau}, {Prunet}, {Ricciardi}, {Roman},
  {Schaefer}, \& {Toffolatti}}]{delabrouille2012}
{Delabrouille}, J., {Betoule}, M., {Melin}, J.-B., {et~al.} 2013, \aap, 553,
  A96.
\newblock {The pre-launch Planck Sky Model: a model of sky emission at
  submillimetre to centimetre wavelengths}

\bibitem[{{Delabrouille} {et~al.}(2009){Delabrouille}, {Cardoso}, {Le Jeune},
  {Betoule}, {Fay}, \& {Guilloux}}]{delabrouille2009}
{Delabrouille}, J., {Cardoso}, J., {Le Jeune}, M., {et~al.} 2009, \aap, 493,
  835.
\newblock {A full sky, low foreground, high resolution CMB map from WMAP}

\bibitem[{{Delabrouille} {et~al.}(2003){Delabrouille}, {Cardoso}, \&
  {Patanchon}}]{delabrouille2003}
{Delabrouille}, J., {Cardoso}, J.-F., \& {Patanchon}, G. 2003, \mnras, 346,
  1089.
\newblock {Multidetector multicomponent spectral matching and applications for
  cosmic microwave background data analysis}

\bibitem[{{Dunkley} {et~al.}(2009){Dunkley}, {Komatsu}, {Nolta}, {Spergel},
  {Larson}, {Hinshaw}, {Page}, {Bennett}, {Gold}, {Jarosik}, {Weiland},
  {Halpern}, {Hill}, {Kogut}, {Limon}, {Meyer}, {Tucker}, {Wollack}, \&
  {Wright}}]{dunkley2009}
{Dunkley}, J., {Komatsu}, E., {Nolta}, M.~R., {et~al.} 2009, \apjs, 180, 306.
\newblock {Five-Year Wilkinson Microwave Anisotropy Probe (WMAP) Observations:
  Likelihoods and Parameters from the WMAP data}

\bibitem[{{Dupac} \& {Tauber}(2005)}]{dupac2005}
{Dupac}, X. \& {Tauber}, J. 2005, \aap, 430, 363.
\newblock {Scanning strategy for mapping the Cosmic Microwave Background
  anisotropies with Planck}

\bibitem[{{Efstathiou}(2007)}]{efstathiou2007}
{Efstathiou}, G. 2007, \mnras, 380, 1621.
\newblock {Effects of destriping errors on cosmic microwave background
  polarization power spectra and pixel noise covariances}

\bibitem[{{Eriksen} {et~al.}(2006){Eriksen}, {Dickinson}, {Lawrence},
  {Baccigalupi}, {Banday}, {G{\'o}rski}, {Hansen}, {Lilje}, {Pierpaoli},
  {Seiffert}, {Smith}, \& {Vanderlinde}}]{eriksen2006}
{Eriksen}, H.~K., {Dickinson}, C., {Lawrence}, C.~R., {et~al.} 2006, \apj, 641,
  665.
\newblock {Cosmic Microwave Background Component Separation by Parameter
  Estimation}

\bibitem[{{Eriksen} {et~al.}(2008){Eriksen}, {Jewell}, {Dickinson}, {Banday},
  {G{\'o}rski}, \& {Lawrence}}]{eriksen2008}
{Eriksen}, H.~K., {Jewell}, J.~B., {Dickinson}, C., {et~al.} 2008, \apj, 676,
  10.
\newblock {Joint Bayesian Component Separation and CMB Power Spectrum
  Estimation}

\bibitem[{{Fern{\'a}ndez-Cobos} {et~al.}(2012){Fern{\'a}ndez-Cobos}, {Vielva},
  {Barreiro}, \& {Mart{\'{\i}}nez-Gonz{\'a}lez}}]{fernandez2012}
{Fern{\'a}ndez-Cobos}, R., {Vielva}, P., {Barreiro}, R.~B., \&
  {Mart{\'{\i}}nez-Gonz{\'a}lez}, E. 2012, \mnras, 420, 2162.
\newblock {Multiresolution internal template cleaning: an application to the
  Wilkinson Microwave Anisotropy Probe 7-yr polarization data}

\bibitem[{{Ferreira} \& {Jaffe}(2000)}]{ferreira2000}
{Ferreira}, P.~G. \& {Jaffe}, A.~H. 2000, \mnras, 312, 89.
\newblock {Simultaneous estimation of noise and signal in cosmic microwave
  background experiments}

\bibitem[{{Finkbeiner} {et~al.}(1999){Finkbeiner}, {Davis}, \&
  {Schlegel}}]{finkbeiner1999}
{Finkbeiner}, D.~P., {Davis}, M., \& {Schlegel}, D.~J. 1999, \apj, 524, 867.
\newblock {Extrapolation of Galactic Dust Emission at 100 Microns to Cosmic
  Microwave Background Radiation Frequencies Using FIRAS}

\bibitem[{{Fixsen}(2009)}]{fixsen2009}
{Fixsen}, D.~J. 2009, \apj, 707, 916.
\newblock {The Temperature of the Cosmic Microwave Background}

\bibitem[{{Fixsen} {et~al.}(1997){Fixsen}, {Weiland}, {Brodd}, {Hauser},
  {Kelsall}, {Leisawitz}, {Mather}, {Jensen}, {Schafer}, \&
  {Silverberg}}]{fixsen1997}
{Fixsen}, D.~J., {Weiland}, J.~L., {Brodd}, S., {et~al.} 1997, \apj, 490, 482.
\newblock {Comparison of the COBE FIRAS and DIRBE Calibrations}

\bibitem[{{Fosalba} {et~al.}(2002){Fosalba}, {Dor{\'e}}, \&
  {Bouchet}}]{fosalba2002}
{Fosalba}, P., {Dor{\'e}}, O., \& {Bouchet}, F.~R. 2002, \prd, 65, 063003.
\newblock {Elliptical beams in CMB temperature and polarization anisotropy
  experiments: An analytic approach}

\bibitem[{{Frailis} {et~al.}(2009){Frailis}, {Maris}, {Zacchei}, {Morisset},
  {Rohlfs}, {Meharga}, {Binko}, {T{\"u}rler}, {Galeotta}, {Gasparo},
  {Franceschi}, {Butler}, {D'Arcangelo}, {Fogliani}, {Gregorio}, {Lowe},
  {Maggio}, {Malaspina}, {Mandolesi}, {Manzato}, {Pasian}, {Perrotta},
  {Sandri}, {Terenzi}, {Tomasi}, \& {Zonca}}]{frailis2009}
{Frailis}, M., {Maris}, M., {Zacchei}, A., {et~al.} 2009, Journal of
  Instrumentation, 4, 2021.
\newblock {A systematic approach to the Planck LFI end-to-end test and its
  application to the DPC Level 1 pipeline}

\bibitem[{{Giommi} {et~al.}(2012){Giommi}, {Polenta}, {L{\"a}hteenm{\"a}ki},
  {Thompson}, {Capalbi}, {Cutini}, {Gasparrini}, {Gonz{\'a}lez-Nuevo},
  {Le{\'o}n-Tavares}, {L{\'o}pez-Caniego}, {Mazziotta}, {Monte}, {Perri},
  {Rain{\`o}}, {Tosti}, {Tramacere}, {Verrecchia}, {Aller}, {Aller},
  {Angelakis}, {Bastieri}, {Berdyugin}, {Bonaldi}, {Bonavera}, {Burigana},
  {Burrows}, {Buson}, {Cavazzuti}, {Chincarini}, {Colafrancesco}, {Costamante},
  {Cuttaia}, {D'Ammando}, {de Zotti}, {Frailis}, {Fuhrmann}, {Galeotta},
  {Gargano}, {Gehrels}, {Giglietto}, {Giordano}, {Giroletti}, {Keih{\"a}nen},
  {King}, {Krichbaum}, {Lasenby}, {Lavonen}, {Lawrence}, {Leto}, {Lindfors},
  {Mandolesi}, {Massardi}, {Max-Moerbeck}, {Michelson}, {Mingaliev}, {Natoli},
  {Nestoras}, {Nieppola}, {Nilsson}, {Partridge}, {Pavlidou}, {Pearson},
  {Procopio}, {Rachen}, {Readhead}, {Reeves}, {Reimer}, {Reinthal},
  {Ricciardi}, {Richards}, {Riquelme}, {Saarinen}, {Sajina}, {Sandri},
  {Savolainen}, {Sievers}, {Sillanp{\"a}{\"a}}, {Sotnikova}, {Stevenson},
  {Tagliaferri}, {Takalo}, {Tammi}, {Tavagnacco}, {Terenzi}, {Toffolatti},
  {Tornikoski}, {Trigilio}, {Turunen}, {Umana}, {Ungerechts}, {Villa}, {Wu},
  {Zacchei}, {Zensus}, \& {Zhou}}]{planck2011-6.3b}
{Giommi}, P., {Polenta}, G., {L{\"a}hteenm{\"a}ki}, A., {et~al.} 2012, \aap,
  541, A160.
\newblock {Simultaneous \textit{Planck}, \textit{Swift}, and \textit{Fermi}
  observations of X-ray and $\gamma$-ray selected blazars}

\bibitem[{{Giorgini} {et~al.}(1996){Giorgini}, {Yeomans}, {Chamberlin},
  {Chodas}, {Jacobson}, {Keesey}, {Lieske}, {Ostro}, {Standish}, \&
  {Wimberly}}]{giorgini1996}
{Giorgini}, J.~D., {Yeomans}, D.~K., {Chamberlin}, A.~B., {et~al.} 1996, in
  Bulletin of the American Astronomical Society, Vol.~28, Bulletin of the
  American Astronomical Society, 1158

\bibitem[{{Gold} {et~al.}(2009){Gold}, {Bennett}, {Hill}, {Hinshaw}, {Odegard},
  {Spergel}, {Weiland}, {Dunkley}, {Halpern}, {Jarosik}, {Kogut}, {Komatsu},
  {Larson}, {Meyer}, {Nolta}, {Wollack}, \& {Wright}}]{gold2009}
{Gold}, B., {Bennett}, C.~L., {Hill}, R.~S., {et~al.} 2009, \apjs, 180, 265.
\newblock {Five-Year Wilkinson Microwave Anisotropy Probe (WMAP) Observations:
  Galactic Foreground Emission}

\bibitem[{{Gold} {et~al.}(2011){Gold}, {Odegard}, {Weiland}, {Hill}, {Kogut},
  {Bennett}, {Hinshaw}, {Chen}, {Dunkley}, {Halpern}, {Jarosik}, {Komatsu},
  {Larson}, {Limon}, {Meyer}, {Nolta}, {Page}, {Smith}, {Spergel}, {Tucker},
  {Wollack}, \& {Wright}}]{gold2010}
{Gold}, B., {Odegard}, N., {Weiland}, J.~L., {et~al.} 2011, \apjs, 192, 15.
\newblock {Seven-year Wilkinson Microwave Anisotropy Probe (WMAP) Observations:
  Galactic Foreground Emission}

\bibitem[{{Gonz{\'a}lez-Nuevo} {et~al.}(2008){Gonz{\'a}lez-Nuevo}, {Massardi},
  {Arg{\"u}eso}, {Herranz}, {Toffolatti}, {Sanz}, {L{\'o}pez-Caniego}, \& {de
  Zotti}}]{gonzalez2008}
{Gonz{\'a}lez-Nuevo}, J., {Massardi}, M., {Arg{\"u}eso}, F., {et~al.} 2008,
  \mnras, 384, 711.
\newblock {Statistical properties of extragalactic sources in the New
  Extragalactic WMAP Point Source (NEWPS) catalogue}

\bibitem[{{G{\'o}rski} {et~al.}(2005){G{\'o}rski}, {Hivon}, {Banday},
  {Wandelt}, {Hansen}, {Reinecke}, \& {Bartelmann}}]{gorski2005}
{G{\'o}rski}, K.~M., {Hivon}, E., {Banday}, A.~J., {et~al.} 2005, \apj, 622,
  759.
\newblock {HEALPix: A Framework for High-Resolution Discretization and Fast
  Analysis of Data Distributed on the Sphere}

\bibitem[{{Gregorio} {et~al.}(2013){Gregorio}, {Cuttaia}, {Mennella},
  {Bersanelli}, {Maris}, \& {Meinhold}}]{gregorio2013}
{Gregorio}, A., {Cuttaia}, F., {Mennella}, A., {et~al.} 2013, Journal of
  Instrumentation, 8, 7001.
\newblock {In-flight calibration and verification of the Planck-LFI instrument}

\bibitem[{{Hanson} {et~al.}(2009){Hanson}, {Rocha}, \&
  {G{\'o}rski}}]{hanson2009}
{Hanson}, D., {Rocha}, G., \& {G{\'o}rski}, K. 2009, \mnras, 400, 2169.
\newblock {Lensing reconstruction from Planck sky maps: inhomogeneous noise}

\bibitem[{{Haslam} {et~al.}(1982){Haslam}, {Salter}, {Stoffel}, \&
  {Wilson}}]{haslam1982}
{Haslam}, C.~G.~T., {Salter}, C.~J., {Stoffel}, H., \& {Wilson}, W.~E. 1982,
  \aaps, 47, 1.
\newblock {A 408 MHz all-sky continuum survey. II - The atlas of contour maps}

\bibitem[{{Herranz} {et~al.}(2009){Herranz}, {L{\'o}pez-Caniego}, {Sanz}, \&
  {Gonz{\'a}lez-Nuevo}}]{herranz2009}
{Herranz}, D., {L{\'o}pez-Caniego}, M., {Sanz}, J.~L., \& {Gonz{\'a}lez-Nuevo},
  J. 2009, \mnras, 394, 510.
\newblock {A novel multifrequency technique for the detection of point sources
  in cosmic microwave background maps}

\bibitem[{{Herranz} \& {Sanz}(2008)}]{herranz2008}
{Herranz}, D. \& {Sanz}, J.~L. 2008, IEEE Journal of Selected Topics in Signal
  Processing, 2, 727.
\newblock {Matrix Filters for the Detection of Extragalactic Point Sources in
  Cosmic Microwave Background Images}

\bibitem[{{Herreros} {et~al.}(2009){Herreros}, {G{\'o}mez}, {Rebolo},
  {Chulani}, {Rubi{\~n}o-Martin}, {Hildebrandt}, {Bersanelli}, {Butler},
  {Miccolis}, {Pe{\~n}a}, {Pereira}, {Torrero}, {Franceschet}, {L{\'o}pez}, \&
  {Alcal{\'a}}}]{herreros2009}
{Herreros}, J.~M., {G{\'o}mez}, M.~F., {Rebolo}, R., {et~al.} 2009, Journal of
  Instrumentation, 4, 2008.
\newblock {The Planck-LFI Radiometer Electronics Box Assembly}

\bibitem[{{Hill} {et~al.}(2009){Hill}, {Weiland}, {Odegard}, {Wollack},
  {Hinshaw}, {Larson}, {Bennett}, {Halpern}, {Page}, {Dunkley}, {Gold},
  {Jarosik}, {Kogut}, {Limon}, {Nolta}, {Spergel}, {Tucker}, \&
  {Wright}}]{hill2009}
{Hill}, R.~S., {Weiland}, J.~L., {Odegard}, N., {et~al.} 2009, \apjs, 180, 246.
\newblock {Five-Year Wilkinson Microwave Anisotropy Probe (WMAP) Observations:
  Beam Maps and Window Functions}

\bibitem[{{Hinshaw} {et~al.}(2003{\natexlab{a}}){Hinshaw}, {Barnes}, {Bennett},
  {Greason}, {Halpern}, {Hill}, {Jarosik}, {Kogut}, {Limon}, {Meyer},
  {Odegard}, {Page}, {Spergel}, {Tucker}, {Weiland}, {Wollack}, \&
  {Wright}}]{hinshaw2003a}
{Hinshaw}, G., {Barnes}, C., {Bennett}, C.~L., {et~al.} 2003{\natexlab{a}},
  \apjs, 148, 63.
\newblock {First-Year Wilkinson Microwave Anisotropy Probe (WMAP) Observations:
  Data Processing Methods and Systematic Error Limits}

\bibitem[{{Hinshaw} {et~al.}(2013){Hinshaw}, {Larson}, {Komatsu}, {Spergel},
  {Bennett}, {Dunkley}, {Nolta}, {Halpern}, {Hill}, {Odegard}, {Page}, {Smith},
  {Weiland}, {Gold}, {Jarosik}, {Kogut}, {Limon}, {Meyer}, {Tucker}, {Wollack},
  \& {Wright}}]{hinshaw2012}
{Hinshaw}, G., {Larson}, D., {Komatsu}, E., {et~al.} 2013, \apjs, 208, 19.
\newblock {Nine-year Wilkinson Microwave Anisotropy Probe (WMAP) Observations:
  Cosmological Parameter Results}

\bibitem[{{Hinshaw} {et~al.}(2007){Hinshaw}, {Nolta}, {Bennett}, {Bean},
  {Dor{\'e}}, {Greason}, {Halpern}, {Hill}, {Jarosik}, {Kogut}, {Komatsu},
  {Limon}, {Odegard}, {Meyer}, {Page}, {Peiris}, {Spergel}, {Tucker}, {Verde},
  {Weiland}, {Wollack}, \& {Wright}}]{hinshaw2007}
{Hinshaw}, G., {Nolta}, M.~R., {Bennett}, C.~L., {et~al.} 2007, \apjs, 170,
  288.
\newblock {Three-Year Wilkinson Microwave Anisotropy Probe (WMAP) Observations:
  Temperature Analysis}

\bibitem[{{Hinshaw} {et~al.}(2003{\natexlab{b}}){Hinshaw}, {Spergel}, {Verde},
  {Hill}, {Meyer}, {Barnes}, {Bennett}, {Halpern}, {Jarosik}, {Kogut},
  {Komatsu}, {Limon}, {Page}, {Tucker}, {Weiland}, {Wollack}, \&
  {Wright}}]{hinshaw2003b}
{Hinshaw}, G., {Spergel}, D.~N., {Verde}, L., {et~al.} 2003{\natexlab{b}},
  \apjs, 148, 135.
\newblock {First-Year Wilkinson Microwave Anisotropy Probe (WMAP) Observations:
  The Angular Power Spectrum}

\bibitem[{{Hinshaw} {et~al.}(2009){Hinshaw}, {Weiland}, {Hill}, {Odegard},
  {Larson}, {Bennett}, {Dunkley}, {Gold}, {Greason}, {Jarosik}, {Komatsu},
  {Nolta}, {Page}, {Spergel}, {Wollack}, {Halpern}, {Kogut}, {Limon}, {Meyer},
  {Tucker}, \& {Wright}}]{hinshaw2009}
{Hinshaw}, G., {Weiland}, J.~L., {Hill}, R.~S., {et~al.} 2009, \apjs, 180, 225.
\newblock {Five-Year Wilkinson Microwave Anisotropy Probe (WMAP) Observations:
  Data Processing, Sky Maps, and Basic Results}

\bibitem[{{Holmes} {et~al.}(2008){Holmes}, {Bock}, {Crill}, {Koch}, {Jones},
  {Lange}, \& {Paine}}]{holmes2008}
{Holmes}, W.~A., {Bock}, J.~J., {Crill}, B.~P., {et~al.} 2008, \ao, 47, 5996.
\newblock {Initial test results on bolometers for the Planck high frequency
  instrument}

\bibitem[{{Huffenberger} {et~al.}(2010){Huffenberger}, {Crill}, {Lange},
  {G{\'o}rski}, \& {Lawrence}}]{huffenberger2010}
{Huffenberger}, K.~M., {Crill}, B.~P., {Lange}, A.~E., {G{\'o}rski}, K.~M., \&
  {Lawrence}, C.~R. 2010, \aap, 510, A58.
\newblock {Measuring Planck beams with planets}

\bibitem[{{Jarosik} {et~al.}(2003){Jarosik}, {Barnes}, {Bennett}, {Halpern},
  {Hinshaw}, {Kogut}, {Limon}, {Meyer}, {Page}, {Spergel}, {Tucker}, {Weiland},
  {Wollack}, \& {Wright}}]{jarosik2003}
{Jarosik}, N., {Barnes}, C., {Bennett}, C.~L., {et~al.} 2003, \apjs, 148, 29.
\newblock {First-Year Wilkinson Microwave Anisotropy Probe (WMAP) Observations:
  On-Orbit Radiometer Characterization}

\bibitem[{{Jarosik} {et~al.}(2007){Jarosik}, {Barnes}, {Greason}, {Hill},
  {Nolta}, {Odegard}, {Weiland}, {Bean}, {Bennett}, {Dor{\'e}}, {Halpern},
  {Hinshaw}, {Kogut}, {Komatsu}, {Limon}, {Meyer}, {Page}, {Spergel}, {Tucker},
  {Wollack}, \& {Wright}}]{jarosik2007}
{Jarosik}, N., {Barnes}, C., {Greason}, M.~R., {et~al.} 2007, \apjs, 170, 263.
\newblock {Three-Year Wilkinson Microwave Anisotropy Probe (WMAP) Observations:
  Beam Profiles, Data Processing, Radiometer Characterization, and Systematic
  Error Limits}

\bibitem[{{Jarosik} {et~al.}(2011){Jarosik}, {Bennett}, {Dunkley}, {Gold},
  {Greason}, {Halpern}, {Hill}, {Hinshaw}, {Kogut}, {Komatsu}, {Larson},
  {Limon}, {Meyer}, {Nolta}, {Odegard}, {Page}, {Smith}, {Spergel}, {Tucker},
  {Weiland}, {Wollack}, \& {Wright}}]{jarosik2010}
{Jarosik}, N., {Bennett}, C.~L., {Dunkley}, J., {et~al.} 2011, \apjs, 192, 14.
\newblock {Seven-year Wilkinson Microwave Anisotropy Probe (WMAP) Observations:
  Sky Maps, Systematic Errors, and Basic Results}

\bibitem[{{Kaiser}(1974)}]{kaiser1974}
{Kaiser}, J.~F. 1974, Proc. 1974 IEEE Int. Symp. Circuit Theory, 20.
\newblock {Nonrecursive Digital Filter Design Using the Io-sinh Window
  Function}

\bibitem[{{Keih{\"a}nen} {et~al.}(2010){Keih{\"a}nen}, {Keskitalo},
  {Kurki-Suonio}, {Poutanen}, \& {Sirvi{\"o}}}]{keihanen2010}
{Keih{\"a}nen}, E., {Keskitalo}, R., {Kurki-Suonio}, H., {Poutanen}, T., \&
  {Sirvi{\"o}}, A. 2010, \aap, 510, A57.
\newblock {Making cosmic microwave background temperature and polarization maps
  with MADAM}

\bibitem[{{Keih{\"a}nen} {et~al.}(2005){Keih{\"a}nen}, {Kurki-Suonio}, \&
  {Poutanen}}]{keihanen2005}
{Keih{\"a}nen}, E., {Kurki-Suonio}, H., \& {Poutanen}, T. 2005, \mnras, 360,
  390.
\newblock {MADAM- a map-making method for CMB experiments}

\bibitem[{{Keih{\"a}nen} {et~al.}(2004){Keih{\"a}nen}, {Kurki-Suonio},
  {Poutanen}, {Maino}, \& {Burigana}}]{keihanen2004}
{Keih{\"a}nen}, E., {Kurki-Suonio}, H., {Poutanen}, T., {Maino}, D., \&
  {Burigana}, C. 2004, \aap, 428, 287.
\newblock {A maximum likelihood approach to the destriping technique}

\bibitem[{Keihanen \& Reinecke(2012)}]{keihanen2012}
Keihanen, E. \& Reinecke, M. 2012, Astronomy and Astrophysics, 548.
\newblock {ArtDeco: A beam deconvolution code for absolute CMB measurements}

\bibitem[{{Kelsall} {et~al.}(1998){Kelsall}, {Weiland}, {Franz}, {Reach},
  {Arendt}, {Dwek}, {Freudenreich}, {Hauser}, {Moseley}, {Odegard},
  {Silverberg}, \& {Wright}}]{kelsall1998}
{Kelsall}, T., {Weiland}, J.~L., {Franz}, B.~A., {et~al.} 1998, \apj, 508, 44.
\newblock {The COBE Diffuse Infrared Background Experiment Search for the
  Cosmic Infrared Background. II. Model of the Interplanetary Dust Cloud}

\bibitem[{{Keskitalo} {et~al.}(2010){Keskitalo}, {Ashdown}, {Cabella},
  {Kisner}, {Poutanen}, {Stompor}, {Bartlett}, {Borrill}, {Cantalupo}, {de
  Gasperis}, {de Rosa}, {de Troia}, {Eriksen}, {Finelli}, {G{\'o}rski},
  {Gruppuso}, {Hivon}, {Jaffe}, {Keih{\"a}nen}, {Kurki-Suonio}, {Lawrence},
  {Natoli}, {Paci}, {Polenta}, \& {Rocha}}]{keskitalo2010}
{Keskitalo}, R., {Ashdown}, M.~A.~J., {Cabella}, P., {et~al.} 2010, \aap, 522,
  A94.
\newblock {Residual noise covariance for Planck low-resolution data analysis}

\bibitem[{{Kogut} {et~al.}(2007){Kogut}, {Dunkley}, {Bennett}, {Dor{\'e}},
  {Gold}, {Halpern}, {Hinshaw}, {Jarosik}, {Komatsu}, {Nolta}, {Odegard},
  {Page}, {Spergel}, {Tucker}, {Weiland}, {Wollack}, \& {Wright}}]{kogut2007}
{Kogut}, A., {Dunkley}, J., {Bennett}, C.~L., {et~al.} 2007, \apj, 665, 355.
\newblock {Three-Year Wilkinson Microwave Anisotropy Probe (WMAP) Observations:
  Foreground Polarization}

\bibitem[{{Kogut} {et~al.}(2003){Kogut}, {Spergel}, {Barnes}, {Bennett},
  {Halpern}, {Hinshaw}, {Jarosik}, {Limon}, {Meyer}, {Page}, {Tucker},
  {Wollack}, \& {Wright}}]{kogut2003}
{Kogut}, A., {Spergel}, D.~N., {Barnes}, C., {et~al.} 2003, \apjs, 148, 161.
\newblock {First-Year Wilkinson Microwave Anisotropy Probe (WMAP) Observations:
  Temperature-Polarization Correlation}

\bibitem[{{Komatsu} {et~al.}(2009){Komatsu}, {Dunkley}, {Nolta}, {Bennett},
  {Gold}, {Hinshaw}, {Jarosik}, {Larson}, {Limon}, {Page}, {Spergel},
  {Halpern}, {Hill}, {Kogut}, {Meyer}, {Tucker}, {Weiland}, {Wollack}, \&
  {Wright}}]{komatsu2009}
{Komatsu}, E., {Dunkley}, J., {Nolta}, M.~R., {et~al.} 2009, \apjs, 180, 330.
\newblock {Five-Year Wilkinson Microwave Anisotropy Probe (WMAP) Observations:
  Cosmological Interpretation}

\bibitem[{{Komatsu} {et~al.}(2003){Komatsu}, {Kogut}, {Nolta}, {Bennett},
  {Halpern}, {Hinshaw}, {Jarosik}, {Limon}, {Meyer}, {Page}, {Spergel},
  {Tucker}, {Verde}, {Wollack}, \& {Wright}}]{komatsu2003}
{Komatsu}, E., {Kogut}, A., {Nolta}, M.~R., {et~al.} 2003, \apjs, 148, 119.
\newblock {First-Year Wilkinson Microwave Anisotropy Probe (WMAP) Observations:
  Tests of Gaussianity}

\bibitem[{{Komatsu} {et~al.}(2011){Komatsu}, {Smith}, {Dunkley}, {Bennett},
  {Gold}, {Hinshaw}, {Jarosik}, {Larson}, {Nolta}, {Page}, {Spergel},
  {Halpern}, {Hill}, {Kogut}, {Limon}, {Meyer}, {Odegard}, {Tucker}, {Weiland},
  {Wollack}, \& {Wright}}]{komatsu2010}
{Komatsu}, E., {Smith}, K.~M., {Dunkley}, J., {et~al.} 2011, \apjs, 192, 18.
\newblock {Seven-year Wilkinson Microwave Anisotropy Probe (WMAP) Observations:
  Cosmological Interpretation}

\bibitem[{{Kurki-Suonio} {et~al.}(2009){Kurki-Suonio}, {Keih{\"a}nen},
  {Keskitalo}, {Poutanen}, {Sirvi{\"o}}, {Maino}, \&
  {Burigana}}]{kurki-suonio2009}
{Kurki-Suonio}, H., {Keih{\"a}nen}, E., {Keskitalo}, R., {et~al.} 2009, \aap,
  506, 1511.
\newblock {Destriping CMB temperature and polarization maps}

\bibitem[{{Lamarre} {et~al.}(2010){Lamarre}, {Puget}, {Ade}, {Bouchet},
  {Guyot}, {Lange}, {Pajot}, {Arondel}, {Benabed}, {Beney}, {Beno{\^i}t},
  {Bernard}, {Bhatia}, {Blanc}, {Bock}, {Br{\'e}elle}, {Bradshaw}, {Camus},
  {Catalano}, {Charra}, {Charra}, {Church}, {Couchot}, {Coulais}, {Crill},
  {Crook}, {Dassas}, {de Bernardis}, {Delabrouille}, {de Marcillac}, {Delouis},
  {D{\'e}sert}, {Dumesnil}, {Dupac}, {Efstathiou}, {Eng}, {Evesque},
  {Fourmond}, {Ganga}, {Giard}, {Gispert}, {Guglielmi}, {Haissinski},
  {Henrot-Versill{\'e}}, {Hivon}, {Holmes}, {Jones}, {Koch}, {Lagard{\`e}re},
  {Lami}, {Land{\'e}}, {Leriche}, {Leroy}, {Longval},
  {Mac{\'{\i}}as-P{\'e}rez}, {Maciaszek}, {Maffei}, {Mansoux}, {Marty}, {Masi},
  {Mercier}, {Miville-Desch{\^e}nes}, {Moneti}, {Montier}, {Murphy},
  {Narbonne}, {Nexon}, {Paine}, {Pahn}, {Perdereau}, {Piacentini}, {Piat},
  {Plaszczynski}, {Pointecouteau}, {Pons}, {Ponthieu}, {Prunet}, {Rambaud},
  {Recouvreur}, {Renault}, {Ristorcelli}, {Rosset}, {Santos}, {Savini},
  {Serra}, {Stassi}, {Sudiwala}, {Sygnet}, {Tauber}, {Torre}, {Tristram},
  {Vibert}, {Woodcraft}, {Yurchenko}, \& {Yvon}}]{lamarre2010}
{Lamarre}, J., {Puget}, J., {Ade}, P.~A.~R., {et~al.} 2010, \aap, 520, A9.
\newblock {\textit{Planck} pre-launch status: The HFI instrument, from
  specification to actual performance}

\bibitem[{{Larson} {et~al.}(2011){Larson}, {Dunkley}, {Hinshaw}, {Komatsu},
  {Nolta}, {Bennett}, {Gold}, {Halpern}, {Hill}, {Jarosik}, {Kogut}, {Limon},
  {Meyer}, {Odegard}, {Page}, {Smith}, {Spergel}, {Tucker}, {Weiland},
  {Wollack}, \& {Wright}}]{larson2010}
{Larson}, D., {Dunkley}, J., {Hinshaw}, G., {et~al.} 2011, \apjs, 192, 16.
\newblock {Seven-year Wilkinson Microwave Anisotropy Probe (WMAP) Observations:
  Power Spectra and WMAP-derived Parameters}

\bibitem[{{Leach} {et~al.}(2008){Leach}, {Cardoso}, {Baccigalupi}, {Barreiro},
  {Betoule}, {Bobin}, {Bonaldi}, {Delabrouille}, {de Zotti}, {Dickinson},
  {Eriksen}, {Gonz{\'a}lez-Nuevo}, {Hansen}, {Herranz}, {Le Jeune},
  {L{\'o}pez-Caniego}, {Mart{\'{\i}}nez-Gonz{\'a}lez}, {Massardi}, {Melin},
  {Miville-Desch{\^e}nes}, {Patanchon}, {Prunet}, {Ricciardi}, {Salerno},
  {Sanz}, {Starck}, {Stivoli}, {Stolyarov}, {Stompor}, \& {Vielva}}]{leach2008}
{Leach}, S.~M., {Cardoso}, J., {Baccigalupi}, C., {et~al.} 2008, \aap, 491,
  597.
\newblock {Component separation methods for the PLANCK mission}

\bibitem[{{Leahy} {et~al.}(2010){Leahy}, {Bersanelli}, {D'Arcangelo}, {Ganga},
  {Leach}, {Moss}, {Keih{\"a}nen}, {Keskitalo}, {Kurki-Suonio}, {Poutanen},
  {Sandri}, {Scott}, {Tauber}, {Valenziano}, {Villa}, {Wilkinson}, {Zonca},
  {Baccigalupi}, {Borrill}, {Butler}, {Cuttaia}, {Davis}, {Frailis},
  {Francheschi}, {Galeotta}, {Gregorio}, {Leonardi}, {Mandolesi}, {Maris},
  {Meinhold}, {Mendes}, {Mennella}, {Morgante}, {Prezeau}, {Rocha},
  {Stringhetti}, {Terenzi}, \& {Tomasi}}]{leahy2010}
{Leahy}, J.~P., {Bersanelli}, M., {D'Arcangelo}, O., {et~al.} 2010, \aap, 520,
  A8.
\newblock {\textit{Planck} pre-launch status: Expected LFI polarisation
  capability}

\bibitem[{{Mac{\'{\i}}as-P{\'e}rez} {et~al.}(2007){Mac{\'{\i}}as-P{\'e}rez},
  {Lagache}, {Maffei}, {Ganga}, {Bourrachot}, {Ade}, {Amblard}, {Ansari},
  {Aubourg}, {Aumont}, {Bargot}, {Bartlett}, {Beno{\^i}t}, {Bernard}, {Bhatia},
  {Blanchard}, {Bock}, {Boscaleri}, {Bouchet}, {Camus}, {Cardoso}, {Couchot},
  {de Bernardis}, {Delabrouille}, {D{\'e}sert}, {Dor{\'e}}, {Douspis},
  {Dumoulin}, {Dupac}, {Filliatre}, {Fosalba}, {Gannaway}, {Gautier}, {Giard},
  {Giraud-H{\'e}raud}, {Gispert}, {Guglielmi}, {Hamilton}, {Hanany},
  {Henrot-Versill{\'e}}, {Hristov}, {Kaplan}, {Lamarre}, {Lange}, {Madet},
  {Magneville}, {Marrone}, {Masi}, {Mayet}, {Murphy}, {Naraghi}, {Nati},
  {Patanchon}, {Perdereau}, {Perrin}, {Plaszczynski}, {Piat}, {Ponthieu},
  {Prunet}, {Puget}, {Renault}, {Rosset}, {Santos}, {Starobinsky}, {Strukov},
  {Sudiwala}, {Teyssier}, {Tristram}, {Tucker}, {Vanel}, {Vibert}, {Wakui}, \&
  {Yvon}}]{macias2007}
{Mac{\'{\i}}as-P{\'e}rez}, J.~F., {Lagache}, G., {Maffei}, B., {et~al.} 2007,
  \aap, 467, 1313.
\newblock {Archeops in-flight performance, data processing, and map making}

\bibitem[{{Maffei} {et~al.}(2010){Maffei}, {Noviello}, {Murphy}, {Ade},
  {Lamarre}, {Bouchet}, {Brossard}, {Catalano}, {Colgan}, {Gispert}, {Gleeson},
  {Haynes}, {Jones}, {Lange}, {Longval}, {McAuley}, {Pajot}, {Peacocke},
  {Pisano}, {Puget}, {Ristorcelli}, {Savini}, {Sudiwala}, {Wylde}, \&
  {Yurchenko}}]{maffei2010}
{Maffei}, B., {Noviello}, F., {Murphy}, J.~A., {et~al.} 2010, \aap, 520, A12.
\newblock {\textit{Planck} pre-launch status: HFI beam expectations from the
  optical optimisation of the focal plane}

\bibitem[{{Maino} {et~al.}(2002){Maino}, {Burigana}, {G{\'o}rski}, {Mandolesi},
  \& {Bersanelli}}]{maino2002}
{Maino}, D., {Burigana}, C., {G{\'o}rski}, K.~M., {Mandolesi}, N., \&
  {Bersanelli}, M. 2002, \aap, 387, 356.
\newblock {Removing 1/f noise stripes in cosmic microwave background anisotropy
  observations}

\bibitem[{{Malaspina} {et~al.}(2009){Malaspina}, {Franceschi}, {Battaglia},
  {Binko}, {Butler}, {D'Arcangelo}, {Fogliani}, {Frailis}, {Franceschet},
  {Galeotta}, {Gasparo}, {Gregorio}, {Lapolla}, {Leonardi}, {Maggio},
  {Mandolesi}, {Manzato}, {Maris}, {Meharga}, {Meinhold}, {Morisset}, {Pasian},
  {Perrotta}, {Rohlfs}, {Sandri}, {Tomasi}, {T{\"u}rler}, {Zacchei}, \&
  {Zonca}}]{malaspina2009}
{Malaspina}, M., {Franceschi}, E., {Battaglia}, P., {et~al.} 2009, Journal of
  Instrumentation, 4, 2017.
\newblock {LFI Radiometric Chain Assembly (RCA) data handling ``Rachel''}

\bibitem[{{Mandolesi} {et~al.}(2010){Mandolesi}, {Bersanelli}, {Butler},
  {Artal}, {Baccigalupi}, {Balbi}, {Banday}, {Barreiro}, {Bartelmann},
  {Bennett}, {Bhandari}, {Bonaldi}, {Borrill}, {Bremer}, {Burigana}, {Bowman},
  {Cabella}, {Cantalupo}, {Cappellini}, {Courvoisier}, {Crone}, {Cuttaia},
  {Danese}, {D'Arcangelo}, {Davies}, {Davis}, {de Angelis}, {de Gasperis}, {de
  Rosa}, {de Troia}, {de Zotti}, {Dick}, {Dickinson}, {Diego}, {Donzelli},
  {D{\"o}rl}, {Dupac}, {En{\ss}lin}, {Eriksen}, {Falvella}, {Finelli},
  {Frailis}, {Franceschi}, {Gaier}, {Galeotta}, {Gasparo}, {Giardino}, {Gomez},
  {Gonzalez-Nuevo}, {G{\'o}rski}, {Gregorio}, {Gruppuso}, {Hansen}, {Hell},
  {Herranz}, {Herreros}, {Hildebrandt}, {Hovest}, {Hoyland}, {Huffenberger},
  {Janssen}, {Jaffe}, {Keih{\"a}nen}, {Keskitalo}, {Kisner}, {Kurki-Suonio},
  {L{\"a}hteenm{\"a}ki}, {Lawrence}, {Leach}, {Leahy}, {Leonardi}, {Levin},
  {Lilje}, {L{\'o}pez-Caniego}, {Lowe}, {Lubin}, {Maino}, {Malaspina}, {Maris},
  {Marti-Canales}, {Martinez-Gonzalez}, {Massardi}, {Matarrese}, {Matthai},
  {Meinhold}, {Melchiorri}, {Mendes}, {Mennella}, {Morgante}, {Morigi},
  {Morisset}, {Moss}, {Nash}, {Natoli}, {Nesti}, {Paine}, {Partridge},
  {Pasian}, {Passvogel}, {Pearson}, {P{\'e}rez-Cuevas}, {Perrotta}, {Polenta},
  {Popa}, {Poutanen}, {Prezeau}, {Prina}, {Rachen}, {Rebolo}, {Reinecke},
  {Ricciardi}, {Riller}, {Rocha}, {Roddis}, {Rohlfs}, {Rubi{\~n}o-Martin},
  {Salerno}, {Sandri}, {Scott}, {Seiffert}, {Silk}, {Simonetto}, {Smoot},
  {Sozzi}, {Sternberg}, {Stivoli}, {Stringhetti}, {Tauber}, {Terenzi},
  {Tomasi}, {Tuovinen}, {T{\"u}rler}, {Valenziano}, {Varis}, {Vielva}, {Villa},
  {Vittorio}, {Wade}, {White}, {White}, {Wilkinson}, {Zacchei}, \&
  {Zonca}}]{mandolesi2010}
{Mandolesi}, N., {Bersanelli}, M., {Butler}, R.~C., {et~al.} 2010, \aap, 520,
  A3.
\newblock {\textit{Planck} pre-launch status: The Planck-LFI programme}

\bibitem[{{Maris} {et~al.}(2009){Maris}, {Tomasi}, {Galeotta}, {Miccolis},
  {Hildebrandt}, {Frailis}, {Rohlfs}, {Morisset}, {Zacchei}, {Bersanelli},
  {Binko}, {Burigana}, {Butler}, {Cuttaia}, {Chulani}, {D'Arcangelo},
  {Fogliani}, {Franceschi}, {Gasparo}, {Gomez}, {Gregorio}, {Herreros},
  {Leonardi}, {Leutenegger}, {Maggio}, {Maino}, {Malaspina}, {Mandolesi},
  {Manzato}, {Meharga}, {Meinhold}, {Mennella}, {Pasian}, {Perrotta}, {Rebolo},
  {T{\"u}rler}, \& {Zonca}}]{maris2009}
{Maris}, M., {Tomasi}, M., {Galeotta}, S., {et~al.} 2009, Journal of
  Instrumentation, 4, 2018.
\newblock {Optimization of Planck-LFI on-board data handling}

\bibitem[{{Massardi}(2006)}]{massardi2006}
{Massardi}, M. 2006, in CMB and Physics of the Early Universe, 45

\bibitem[{{Meinhold} {et~al.}(2009){Meinhold}, {Leonardi}, {Aja}, {Artal},
  {Battaglia}, {Bersanelli}, {Blackhurst}, {Butler}, {Cuevas}, {Cuttaia},
  {D'Arcangelo}, {Davis}, {de la Fuente}, {Frailis}, {Franceschet},
  {Franceschi}, {Gaier}, {Galeotta}, {Gregorio}, {Hoyland}, {Hughes},
  {Jukkala}, {Kettle}, {Laaninen}, {Leutenegger}, {Lowe}, {Malaspina},
  {Mandolesi}, {Maris}, {Mart{\'{\i}}nez-Gonz{\'a}lez}, {Mendes}, {Mennella},
  {Miccolis}, {Morgante}, {Roddis}, {Sandri}, {Seiffert}, {Salm{\'o}n},
  {Stringhetti}, {Poutanen}, {Terenzi}, {Tomasi}, {Tuovinen}, {Varis},
  {Valenziano}, {Villa}, {Wilkinson}, {Winder}, {Zacchei}, \&
  {Zonca}}]{meinhold2009}
{Meinhold}, P., {Leonardi}, R., {Aja}, B., {et~al.} 2009, Journal of
  Instrumentation, 4, 2009.
\newblock {Noise properties of the Planck-LFI receivers}

\bibitem[{{Melin} {et~al.}(2006){Melin}, {Bartlett}, \&
  {Delabrouille}}]{melin2006}
{Melin}, J., {Bartlett}, J.~G., \& {Delabrouille}, J. 2006, \aap, 459, 341.
\newblock {Catalog extraction in SZ cluster surveys: a matched filter approach}

\bibitem[{{Mennella} {et~al.}(2002){Mennella}, {Bersanelli}, {Burigana},
  {Maino}, {Mandolesi}, {Morgante}, \& {Stanghellini}}]{mennella2002}
{Mennella}, A., {Bersanelli}, M., {Burigana}, C., {et~al.} 2002, \aap, 384,
  736.
\newblock {PLANCK: Systematic effects induced by periodic fluctuations of
  arbitrary shape}

\bibitem[{{Mennella} {et~al.}(2010){Mennella}, {Bersanelli}, {Butler},
  {Cuttaia}, {D'Arcangelo}, {Davis}, {Frailis}, {Galeotta}, {Gregorio},
  {Lawrence}, {Leonardi}, {Lowe}, {Mandolesi}, {Maris}, {Meinhold}, {Mendes},
  {Morgante}, {Sandri}, {Stringhetti}, {Terenzi}, {Tomasi}, {Valenziano},
  {Villa}, {Zacchei}, {Zonca}, {Balasini}, {Franceschet}, {Battaglia},
  {Lapolla}, {Leutenegger}, {Miccolis}, {Pagan}, {Silvestri}, {Aja}, {Artal},
  {Baldan}, {Bastia}, {Bernardino}, {Boschini}, {Cafagna}, {Cappellini},
  {Cavaliere}, {Colombo}, {de La Fuente}, {Edgeley}, {Falvella}, {Ferrari},
  {Fogliani}, {Franceschi}, {Gaier}, {Gomez}, {Herreros}, {Hildebrandt},
  {Hoyland}, {Hughes}, {Jukkala}, {Kettle}, {Laaninen}, {Lawson}, {Leahy},
  {Levin}, {Lilje}, {Maino}, {Malaspina}, {Manzato}, {Marti-Canales},
  {Martinez-Gonzalez}, {Mediavilla}, {Pasian}, {Pascual}, {Pecora},
  {Peres-Cuevas}, {Platania}, {Pospieszalsky}, {Poutanen}, {Rebolo}, {Roddis},
  {Salmon}, {Seiffert}, {Simonetto}, {Sozzi}, {Tauber}, {Tuovinen}, {Varis},
  {Wilkinson}, \& {Winder}}]{mennella2010}
{Mennella}, A., {Bersanelli}, M., {Butler}, R.~C., {et~al.} 2010, \aap, 520,
  A5.
\newblock {\textit{Planck} pre-launch status: Low Frequency Instrument
  calibration and expected scientific performance}

\bibitem[{{Mennella} {et~al.}(2003){Mennella}, {Bersanelli}, {Seiffert},
  {Kettle}, {Roddis}, {Wilkinson}, \& {Meinhold}}]{mennella2003}
{Mennella}, A., {Bersanelli}, M., {Seiffert}, M., {et~al.} 2003, \aap, 410,
  1089.
\newblock {Offset balancing in pseudo-correlation radiometers for CMB
  measurements}

\bibitem[{{Mennella} {et~al.}(2011){Mennella}, {Butler}, {Curto}, {Cuttaia},
  {Davis}, {Dick}, {Frailis}, {Galeotta}, {Gregorio}, {Kurki-Suonio},
  {Lawrence}, {Leach}, {Leahy}, {Lowe}, {Maino}, {Mandolesi}, {Maris},
  {Mart{\'{\i}}nez-Gonz{\'a}lez}, {Meinhold}, {Morgante}, {Pearson},
  {Perrotta}, {Polenta}, {Poutanen}, {Sandri}, {Seiffert}, {Suur-Uski},
  {Tavagnacco}, {Terenzi}, {Tomasi}, {Valiviita}, {Villa}, {Watson},
  {Wilkinson}, {Zacchei}, {Zonca}, {Aja}, {Artal}, {Baccigalupi}, {Banday},
  {Barreiro}, {Bartlett}, {Bartolo}, {Battaglia}, {Bennett}, {Bonaldi},
  {Bonavera}, {Borrill}, {Bouchet}, {Burigana}, {Cabella}, {Cappellini},
  {Chen}, {Colombo}, {Cruz}, {Danese}, {D'Arcangelo}, {Davies}, {de Gasperis},
  {de Rosa}, {de Zotti}, {Dickinson}, {Diego}, {Donzelli}, {Efstathiou},
  {En{\ss}lin}, {Eriksen}, {Falvella}, {Finelli}, {Foley}, {Franceschet},
  {Franceschi}, {Gaier}, {G{\'e}nova-Santos}, {George}, {G{\'o}mez},
  {Gonz{\'a}lez-Nuevo}, {G{\'o}rski}, {Gruppuso}, {Hansen}, {Herranz},
  {Herreros}, {Hoyland}, {Hughes}, {Jewell}, {Jukkala}, {Juvela},
  {Kangaslahti}, {Keih{\"a}nen}, {Keskitalo}, {Kilpia}, {Kisner}, {Knoche},
  {Knox}, {Laaninen}, {L{\"a}hteenm{\"a}ki}, {Lamarre}, {Leonardi},
  {Le{\'o}n-Tavares}, {Leutenegger}, {Lilje}, {L{\'o}pez-Caniego}, {Lubin},
  {Malaspina}, {Marinucci}, {Massardi}, {Matarrese}, {Matthai}, {Melchiorri},
  {Mendes}, {Miccolis}, {Migliaccio}, {Mitra}, {Moss}, {Natoli}, {Nesti},
  {N{\o}rgaard-Nielsen}, {Pagano}, {Paladini}, {Paoletti}, {Partridge},
  {Pasian}, {Pettorino}, {Pietrobon}, {Pospieszalski}, {Pr{\'e}zeau}, {Prina},
  {Procopio}, {Puget}, {Quercellini}, {Rachen}, {Rebolo}, {Reinecke},
  {Ricciardi}, {Robbers}, {Rocha}, {Roddis}, {Rubino-Mart{\'{\i}}n},
  {Savelainen}, {Scott}, {Silvestri}, {Simonetto}, {Sjoman}, {Smoot}, {Sozzi},
  {Stringhetti}, {Tauber}, {Tofani}, {Toffolatti}, {Tuovinen}, {T{\"u}rler},
  {Umana}, {Valenziano}, {Varis}, {Vielva}, {Vittorio}, {Wade}, {Watson},
  {White}, \& {Winder}}]{planck2011-1.4}
{Mennella}, A., {Butler}, R.~C., {Curto}, A., {et~al.} 2011, \aap, 536, A3.
\newblock {\textit{Planck} early results. III. First assessment of the Low
  Frequency Instrument in-flight performance}

\bibitem[{{Mennella} {et~al.}(2009){Mennella}, {Villa}, {Terenzi}, {Cuttaia},
  {Battaglia}, {Bersanelli}, {Butler}, {D'Arcangelo}, {Artal}, {Davis},
  {Frailis}, {Franceschet}, {Galeotta}, {Gregorio}, {Hughes}, {Jukkala},
  {Kettle}, {Kilpi{\"a}}, {Laaninen}, {Lapolla}, {Leonardi}, {Leutenegger},
  {Lowe}, {Mandolesi}, {Maris}, {Meinhold}, {Mendes}, {Miccolis}, {Morgante},
  {Roddis}, {Sandri}, {Silvestri}, {Stringhetti}, {Tomasi}, {Tuovinen},
  {Valenziano}, {Zacchei}, {Varis}, {Wilkinson}, \& {Zonca}}]{mennella2009}
{Mennella}, A., {Villa}, F., {Terenzi}, L., {et~al.} 2009, Journal of
  Instrumentation, 4, 2011.
\newblock {The linearity response of the Planck-LFI flight model receivers}

\bibitem[{{Mitra} {et~al.}(2011){Mitra}, {Rocha}, {G{\'o}rski}, {Huffenberger},
  {Eriksen}, {Ashdown}, \& {Lawrence}}]{mitra2010}
{Mitra}, S., {Rocha}, G., {G{\'o}rski}, K.~M., {et~al.} 2011, \apjs, 193, 5.
\newblock {Fast Pixel Space Convolution for Cosmic Microwave Background Surveys
  with Asymmetric Beams and Complex Scan Strategies: FEBeCoP}

\bibitem[{{Morgante} {et~al.}(2009){Morgante}, {Pearson}, {Melot}, {Stassi},
  {Terenzi}, {Wilson}, {Hernandez}, {Wade}, {Gregorio}, {Bersanelli}, {Butler},
  \& {Mandolesi}}]{morgante2009}
{Morgante}, G., {Pearson}, D., {Melot}, F., {et~al.} 2009, Journal of
  Instrumentation, 4, 2016.
\newblock {Cryogenic characterization of the Planck sorption cooler system
  flight model}

\bibitem[{{Natoli} {et~al.}(2001){Natoli}, {de Gasperis}, {Gheller}, \&
  {Vittorio}}]{natoli2001}
{Natoli}, P., {de Gasperis}, G., {Gheller}, C., \& {Vittorio}, N. 2001, \aap,
  372, 346.
\newblock {A Map-Making algorithm for the Planck Surveyor.}

\bibitem[{{Nolta} {et~al.}(2009){Nolta}, {Dunkley}, {Hill}, {Hinshaw},
  {Komatsu}, {Larson}, {Page}, {Spergel}, {Bennett}, {Gold}, {Jarosik},
  {Odegard}, {Weiland}, {Wollack}, {Halpern}, {Kogut}, {Limon}, {Meyer},
  {Tucker}, \& {Wright}}]{nolta2009}
{Nolta}, M.~R., {Dunkley}, J., {Hill}, R.~S., {et~al.} 2009, \apjs, 180, 296.
\newblock {Five-Year Wilkinson Microwave Anisotropy Probe (WMAP) Observations:
  Angular Power Spectra}

\bibitem[{{Nolta} {et~al.}(2004){Nolta}, {Wright}, {Page}, {Bennett},
  {Halpern}, {Hinshaw}, {Jarosik}, {Kogut}, {Limon}, {Meyer}, {Spergel},
  {Tucker}, \& {Wollack}}]{nolta2004}
{Nolta}, M.~R., {Wright}, E.~L., {Page}, L., {et~al.} 2004, \apj, 608, 10.
\newblock {First Year Wilkinson Microwave Anisotropy Probe Observations: Dark
  Energy Induced Correlation with Radio Sources}

\bibitem[{{Onishi} {et~al.}(2002){Onishi}, {Mizuno}, {Kawamura}, {Tachihara},
  \& {Fukui}}]{onishi2002}
{Onishi}, T., {Mizuno}, A., {Kawamura}, A., {Tachihara}, K., \& {Fukui}, Y.
  2002, \apj, 575, 950.
\newblock {A Complete Search for Dense Cloud Cores in Taurus}

\bibitem[{{Page} {et~al.}(2003{\natexlab{a}}){Page}, {Barnes}, {Hinshaw},
  {Spergel}, {Weiland}, {Wollack}, {Bennett}, {Halpern}, {Jarosik}, {Kogut},
  {Limon}, {Meyer}, {Tucker}, \& {Wright}}]{page2003a}
{Page}, L., {Barnes}, C., {Hinshaw}, G., {et~al.} 2003{\natexlab{a}}, \apjs,
  148, 39.
\newblock {First-Year Wilkinson Microwave Anisotropy Probe (WMAP) Observations:
  Beam Profiles and Window Functions}

\bibitem[{{Page} {et~al.}(2007){Page}, {Hinshaw}, {Komatsu}, {Nolta},
  {Spergel}, {Bennett}, {Barnes}, {Bean}, {Dor{\'e}}, {Dunkley}, {Halpern},
  {Hill}, {Jarosik}, {Kogut}, {Limon}, {Meyer}, {Odegard}, {Peiris}, {Tucker},
  {Verde}, {Weiland}, {Wollack}, \& {Wright}}]{page2007}
{Page}, L., {Hinshaw}, G., {Komatsu}, E., {et~al.} 2007, \apjs, 170, 335.
\newblock {Three-Year Wilkinson Microwave Anisotropy Probe (WMAP) Observations:
  Polarization Analysis}

\bibitem[{{Page} {et~al.}(2003{\natexlab{b}}){Page}, {Nolta}, {Barnes},
  {Bennett}, {Halpern}, {Hinshaw}, {Jarosik}, {Kogut}, {Limon}, {Meyer},
  {Peiris}, {Spergel}, {Tucker}, {Wollack}, \& {Wright}}]{page2003b}
{Page}, L., {Nolta}, M.~R., {Barnes}, C., {et~al.} 2003{\natexlab{b}}, \apjs,
  148, 233.
\newblock {First-Year Wilkinson Microwave Anisotropy Probe (WMAP) Observations:
  Interpretation of the TT and TE Angular Power Spectrum Peaks}

\bibitem[{{Pajot} {et~al.}(2010){Pajot}, {Ade}, {Beney}, {Br{\'e}elle},
  {Broszkiewicz}, {Camus}, {Carab{\'e}tian}, {Catalano}, {Chardin}, {Charra},
  {Charra}, {Cizeron}, {Couchot}, {Coulais}, {Crill}, {Dassas}, {Daubin}, {de
  Bernardis}, {de Marcillac}, {Delouis}, {D{\'e}sert}, {Duret}, {Eng},
  {Evesque}, {Fourmond}, {Fran{\c c}ois}, {Giard}, {Giraud-H{\'e}raud},
  {Guglielmi}, {Guyot}, {Haissinski}, {Henrot-Versill{\'e}}, {Hervier},
  {Holmes}, {Jones}, {Lamarre}, {Lami}, {Lange}, {Lefebvre}, {Leriche},
  {Leroy}, {Macias-Perez}, {Maciaszek}, {Maffei}, {Mahendran}, {Mansoux},
  {Marty}, {Masi}, {Mercier}, {Miville-Deschenes}, {Montier}, {Nicolas},
  {Noviello}, {Perdereau}, {Piacentini}, {Piat}, {Plaszczynski},
  {Pointecouteau}, {Pons}, {Ponthieu}, {Puget}, {Rambaud}, {Renault},
  {Renault}, {Rioux}, {Ristorcelli}, {Rosset}, {Savini}, {Sudiwala}, {Torre},
  {Tristram}, {Vall{\'e}e}, {Veneziani}, \& {Yvon}}]{pajot2010}
{Pajot}, F., {Ade}, P.~A.~R., {Beney}, J., {et~al.} 2010, \aap, 520, A10.
\newblock {\textit{Planck} pre-launch status: HFI ground calibration}

\bibitem[{{Peiris} {et~al.}(2003){Peiris}, {Komatsu}, {Verde}, {Spergel},
  {Bennett}, {Halpern}, {Hinshaw}, {Jarosik}, {Kogut}, {Limon}, {Meyer},
  {Page}, {Tucker}, {Wollack}, \& {Wright}}]{peiris2003}
{Peiris}, H.~V., {Komatsu}, E., {Verde}, L., {et~al.} 2003, \apjs, 148, 213.
\newblock {First-Year Wilkinson Microwave Anisotropy Probe (WMAP) Observations:
  Implications For Inflation}

\bibitem[{{Planck Collaboration}(2005)}]{planck2005-bluebook}
{Planck Collaboration}. 2005, ESA publication ESA-SCI(2005)/01.
\newblock {The Scientific Programme of Planck}

\bibitem[{{Planck Collaboration}(2011)}]{planck2011-1.10sup}
{Planck Collaboration}. 2011, {The Explanatory Supplement to the Planck Early
  Release Compact Source Catalogue} ({ESA})

\bibitem[{{Planck Collaboration ES}(2013)}]{planck2013-p28}
{Planck Collaboration ES}. 2013, {The Explanatory Supplement to the
  \textit{Planck} 2013 results, http://pla.esac.esa.int/pla/index.html} ({ESA})

\bibitem[{{Planck Collaboration ES}(2015)}]{planck2014-ES}
{Planck Collaboration ES}. 2015, {The Explanatory Supplement to the \Planck\
  2015 results, \url{http://wiki.cosmos.esa.int/planckpla/index.php/Main_Page}}
  ({ESA})

\bibitem[{{Planck Collaboration XXXII}(2015)}]{planck2013-p05a-addendum}
{Planck Collaboration XXXII}, {\sorthelp{Planck Collaboration 2014ZG}}. 2015,
  \aap, 581, A14.
\newblock {\textit{Planck} 2013 results. XXXII. The updated Planck catalogue of
  Sunyaev-Zeldovich sources}

\bibitem[{{Planck HFI Core Team}(2011{\natexlab{a}})}]{planck2011-1.5}
{Planck HFI Core Team}. 2011{\natexlab{a}}, \aap, 536, A4.
\newblock {\textit{Planck} early results, IV. First assessment of the High
  Frequency Instrument in-flight performance}

\bibitem[{{Planck HFI Core Team}(2011{\natexlab{b}})}]{planck2011-1.7}
{Planck HFI Core Team}. 2011{\natexlab{b}}, \aap, 536, A6.
\newblock {\textit{Planck} early results. VI. The High Frequency Instrument
  data processing}

\bibitem[{{\sorthelp{Planck Collaboration 2011A}}{Planck Collaboration
  I}(2011)}]{planck2011-1.1}
{\sorthelp{Planck Collaboration 2011A}}{Planck Collaboration I}. 2011, \aap,
  536, A1.
\newblock {\textit{Planck} early results. I. The Planck mission}

\bibitem[{{\sorthelp{Planck Collaboration 2011B}}{Planck Collaboration
  II}(2011)}]{planck2011-1.3}
{\sorthelp{Planck Collaboration 2011B}}{Planck Collaboration II}. 2011, \aap,
  536, A2.
\newblock {\textit{Planck} early results. II. The thermal performance of
  \textit{Planck}}

\bibitem[{{\sorthelp{Planck Collaboration 2011G}}{Planck Collaboration
  VII}(2011)}]{planck2011-1.10}
{\sorthelp{Planck Collaboration 2011G}}{Planck Collaboration VII}. 2011, \aap,
  536, A7.
\newblock {\textit{Planck} early results. VII. The Early Release Compact Source
  Catalogue}

\bibitem[{{\sorthelp{Planck Collaboration 2011H}}{Planck Collaboration
  VIII}(2011)}]{planck2011-5.1a}
{\sorthelp{Planck Collaboration 2011H}}{Planck Collaboration VIII}. 2011, \aap,
  536, A8.
\newblock {\textit{Planck} early results. VIII. The all-sky early
  Sunyaev-Zeldovich cluster sample}

\bibitem[{{\sorthelp{Planck Collaboration 2011I}}{Planck Collaboration
  IX}(2011)}]{planck2011-5.1b}
{\sorthelp{Planck Collaboration 2011I}}{Planck Collaboration IX}. 2011, \aap,
  536, A9.
\newblock {\textit{Planck} early results. IX. XMM-\textit{Newton} follow-up
  validation programme of \textit{Planck} cluster candidates}

\bibitem[{{\sorthelp{Planck Collaboration 2011J}}{Planck Collaboration
  X}(2011)}]{planck2011-5.2a}
{\sorthelp{Planck Collaboration 2011J}}{Planck Collaboration X}. 2011, \aap,
  536, A10.
\newblock {\textit{Planck} early results. X. Statistical analysis of
  Sunyaev-Zeldovich scaling relations for X-ray galaxy clusters}

\bibitem[{{\sorthelp{Planck Collaboration 2011K}}{Planck Collaboration
  XI}(2011)}]{planck2011-5.2b}
{\sorthelp{Planck Collaboration 2011K}}{Planck Collaboration XI}. 2011, \aap,
  536, A11.
\newblock {\textit{Planck} early results. XI. Calibration of the local galaxy
  cluster Sunyaev-Zeldovich scaling relations}

\bibitem[{{\sorthelp{Planck Collaboration 2011L}}{Planck Collaboration
  XII}(2011)}]{planck2011-5.2c}
{\sorthelp{Planck Collaboration 2011L}}{Planck Collaboration XII}. 2011, \aap,
  536, A12.
\newblock {\textit{Planck} early results. XII. Cluster Sunyaev-Zeldovich
  optical scaling relations}

\bibitem[{{\sorthelp{Planck Collaboration 2011M}}{Planck Collaboration
  XIII}(2011)}]{planck2011-6.1}
{\sorthelp{Planck Collaboration 2011M}}{Planck Collaboration XIII}. 2011, \aap,
  536, A13.
\newblock {\textit{Planck} early results. XIII. Statistical properties of
  extragalactic radio sources in the Planck Early Release Compact Source
  Catalogue}

\bibitem[{{\sorthelp{Planck Collaboration 2011N}}{Planck Collaboration
  XIV}(2011)}]{planck2011-6.2}
{\sorthelp{Planck Collaboration 2011N}}{Planck Collaboration XIV}. 2011, \aap,
  536, A14.
\newblock {\textit{Planck} early results. XIV. ERCSC validation and extreme
  radio sources}

\bibitem[{{\sorthelp{Planck Collaboration 2011O}}{Planck Collaboration
  XV}(2011)}]{planck2011-6.3a}
{\sorthelp{Planck Collaboration 2011O}}{Planck Collaboration XV}. 2011, \aap,
  536, A15.
\newblock {\textit{Planck} early results. XV. Spectral energy distributions and
  radio continuum spectra of northern extragalactic radio sources}

\bibitem[{{\sorthelp{Planck Collaboration 2011P}}{Planck Collaboration
  XVI}(2011)}]{planck2011-6.4a}
{\sorthelp{Planck Collaboration 2011P}}{Planck Collaboration XVI}. 2011, \aap,
  536, A16.
\newblock {\textit{Planck} early results. XVI. The \textit{Planck} view of
  nearby galaxies}

\bibitem[{{\sorthelp{Planck Collaboration 2011Q}}{Planck Collaboration
  XVII}(2011)}]{planck2011-6.4b}
{\sorthelp{Planck Collaboration 2011Q}}{Planck Collaboration XVII}. 2011, \aap,
  536, A17.
\newblock {\textit{Planck} early results. XVII. Origin of the submillimetre
  excess dust emission in the Magellanic Clouds}

\bibitem[{{\sorthelp{Planck Collaboration 2011R}}{Planck Collaboration
  XVIII}(2011)}]{planck2011-6.6}
{\sorthelp{Planck Collaboration 2011R}}{Planck Collaboration XVIII}. 2011,
  \aap, 536, A18.
\newblock {\textit{Planck} early results. XVIII. The power spectrum of cosmic
  infrared background anisotropies}

\bibitem[{{\sorthelp{Planck Collaboration 2011S}}{Planck Collaboration
  XIX}(2011)}]{planck2011-7.0}
{\sorthelp{Planck Collaboration 2011S}}{Planck Collaboration XIX}. 2011, \aap,
  536, A19.
\newblock {\textit{Planck} early results. XIX. All-sky temperature and dust
  optical depth from \textit{Planck} and IRAS. Constraints on the ``dark gas''
  in our Galaxy}

\bibitem[{{\sorthelp{Planck Collaboration 2011T}}{Planck Collaboration
  XX}(2011)}]{planck2011-7.2}
{\sorthelp{Planck Collaboration 2011T}}{Planck Collaboration XX}. 2011, \aap,
  536, A20.
\newblock {\textit{Planck} early results. XX. New light on anomalous microwave
  emission from spinning dust grains}

\bibitem[{{\sorthelp{Planck Collaboration 2011U}}{Planck Collaboration
  XXI}(2011)}]{planck2011-7.3}
{\sorthelp{Planck Collaboration 2011U}}{Planck Collaboration XXI}. 2011, \aap,
  536, A21.
\newblock {\textit{Planck} early results. XXI. Properties of the interstellar
  medium in the Galactic plane}

\bibitem[{{\sorthelp{Planck Collaboration 2011V}}{Planck Collaboration
  XXII}(2011)}]{planck2011-7.7a}
{\sorthelp{Planck Collaboration 2011V}}{Planck Collaboration XXII}. 2011, \aap,
  536, A22.
\newblock {\textit{Planck} early results. XXII. The submillimetre properties of
  a sample of Galactic cold clumps}

\bibitem[{{\sorthelp{Planck Collaboration 2011W}}{Planck Collaboration
  XXIII}(2011)}]{planck2011-7.7b}
{\sorthelp{Planck Collaboration 2011W}}{Planck Collaboration XXIII}. 2011,
  \aap, 536, A23.
\newblock {\textit{Planck} early results. XXIII. The Galactic cold core
  population revealed by the first all-sky survey}

\bibitem[{{\sorthelp{Planck Collaboration 2011X}}{Planck Collaboration
  XXIV}(2011)}]{planck2011-7.12}
{\sorthelp{Planck Collaboration 2011X}}{Planck Collaboration XXIV}. 2011, \aap,
  536, A24.
\newblock {\textit{Planck} early results. XXIV. Dust in the diffuse
  interstellar medium and the Galactic halo}

\bibitem[{{\sorthelp{Planck Collaboration 2011Y}}{Planck Collaboration
  XXV}(2011)}]{planck2011-7.13}
{\sorthelp{Planck Collaboration 2011Y}}{Planck Collaboration XXV}. 2011, \aap,
  536, A25.
\newblock {\textit{Planck} early results. XXV. Thermal dust in nearby molecular
  clouds}

\bibitem[{{\sorthelp{Planck Collaboration 2011Z}}{Planck Collaboration
  XXVI}(2011)}]{planck2011-5.1c}
{\sorthelp{Planck Collaboration 2011Z}}{Planck Collaboration XXVI}. 2011, \aap,
  536, A26.
\newblock {\textit{Planck} early results. XXVI. Detection with \textit{Planck}
  and confirmation by XMM-\textit{Newton} of PLCK G266.6-27.3, an exceptionally
  X-ray luminous and massive galaxy cluster at \textit{z}$\sim$1}

\bibitem[{{\sorthelp{Planck Collaboration 2014A}}{Planck Collaboration
  I}(2014)}]{planck2013-p01}
{\sorthelp{Planck Collaboration 2014A}}{Planck Collaboration I}. 2014, \aap,
  571, A1.
\newblock {\textit{Planck} 2013 results. I. Overview of products and scientific
  results}

\bibitem[{{\sorthelp{Planck Collaboration 2014B}}{Planck Collaboration
  II}(2014)}]{planck2013-p02}
{\sorthelp{Planck Collaboration 2014B}}{Planck Collaboration II}. 2014, \aap,
  571, A2.
\newblock {\textit{Planck} 2013 results. II. Low Frequency Instrument data
  processing}

\bibitem[{{\sorthelp{Planck Collaboration 2014C}}{Planck Collaboration
  III}(2014)}]{planck2013-p02a}
{\sorthelp{Planck Collaboration 2014C}}{Planck Collaboration III}. 2014, \aap,
  571, A3.
\newblock {\textit{Planck} 2013 results. III. LFI systematic uncertainties}

\bibitem[{{\sorthelp{Planck Collaboration 2014D}}{Planck Collaboration
  IV}(2014)}]{planck2013-p02d}
{\sorthelp{Planck Collaboration 2014D}}{Planck Collaboration IV}. 2014, \aap,
  571, A4.
\newblock {\textit{Planck} 2013 results. IV. LFI Beams and window functions}

\bibitem[{{\sorthelp{Planck Collaboration 2014E}}{Planck Collaboration
  V}(2014)}]{planck2013-p02b}
{\sorthelp{Planck Collaboration 2014E}}{Planck Collaboration V}. 2014, \aap,
  571, A5.
\newblock {\textit{Planck} 2013 results. V. LFI Calibration}

\bibitem[{{\sorthelp{Planck Collaboration 2014F}}{Planck Collaboration
  VI}(2014)}]{planck2013-p03}
{\sorthelp{Planck Collaboration 2014F}}{Planck Collaboration VI}. 2014, \aap,
  571, A6.
\newblock {\textit{Planck} 2013 results. VI. High Frequency Instrument data
  processing}

\bibitem[{{\sorthelp{Planck Collaboration 2014G}}{Planck Collaboration
  VII}(2014)}]{planck2013-p03c}
{\sorthelp{Planck Collaboration 2014G}}{Planck Collaboration VII}. 2014, \aap,
  571, A7.
\newblock {\textit{Planck} 2013 results. VII. HFI time response and beams}

\bibitem[{{\sorthelp{Planck Collaboration 2014H}}{Planck Collaboration
  VIII}(2014)}]{planck2013-p03f}
{\sorthelp{Planck Collaboration 2014H}}{Planck Collaboration VIII}. 2014, \aap,
  571, A8.
\newblock {\textit{Planck} 2013 results. VIII. HFI photometric calibration and
  mapmaking}

\bibitem[{{\sorthelp{Planck Collaboration 2014I}}{Planck Collaboration
  IX}(2014)}]{planck2013-p03d}
{\sorthelp{Planck Collaboration 2014I}}{Planck Collaboration IX}. 2014, \aap,
  571, A9.
\newblock {\textit{Planck} 2013 results. IX. HFI spectral response}

\bibitem[{{\sorthelp{Planck Collaboration 2014J}}{Planck Collaboration
  X}(2014)}]{planck2013-p03e}
{\sorthelp{Planck Collaboration 2014J}}{Planck Collaboration X}. 2014, \aap,
  571, A10.
\newblock {\textit{Planck} 2013 results. X. HFI energetic particle effects:
  characterization, removal, and simulation}

\bibitem[{{\sorthelp{Planck Collaboration 2014K}}{Planck Collaboration
  XI}(2014)}]{planck2013-p06b}
{\sorthelp{Planck Collaboration 2014K}}{Planck Collaboration XI}. 2014, \aap,
  571, A11.
\newblock {\textit{Planck} 2013 results. XI. All-sky model of thermal dust
  emission}

\bibitem[{{\sorthelp{Planck Collaboration 2014L}}{Planck Collaboration
  XII}(2014)}]{planck2013-p06}
{\sorthelp{Planck Collaboration 2014L}}{Planck Collaboration XII}. 2014, \aap,
  571, A12.
\newblock {\textit{Planck} 2013 results. XII. Diffuse component separation}

\bibitem[{{\sorthelp{Planck Collaboration 2014M}}{Planck Collaboration
  XIII}(2014)}]{planck2013-p03a}
{\sorthelp{Planck Collaboration 2014M}}{Planck Collaboration XIII}. 2014, \aap,
  571, A13.
\newblock {\textit{Planck} 2013 results. XIII. Galactic CO emission}

\bibitem[{{\sorthelp{Planck Collaboration 2014N}}{Planck Collaboration
  XIV}(2014)}]{planck2013-pip88}
{\sorthelp{Planck Collaboration 2014N}}{Planck Collaboration XIV}. 2014, \aap,
  571, A14.
\newblock {\textit{Planck} 2013 results. XIV. Zodiacal emission}

\bibitem[{{\sorthelp{Planck Collaboration 2014O}}{Planck Collaboration
  XV}(2014)}]{planck2013-p08}
{\sorthelp{Planck Collaboration 2014O}}{Planck Collaboration XV}. 2014, \aap,
  571, A15.
\newblock {\textit{Planck} 2013 results. XV. CMB power spectra and likelihood}

\bibitem[{{\sorthelp{Planck Collaboration 2014P}}{Planck Collaboration
  XVI}(2014)}]{planck2013-p11}
{\sorthelp{Planck Collaboration 2014P}}{Planck Collaboration XVI}. 2014, \aap,
  571, A16.
\newblock {\textit{Planck} 2013 results. XVI. Cosmological parameters}

\bibitem[{{\sorthelp{Planck Collaboration 2014Q}}{Planck Collaboration
  XVII}(2014)}]{planck2013-p12}
{\sorthelp{Planck Collaboration 2014Q}}{Planck Collaboration XVII}. 2014, \aap,
  571, A17.
\newblock {\textit{Planck} 2013 results. XVII. Gravitational lensing by
  large-scale structure}

\bibitem[{{\sorthelp{Planck Collaboration 2014R}}{Planck Collaboration
  XVIII}(2014)}]{planck2013-p13}
{\sorthelp{Planck Collaboration 2014R}}{Planck Collaboration XVIII}. 2014,
  \aap, 571, A18.
\newblock {\textit{Planck} 2013 results. XVIII. The gravitational
  lensing-infrared background correlation}

\bibitem[{{\sorthelp{Planck Collaboration 2014S}}{Planck Collaboration
  XIX}(2014)}]{planck2013-p14}
{\sorthelp{Planck Collaboration 2014S}}{Planck Collaboration XIX}. 2014, \aap,
  571, A19.
\newblock {\textit{Planck} 2013 results. XIX. The integrated Sachs-Wolfe
  effect}

\bibitem[{{\sorthelp{Planck Collaboration 2014T}}{Planck Collaboration
  XX}(2014)}]{planck2013-p15}
{\sorthelp{Planck Collaboration 2014T}}{Planck Collaboration XX}. 2014, \aap,
  571, A20.
\newblock {\textit{Planck} 2013 results. XX. Cosmology from Sunyaev-Zeldovich
  cluster counts}

\bibitem[{{\sorthelp{Planck Collaboration 2014U}}{Planck Collaboration
  XXI}(2014)}]{planck2013-p05b}
{\sorthelp{Planck Collaboration 2014U}}{Planck Collaboration XXI}. 2014, \aap,
  571, A21.
\newblock {\textit{Planck} 2013 results. XXI. Power spectrum and high-order
  statistics of the \textit{Planck} all-sky Compton parameter map}

\bibitem[{{\sorthelp{Planck Collaboration 2014V}}{Planck Collaboration
  XXII}(2014)}]{planck2013-p17}
{\sorthelp{Planck Collaboration 2014V}}{Planck Collaboration XXII}. 2014, \aap,
  571, A22.
\newblock {\textit{Planck} 2013 results. XXII. Constraints on inflation}

\bibitem[{{\sorthelp{Planck Collaboration 2014W}}{Planck Collaboration
  XXIII}(2014)}]{planck2013-p09}
{\sorthelp{Planck Collaboration 2014W}}{Planck Collaboration XXIII}. 2014,
  \aap, 571, A23.
\newblock {\textit{Planck} 2013 results. XXIII. Isotropy and statistics of the
  CMB}

\bibitem[{{\sorthelp{Planck Collaboration 2014X}}{Planck Collaboration
  XXIV}(2014)}]{planck2013-p09a}
{\sorthelp{Planck Collaboration 2014X}}{Planck Collaboration XXIV}. 2014, \aap,
  571, A24.
\newblock {\textit{Planck} 2013 results. XXIV. Constraints on primordial
  non-Gaussianity}

\bibitem[{{\sorthelp{Planck Collaboration 2014Y}}{Planck Collaboration
  XXV}(2014)}]{planck2013-p20}
{\sorthelp{Planck Collaboration 2014Y}}{Planck Collaboration XXV}. 2014, \aap,
  571, A25.
\newblock {\textit{Planck} 2013 results. XXV. Searches for cosmic strings and
  other topological defects}

\bibitem[{{\sorthelp{Planck Collaboration 2014ZA}}{Planck Collaboration
  XXVI}(2014)}]{planck2013-p19}
{\sorthelp{Planck Collaboration 2014ZA}}{Planck Collaboration XXVI}. 2014,
  \aap, 571, A26.
\newblock {\textit{Planck} 2013 results. XXVI. Background geometry and topology
  of the Universe}

\bibitem[{{\sorthelp{Planck Collaboration 2014ZB}}{Planck Collaboration
  XXVII}(2014)}]{planck2013-pipaberration}
{\sorthelp{Planck Collaboration 2014ZB}}{Planck Collaboration XXVII}. 2014,
  \aap, 571, A27.
\newblock {\textit{Planck} 2013 results. XXVII. Doppler boosting of the CMB:
  Eppur si muove}

\bibitem[{{\sorthelp{Planck Collaboration 2014ZC}}{Planck Collaboration
  XXVIII}(2014)}]{planck2013-p05}
{\sorthelp{Planck Collaboration 2014ZC}}{Planck Collaboration XXVIII}. 2014,
  \aap, 571, A28.
\newblock {\textit{Planck} 2013 results. XXVIII. The Planck Catalogue of
  Compact Sources}

\bibitem[{{\sorthelp{Planck Collaboration 2014ZD}}{Planck Collaboration
  XXIX}(2014)}]{planck2013-p05a}
{\sorthelp{Planck Collaboration 2014ZD}}{Planck Collaboration XXIX}. 2014,
  \aap, 571, A29.
\newblock {\textit{Planck} 2013 results. XXIX. The Planck catalogue of
  Sunyaev-Zeldovich sources}

\bibitem[{{\sorthelp{Planck Collaboration 2014ZE}}{Planck Collaboration
  XXX}(2014)}]{planck2013-pip56}
{\sorthelp{Planck Collaboration 2014ZE}}{Planck Collaboration XXX}. 2014, \aap,
  571, A30.
\newblock {\textit{Planck} 2013 results. XXX. Cosmic infrared background
  measurements and implications for star formation}

\bibitem[{{\sorthelp{Planck Collaboration 2014ZF}}{Planck Collaboration
  XXXI}(2014)}]{planck2013-p01a}
{\sorthelp{Planck Collaboration 2014ZF}}{Planck Collaboration XXXI}. 2014,
  \aap, 571, A31.
\newblock {\textit{Planck} 2013 results. XXXI. Consistency of the
  \textit{Planck} data}

\bibitem[{{\sorthelp{Planck Collaboration 2015A}}{Planck Collaboration
  I}(2016)}]{planck2014-a01}
{\sorthelp{Planck Collaboration 2015A}}{Planck Collaboration I}. 2016, \aap,
  submitted.
\newblock {\textit{Planck} 2015 results. I. Overview of products and results}

\bibitem[{{\sorthelp{Planck Collaboration 2015B}}{Planck Collaboration
  II}(2016)}]{planck2014-a03}
{\sorthelp{Planck Collaboration 2015B}}{Planck Collaboration II}. 2016, \aap,
  submitted.
\newblock {\textit{Planck} 2015 results. II. Low Frequency Instrument data
  processing}

\bibitem[{{\sorthelp{Planck Collaboration 2015C}}{Planck Collaboration
  III}(2016)}]{planck2014-a04}
{\sorthelp{Planck Collaboration 2015C}}{Planck Collaboration III}. 2016, \aap,
  submitted.
\newblock {\textit{Planck} 2015 results. III. LFI systematic uncertainties}

\bibitem[{{\sorthelp{Planck Collaboration 2015D}}{Planck Collaboration
  IV}(2016)}]{planck2014-a05}
{\sorthelp{Planck Collaboration 2015D}}{Planck Collaboration IV}. 2016, \aap,
  in press.
\newblock {\textit{Planck} 2015 results. IV. LFI beams and window functions}

\bibitem[{{\sorthelp{Planck Collaboration 2015E}}{Planck Collaboration
  V}(2016)}]{planck2014-a06}
{\sorthelp{Planck Collaboration 2015E}}{Planck Collaboration V}. 2016, \aap, in
  press.
\newblock {\textit{Planck} 2015 results. V. LFI calibration}

\bibitem[{{\sorthelp{Planck Collaboration 2015F}}{Planck Collaboration
  VI}(2016)}]{planck2014-a07}
{\sorthelp{Planck Collaboration 2015F}}{Planck Collaboration VI}. 2016, \aap,
  submitted.
\newblock {\textit{Planck} 2015 results. VI. LFI maps}

\bibitem[{{\sorthelp{Planck Collaboration 2015G}}{Planck Collaboration
  VII}(2016)}]{planck2014-a08}
{\sorthelp{Planck Collaboration 2015G}}{Planck Collaboration VII}. 2016, \aap,
  in press.
\newblock {\textit{Planck} 2015 results. VII. High Frequency Instrument data
  processing: Time-ordered information and beam processing}

\bibitem[{{\sorthelp{Planck Collaboration 2015H}}{Planck Collaboration
  VIII}(2016)}]{planck2014-a09}
{\sorthelp{Planck Collaboration 2015H}}{Planck Collaboration VIII}. 2016, \aap,
  in press.
\newblock {\textit{Planck} 2015 results. VIII. High Frequency Instrument data
  processing: Calibration and maps}

\bibitem[{{\sorthelp{Planck Collaboration 2015I}}{Planck Collaboration
  IX}(2016)}]{planck2014-a11}
{\sorthelp{Planck Collaboration 2015I}}{Planck Collaboration IX}. 2016, \aap,
  submitted.
\newblock {\textit{Planck} 2015 results. IX. Diffuse component separation: CMB
  maps}

\bibitem[{{\sorthelp{Planck Collaboration 2015J}}{Planck Collaboration
  X}(2016)}]{planck2014-a12}
{\sorthelp{Planck Collaboration 2015J}}{Planck Collaboration X}. 2016, \aap,
  submitted.
\newblock {\textit{Planck} 2015 results. X. Diffuse component separation:
  Foreground maps}

\bibitem[{{\sorthelp{Planck Collaboration 2015K}}{Planck Collaboration
  XI}(2016)}]{planck2014-a13}
{\sorthelp{Planck Collaboration 2015K}}{Planck Collaboration XI}. 2016, \aap,
  submitted.
\newblock {\textit{Planck} 2015 results. XI. CMB power spectra, likelihoods,
  and robustness of parameters}

\bibitem[{{\sorthelp{Planck Collaboration 2015L}}{Planck Collaboration
  XII}(2016)}]{planck2014-a14}
{\sorthelp{Planck Collaboration 2015L}}{Planck Collaboration XII}. 2016, \aap,
  submitted.
\newblock {\textit{Planck} 2015 results. XII. Full Focal Plane simulations}

\bibitem[{{\sorthelp{Planck Collaboration 2015M}}{Planck Collaboration
  XIII}(2016)}]{planck2014-a15}
{\sorthelp{Planck Collaboration 2015M}}{Planck Collaboration XIII}. 2016, \aap,
  submitted.
\newblock {\textit{Planck} 2015 results. XIII. Cosmological parameters}

\bibitem[{{\sorthelp{Planck Collaboration 2015N}}{Planck Collaboration
  XIV}(2016)}]{planck2014-a16}
{\sorthelp{Planck Collaboration 2015N}}{Planck Collaboration XIV}. 2016, \aap,
  submitted.
\newblock {\textit{Planck} 2015 results. XIV. Dark energy and modified gravity}

\bibitem[{{\sorthelp{Planck Collaboration 2015O}}{Planck Collaboration
  XV}(2016)}]{planck2014-a17}
{\sorthelp{Planck Collaboration 2015O}}{Planck Collaboration XV}. 2016, \aap,
  submitted.
\newblock {\textit{Planck} 2015 results. XV. Gravitational lensing}

\bibitem[{{\sorthelp{Planck Collaboration 2015P}}{Planck Collaboration
  XVI}(2016)}]{planck2014-a18}
{\sorthelp{Planck Collaboration 2015P}}{Planck Collaboration XVI}. 2016, \aap,
  in press.
\newblock {\textit{Planck} 2015 results. XVI. Isotropy and statistics of the
  CMB}

\bibitem[{{\sorthelp{Planck Collaboration 2015Q}}{Planck Collaboration
  XVII}(2016)}]{planck2014-a19}
{\sorthelp{Planck Collaboration 2015Q}}{Planck Collaboration XVII}. 2016, \aap,
  submitted.
\newblock {\textit{Planck} 2015 results. XVII. Constraints on primordial
  non-Gaussianity}

\bibitem[{{\sorthelp{Planck Collaboration 2015R}}{Planck Collaboration
  XVIII}(2016)}]{planck2014-a20}
{\sorthelp{Planck Collaboration 2015R}}{Planck Collaboration XVIII}. 2016,
  \aap, submitted.
\newblock {\textit{Planck} 2015 results. XVIII. Background geometry and
  topology of the Universe}

\bibitem[{{\sorthelp{Planck Collaboration 2015S}}{Planck Collaboration
  XIX}(2016)}]{planck2014-a22}
{\sorthelp{Planck Collaboration 2015S}}{Planck Collaboration XIX}. 2016, \aap,
  submitted.
\newblock {\textit{Planck} 2015 results. XIX. Constraints on primordial
  magnetic fields}

\bibitem[{{\sorthelp{Planck Collaboration 2015T}}{Planck Collaboration
  XX}(2016)}]{planck2014-a24}
{\sorthelp{Planck Collaboration 2015T}}{Planck Collaboration XX}. 2016, \aap,
  submitted.
\newblock {\textit{Planck} 2015 results. XX. Constraints on inflation}

\bibitem[{{\sorthelp{Planck Collaboration 2015U}}{Planck Collaboration
  XXI}(2016)}]{planck2014-a26}
{\sorthelp{Planck Collaboration 2015U}}{Planck Collaboration XXI}. 2016, \aap,
  submitted.
\newblock {\textit{Planck} 2015 results. XXI. The integrated Sachs-Wolfe
  effect}

\bibitem[{{\sorthelp{Planck Collaboration 2015V}}{Planck Collaboration
  XXII}(2016)}]{planck2014-a28}
{\sorthelp{Planck Collaboration 2015V}}{Planck Collaboration XXII}. 2016, \aap,
  submitted.
\newblock {\textit{Planck} 2015 results. XXII. A map of the thermal
  Sunyaev-Zeldovich effect}

\bibitem[{{\sorthelp{Planck Collaboration 2015W}}{Planck Collaboration
  XXIII}(2016)}]{planck2014-a29}
{\sorthelp{Planck Collaboration 2015W}}{Planck Collaboration XXIII}. 2016,
  \aap, submitted.
\newblock {\textit{Planck} 2015 results. XXIII. The thermal Sunyaev-Zeldovich
  effect--cosmic infrared background correlation}

\bibitem[{{\sorthelp{Planck Collaboration 2015X}}{Planck Collaboration
  XXIV}(2016)}]{planck2014-a30}
{\sorthelp{Planck Collaboration 2015X}}{Planck Collaboration XXIV}. 2016, \aap,
  submitted.
\newblock {\textit{Planck} 2015 results. XXIV. Cosmology from Sunyaev-Zeldovich
  cluster counts}

\bibitem[{{\sorthelp{Planck Collaboration 2015Y}}{Planck Collaboration
  XXV}(2016)}]{planck2014-a31}
{\sorthelp{Planck Collaboration 2015Y}}{Planck Collaboration XXV}. 2016, \aap,
  submitted.
\newblock {\textit{Planck} 2015 results. XXV. Diffuse, low-frequency Galactic
  foregrounds}

\bibitem[{{\sorthelp{Planck Collaboration 2015ZA}}{Planck Collaboration
  XXVI}(2016)}]{planck2014-a35}
{\sorthelp{Planck Collaboration 2015ZA}}{Planck Collaboration XXVI}. 2016,
  \aap, submitted.
\newblock {\textit{Planck} 2015 results. XXVI. The Second Planck Catalogue of
  Compact Sources}

\bibitem[{{\sorthelp{Planck Collaboration 2015ZB}}{Planck Collaboration
  XXVII}(2016)}]{planck2014-a36}
{\sorthelp{Planck Collaboration 2015ZB}}{Planck Collaboration XXVII}. 2016,
  \aap, in press.
\newblock {\textit{Planck} 2015 results. XXVII. The Second Planck Catalogue of
  Sunyaev-Zeldovich Sources}

\bibitem[{{\sorthelp{Planck Collaboration 2015ZC}}{Planck Collaboration
  XXVIII}(2016)}]{planck2014-a37}
{\sorthelp{Planck Collaboration 2015ZC}}{Planck Collaboration XXVIII}. 2016,
  \aap, in press.
\newblock {\textit{Planck} 2015 results. XXVIII. The Planck Catalogue of
  Galactic Cold Clumps}

\bibitem[{{\sorthelp{Planck Collaboration 2015ZD}}{Planck Collaboration
  XXIX}(2016)}]{planck2014-a02}
{\sorthelp{Planck Collaboration 2015ZD}}{Planck Collaboration XXIX}. 2016, in
  preparation.
\newblock {\textit{Planck} 2015 results. The Planck Telescope}

\bibitem[{{\sorthelp{Planck Collaboration 2015ZE}}{Planck Collaboration
  XXX}(2016)}]{planck2014-a10}
{\sorthelp{Planck Collaboration 2015ZE}}{Planck Collaboration XXX}. 2016, in
  preparation.
\newblock {\textit{Planck} 2015 results. CMB polarization at low multipoles}

\bibitem[{{\sorthelp{Planck Collaboration 2015ZF}}{Planck Collaboration
  XXXI}(2016)}]{planck2014-a23}
{\sorthelp{Planck Collaboration 2015ZF}}{Planck Collaboration XXXI}. 2016, in
  preparation.
\newblock {\textit{Planck} 2015 results. Constraints on parity and
  birefringence}

\bibitem[{{\sorthelp{Planck Collaboration 2015ZG}}{Planck Collaboration
  XXXII}(2016)}]{planck2014-a25}
{\sorthelp{Planck Collaboration 2015ZG}}{Planck Collaboration XXXII}. 2016, in
  preparation.
\newblock {\textit{Planck} 2015 results. Reionization}

\bibitem[{{\sorthelp{Planck Collaboration 2015ZH}}{Planck Collaboration
  XXXIII}(2016)}]{planck2014-a32}
{\sorthelp{Planck Collaboration 2015ZH}}{Planck Collaboration XXXIII}. 2016, in
  preparation.
\newblock {\textit{Planck} 2015 results. A model of dust emission in
  temperature and polarization}

\bibitem[{{\sorthelp{Planck Collaboration 2015ZI}}{Planck Collaboration
  XXXIV}(2016)}]{planck2014-a33}
{\sorthelp{Planck Collaboration 2015ZI}}{Planck Collaboration XXXIV}. 2016, in
  preparation.
\newblock {\textit{Planck} 2015 results. The zodiacal light}

\bibitem[{{\sorthelp{Planck Collaboration 2015ZJ}}{Planck Collaboration
  XXXV}(2016)}]{planck2014-a34}
{\sorthelp{Planck Collaboration 2015ZJ}}{Planck Collaboration XXXV}. 2016, in
  preparation.
\newblock {\textit{Planck} 2015 results. The non-CMB \Planck\ sky}

\bibitem[{{\sorthelp{Planck Collaboration IntA}}{Planck Collaboration Int.
  I}(2012)}]{planck2012-I}
{\sorthelp{Planck Collaboration IntA}}{Planck Collaboration Int. I}. 2012,
  \aap, 543, A102.
\newblock {\textit{Planck} intermediate results. I. Further validation of new
  \textit{Planck} clusters with XMM-\textit{Newton}}

\bibitem[{{\sorthelp{Planck Collaboration IntB}}{Planck and AMI
  Collaborations}(2013)}]{planck2012-II}
{\sorthelp{Planck Collaboration IntB}}{Planck and AMI Collaborations}. 2013,
  \aap, 550, A128.
\newblock {\textit{Planck} intermediate results. II. Comparison of
  Sunyaev-Zeldovich measurements from \textit{Planck} and from the Arcminute
  Microkelvin Imager for 11 galaxy clusters}

\bibitem[{{\sorthelp{Planck Collaboration IntC}}{Planck Collaboration Int.
  III}(2013)}]{planck2012-III}
{\sorthelp{Planck Collaboration IntC}}{Planck Collaboration Int. III}. 2013,
  \aap, 550, A129.
\newblock {\textit{Planck} intermediate results. III. The relation between
  galaxy cluster mass and Sunyaev-Zeldovich signal}

\bibitem[{{\sorthelp{Planck Collaboration IntD}}{Planck Collaboration Int.
  IV}(2013)}]{planck2012-IV}
{\sorthelp{Planck Collaboration IntD}}{Planck Collaboration Int. IV}. 2013,
  \aap, 550, A130.
\newblock {\textit{Planck} intermediate results. IV. The XMM-\textit{Newton}
  validation programme for new \textit{Planck} clusters}

\bibitem[{{\sorthelp{Planck Collaboration IntEe}}{Planck Collaboration Int. V
  (Erratum)}(2013)}]{planck2012-Ve}
{\sorthelp{Planck Collaboration IntEe}}{Planck Collaboration Int. V (Erratum)}.
  2013, \aap, 558, C2.
\newblock {Erratum: \textit{Planck} intermediate results. V. Pressure profiles
  of galaxy clusters from the Sunyaev-Zeldovich effect}

\bibitem[{{\sorthelp{Planck Collaboration IntE}}{Planck Collaboration Int.
  V}(2013)}]{planck2012-V}
{\sorthelp{Planck Collaboration IntE}}{Planck Collaboration Int. V}. 2013,
  \aap, 550, A131.
\newblock {\textit{Planck} intermediate results. V. Pressure profiles of galaxy
  clusters from the Sunyaev-Zeldovich effect}

\bibitem[{{\sorthelp{Planck Collaboration IntF}}{Planck Collaboration Int.
  VI}(2013)}]{planck2012-VI}
{\sorthelp{Planck Collaboration IntF}}{Planck Collaboration Int. VI}. 2013,
  \aap, 550, A132.
\newblock {\textit{Planck} intermediate results. VI. The dynamical structure of
  PLCKG214.6+37.0, a \textit{Planck} discovered triple system of galaxy
  clusters}

\bibitem[{{\sorthelp{Planck Collaboration IntG}}{Planck Collaboration Int.
  VII}(2013)}]{planck2012-VII}
{\sorthelp{Planck Collaboration IntG}}{Planck Collaboration Int. VII}. 2013,
  \aap, 550, A133.
\newblock {\textit{Planck} intermediate results. VII. Statistical properties of
  infrared and radio extragalactic sources from the Planck Early Release
  Compact Source Catalogue at frequencies between 100 and 857\,GHz}

\bibitem[{{\sorthelp{Planck Collaboration IntH}}{Planck Collaboration Int.
  VIII}(2013)}]{planck2012-VIII}
{\sorthelp{Planck Collaboration IntH}}{Planck Collaboration Int. VIII}. 2013,
  \aap, 550, A134.
\newblock {\textit{Planck} intermediate results. VIII. Filaments between
  interacting clusters}

\bibitem[{{\sorthelp{Planck Collaboration IntI}}{Planck Collaboration Int.
  IX}(2013)}]{planck2012-IX}
{\sorthelp{Planck Collaboration IntI}}{Planck Collaboration Int. IX}. 2013,
  \aap, 554, A139.
\newblock {\textit{Planck} intermediate results. IX. Detection of the Galactic
  haze with Planck}

\bibitem[{{\sorthelp{Planck Collaboration IntJ}}{Planck Collaboration Int.
  X}(2013)}]{planck2012-X}
{\sorthelp{Planck Collaboration IntJ}}{Planck Collaboration Int. X}. 2013,
  \aap, 554, A140.
\newblock {\textit{Planck} intermediate results. X. Physics of the hot gas in
  the Coma cluster}

\bibitem[{{\sorthelp{Planck Collaboration IntK}}{Planck Collaboration Int.
  XI}(2013)}]{planck2012-XI}
{\sorthelp{Planck Collaboration IntK}}{Planck Collaboration Int. XI}. 2013,
  \aap, 557, A52.
\newblock {\textit{Planck} intermediate results. XI. The gas content of dark
  matter halos: the Sunyaev-Zeldovich-stellar mass relation for locally
  brightest galaxies}

\bibitem[{{\sorthelp{Planck Collaboration IntL}}{Planck Collaboration Int.
  XII}(2013)}]{planck2013-XII}
{\sorthelp{Planck Collaboration IntL}}{Planck Collaboration Int. XII}. 2013,
  \aap, 557, A53.
\newblock {\textit{Planck} intermediate results. XII. Diffuse Galactic
  components in the Gould Belt System}

\bibitem[{{\sorthelp{Planck Collaboration IntM}}{Planck Collaboration Int.
  XIII}(2014)}]{planck2013-XIII}
{\sorthelp{Planck Collaboration IntM}}{Planck Collaboration Int. XIII}. 2014,
  \aap, 561, A97.
\newblock {\textit{Planck} intermediate results. XIII. Constraints on peculiar
  velocities}

\bibitem[{{\sorthelp{Planck Collaboration IntN}}{Planck Collaboration Int.
  XIV}(2014)}]{planck2013-XIV}
{\sorthelp{Planck Collaboration IntN}}{Planck Collaboration Int. XIV}. 2014,
  \aap, 564, A45.
\newblock {\textit{Planck} intermediate results. XIV. Dust emission at
  millimetre wavelengths in the Galactic plane}

\bibitem[{{\sorthelp{Planck Collaboration IntO}}{Planck Collaboration Int.
  XV}(2014)}]{planck2013-XV}
{\sorthelp{Planck Collaboration IntO}}{Planck Collaboration Int. XV}. 2014,
  \aap, 565, A103.
\newblock {\textit{Planck} intermediate results. XV. A study of anomalous
  microwave emission in Galactic clouds}

\bibitem[{{\sorthelp{Planck Collaboration IntP}}{Planck Collaboration Int.
  XVI}(2014)}]{planck2013-XVI}
{\sorthelp{Planck Collaboration IntP}}{Planck Collaboration Int. XVI}. 2014,
  \aap, 566, A54.
\newblock {\textit{Planck} intermediate results. XVI. Profile likelihoods for
  cosmological parameters}

\bibitem[{{\sorthelp{Planck Collaboration IntQ}}{Planck Collaboration Int.
  XVII}(2014)}]{planck2013-XVII}
{\sorthelp{Planck Collaboration IntQ}}{Planck Collaboration Int. XVII}. 2014,
  \aap, 566, A55.
\newblock {\textit{Planck} intermediate results. XVII. Emission of dust in the
  diffuse interstellar medium from the far-infrared to microwave frequencies}

\bibitem[{{\sorthelp{Planck Collaboration IntR}}{Planck Collaboration Int.
  XVIII}(2015)}]{planck2014-XVIII}
{\sorthelp{Planck Collaboration IntR}}{Planck Collaboration Int. XVIII}. 2015,
  \aap, 573, A6.
\newblock {\textit{Planck} intermediate results. XVIII. The millimetre and
  submillimetre emission from planetary nebulae}

\bibitem[{{\sorthelp{Planck Collaboration IntS}}{Planck Collaboration Int.
  XIX}(2015)}]{planck2014-XIX}
{\sorthelp{Planck Collaboration IntS}}{Planck Collaboration Int. XIX}. 2015,
  \aap, 576, A104.
\newblock {\textit{Planck} intermediate results. XIX. An overview of the
  polarized thermal emission from Galactic dust}

\bibitem[{{\sorthelp{Planck Collaboration IntT}}{Planck Collaboration Int.
  XX}(2015)}]{planck2014-XX}
{\sorthelp{Planck Collaboration IntT}}{Planck Collaboration Int. XX}. 2015,
  \aap, 576, A105.
\newblock {\textit{Planck} intermediate results. XX. Comparison of polarized
  thermal emission from Galactic dust with simulations of MHD turbulence}

\bibitem[{{\sorthelp{Planck Collaboration IntU}}{Planck Collaboration Int.
  XXI}(2015)}]{planck2014-XXI}
{\sorthelp{Planck Collaboration IntU}}{Planck Collaboration Int. XXI}. 2015,
  \aap, 576, A106.
\newblock {\textit{Planck} intermediate results. XXI. Comparison of polarized
  thermal emission from Galactic dust at 353\,GHz with optical interstellar
  polarization}

\bibitem[{{\sorthelp{Planck Collaboration IntV}}{Planck Collaboration Int.
  XXII}(2015)}]{planck2014-XXII}
{\sorthelp{Planck Collaboration IntV}}{Planck Collaboration Int. XXII}. 2015,
  \aap, submitted, 576, A107.
\newblock {\textit{Planck} intermediate results. XXII. Frequency dependence of
  thermal emission from Galactic dust in intensity and polarization}

\bibitem[{{\sorthelp{Planck Collaboration IntW}}{Planck Collaboration Int.
  XXIII}(2015)}]{planck2014-XXIII}
{\sorthelp{Planck Collaboration IntW}}{Planck Collaboration Int. XXIII}. 2015,
  \aap, 580, A13.
\newblock {\textit{Planck} intermediate results. XXIII. Galactic plane emission
  components derived from Planck with ancillary data}

\bibitem[{{\sorthelp{Planck Collaboration IntX}}{Planck Collaboration Int.
  XXIV}(2015)}]{planck2014-XXIV}
{\sorthelp{Planck Collaboration IntX}}{Planck Collaboration Int. XXIV}. 2015,
  \aap, 580, A22.
\newblock {\textit{Planck} intermediate results. XXIV. Constraints on variation
  of fundamental constants}

\bibitem[{{\sorthelp{Planck Collaboration IntY}}{Planck Collaboration Int.
  XXV}(2015)}]{planck2014-XXV}
{\sorthelp{Planck Collaboration IntY}}{Planck Collaboration Int. XXV}. 2015,
  \aap, 582, A28.
\newblock {\textit{Planck} intermediate results. XXV. The Andromeda Galaxy as
  seen by \textit{Planck}}

\bibitem[{{\sorthelp{Planck Collaboration IntZA}}{Planck Collaboration Int.
  XXVI}(2015)}]{planck2014-XXVI}
{\sorthelp{Planck Collaboration IntZA}}{Planck Collaboration Int. XXVI}. 2015,
  \aap, 582, A29.
\newblock {\textit{Planck} intermediate results. XXVI. Optical identification
  and redshifts of \textit{Planck} clusters with the RTT150 telescope}

\bibitem[{{\sorthelp{Planck Collaboration IntZB}}{Planck Collaboration Int.
  XXVII}(2015)}]{planck2014-XXVII}
{\sorthelp{Planck Collaboration IntZB}}{Planck Collaboration Int. XXVII}. 2015,
  \aap, 582, A30.
\newblock {\textit{Planck} intermediate results. XXVII. High-redshift infrared
  galaxy overdensity candidates and lensed sources discovered by
  \textit{Planck} and confirmed by \textit{Herschel}-SPIRE}

\bibitem[{{\sorthelp{Planck Collaboration IntZC}}{Planck Collaboration Int.
  XXVIII}(2015)}]{planck2014-XXVIII}
{\sorthelp{Planck Collaboration IntZC}}{Planck Collaboration Int. XXVIII}.
  2015, \aap, 582, A31.
\newblock {\textit{Planck} intermediate results. XXVIII. Interstellar gas and
  dust in the Chamaeleon clouds as seen by \textit{Fermi} LAT and
  \textit{Planck}}

\bibitem[{{\sorthelp{Planck Collaboration IntZD}}{Planck Collaboration Int.
  XXIX}(2016)}]{planck2014-XXIX}
{\sorthelp{Planck Collaboration IntZD}}{Planck Collaboration Int. XXIX}. 2016,
  \aap, 586, A132.
\newblock {\textit{Planck} intermediate results. XXIX. All-sky dust modelling
  with \textit{Planck}, IRAS, and WISE observations}

\bibitem[{{\sorthelp{Planck Collaboration IntZE}}{Planck Collaboration Int.
  XXX}(2016)}]{planck2014-XXX}
{\sorthelp{Planck Collaboration IntZE}}{Planck Collaboration Int. XXX}. 2016,
  \aap, 586, A133.
\newblock {\textit{Planck} intermediate results. XXX. The angular power
  spectrum of polarized dust emission at intermediate and high Galactic
  latitudes}

\bibitem[{{\sorthelp{Planck Collaboration IntZF}}{Planck Collaboration Int.
  XXXI}(2016)}]{planck2014-XXXI}
{\sorthelp{Planck Collaboration IntZF}}{Planck Collaboration Int. XXXI}. 2016,
  \aap, 586, A134.
\newblock {\textit{Planck} intermediate results. XXXI. Microwave survey of
  Galactic supernova remnants}

\bibitem[{{\sorthelp{Planck Collaboration IntZG}}{Planck Collaboration Int.
  XXXII}(2016)}]{planck2014-XXXII}
{\sorthelp{Planck Collaboration IntZG}}{Planck Collaboration Int. XXXII}. 2016,
  \aap, 586, A135.
\newblock {\textit{Planck} intermediate results. XXXII. The relative
  orientation between the magnetic field and structures traced by interstellar
  dust}

\bibitem[{{\sorthelp{Planck Collaboration IntZH}}{Planck Collaboration Int.
  XXXIII}(2016)}]{planck2014-XXXIII}
{\sorthelp{Planck Collaboration IntZH}}{Planck Collaboration Int. XXXIII}.
  2016, \aap, 586, A136.
\newblock {\textit{Planck} intermediate results. XXXIII. Signature of the
  magnetic field geometry of interstellar filaments in dust polarization maps}

\bibitem[{{\sorthelp{Planck Collaboration IntZI}}{Planck Collaboration Int.
  XXXIV}(2016)}]{planck2015-XXXIV}
{\sorthelp{Planck Collaboration IntZI}}{Planck Collaboration Int. XXXIV}. 2016,
  \aap, 586, A137.
\newblock {\textit{Planck} intermediate results. XXXIV. The magnetic field
  structure in the Rosette Nebula}

\bibitem[{{\sorthelp{Planck Collaboration IntZJ}}{Planck Collaboration Int.
  XXXV}(2016)}]{planck2015-XXXV}
{\sorthelp{Planck Collaboration IntZJ}}{Planck Collaboration Int. XXXV}. 2016,
  \aap, 586, A138.
\newblock {\textit{Planck} intermediate results. XXXV. Probing the role of the
  magnetic field in the formation of structure in molecular clouds}

\bibitem[{{\sorthelp{Planck Collaboration IntZK}}{Planck Collaboration Int.
  XXXVI}(2016)}]{planck2015-XXXVI}
{\sorthelp{Planck Collaboration IntZK}}{Planck Collaboration Int. XXXVI}. 2016,
  \aap, 586, A139.
\newblock {\textit{Planck} intermediate results. XXXVI. Optical identification
  and redshifts of Planck SZ sources with telescopes in the Canary Islands
  Observatories}

\bibitem[{{\sorthelp{Planck Collaboration IntZL}}{Planck Collaboration Int.
  XXXVII}(2016)}]{planck2015-XXXVII}
{\sorthelp{Planck Collaboration IntZL}}{Planck Collaboration Int. XXXVII}.
  2016, \aap, 586, A140.
\newblock {\textit{Planck} intermediate results. XXXVII. Evidence of unbound
  gas from the kinetic Sunyaev-Zeldovich effect}

\bibitem[{{\sorthelp{Planck Collaboration IntZM}}{Planck Collaboration Int.
  XXXVIII}(2016)}]{planck2015-XXXVIII}
{\sorthelp{Planck Collaboration IntZM}}{Planck Collaboration Int. XXXVIII}.
  2016, \aap, 586, A141.
\newblock {\textit{Planck} intermediate results. XXXVIII. E- and B-modes of
  dust polarization from the magnetized filamentary structure of the
  interstellar medium}

\bibitem[{{\sorthelp{Planck Collaboration IntZN}}{Planck Collaboration Int.
  XXXIX}(2015)}]{planck2015-XXXIX}
{\sorthelp{Planck Collaboration IntZN}}{Planck Collaboration Int. XXXIX}. 2015,
  \aap, submitted.
\newblock {\textit{Planck} intermediate results. XXXIX. The Planck List of
  High-redshift Source Candidates}

\bibitem[{{\sorthelp{Planck Collaboration IntZO}}{Planck Collaboration Int.
  XL}(2015)}]{planck2015-XL}
{\sorthelp{Planck Collaboration IntZO}}{Planck Collaboration Int. XL}. 2015,
  \aap, submitted.
\newblock {\textit{Planck} intermediate results. XL. The Sunyaev-Zeldovich
  signal from the Virgo cluster}

\bibitem[{{\sorthelp{Planck Collaboration IntZP}}{Planck Collaboration Int.
  XLI}(2015)}]{planck2015-XLI}
{\sorthelp{Planck Collaboration IntZP}}{Planck Collaboration Int. XLI}. 2015,
  \aap, submitted.
\newblock {\textit{Planck} intermediate results. XLI. A map of lensing-induced
  \textit{B}-modes}

\bibitem[{{\sorthelp{Planck Collaboration IntZQ}}{Planck Collaboration Int.
  XLII}(2016)}]{planck2016-XLII}
{\sorthelp{Planck Collaboration IntZQ}}{Planck Collaboration Int. XLII}. 2016,
  \aap, submitted.
\newblock {\textit{Planck} intermediate results. XLII. Large-scale Galactic
  magnetic fields}

\bibitem[{{Poutanen} {et~al.}(2006){Poutanen}, {de Gasperis}, {Hivon},
  {Kurki-Suonio}, {Balbi}, {Borrill}, {Cantalupo}, {Dor{\'e}}, {Keih{\"a}nen},
  {Lawrence}, {Maino}, {Natoli}, {Prunet}, {Stompor}, \&
  {Teyssier}}]{poutanen2006}
{Poutanen}, T., {de Gasperis}, G., {Hivon}, E., {et~al.} 2006, \aap, 449, 1311.
\newblock {Comparison of map-making algorithms for CMB experiments}

\bibitem[{{Pratt}(1978)}]{pratt1978}
{Pratt}, W.~K. 1978, {Digital image processing} (New York: Wiley)

\bibitem[{{Prunet} {et~al.}(2001){Prunet}, {Ade}, {Bock}, {Bond}, {Borrill},
  {Boscaleri}, {Coble}, {Crill}, {de Bernardis}, {De Gasperis}, {De Troia},
  {Farese}, {Ferreira}, {Ganga}, {Giacometti}, {Hivon}, {Hristov},
  {Iacoangeli}, {Jaffe}, {Lange}, {Martinis}, {Masi}, {Mason}, {Mauskopf},
  {Melchiorri}, {Miglio}, {Montroy}, {Netterfield}, {Pascale}, {Piacentini},
  {Pogosyan}, {Pongetti}, {Prunet}, {Rao}, {Romeo}, {Ruhl}, {Scaramuzzi},
  {Sforna}, \& {Vittorio}}]{prunet2001}
{Prunet}, S., {Ade}, P.~A.~R., {Bock}, J.~J., {et~al.} 2001, e-print.
\newblock {Noise estimation in CMB time-streams and fast map-making.
  Application to the BOOMERanG98 data}

\bibitem[{{Reinecke}(2011)}]{reinecke2011}
{Reinecke}, M. 2011, \aap, 526, A108, accepted.
\newblock {Libpsht - algorithms for efficient spherical harmonic transforms}

\bibitem[{{Reinecke} {et~al.}(2006){Reinecke}, {Dolag}, {Hell}, {Bartelmann},
  \& {En{\ss}lin}}]{reinecke2006}
{Reinecke}, M., {Dolag}, K., {Hell}, R., {Bartelmann}, M., \& {En{\ss}lin},
  T.~A. 2006, \aap, 445, 373.
\newblock {A simulation pipeline for the Planck mission}

\bibitem[{{Revenu} {et~al.}(2000){Revenu}, {Kim}, {Ansari}, {Couchot},
  {Delabrouille}, \& {Kaplan}}]{revenu2000}
{Revenu}, B., {Kim}, A., {Ansari}, R., {et~al.} 2000, \aaps, 142, 499.
\newblock {Destriping of polarized data in a CMB mission with a circular
  scanning strategy}

\bibitem[{{Ricciardi}(2007)}]{ricciardi2007}
{Ricciardi}, S. 2007, New Astronomy Review, 51, 310.
\newblock {Planck Reference Sky versus WMAP}

\bibitem[{{Rocha} {et~al.}(2011){Rocha}, {Contaldi}, {Bond}, \&
  {Gorski}}]{rocha2009}
{Rocha}, G., {Contaldi}, C.~R., {Bond}, J.~R., \& {Gorski}, K.~M. 2011, \mnras,
  414, 823.
\newblock {Application of XFaster power spectrum and likelihood estimator to
  Planck}

\bibitem[{{Rocha} {et~al.}(2010{\natexlab{a}}){Rocha}, {Contaldi}, {Colombo},
  {Bond}, {Gorski}, \& {Lawrence}}]{rocha2010b}
{Rocha}, G., {Contaldi}, C.~R., {Colombo}, L.~P.~L., {et~al.}
  2010{\natexlab{a}}, e-print.
\newblock {Performance of XFaster likelihood in real CMB experiments}

\bibitem[{{Rocha} {et~al.}(2010{\natexlab{b}}){Rocha}, {Pagano}, {G{\'o}rski},
  {Huffenberger}, {Lawrence}, \& {Lange}}]{rocha2010a}
{Rocha}, G., {Pagano}, L., {G{\'o}rski}, K.~M., {et~al.} 2010{\natexlab{b}},
  \aap, 513, A23.
\newblock {Markov chain beam randomization: a study of the impact of PLANCK
  beam measurement errors on cosmological parameter estimation}

\bibitem[{{Rosset} {et~al.}(2010){Rosset}, {Tristram}, {Ponthieu}, {Ade},
  {Aumont}, {Catalano}, {Conversi}, {Couchot}, {Crill}, {D{\'e}sert}, {Ganga},
  {Giard}, {Giraud-H{\'e}raud}, {Ha{\"i}ssinski}, {Henrot-Versill{\'e}},
  {Holmes}, {Jones}, {Lamarre}, {Lange}, {Leroy}, {Mac{\'{\i}}as-P{\'e}rez},
  {Maffei}, {de Marcillac}, {Miville-Desch{\^e}nes}, {Montier}, {Noviello},
  {Pajot}, {Perdereau}, {Piacentini}, {Piat}, {Plaszczynski}, {Pointecouteau},
  {Puget}, {Ristorcelli}, {Savini}, {Sudiwala}, {Veneziani}, \&
  {Yvon}}]{rosset2010}
{Rosset}, C., {Tristram}, M., {Ponthieu}, N., {et~al.} 2010, \aap, 520, A13.
\newblock {\textit{Planck} pre-launch status: High Frequency Instrument
  polarization calibration}

\bibitem[{{Sandri} {et~al.}(2010){Sandri}, {Villa}, {Bersanelli}, {Burigana},
  {Butler}, {D'Arcangelo}, {Figini}, {Gregorio}, {Lawrence}, {Maino},
  {Mandolesi}, {Maris}, {Nesti}, {Perrotta}, {Platania}, {Simonetto}, {Sozzi},
  {Tauber}, \& {Valenziano}}]{sandri2010}
{Sandri}, M., {Villa}, F., {Bersanelli}, M., {et~al.} 2010, \aap, 520, A7.
\newblock {\textit{Planck} pre-launch status: Low Frequency Instrument optics}

\bibitem[{{Schlegel} {et~al.}(1998){Schlegel}, {Finkbeiner}, \&
  {Davis}}]{schlegel1998}
{Schlegel}, D.~J., {Finkbeiner}, D.~P., \& {Davis}, M. 1998, \apj, 500, 525.
\newblock {Maps of Dust Infrared Emission for Use in Estimation of Reddening
  and Cosmic Microwave Background Radiation Foregrounds}

\bibitem[{{Seiffert} {et~al.}(2002){Seiffert}, {Mennella}, {Burigana},
  {Mandolesi}, {Bersanelli}, {Meinhold}, \& {Lubin}}]{seiffert2002}
{Seiffert}, M., {Mennella}, A., {Burigana}, C., {et~al.} 2002, \aap, 391, 1185.
\newblock {1/f noise and other systematic effects in the Planck-LFI
  radiometers}

\bibitem[{Shoemake(1985)}]{shoemake1985}
Shoemake, K. 1985, in SIGGRAPH '85: Proceedings of the 12th annual conference
  on Computer graphics and interactive techniques (New York, NY, USA: ACM),
  245--254

\bibitem[{{Shuster}(1993)}]{shuster1993}
{Shuster}, M.~D. 1993, Journal of the Astronautical Sciences, 41, 439.
\newblock {Survey of attitude representations}

\bibitem[{{Smith} {et~al.}(2007){Smith}, {Zahn}, \& {Dor{\'e}}}]{smith2007}
{Smith}, K.~M., {Zahn}, O., \& {Dor{\'e}}, O. 2007, \prd, 76, 043510.
\newblock {Detection of gravitational lensing in the cosmic microwave
  background}

\bibitem[{{Spencer} {et~al.}(2010){Spencer}, {Naylor}, \&
  {Swinyard}}]{spencer2010}
{Spencer}, L.~D., {Naylor}, D.~A., \& {Swinyard}, B.~M. 2010, Measurement
  Science and Technology, 21, 065601.
\newblock {Performance evaluation of the Herschel/SPIRE imaging Fourier
  transform spectrometer through ground-based measurements}

\bibitem[{{Spergel} {et~al.}(2007){Spergel}, {Bean}, {Dor{\'e}}, {Nolta},
  {Bennett}, {Dunkley}, {Hinshaw}, {Jarosik}, {Komatsu}, {Page}, {Peiris},
  {Verde}, {Halpern}, {Hill}, {Kogut}, {Limon}, {Meyer}, {Odegard}, {Tucker},
  {Weiland}, {Wollack}, \& {Wright}}]{spergel2007}
{Spergel}, D.~N., {Bean}, R., {Dor{\'e}}, O., {et~al.} 2007, \apjs, 170, 377.
\newblock {Three-Year Wilkinson Microwave Anisotropy Probe (WMAP) Observations:
  Implications for Cosmology}

\bibitem[{{Spergel} {et~al.}(2003){Spergel}, {Verde}, {Peiris}, {Komatsu},
  {Nolta}, {Bennett}, {Halpern}, {Hinshaw}, {Jarosik}, {Kogut}, {Limon},
  {Meyer}, {Page}, {Tucker}, {Weiland}, {Wollack}, \& {Wright}}]{spergel2003}
{Spergel}, D.~N., {Verde}, L., {Peiris}, H.~V., {et~al.} 2003, \apjs, 148, 175.
\newblock {First-Year Wilkinson Microwave Anisotropy Probe (WMAP) Observations:
  Determination of Cosmological Parameters}

\bibitem[{{Stute}(2004)}]{stute2004}
{Stute}, T. 2004, in Society of Photo-Optical Instrumentation Engineers (SPIE)
  Conference Series, Vol. 5495, Society of Photo-Optical Instrumentation
  Engineers (SPIE) Conference Series, ed. {J.~Antebi \& D.~Lemke}, 1--10

\bibitem[{{Swinyard} {et~al.}(2010){Swinyard}, {Hartogh}, {Sidher}, {Fulton},
  {Lellouch}, {Jarchow}, {Griffin}, {Moreno}, {Sagawa}, {Portyankina},
  {Blecka}, {Banaszkiewicz}, {Bockelee-Morvan}, {Crovisier}, {Encrenaz},
  {Kueppers}, {Lara}, {Lis}, {Medvedev}, {Rengel}, {Szutowicz},
  {Vandenbussche}, {Bensch}, {Bergin}, {Billebaud}, {Biver}, {Blake},
  {Blommaert}, {de Val-Borro}, {Cernicharo}, {Cavalie}, {Courtin}, {Davis},
  {Decin}, {Encrenaz}, {de Graauw}, {Jehin}, {Kidger}, {Leeks}, {Orton},
  {Naylor}, {Schieder}, {Stam}, {Thomas}, {Verdugo}, {Waelkens}, \&
  {Walker}}]{swinyard2010}
{Swinyard}, B.~M., {Hartogh}, P., {Sidher}, S., {et~al.} 2010, \aap, 518, L151.
\newblock {The Herschel-SPIRE submillimetre spectrum of Mars}

\bibitem[{{Tauber} {et~al.}(2010{\natexlab{a}}){Tauber}, {Mandolesi}, {Puget},
  {Banos}, {Bersanelli}, {Bouchet}, {Butler}, {Charra}, {Crone}, {Dodsworth},
  \& et~al.}]{tauber2010a}
{Tauber}, J.~A., {Mandolesi}, N., {Puget}, J., {et~al.} 2010{\natexlab{a}},
  \aap, 520, A1.
\newblock {\textit{Planck} pre-launch status: The Planck mission}

\bibitem[{{Tauber} {et~al.}(2010{\natexlab{b}}){Tauber}, {Norgaard-Nielsen},
  {Ade}, {Amiri Parian}, {Banos}, {Bersanelli}, {Burigana}, {Chamballu}, {de
  Chambure}, {Christensen}, {Corre}, {Cozzani}, {Crill}, {Crone},
  {D'Arcangelo}, {Daddato}, {Doyle}, {Dubruel}, {Forma}, {Hills},
  {Huffenberger}, {Jaffe}, {Jessen}, {Kletzkine}, {Lamarre}, {Leahy},
  {Longval}, {de Maagt}, {Maffei}, {Mandolesi}, {Mart{\'{\i}}-Canales},
  {Mart{\'{\i}}n-Polegre}, {Martin}, {Mendes}, {Murphy}, {Nielsen}, {Noviello},
  {Paquay}, {Peacocke}, {Ponthieu}, {Pontoppidan}, {Ristorcelli}, {Riti},
  {Rolo}, {Rosset}, {Sandri}, {Savini}, {Sudiwala}, {Tristram}, {Valenziano},
  {van der Vorst}, {van't Klooster}, {Villa}, \& {Yurchenko}}]{tauber2010b}
{Tauber}, J.~A., {Norgaard-Nielsen}, H.~U., {Ade}, P.~A.~R., {et~al.}
  2010{\natexlab{b}}, \aap, 520, A2.
\newblock {\textit{Planck} pre-launch status: The optical system}

\bibitem[{{Terenzi} {et~al.}(2009{\natexlab{a}}){Terenzi}, {Lapolla},
  {Laaninen}, {Battaglia}, {Cavaliere}, {De Rosa}, {Hughes}, {Jukkala},
  {Kilpi{\"a}}, {Morgante}, {Tomasi}, {Varis}, {Bersanelli}, {Butler},
  {Ferrari}, {Franceschet}, {Leutenegger}, {Mandolesi}, {Mennella},
  {Silvestri}, {Stringhetti}, {Tuovinen}, {Valenziano}, \&
  {Villa}}]{terenzi2009a}
{Terenzi}, L., {Lapolla}, M., {Laaninen}, M., {et~al.} 2009{\natexlab{a}},
  Journal of Instrumentation, 4, 2015.
\newblock {Cryogenic environment and performance for testing the Planck
  radiometers}

\bibitem[{{Terenzi} {et~al.}(2009{\natexlab{b}}){Terenzi}, {Salmon}, {Colin},
  {Mennella}, {Morgante}, {Tomasi}, {Battaglia}, {Lapolla}, {Bersanelli},
  {Butler}, {Cuttaia}, {D'Arcangelo}, {Davis}, {Franceschet}, {Galeotta},
  {Gregorio}, {Hughes}, {Jukkala}, {Kettle}, {Laaninen}, {Leutenegger},
  {Leonardi}, {Mandolesi}, {Maris}, {Meinhold}, {Miccolis}, {Roddis}, {Sambo},
  {Sandri}, {Silvestri}, {Tuovinen}, {Valenziano}, {Varis}, {Villa},
  {Wilkinson}, \& {Zonca}}]{terenzi2009b}
{Terenzi}, L., {Salmon}, M.~J., {Colin}, A., {et~al.} 2009{\natexlab{b}},
  Journal of Instrumentation, 4, 2012.
\newblock {Thermal susceptibility of the Planck-LFI receivers}

\bibitem[{{Tomasi} {et~al.}(2010){Tomasi}, {Cappellini}, {Gregorio}, {Colombo},
  {Lapolla}, {Terenzi}, {Morgante}, {Bersanelli}, {Butler}, {Galeotta},
  {Mandolesi}, {Maris}, {Mennella}, {Valenziano}, \& {Zacchei}}]{tomasi2010}
{Tomasi}, M., {Cappellini}, B., {Gregorio}, A., {et~al.} 2010, Journal of
  Instrumentation, 5, 1002.
\newblock {Dynamic validation of the Planck-LFI thermal model}

\bibitem[{{Tomasi} {et~al.}(2009){Tomasi}, {Mennella}, {Galeotta}, {Lowe},
  {Mendes}, {Leonardi}, {Villa}, {Cappellini}, {Gregorio}, {Meinhold},
  {Sandri}, {Cuttaia}, {Terenzi}, {Maris}, {Valenziano}, {Salmon},
  {Bersanelli}, {Binko}, {Butler}, {D'Arcangelo}, {Fogliani}, {Frailis},
  {Franceschi}, {Gasparo}, {Maggio}, {Maino}, {Malaspina}, {Mandolesi},
  {Manzato}, {Meharga}, {Morgante}, {Morisset}, {Pasian}, {Perrotta}, {Rohlfs},
  {T{\"u}rler}, {Zacchei}, \& {Zonca}}]{tomasi2009}
{Tomasi}, M., {Mennella}, A., {Galeotta}, S., {et~al.} 2009, Journal of
  Instrumentation, 4, 2020.
\newblock {Off-line radiometric analysis of Planck-LFI data}

\bibitem[{{Tristram} {et~al.}(2005){Tristram}, {Mac{\'{\i}}as-P{\'e}rez},
  {Renault}, \& {Santos}}]{tristram2005}
{Tristram}, M., {Mac{\'{\i}}as-P{\'e}rez}, J.~F., {Renault}, C., \& {Santos},
  D. 2005, \mnras, 358, 833.
\newblock {XSPECT, estimation of the angular power spectrum by computing
  cross-power spectra with analytical error bars}

\bibitem[{{Valenziano} {et~al.}(2009){Valenziano}, {Cuttaia}, {De Rosa},
  {Terenzi}, {Brighenti}, {Cazzola}, {Garbesi}, {Mariotti}, {Orsi}, {Pagan},
  {Cavaliere}, {Biggi}, {Lapini}, {Panagin}, {Battaglia}, {Butler},
  {Bersanelli}, {D'Arcangelo}, {Levin}, {Mandolesi}, {Mennella}, {Morgante},
  {Morigi}, {Sandri}, {Simonetto}, {Tomasi}, {Villa}, {Frailis}, {Galeotta},
  {Gregorio}, {Leonardi}, {Lowe}, {Maris}, {Meinhold}, {Mendes}, {Stringhetti},
  {Zonca}, \& {Zacchei}}]{valenziano2009}
{Valenziano}, L., {Cuttaia}, F., {De Rosa}, A., {et~al.} 2009, Journal of
  Instrumentation, 4, 2006.
\newblock {Planck-LFI: design and performance of the 4 Kelvin Reference Load
  Unit}

\bibitem[{{Varis} {et~al.}(2009){Varis}, {Hughes}, {Laaninen}, {Kilpi{\"a}},
  {Jukkala}, {Tuovinen}, {Ovaska}, {Sj{\"o}man}, {Kangaslahti}, {Gaier},
  {Hoyland}, {Meinhold}, {Mennella}, {Bersanelli}, {Butler}, {Cuttaia},
  {Franceschi}, {Leonardi}, {Leutenegger}, {Malaspina}, {Mandolesi},
  {Miccolis}, {Poutanen}, {Kurki-Suonio}, {Sandri}, {Stringhetti}, {Terenzi},
  {Tomasi}, \& {Valenziano}}]{varis2009}
{Varis}, J., {Hughes}, N.~J., {Laaninen}, M., {et~al.} 2009, Journal of
  Instrumentation, 4, 2001.
\newblock {Design, development, and verification of the Planck Low Frequency
  Instrument 70 GHz Front-End and Back-End Modules}

\bibitem[{{Verde} {et~al.}(2003){Verde}, {Peiris}, {Spergel}, {Nolta},
  {Bennett}, {Halpern}, {Hinshaw}, {Jarosik}, {Kogut}, {Limon}, {Meyer},
  {Page}, {Tucker}, {Wollack}, \& {Wright}}]{verde2003}
{Verde}, L., {Peiris}, H.~V., {Spergel}, D.~N., {et~al.} 2003, \apjs, 148, 195.
\newblock {First-Year Wilkinson Microwave Anisotropy Probe (WMAP) Observations:
  Parameter Estimation Methodology}

\bibitem[{{Villa} {et~al.}(2009){Villa}, {D'Arcangelo}, {Pecora}, {Figini},
  {Nesti}, {Simonetto}, {Sozzi}, {Sandri}, {Battaglia}, {Guzzi}, {Bersanelli},
  {Butler}, \& {Mandolesi}}]{villa2009}
{Villa}, F., {D'Arcangelo}, O., {Pecora}, M., {et~al.} 2009, Journal of
  Instrumentation, 4, 2004.
\newblock {Planck-LFI flight model feed horns}

\bibitem[{{Villa} {et~al.}(2010){Villa}, {Terenzi}, {Sandri}, {Meinhold},
  {Poutanen}, {Battaglia}, {Franceschet}, {Hughes}, {Laaninen}, {Lapolla},
  {Bersanelli}, {Butler}, {Cuttaia}, {D'Arcangelo}, {Frailis}, {Franceschi},
  {Galeotta}, {Gregorio}, {Leonardi}, {Lowe}, {Mandolesi}, {Maris}, {Mendes},
  {Mennella}, {Morgante}, {Stringhetti}, {Tomasi}, {Valenziano}, {Zacchei},
  {Zonca}, {Aja}, {Artal}, {Balasini}, {Bernardino}, {Blackhurst}, {Boschini},
  {Cappellini}, {Cavaliere}, {Colin}, {Colombo}, {Davis}, {de La Fuente},
  {Edgeley}, {Gaier}, {Galtress}, {Hoyland}, {Jukkala}, {Kettle}, {Kilpia},
  {Lawrence}, {Lawson}, {Leahy}, {Leutenegger}, {Levin}, {Maino}, {Malaspina},
  {Mediavilla}, {Miccolis}, {Pagan}, {Pascual}, {Pasian}, {Pecora},
  {Pospieszalski}, {Roddis}, {Salmon}, {Seiffert}, {Silvestri}, {Simonetto},
  {Sjoman}, {Sozzi}, {Tuovinen}, {Varis}, {Wilkinson}, \& {Winder}}]{villa2010}
{Villa}, F., {Terenzi}, L., {Sandri}, M., {et~al.} 2010, \aap, 520, A6.
\newblock {\textit{Planck} pre-launch status: Calibration of the Low Frequency
  Instrument flight model radiometers}

\bibitem[{Viterbi(1967)}]{viterbi1967}
Viterbi, A. 1967, IEEE Trans. Inform. Theor., 13, 260.
\newblock {Error bounds for convolutional codes and an asymptotically optimum
  decoding algorithm}

\bibitem[{{Walraven}(1984)}]{walraven1984}
{Walraven}, R. 1984, Proc. of the Digital Equipment User's Society, Fall, 1984.
\newblock {Digital Filters}

\bibitem[{{Weiland} {et~al.}(2011){Weiland}, {Odegard}, {Hill}, {Wollack},
  {Hinshaw}, {Greason}, {Jarosik}, {Page}, {Bennett}, {Dunkley}, {Gold},
  {Halpern}, {Kogut}, {Komatsu}, {Larson}, {Limon}, {Meyer}, {Nolta}, {Smith},
  {Spergel}, {Tucker}, \& {Wright}}]{weiland2010}
{Weiland}, J.~L., {Odegard}, N., {Hill}, R.~S., {et~al.} 2011, \apjs, 192, 19.
\newblock {Seven-year Wilkinson Microwave Anisotropy Probe (WMAP) Observations:
  Planets and Celestial Calibration Sources}

\bibitem[{{White}(2006)}]{white2006}
{White}, M. 2006, \nar, 50, 938.
\newblock {Cosmological science enabled by Planck}

\bibitem[{{Wright} {et~al.}(2009){Wright}, {Chen}, {Odegard}, {Bennett},
  {Hill}, {Hinshaw}, {Jarosik}, {Komatsu}, {Nolta}, {Page}, {Spergel},
  {Weiland}, {Wollack}, {Dunkley}, {Gold}, {Halpern}, {Kogut}, {Larson},
  {Limon}, {Meyer}, \& {Tucker}}]{wright2009}
{Wright}, E.~L., {Chen}, X., {Odegard}, N., {et~al.} 2009, \apjs, 180, 283.
\newblock {Five-Year Wilkinson Microwave Anisotropy Probe (WMAP) Observations:
  Source Catalog}

\bibitem[{{Zacchei} {et~al.}(2009){Zacchei}, {Frailis}, {Maris}, {Morisset},
  {Rohlfs}, {Meharga}, {Binko}, {T{\"u}rler}, {Galeotta}, {Gasparo},
  {Franceschi}, {Butler}, {Cuttaia}, {D'Arcangelo}, {Fogliani}, {Gregorio},
  {Leonardi}, {Lowe}, {Maino}, {Maggio}, {Malaspina}, {Mandolesi}, {Manzato},
  {Meinhold}, {Mendes}, {Mennella}, {Morgante}, {Pasian}, {Perrotta}, {Sandri},
  {Stringhetti}, {Terenzi}, {Tomasi}, \& {Zonca}}]{zacchei2009}
{Zacchei}, A., {Frailis}, M., {Maris}, M., {et~al.} 2009, Journal of
  Instrumentation, 4, 2019.
\newblock {Level 1 on-ground telemetry handling in Planck-LFI}

\bibitem[{{Zacchei} {et~al.}(2011){Zacchei}, {Maino}, {Baccigalupi},
  {Bersanelli}, {Bonaldi}, {Bonavera}, {Burigana}, {Butler}, {Cuttaia}, {de
  Zotti}, {Dick}, {Frailis}, {Galeotta}, {Gonz{\'a}lez-Nuevo}, {G{\'o}rski},
  {Gregorio}, {Keih{\"a}nen}, {Keskitalo}, {Knoche}, {Kurki-Suonio},
  {Lawrence}, {Leach}, {Leahy}, {L{\'o}pez-Caniego}, {Mandolesi}, {Maris},
  {Matthai}, {Meinhold}, {Mennella}, {Morgante}, {Morisset}, {Natoli},
  {Pasian}, {Perrotta}, {Polenta}, {Poutanen}, {Reinecke}, {Ricciardi},
  {Rohlfs}, {Sandri}, {Suur-Uski}, {Tauber}, {Tavagnacco}, {Terenzi}, {Tomasi},
  {Valiviita}, {Villa}, {Zonca}, {Banday}, {Barreiro}, {Bartlett}, {Bartolo},
  {Bedini}, {Bennett}, {Binko}, {Borrill}, {Bouchet}, {Bremer}, {Cabella},
  {Cappellini}, {Chen}, {Colombo}, {Cruz}, {Curto}, {Danese}, {Davies},
  {Davis}, {de Gasperis}, {de Rosa}, {de Troia}, {Dickinson}, {Diego},
  {Donzelli}, {D{\"o}rl}, {Efstathiou}, {En{\ss}lin}, {Eriksen}, {Falvella},
  {Finelli}, {Franceschi}, {Gaier}, {Gasparo}, {G{\'e}nova-Santos}, {Giardino},
  {G{\'o}mez}, {Gruppuso}, {Hansen}, {Hell}, {Herranz}, {Hovest}, {Huynh},
  {Jewell}, {Juvela}, {Kisner}, {Knox}, {L{\"a}hteenm{\"a}ki}, {Lamarre},
  {Leonardi}, {Le{\'o}n-Tavares}, {Lilje}, {Lubin}, {Maggio}, {Marinucci},
  {Mart{\'{\i}}nez-Gonz{\'a}lez}, {Massardi}, {Matarrese}, {Meharga},
  {Melchiorri}, {Migliaccio}, {Mitra}, {Moss}, {N{\o}rgaard-Nielsen}, {Pagano},
  {Paladini}, {Paoletti}, {Partridge}, {Pearson}, {Pettorino}, {Pietrobon},
  {Pr{\'e}zeau}, {Procopio}, {Puget}, {Quercellini}, {Rachen}, {Rebolo},
  {Robbers}, {Rocha}, {Rubi{\~n}o-Mart{\'{\i}}n}, {Salerno}, {Savelainen},
  {Scott}, {Seiffert}, {Silk}, {Smoot}, {Sternberg}, {Stivoli}, {Stompor},
  {Tofani}, {Toffolatti}, {Tuovinen}, {T{\"u}rler}, {Umana}, {Vielva},
  {Vittorio}, {Vuerli}, {Wade}, {Watson}, {White}, \&
  {Wilkinson}}]{planck2011-1.6}
{Zacchei}, A., {Maino}, D., {Baccigalupi}, C., {et~al.} 2011, \aap, 536, A5.
\newblock {\textit{Planck} early results. V. The Low Frequency Instrument data
  processing}

\bibitem[{{Zonca} {et~al.}(2009){Zonca}, {Franceschet}, {Battaglia}, {Villa},
  {Mennella}, {D'Arcangelo}, {Silvestri}, {Bersanelli}, {Artal}, {Butler},
  {Cuttaia}, {Davis}, {Galeotta}, {Hughes}, {Jukkala}, {Kilpi{\"a}},
  {Laaninen}, {Mandolesi}, {Maris}, {Mendes}, {Sandri}, {Terenzi}, {Tuovinen},
  {Varis}, \& {Wilkinson}}]{zonca2009}
{Zonca}, A., {Franceschet}, C., {Battaglia}, P., {et~al.} 2009, Journal of
  Instrumentation, 4, 2010.
\newblock {Planck-LFI radiometers' spectral response}

\end{thebibliography}


\def\eprinttmppp@#1arXiv:@{#1}
\providecommand{\arxivlink[1]}{\href{http://arxiv.org/abs/#1}{arXiv:#1}}
\def\eprinttmp@#1arXiv:#2 [#3]#4@{\ifthenelse{\equal{#3}{x}}{\ifthenelse{
\equal{#1}{}}{\arxivlink{\eprinttmppp@#2@}}{\arxivlink{#1}}}{\arxivlink{#2}
  [#3]}}
\providecommand{\eprintlink}[1]{\eprinttmp@#1arXiv: [x]@}
\providecommand{\eprint}[1]{\eprintlink{#1}}
\providecommand{\adsurl}[1]{\href{#1}{ADS}}
\begin{thebibliography}{116}
\expandafter\ifx\csname natexlab\endcsname\relax\def\natexlab#1{#1}\fi

\bibitem[{{Agudo} {et~al.}(2014){Agudo}, {Thum}, {G{\'o}mez}, \&
  {Wiesemeyer}}]{Agudo14}
{Agudo}, I., {Thum}, C., {G{\'o}mez}, J.~L., \& {Wiesemeyer}, H., {A
  simultaneous 3.5 and 1.3 mm polarimetric survey of active galactic nuclei in
  the northern sky}. 2014, \aap, 566, A59, \eprint{1402.0717}

\bibitem[{{Arg{\"u}eso} {et~al.}(2009){Arg{\"u}eso}, {Sanz}, {Herranz},
  {L{\'o}pez-Caniego}, \& {Gonz{\'a}lez-Nuevo}}]{Argueso09}
{Arg{\"u}eso}, F., {Sanz}, J.~L., {Herranz}, D., {L{\'o}pez-Caniego}, M., \&
  {Gonz{\'a}lez-Nuevo}, J., {Detection/estimation of the modulus of a vector.
  Application to point-source detection in polarization data}. 2009, \mnras,
  395, 649, \eprint{0906.0893}

\bibitem[{{Aumont} {et~al.}(2010){Aumont}, {Conversi}, {Thum}, {Wiesemeyer},
  {Falgarone}, {Mac{\'{\i}}as-P{\'e}rez}, {Piacentini}, {Pointecouteau},
  {Ponthieu}, {Puget}, {Rosset}, {Tauber}, \& {Tristram}}]{Aumont10}
{Aumont}, J., {Conversi}, L., {Thum}, C., {et~al.}, {Measurement of the Crab
  nebula polarization at 90 GHz as a calibrator for CMB experiments}. 2010,
  \aap, 514, A70

\bibitem[{{Beichman} {et~al.}(1988){Beichman}, {Neugebauer}, {Habing}, {Clegg},
  \& {Chester}}]{beichman88}
{Beichman}, C.~A., {Neugebauer}, G., {Habing}, H.~J., {Clegg}, P.~E., \&
  {Chester}, T.~J., eds. 1988, {Infrared astronomical satellite (IRAS) catalogs
  and atlases. Volume 1: Explanatory supplement}, Vol.~1

\bibitem[{{Bonavera} {et~al.}(2011){Bonavera}, {Massardi}, {Bonaldi},
  {Gonz{\'a}lez-Nuevo}, {de Zotti}, \& {Ekers}}]{bonavera11}
{Bonavera}, L., {Massardi}, M., {Bonaldi}, A., {et~al.}, {The Planck-ATCA
  Coeval Observations project: the faint sample}. 2011, \mnras, 416, 559,
  \eprint{1106.0614}

\bibitem[{{Bonnarel} {et~al.}(2000){Bonnarel}, {Fernique}, {Bienaym{\'e}},
  {Egret}, {Genova}, {Louys}, {Ochsenbein}, {Wenger}, \&
  {Bartlett}}]{Bonnarel10}
{Bonnarel}, F., {Fernique}, P., {Bienaym{\'e}}, O., {et~al.}, {The ALADIN
  interactive sky atlas. A reference tool for identification of astronomical
  sources}. 2000, \aaps, 143, 33

\bibitem[{{Boselli} {et~al.}(2010){Boselli}, {Eales}, {Cortese}, {Bendo},
  {Chanial}, {Buat}, {Davies}, {Auld}, {Rigby}, {Baes}, {Barlow}, {Bock},
  {Bradford}, {Castro-Rodriguez}, {Charlot}, {Clements}, {Cormier}, {Dwek},
  {Elbaz}, {Galametz}, {Galliano}, {Gear}, {Glenn}, {Gomez}, {Griffin}, {Hony},
  {Isaak}, {Levenson}, {Lu}, {Madden}, {O'Halloran}, {Okamura}, {Oliver},
  {Page}, {Panuzzo}, {Papageorgiou}, {Parkin}, {Perez-Fournon}, {Pohlen},
  {Rangwala}, {Roussel}, {Rykala}, {Sacchi}, {Sauvage}, {Schulz}, {Schirm},
  {Smith}, {Spinoglio}, {Stevens}, {Symeonidis}, {Vaccari}, {Vigroux},
  {Wilson}, {Wozniak}, {Wright}, \& {Zeilinger}}]{boselli10}
{Boselli}, A., {Eales}, S., {Cortese}, L., {et~al.}, {The Herschel Reference
  Survey}. 2010, \pasp, 122, 261, \eprint{1001.5136}

\bibitem[{Box \& Tiao(1992)}]{BoxTiao}
Box, G. \& Tiao, G. 1992, Bayesian Inference in Statistical Analisys (John
  Wiley \& Sons, Wiley-Interscience)

\bibitem[{{Butler et al.}(2015)}]{Butler15}
{Butler et al.}, {Absolute Calibron of the Radio Astronomy Flux Density Scale
  from 22 to 43 GHz using Planck}. 2015, Bull. AAS, 225

\bibitem[{{Carvalho} {et~al.}(2009){Carvalho}, {Rocha}, \& {Hobson}}]{PwSI}
{Carvalho}, P., {Rocha}, G., \& {Hobson}, M.~P., {A fast Bayesian approach to
  discrete object detection in astronomical data sets - PowellSnakes I}. 2009,
  \mnras, 393, 681, \eprint{0802.3916}

\bibitem[{{Carvalho} {et~al.}(2012){Carvalho}, {Rocha}, {Hobson}, \&
  {Lasenby}}]{PwSII}
{Carvalho}, P., {Rocha}, G., {Hobson}, M.~P., \& {Lasenby}, A., {PowellSnakes
  II: a fast Bayesian approach to discrete object detection in multi-frequency
  astronomical data sets}. 2012, \mnras, 427, 1384, \eprint{1112.4886}

\bibitem[{{Ciesla} {et~al.}(2012){Ciesla}, {Boselli}, {Smith}, {Bendo},
  {Cortese}, {Eales}, {Bianchi}, {Boquien}, {Buat}, {Davies}, {Pohlen},
  {Zibetti}, {Baes}, {Cooray}, {de Looze}, {di Serego Alighieri}, {Galametz},
  {Gomez}, {Lebouteiller}, {Madden}, {Pappalardo}, {Remy}, {Spinoglio},
  {Vaccari}, {Auld}, \& {Clements}}]{ciesla12}
{Ciesla}, L., {Boselli}, A., {Smith}, M.~W.~L., {et~al.}, {Submillimetre
  photometry of 323 nearby galaxies from the Herschel Reference Survey}. 2012,
  \aap, 543, A161, \eprint{1204.4726}

\bibitem[{{Clements} {et~al.}(2010){Clements}, {Rigby}, {Maddox}, {Dunne},
  {Mortier}, {Pearson}, {Amblard}, {Auld}, {Baes}, {Bonfield}, {Burgarella},
  {Buttiglione}, {Cava}, {Cooray}, {Dariush}, {de Zotti}, {Dye}, {Eales},
  {Frayer}, {Fritz}, {Gardner}, {Gonzalez-Nuevo}, {Herranz}, {Ibar}, {Ivison},
  {Jarvis}, {Lagache}, {Leeuw}, {Lopez-Caniego}, {Negrello}, {Pascale},
  {Pohlen}, {Rodighiero}, {Samui}, {Serjeant}, {Sibthorpe}, {Scott}, {Smith},
  {Temi}, {Thompson}, {Valtchanov}, {van der Werf}, \& {Verma}}]{clements10}
{Clements}, D.~L., {Rigby}, E., {Maddox}, S., {et~al.}, {Herschel-ATLAS:
  Extragalactic number counts from 250 to 500 microns}. 2010, \aap, 518, L8,
  \eprint{1005.2409}

\bibitem[{{de Zotti} {et~al.}(2005){de Zotti}, {Ricci}, {Mesa}, {Silva},
  {Mazzotta}, {Toffolatti}, \& {Gonz{\'a}lez-Nuevo}}]{deZotti05}
{de Zotti}, G., {Ricci}, R., {Mesa}, D., {et~al.}, {Predictions for
  high-frequency radio surveys of extragalactic sources}. 2005, \aap, 431, 893,
  \eprint{arXiv:astro-ph/0410709}

\bibitem[{{Eales} {et~al.}(2010){Eales}, {Dunne}, {Clements}, {Cooray}, {de
  Zotti}, {Dye}, {Ivison}, {Jarvis}, {Lagache}, {Maddox}, {Negrello},
  {Serjeant}, {Thompson}, {van Kampen}, {Amblard}, {Andreani}, {Baes},
  {Beelen}, {Bendo}, {Benford}, {Bertoldi}, {Bock}, {Bonfield}, {Boselli},
  {Bridge}, {Buat}, {Burgarella}, {Carlberg}, {Cava}, {Chanial}, {Charlot},
  {Christopher}, {Coles}, {Cortese}, {Dariush}, {da Cunha}, {Dalton}, {Danese},
  {Dannerbauer}, {Driver}, {Dunlop}, {Fan}, {Farrah}, {Frayer}, {Frenk},
  {Geach}, {Gardner}, {Gomez}, {Gonz{\'a}lez-Nuevo}, {Gonz{\'a}lez-Solares},
  {Griffin}, {Hardcastle}, {Hatziminaoglou}, {Herranz}, {Hughes}, {Ibar},
  {Jeong}, {Lacey}, {Lapi}, {Lawrence}, {Lee}, {Leeuw}, {Liske},
  {L{\'o}pez-Caniego}, {M{\"u}ller}, {Nandra}, {Panuzzo}, {Papageorgiou},
  {Patanchon}, {Peacock}, {Pearson}, {Phillipps}, {Pohlen}, {Popescu},
  {Rawlings}, {Rigby}, {Rigopoulou}, {Robotham}, {Rodighiero}, {Sansom},
  {Schulz}, {Scott}, {Smith}, {Sibthorpe}, {Smail}, {Stevens}, {Sutherland},
  {Takeuchi}, {Tedds}, {Temi}, {Tuffs}, {Trichas}, {Vaccari}, {Valtchanov},
  {van der Werf}, {Verma}, {Vieria}, {Vlahakis}, \& {White}}]{eales10}
{Eales}, S., {Dunne}, L., {Clements}, D., {et~al.}, {The Herschel ATLAS}. 2010,
  \pasp, 122, 499, \eprint{0910.4279}

\bibitem[{{Fixsen}(2009)}]{Fixsen09}
{Fixsen}, D.~J., {The Temperature of the Cosmic Microwave Background}. 2009,
  \apj, 707, 916, \eprint{0911.1955}

\bibitem[{{Gonz{\'a}lez-Nuevo} {et~al.}(2006){Gonz{\'a}lez-Nuevo},
  {Arg{\"u}eso}, {L{\'o}pez-Caniego}, {Toffolatti}, {Sanz}, {Vielva}, \&
  {Herranz}}]{gnuevo06}
{Gonz{\'a}lez-Nuevo}, J., {Arg{\"u}eso}, F., {L{\'o}pez-Caniego}, M., {et~al.},
  {The Mexican hat wavelet family: application to point-source detection in
  cosmic microwave background maps}. 2006, \mnras, 369, 1603,
  \eprint{arXiv:astro-ph/0604376}

\bibitem[{{Gonz{\'a}lez-Nuevo} {et~al.}(2008){Gonz{\'a}lez-Nuevo}, {Massardi},
  {Arg{\"u}eso}, {Herranz}, {Toffolatti}, {Sanz}, {L{\'o}pez-Caniego}, \& {de
  Zotti}}]{gnuevo08}
{Gonz{\'a}lez-Nuevo}, J., {Massardi}, M., {Arg{\"u}eso}, F., {et~al.},
  {Statistical properties of extragalactic sources in the New Extragalactic
  WMAP Point Source (NEWPS) catalogue}. 2008, \mnras, 384, 711,
  \eprint{0711.2631}

\bibitem[{{G{\'o}rski} {et~al.}(2005){G{\'o}rski}, {Hivon}, {Banday},
  {Wandelt}, {Hansen}, {Reinecke}, \& {Bartelmann}}]{gorski2005}
{G{\'o}rski}, K.~M., {Hivon}, E., {Banday}, A.~J., {et~al.}, {HEALPix: A
  Framework for High-Resolution Discretization and Fast Analysis of Data
  Distributed on the Sphere}. 2005, \apj, 622, 759, \eprint{astro-ph/0409513}

\bibitem[{{Gregory} {et~al.}(1996){Gregory}, {Scott}, {Douglas}, \&
  {Condon}}]{gregory96}
{Gregory}, P.~C., {Scott}, W.~K., {Douglas}, K., \& {Condon}, J.~J., {The GB6
  Catalog of Radio Sources}. 1996, \apjs, 103, 427

\bibitem[{{Griffin} {et~al.}(2010){Griffin}, {Abergel}, {Abreu}, {Ade},
  {Andr{\'e}}, {Augueres}, {Babbedge}, {Bae}, {Baillie}, {Baluteau}, {Barlow},
  {Bendo}, {Benielli}, {Bock}, {Bonhomme}, {Brisbin}, {Brockley-Blatt},
  {Caldwell}, {Cara}, {Castro-Rodriguez}, {Cerulli}, {Chanial}, {Chen},
  {Clark}, {Clements}, {Clerc}, {Coker}, {Communal}, {Conversi}, {Cox},
  {Crumb}, {Cunningham}, {Daly}, {Davis}, {de Antoni}, {Delderfield}, {Devin},
  {di Giorgio}, {Didschuns}, {Dohlen}, {Donati}, {Dowell}, {Dowell}, {Duband},
  {Dumaye}, {Emery}, {Ferlet}, {Ferrand}, {Fontignie}, {Fox}, {Franceschini},
  {Frerking}, {Fulton}, {Garcia}, {Gastaud}, {Gear}, {Glenn}, {Goizel},
  {Griffin}, {Grundy}, {Guest}, {Guillemet}, {Hargrave}, {Harwit}, {Hastings},
  {Hatziminaoglou}, {Herman}, {Hinde}, {Hristov}, {Huang}, {Imhof}, {Isaak},
  {Israelsson}, {Ivison}, {Jennings}, {Kiernan}, {King}, {Lange}, {Latter},
  {Laurent}, {Laurent}, {Leeks}, {Lellouch}, {Levenson}, {Li}, {Li},
  {Lilienthal}, {Lim}, {Liu}, {Lu}, {Madden}, {Mainetti}, {Marliani}, {McKay},
  {Mercier}, {Molinari}, {Morris}, {Moseley}, {Mulder}, {Mur}, {Naylor},
  {Nguyen}, {O'Halloran}, {Oliver}, {Olofsson}, {Olofsson}, {Orfei}, {Page},
  {Pain}, {Panuzzo}, {Papageorgiou}, {Parks}, {Parr-Burman}, {Pearce},
  {Pearson}, {P{\'e}rez-Fournon}, {Pinsard}, {Pisano}, {Podosek}, {Pohlen},
  {Polehampton}, {Pouliquen}, {Rigopoulou}, {Rizzo}, {Roseboom}, {Roussel},
  {Rowan-Robinson}, {Rownd}, {Saraceno}, {Sauvage}, {Savage}, {Savini},
  {Sawyer}, {Scharmberg}, {Schmitt}, {Schneider}, {Schulz}, {Schwartz},
  {Shafer}, {Shupe}, {Sibthorpe}, {Sidher}, {Smith}, {Smith}, {Smith},
  {Spencer}, {Stobie}, {Sudiwala}, {Sukhatme}, {Surace}, {Stevens}, {Swinyard},
  {Trichas}, {Tourette}, {Triou}, {Tseng}, {Tucker}, {Turner}, {Vaccari},
  {Valtchanov}, {Vigroux}, {Virique}, {Voellmer}, {Walker}, {Ward}, {Waskett},
  {Weilert}, {Wesson}, {White}, {Whitehouse}, {Wilson}, {Winter}, {Woodcraft},
  {Wright}, {Xu}, {Zavagno}, {Zemcov}, {Zhang}, \& {Zonca}}]{griffin10}
{Griffin}, M.~J., {Abergel}, A., {Abreu}, A., {et~al.}, {The Herschel-SPIRE
  instrument and its in-flight performance}. 2010, \aap, 518, L3,
  \eprint{1005.5123}

\bibitem[{Hagen \& Dereniak(2008)}]{hagen08}
Hagen, N. \& Dereniak, E.~L., Gaussian profile estimation in two dimensions.
  2008, Appl. Opt., 47, 6842

\bibitem[{{Hamaker} \& {Bregman}(1996)}]{IAUdef}
{Hamaker}, J.~P. \& {Bregman}, J.~D., {Understanding radio polarimetry. III.
  Interpreting the IAU/IEEE definitions of the Stokes parameters.} 1996, \aaps,
  117, 161

\bibitem[{{Healey} {et~al.}(2007){Healey}, {Romani}, {Taylor}, {Sadler},
  {Ricci}, {Murphy}, {Ulvestad}, \& {Winn}}]{healey07}
{Healey}, S.~E., {Romani}, R.~W., {Taylor}, G.~B., {et~al.}, {CRATES: An
  All-Sky Survey of Flat-Spectrum Radio Sources}. 2007, \apjs, 171, 61,
  \eprint{arXiv:astro-ph/0702346}

\bibitem[{{Herranz} {et~al.}(2012){Herranz}, {Arg{\"u}eso}, \&
  {Carvalho}}]{PolLikelihood}
{Herranz}, D., {Arg{\"u}eso}, F., \& {Carvalho}, P., {Compact Source Detection
  in Multichannel Microwave Surveys: From SZ Clusters to Polarized Sources}.
  2012, Advances in Astronomy, 2012, \eprint{1204.3834}

\bibitem[{Hinkley(1969)}]{AngleRatio}
Hinkley, D.~V., On the ratio of two correlated normal random variables. 1969,
  Biometrika, 56, 635,
  \eprint{http://biomet.oxfordjournals.org/content/56/3/635.full.pdf+html}

\bibitem[{{Lanz} {et~al.}(2013){Lanz}, {Herranz}, {L{\'o}pez-Caniego},
  {Gonz{\'a}lez-Nuevo}, {de Zotti}, {Massardi}, \& {Sanz}}]{lanz13}
{Lanz}, L.~F., {Herranz}, D., {L{\'o}pez-Caniego}, M., {et~al.}, {Extragalactic
  point source detection in Wilkinson Microwave Anisotropy Probe 7-year data at
  61 and 94 GHz}. 2013, \mnras, 428, 3048

\bibitem[{{L{\'o}pez-Caniego} {et~al.}(2007){L{\'o}pez-Caniego},
  {Gonz{\'a}lez-Nuevo}, {Herranz}, {Massardi}, {Sanz}, {De Zotti},
  {Toffolatti}, \& {Arg{\"u}eso}}]{caniego07}
{L{\'o}pez-Caniego}, M., {Gonz{\'a}lez-Nuevo}, J., {Herranz}, D., {et~al.},
  {Nonblind Catalog of Extragalactic Point Sources from the Wilkinson Microwave
  Anisotropy Probe (WMAP) First 3 Year Survey Data}. 2007, \apjs, 170, 108,
  \eprint{arXiv:astro-ph/0701473}

\bibitem[{{L{\'o}pez-Caniego} {et~al.}(2006){L{\'o}pez-Caniego}, {Herranz},
  {Gonz{\'a}lez-Nuevo}, {Sanz}, {Barreiro}, {Vielva}, {Arg{\"u}eso}, \&
  {Toffolatti}}]{caniego06}
{L{\'o}pez-Caniego}, M., {Herranz}, D., {Gonz{\'a}lez-Nuevo}, J., {et~al.},
  {Comparison of filters for the detection of point sources in Planck
  simulations}. 2006, \mnras, 370, 2047, \eprint{arXiv:astro-ph/0606199}

\bibitem[{{L{\'o}pez-Caniego} {et~al.}(2009){L{\'o}pez-Caniego}, {Massardi},
  {Gonz{\'a}lez-Nuevo}, {Lanz}, {Herranz}, {De Zotti}, {Sanz}, \&
  {Arg{\"u}eso}}]{LopezCaniego09}
{L{\'o}pez-Caniego}, M., {Massardi}, M., {Gonz{\'a}lez-Nuevo}, J., {et~al.},
  {Polarization of the WMAP Point Sources}. 2009, \apj, 705, 868,
  \eprint{0909.4311}

\bibitem[{{Louis} {et~al.}(2014){Louis}, {Addison}, {Hasselfield}, {Bond},
  {Calabrese}, {Das}, {Devlin}, {Dunkley}, {D{\"u}nner}, {Gralla}, {Hajian},
  {Hincks}, {Hlozek}, {Huffenberger}, {Infante}, {Kosowsky}, {Marriage},
  {Moodley}, {N{\ae}ss}, {Niemack}, {Nolta}, {Page}, {Partridge}, {Sehgal},
  {Sievers}, {Spergel}, {Staggs}, {Walter}, \& {Wollack}}]{Louis14}
{Louis}, T., {Addison}, G.~E., {Hasselfield}, M., {et~al.}, {The Atacama
  Cosmology Telescope: cross correlation with Planck maps}. 2014, \jcap, 7, 16,
  \eprint{1403.0608}

\bibitem[{{MacKenzie} {et~al.}(2011){MacKenzie}, {Braglia}, {Gibb}, {Scott},
  {Jenness}, {Serjeant}, {Thompson}, {Berry}, {Brunt}, {Chapin},
  {Chrysostomou}, {Clements}, {Coppin}, {Economou}, {Evans}, {Friberg},
  {Greaves}, {Hill}, {Holland}, {Ivison}, {Knapen}, {Jackson}, {Joncas},
  {Morgan}, {Mortier}, {Pearson}, {Pestalozzi}, {Pope}, {Richer}, {Urquhart},
  {Vaccari}, {Weferling}, {White}, \& {Zhu}}]{MacKenzie11}
{MacKenzie}, T., {Braglia}, F.~G., {Gibb}, A.~G., {et~al.}, {A pilot study for
  the SCUBA-2 `All-Sky' Survey}. 2011, \mnras, 415, 1950, \eprint{1012.1655}

\bibitem[{{Maddox et al.}(2015)}]{maddox15}
{Maddox et al.} 2015, in prep.

\bibitem[{{Marsden} {et~al.}(2013){Marsden}, {Gralla}, {Marriage}, {Switzer},
  {Partridge}, {Massardi}, {Morales}, {Addison}, {Bond}, {Crichton}, {Das},
  {Devlin}, {Dunner}, {Hajian}, {Hilton}, {Hincks}, {Hughes}, {Irwin},
  {Kosowsky}, {Menanteau}, {Moodley}, {Niemack}, {Page}, {Reese}, {Schmitt},
  {Sehgal}, {Sievers}, {Staggs}, {Swetz}, {Thornton}, \& {Wollack}}]{Mar13}
{Marsden}, D., {Gralla}, M., {Marriage}, T.~A., {et~al.}, {The Atacama
  Cosmology Telescope: Dusty Star-Forming Galaxies and Active Galactic Nuclei
  in the Southern Survey}. 2013, ArXiv:1306.2288, \eprint{1306.2288}

\bibitem[{{Massardi} {et~al.}(2009){Massardi}, {L{\'o}pez-Caniego},
  {Gonz{\'a}lez-Nuevo}, {Herranz}, {de Zotti}, \& {Sanz}}]{massardi09}
{Massardi}, M., {L{\'o}pez-Caniego}, M., {Gonz{\'a}lez-Nuevo}, J., {et~al.},
  {Blind and non-blind source detection in WMAP 5-yr maps}. 2009, \mnras, 392,
  733, \eprint{0810.2338}

\bibitem[{{Mitra} {et~al.}(2011){Mitra}, {Rocha}, {G{\'o}rski}, {Huffenberger},
  {Eriksen}, {Ashdown}, \& {Lawrence}}]{mitra2010}
{Mitra}, S., {Rocha}, G., {G{\'o}rski}, K.~M., {et~al.}, {Fast Pixel Space
  Convolution for Cosmic Microwave Background Surveys with Asymmetric Beams and
  Complex Scan Strategies: FEBeCoP}. 2011, \apjs, 193, 5, \eprint{1005.1929}

\bibitem[{{Miville-Desch{\^e}nes} \& {Lagache}(2005)}]{mdeschenes05}
{Miville-Desch{\^e}nes}, M.-A. \& {Lagache}, G., {IRIS: A New Generation of
  IRAS Maps}. 2005, \apjs, 157, 302, \eprint{arXiv:astro-ph/0412216}

\bibitem[{{Mocanu} {et~al.}(2013{\natexlab{a}}){Mocanu}, {Crawford}, {Vieira},
  {Aird}, {Aravena}, {Austermann}, {Benson}, {B{\'e}thermin}, {Bleem},
  {Bothwell}, {Carlstrom}, {Chang}, {Chapman}, {Cho}, {Crites}, {de Haan},
  {Dobbs}, {Everett}, {George}, {Halverson}, {Harrington}, {Hezaveh}, {Holder},
  {Holzapfel}, {Hoover}, {Hrubes}, {Keisler}, {Knox}, {Lee}, {Leitch},
  {Lueker}, {Luong-Van}, {Marrone}, {McMahon}, {Mehl}, {Meyer}, {Mohr},
  {Montroy}, {Natoli}, {Padin}, {Plagge}, {Pryke}, {Rest}, {Reichardt}, {Ruhl},
  {Sayre}, {Schaffer}, {Shirokoff}, {Spieler}, {Spilker}, {Stalder},
  {Staniszewski}, {Stark}, {Story}, {Switzer}, {Vanderlinde}, \&
  {Williamson}}]{Moc13}
{Mocanu}, L.~M., {Crawford}, T.~M., {Vieira}, J.~D., {et~al.}, {Extragalactic
  millimeter-wave point source catalog, number counts and statistics from 771
  square degrees of the SPT-SZ Survey}. 2013{\natexlab{a}}, ArXiv:1306.3470,
  \eprint{1306.3470}

\bibitem[{{Mocanu} {et~al.}(2013{\natexlab{b}}){Mocanu}, {Crawford}, {Vieira},
  {Aird}, {Aravena}, {Austermann}, {Benson}, {B{\'e}thermin}, {Bleem},
  {Bothwell}, {Carlstrom}, {Chang}, {Chapman}, {Cho}, {Crites}, {de Haan},
  {Dobbs}, {Everett}, {George}, {Halverson}, {Harrington}, {Hezaveh}, {Holder},
  {Holzapfel}, {Hoover}, {Hrubes}, {Keisler}, {Knox}, {Lee}, {Leitch},
  {Lueker}, {Luong-Van}, {Marrone}, {McMahon}, {Mehl}, {Meyer}, {Mohr},
  {Montroy}, {Natoli}, {Padin}, {Plagge}, {Pryke}, {Rest}, {Reichardt}, {Ruhl},
  {Sayre}, {Schaffer}, {Shirokoff}, {Spieler}, {Spilker}, {Stalder},
  {Staniszewski}, {Stark}, {Story}, {Switzer}, {Vanderlinde}, \&
  {Williamson}}]{Mocanu13}
{Mocanu}, L.~M., {Crawford}, T.~M., {Vieira}, J.~D., {et~al.}, {Extragalactic
  Millimeter-wave Point-source Catalog, Number Counts and Statistics from 771
  deg$^{2}$ of the SPT-SZ Survey}. 2013{\natexlab{b}}, \apj, 779, 61,
  \eprint{1306.3470}

\bibitem[{{Montier} {et~al.}(2015){Montier}, {Plaszczynski}, {Levrier},
  {Tristram}, {Alina}, {Ristorcelli}, \& {Bernard}}]{Montier15}
{Montier}, L., {Plaszczynski}, S., {Levrier}, F., {et~al.}, {Polarization
  measurement analysis. I. Impact of the full covariance matrix on polarization
  fraction and angle measurements}. 2015, \aap, 574, A135, \eprint{1406.6536}

\bibitem[{{Murphy} {et~al.}(2010){Murphy}, {Sadler}, {Ekers}, {Massardi},
  {Hancock}, {Mahony}, {Ricci}, {Burke-Spolaor}, {Calabretta}, {Chhetri}, {de
  Zotti}, {Edwards}, {Ekers}, {Jackson}, {Kesteven}, {Lindley}, {Newton-McGee},
  {Phillips}, {Roberts}, {Sault}, {Staveley-Smith}, {Subrahmanyan}, {Walker},
  \& {Wilson}}]{murphy10}
{Murphy}, T., {Sadler}, E.~M., {Ekers}, R.~D., {et~al.}, {The Australia
  Telescope 20 GHz Survey: the source catalogue}. 2010, \mnras, 402, 2403,
  \eprint{0911.0002}

\bibitem[{{Partridge} {et~al.}(2015){Partridge}, {L{\'o}pez-Caniego}, {Perley},
  {Stevens}, {Butler}, {Rocha}, {Walter}, \& {Zacchei}}]{2015arXiv150602892P}
{Partridge}, B., {L{\'o}pez-Caniego}, M., {Perley}, R.~A., {et~al.}, {Absolute
  Calibration of the Radio Astronomy Flux Density Scale at 22 to 43 GHz Using
  Planck}. 2015, ArXiv e-prints, \eprint{1506.02892}

\bibitem[{{Perley} \& {Butler}(2013)}]{Perley13}
{Perley}, R.~A. \& {Butler}, B.~J., {An Accurate Flux Density Scale from 1 to
  50 GHz}. 2013, \apjs, 204, 19, \eprint{1211.1300}

\bibitem[{{Pilbratt} {et~al.}(2010){Pilbratt}, {Riedinger}, {Passvogel},
  {Crone}, {Doyle}, {Gageur}, {Heras}, {Jewell}, {Metcalfe}, {Ott}, \&
  {Schmidt}}]{pilbratt10}
{Pilbratt}, G.~L., {Riedinger}, J.~R., {Passvogel}, T., {et~al.}, {Herschel
  Space Observatory. An ESA facility for far-infrared and submillimetre
  astronomy}. 2010, \aap, 518, L1, \eprint{1005.5331}

\bibitem[{{Planck Collaboration ES}(2015)}]{planck2014-ES}
{Planck Collaboration ES}. 2015, {The Explanatory Supplement to the \Planck\
  2015 results, \url{http://wiki.cosmos.esa.int/planckpla/index.php/Main_Page}}
  ({ESA})

\bibitem[{{\sorthelp{Planck Collaboration 2011G}}{Planck Collaboration
  VII}(2011)}]{planck2011-1.10}
{\sorthelp{Planck Collaboration 2011G}}{Planck Collaboration VII},
  {\textit{Planck} early results. VII. The Early Release Compact Source
  Catalogue}. 2011, \aap, 536, A7, \eprint{1101.2041}

\bibitem[{{\sorthelp{Planck Collaboration 2011H}}{Planck Collaboration
  VIII}(2011)}]{planck2011-5.1a}
{\sorthelp{Planck Collaboration 2011H}}{Planck Collaboration VIII},
  {\textit{Planck} early results. VIII. The all-sky early Sunyaev-Zeldovich
  cluster sample}. 2011, \aap, 536, A8, \eprint{1101.2024}

\bibitem[{{\sorthelp{Planck Collaboration 2011M}}{Planck Collaboration
  XIII}(2011)}]{planck2011-6.1}
{\sorthelp{Planck Collaboration 2011M}}{Planck Collaboration XIII},
  {\textit{Planck} early results. XIII. Statistical properties of extragalactic
  radio sources in the Planck Early Release Compact Source Catalogue}. 2011,
  \aap, 536, A13, \eprint{1101.2044}

\bibitem[{{\sorthelp{Planck Collaboration 2011N}}{Planck Collaboration
  XIV}(2011)}]{planck2011-6.2}
{\sorthelp{Planck Collaboration 2011N}}{Planck Collaboration XIV},
  {\textit{Planck} early results. XIV. ERCSC validation and extreme radio
  sources}. 2011, \aap, 536, A14, \eprint{1101.1721}

\bibitem[{{\sorthelp{Planck Collaboration 2014A}}{Planck Collaboration
  I}(2014)}]{planck2013-p01}
{\sorthelp{Planck Collaboration 2014A}}{Planck Collaboration I},
  {\textit{Planck} 2013 results. I. Overview of products and scientific
  results}. 2014, \aap, 571, A1, \eprint{1303.5062}

\bibitem[{{\sorthelp{Planck Collaboration 2014B}}{Planck Collaboration
  II}(2014)}]{planck2013-p02}
{\sorthelp{Planck Collaboration 2014B}}{Planck Collaboration II},
  {\textit{Planck} 2013 results. II. Low Frequency Instrument data processing}.
  2014, \aap, 571, A2, \eprint{1303.5063}

\bibitem[{{\sorthelp{Planck Collaboration 2014C}}{Planck Collaboration
  III}(2014)}]{planck2013-p02a}
{\sorthelp{Planck Collaboration 2014C}}{Planck Collaboration III},
  {\textit{Planck} 2013 results. III. LFI systematic uncertainties}. 2014,
  \aap, 571, A3, \eprint{1303.5064}

\bibitem[{{\sorthelp{Planck Collaboration 2014D}}{Planck Collaboration
  IV}(2014)}]{planck2013-p02d}
{\sorthelp{Planck Collaboration 2014D}}{Planck Collaboration IV},
  {\textit{Planck} 2013 results. IV. LFI Beams and window functions}. 2014,
  \aap, 571, A4, \eprint{1303.5065}

\bibitem[{{\sorthelp{Planck Collaboration 2014E}}{Planck Collaboration
  V}(2014)}]{planck2013-p02b}
{\sorthelp{Planck Collaboration 2014E}}{Planck Collaboration V},
  {\textit{Planck} 2013 results. V. LFI Calibration}. 2014, \aap, 571, A5,
  \eprint{1303.5066}

\bibitem[{{\sorthelp{Planck Collaboration 2014F}}{Planck Collaboration
  VI}(2014)}]{planck2013-p03}
{\sorthelp{Planck Collaboration 2014F}}{Planck Collaboration VI},
  {\textit{Planck} 2013 results. VI. High Frequency Instrument data
  processing}. 2014, \aap, 571, A6, \eprint{1303.5067}

\bibitem[{{\sorthelp{Planck Collaboration 2014G}}{Planck Collaboration
  VII}(2014)}]{planck2013-p03c}
{\sorthelp{Planck Collaboration 2014G}}{Planck Collaboration VII},
  {\textit{Planck} 2013 results. VII. HFI time response and beams}. 2014, \aap,
  571, A7, \eprint{1303.5068}

\bibitem[{{\sorthelp{Planck Collaboration 2014H}}{Planck Collaboration
  VIII}(2014)}]{planck2013-p03f}
{\sorthelp{Planck Collaboration 2014H}}{Planck Collaboration VIII},
  {\textit{Planck} 2013 results. VIII. HFI photometric calibration and
  mapmaking}. 2014, \aap, 571, A8, \eprint{1303.5069}

\bibitem[{{\sorthelp{Planck Collaboration 2014I}}{Planck Collaboration
  IX}(2014)}]{planck2013-p03d}
{\sorthelp{Planck Collaboration 2014I}}{Planck Collaboration IX},
  {\textit{Planck} 2013 results. IX. HFI spectral response}. 2014, \aap, 571,
  A9, \eprint{1303.5070}

\bibitem[{{\sorthelp{Planck Collaboration 2014J}}{Planck Collaboration
  X}(2014)}]{planck2013-p03e}
{\sorthelp{Planck Collaboration 2014J}}{Planck Collaboration X},
  {\textit{Planck} 2013 results. X. HFI energetic particle effects:
  characterization, removal, and simulation}. 2014, \aap, 571, A10,
  \eprint{1303.5071}

\bibitem[{{\sorthelp{Planck Collaboration 2014K}}{Planck Collaboration
  XI}(2014)}]{planck2013-p06b}
{\sorthelp{Planck Collaboration 2014K}}{Planck Collaboration XI},
  {\textit{Planck} 2013 results. XI. All-sky model of thermal dust emission}.
  2014, \aap, 571, A11, \eprint{1312.1300}

\bibitem[{{\sorthelp{Planck Collaboration 2014L}}{Planck Collaboration
  XII}(2014)}]{planck2013-p06}
{\sorthelp{Planck Collaboration 2014L}}{Planck Collaboration XII},
  {\textit{Planck} 2013 results. XII. Diffuse component separation}. 2014,
  \aap, 571, A12, \eprint{1303.5072}

\bibitem[{{\sorthelp{Planck Collaboration 2014M}}{Planck Collaboration
  XIII}(2014)}]{planck2013-p03a}
{\sorthelp{Planck Collaboration 2014M}}{Planck Collaboration XIII},
  {\textit{Planck} 2013 results. XIII. Galactic CO emission}. 2014, \aap, 571,
  A13, \eprint{1303.5073}

\bibitem[{{\sorthelp{Planck Collaboration 2014N}}{Planck Collaboration
  XIV}(2014)}]{planck2013-pip88}
{\sorthelp{Planck Collaboration 2014N}}{Planck Collaboration XIV},
  {\textit{Planck} 2013 results. XIV. Zodiacal emission}. 2014, \aap, 571, A14,
  \eprint{1303.5074}

\bibitem[{{\sorthelp{Planck Collaboration 2014O}}{Planck Collaboration
  XV}(2014)}]{planck2013-p08}
{\sorthelp{Planck Collaboration 2014O}}{Planck Collaboration XV},
  {\textit{Planck} 2013 results. XV. CMB power spectra and likelihood}. 2014,
  \aap, 571, A15, \eprint{1303.5075}

\bibitem[{{\sorthelp{Planck Collaboration 2014P}}{Planck Collaboration
  XVI}(2014)}]{planck2013-p11}
{\sorthelp{Planck Collaboration 2014P}}{Planck Collaboration XVI},
  {\textit{Planck} 2013 results. XVI. Cosmological parameters}. 2014, \aap,
  571, A16, \eprint{1303.5076}

\bibitem[{{\sorthelp{Planck Collaboration 2014Q}}{Planck Collaboration
  XVII}(2014)}]{planck2013-p12}
{\sorthelp{Planck Collaboration 2014Q}}{Planck Collaboration XVII},
  {\textit{Planck} 2013 results. XVII. Gravitational lensing by large-scale
  structure}. 2014, \aap, 571, A17, \eprint{1303.5077}

\bibitem[{{\sorthelp{Planck Collaboration 2014R}}{Planck Collaboration
  XVIII}(2014)}]{planck2013-p13}
{\sorthelp{Planck Collaboration 2014R}}{Planck Collaboration XVIII},
  {\textit{Planck} 2013 results. XVIII. The gravitational lensing-infrared
  background correlation}. 2014, \aap, 571, A18, \eprint{1303.5078}

\bibitem[{{\sorthelp{Planck Collaboration 2014S}}{Planck Collaboration
  XIX}(2014)}]{planck2013-p14}
{\sorthelp{Planck Collaboration 2014S}}{Planck Collaboration XIX},
  {\textit{Planck} 2013 results. XIX. The integrated Sachs-Wolfe effect}. 2014,
  \aap, 571, A19, \eprint{1303.5079}

\bibitem[{{\sorthelp{Planck Collaboration 2014T}}{Planck Collaboration
  XX}(2014)}]{planck2013-p15}
{\sorthelp{Planck Collaboration 2014T}}{Planck Collaboration XX},
  {\textit{Planck} 2013 results. XX. Cosmology from Sunyaev-Zeldovich cluster
  counts}. 2014, \aap, 571, A20, \eprint{1303.5080}

\bibitem[{{\sorthelp{Planck Collaboration 2014U}}{Planck Collaboration
  XXI}(2014)}]{planck2013-p05b}
{\sorthelp{Planck Collaboration 2014U}}{Planck Collaboration XXI},
  {\textit{Planck} 2013 results. XXI. Power spectrum and high-order statistics
  of the \textit{Planck} all-sky Compton parameter map}. 2014, \aap, 571, A21,
  \eprint{1303.5081}

\bibitem[{{\sorthelp{Planck Collaboration 2014V}}{Planck Collaboration
  XXII}(2014)}]{planck2013-p17}
{\sorthelp{Planck Collaboration 2014V}}{Planck Collaboration XXII},
  {\textit{Planck} 2013 results. XXII. Constraints on inflation}. 2014, \aap,
  571, A22, \eprint{1303.5082}

\bibitem[{{\sorthelp{Planck Collaboration 2014W}}{Planck Collaboration
  XXIII}(2014)}]{planck2013-p09}
{\sorthelp{Planck Collaboration 2014W}}{Planck Collaboration XXIII},
  {\textit{Planck} 2013 results. XXIII. Isotropy and statistics of the CMB}.
  2014, \aap, 571, A23, \eprint{1303.5083}

\bibitem[{{\sorthelp{Planck Collaboration 2014X}}{Planck Collaboration
  XXIV}(2014)}]{planck2013-p09a}
{\sorthelp{Planck Collaboration 2014X}}{Planck Collaboration XXIV},
  {\textit{Planck} 2013 results. XXIV. Constraints on primordial
  non-Gaussianity}. 2014, \aap, 571, A24, \eprint{1303.5084}

\bibitem[{{\sorthelp{Planck Collaboration 2014Y}}{Planck Collaboration
  XXV}(2014)}]{planck2013-p20}
{\sorthelp{Planck Collaboration 2014Y}}{Planck Collaboration XXV},
  {\textit{Planck} 2013 results. XXV. Searches for cosmic strings and other
  topological defects}. 2014, \aap, 571, A25, \eprint{1303.5085}

\bibitem[{{\sorthelp{Planck Collaboration 2014ZA}}{Planck Collaboration
  XXVI}(2014)}]{planck2013-p19}
{\sorthelp{Planck Collaboration 2014ZA}}{Planck Collaboration XXVI},
  {\textit{Planck} 2013 results. XXVI. Background geometry and topology of the
  Universe}. 2014, \aap, 571, A26, \eprint{1303.5086}

\bibitem[{{\sorthelp{Planck Collaboration 2014ZB}}{Planck Collaboration
  XXVII}(2014)}]{planck2013-pipaberration}
{\sorthelp{Planck Collaboration 2014ZB}}{Planck Collaboration XXVII},
  {\textit{Planck} 2013 results. XXVII. Doppler boosting of the CMB: Eppur si
  muove}. 2014, \aap, 571, A27, \eprint{1303.5087}

\bibitem[{{\sorthelp{Planck Collaboration 2014ZC}}{Planck Collaboration
  XXVIII}(2014)}]{planck2013-p05}
{\sorthelp{Planck Collaboration 2014ZC}}{Planck Collaboration XXVIII},
  {\textit{Planck} 2013 results. XXVIII. The Planck Catalogue of Compact
  Sources}. 2014, \aap, 571, A28, \eprint{1303.5088}

\bibitem[{{\sorthelp{Planck Collaboration 2014ZD}}{Planck Collaboration
  XXIX}(2014)}]{planck2013-p05a}
{\sorthelp{Planck Collaboration 2014ZD}}{Planck Collaboration XXIX},
  {\textit{Planck} 2013 results. XXIX. The Planck catalogue of
  Sunyaev-Zeldovich sources}. 2014, \aap, 571, A29, \eprint{1303.5089}

\bibitem[{{\sorthelp{Planck Collaboration 2014ZE}}{Planck Collaboration
  XXX}(2014)}]{planck2013-pip56}
{\sorthelp{Planck Collaboration 2014ZE}}{Planck Collaboration XXX},
  {\textit{Planck} 2013 results. XXX. Cosmic infrared background measurements
  and implications for star formation}. 2014, \aap, 571, A30,
  \eprint{1309.0382}

\bibitem[{{\sorthelp{Planck Collaboration 2014ZF}}{Planck Collaboration
  XXXI}(2014)}]{planck2013-p01a}
{\sorthelp{Planck Collaboration 2014ZF}}{Planck Collaboration XXXI},
  {\textit{Planck} 2013 results. XXXI. Consistency of the \textit{Planck}
  data}. 2014, \aap, 571, A31, \eprint{1508.03375}

\bibitem[{{\sorthelp{Planck Collaboration 2015A}}{Planck Collaboration
  I}(2016)}]{planck2014-a01}
{\sorthelp{Planck Collaboration 2015A}}{Planck Collaboration I},
  {\textit{Planck} 2015 results. I. Overview of products and results}. 2016,
  \aap, submitted, \eprint{1502.01582}

\bibitem[{{\sorthelp{Planck Collaboration 2015B}}{Planck Collaboration
  II}(2016)}]{planck2014-a03}
{\sorthelp{Planck Collaboration 2015B}}{Planck Collaboration II},
  {\textit{Planck} 2015 results. II. Low Frequency Instrument data processing}.
  2016, \aap, submitted, \eprint{1502.01583}

\bibitem[{{\sorthelp{Planck Collaboration 2015C}}{Planck Collaboration
  III}(2016)}]{planck2014-a04}
{\sorthelp{Planck Collaboration 2015C}}{Planck Collaboration III},
  {\textit{Planck} 2015 results. III. LFI systematic uncertainties}. 2016,
  \aap, submitted, \eprint{1507.08853}

\bibitem[{{\sorthelp{Planck Collaboration 2015D}}{Planck Collaboration
  IV}(2016)}]{planck2014-a05}
{\sorthelp{Planck Collaboration 2015D}}{Planck Collaboration IV},
  {\textit{Planck} 2015 results. IV. LFI beams and window functions}. 2016,
  \aap, in press, \eprint{1502.01584}

\bibitem[{{\sorthelp{Planck Collaboration 2015E}}{Planck Collaboration
  V}(2016)}]{planck2014-a06}
{\sorthelp{Planck Collaboration 2015E}}{Planck Collaboration V},
  {\textit{Planck} 2015 results. V. LFI calibration}. 2016, \aap, in press,
  \eprint{1505.08022}

\bibitem[{{\sorthelp{Planck Collaboration 2015F}}{Planck Collaboration
  VI}(2016)}]{planck2014-a07}
{\sorthelp{Planck Collaboration 2015F}}{Planck Collaboration VI},
  {\textit{Planck} 2015 results. VI. LFI maps}. 2016, \aap, submitted,
  \eprint{1502.01585}

\bibitem[{{\sorthelp{Planck Collaboration 2015G}}{Planck Collaboration
  VII}(2016)}]{planck2014-a08}
{\sorthelp{Planck Collaboration 2015G}}{Planck Collaboration VII},
  {\textit{Planck} 2015 results. VII. High Frequency Instrument data
  processing: Time-ordered information and beam processing}. 2016, \aap, in
  press, \eprint{1502.01586}

\bibitem[{{\sorthelp{Planck Collaboration 2015H}}{Planck Collaboration
  VIII}(2016)}]{planck2014-a09}
{\sorthelp{Planck Collaboration 2015H}}{Planck Collaboration VIII},
  {\textit{Planck} 2015 results. VIII. High Frequency Instrument data
  processing: Calibration and maps}. 2016, \aap, in press, \eprint{1502.01587}

\bibitem[{{\sorthelp{Planck Collaboration 2015I}}{Planck Collaboration
  IX}(2016)}]{planck2014-a11}
{\sorthelp{Planck Collaboration 2015I}}{Planck Collaboration IX},
  {\textit{Planck} 2015 results. IX. Diffuse component separation: CMB maps}.
  2016, \aap, submitted, \eprint{1502.05956}

\bibitem[{{\sorthelp{Planck Collaboration 2015J}}{Planck Collaboration
  X}(2016)}]{planck2014-a12}
{\sorthelp{Planck Collaboration 2015J}}{Planck Collaboration X},
  {\textit{Planck} 2015 results. X. Diffuse component separation: Foreground
  maps}. 2016, \aap, submitted, \eprint{1502.01588}

\bibitem[{{\sorthelp{Planck Collaboration 2015K}}{Planck Collaboration
  XI}(2016)}]{planck2014-a13}
{\sorthelp{Planck Collaboration 2015K}}{Planck Collaboration XI},
  {\textit{Planck} 2015 results. XI. CMB power spectra, likelihoods, and
  robustness of parameters}. 2016, \aap, submitted, \eprint{1507.02704}

\bibitem[{{\sorthelp{Planck Collaboration 2015L}}{Planck Collaboration
  XII}(2016)}]{planck2014-a14}
{\sorthelp{Planck Collaboration 2015L}}{Planck Collaboration XII},
  {\textit{Planck} 2015 results. XII. Full Focal Plane simulations}. 2016,
  \aap, submitted, \eprint{1509.06348}

\bibitem[{{\sorthelp{Planck Collaboration 2015M}}{Planck Collaboration
  XIII}(2016)}]{planck2014-a15}
{\sorthelp{Planck Collaboration 2015M}}{Planck Collaboration XIII},
  {\textit{Planck} 2015 results. XIII. Cosmological parameters}. 2016, \aap,
  submitted, \eprint{1502.01589}

\bibitem[{{\sorthelp{Planck Collaboration 2015N}}{Planck Collaboration
  XIV}(2016)}]{planck2014-a16}
{\sorthelp{Planck Collaboration 2015N}}{Planck Collaboration XIV},
  {\textit{Planck} 2015 results. XIV. Dark energy and modified gravity}. 2016,
  \aap, submitted, \eprint{1502.01590}

\bibitem[{{\sorthelp{Planck Collaboration 2015O}}{Planck Collaboration
  XV}(2016)}]{planck2014-a17}
{\sorthelp{Planck Collaboration 2015O}}{Planck Collaboration XV},
  {\textit{Planck} 2015 results. XV. Gravitational lensing}. 2016, \aap,
  submitted, \eprint{1502.01591}

\bibitem[{{\sorthelp{Planck Collaboration 2015P}}{Planck Collaboration
  XVI}(2016)}]{planck2014-a18}
{\sorthelp{Planck Collaboration 2015P}}{Planck Collaboration XVI},
  {\textit{Planck} 2015 results. XVI. Isotropy and statistics of the CMB}.
  2016, \aap, in press, \eprint{1506.07135}

\bibitem[{{\sorthelp{Planck Collaboration 2015Q}}{Planck Collaboration
  XVII}(2016)}]{planck2014-a19}
{\sorthelp{Planck Collaboration 2015Q}}{Planck Collaboration XVII},
  {\textit{Planck} 2015 results. XVII. Constraints on primordial
  non-Gaussianity}. 2016, \aap, submitted, \eprint{1502.01592}

\bibitem[{{\sorthelp{Planck Collaboration 2015R}}{Planck Collaboration
  XVIII}(2016)}]{planck2014-a20}
{\sorthelp{Planck Collaboration 2015R}}{Planck Collaboration XVIII},
  {\textit{Planck} 2015 results. XVIII. Background geometry and topology of the
  Universe}. 2016, \aap, submitted, \eprint{1502.01593}

\bibitem[{{\sorthelp{Planck Collaboration 2015S}}{Planck Collaboration
  XIX}(2016)}]{planck2014-a22}
{\sorthelp{Planck Collaboration 2015S}}{Planck Collaboration XIX},
  {\textit{Planck} 2015 results. XIX. Constraints on primordial magnetic
  fields}. 2016, \aap, submitted, \eprint{1502.01594}

\bibitem[{{\sorthelp{Planck Collaboration 2015T}}{Planck Collaboration
  XX}(2016)}]{planck2014-a24}
{\sorthelp{Planck Collaboration 2015T}}{Planck Collaboration XX},
  {\textit{Planck} 2015 results. XX. Constraints on inflation}. 2016, \aap,
  submitted, \eprint{1502.02114}

\bibitem[{{\sorthelp{Planck Collaboration 2015U}}{Planck Collaboration
  XXI}(2016)}]{planck2014-a26}
{\sorthelp{Planck Collaboration 2015U}}{Planck Collaboration XXI},
  {\textit{Planck} 2015 results. XXI. The integrated Sachs-Wolfe effect}. 2016,
  \aap, submitted, \eprint{1502.01595}

\bibitem[{{\sorthelp{Planck Collaboration 2015V}}{Planck Collaboration
  XXII}(2016)}]{planck2014-a28}
{\sorthelp{Planck Collaboration 2015V}}{Planck Collaboration XXII},
  {\textit{Planck} 2015 results. XXII. A map of the thermal Sunyaev-Zeldovich
  effect}. 2016, \aap, submitted, \eprint{1502.01596}

\bibitem[{{\sorthelp{Planck Collaboration 2015W}}{Planck Collaboration
  XXIII}(2016)}]{planck2014-a29}
{\sorthelp{Planck Collaboration 2015W}}{Planck Collaboration XXIII},
  {\textit{Planck} 2015 results. XXIII. The thermal Sunyaev-Zeldovich
  effect--cosmic infrared background correlation}. 2016, \aap, submitted,
  \eprint{1509.06555}

\bibitem[{{\sorthelp{Planck Collaboration 2015X}}{Planck Collaboration
  XXIV}(2016)}]{planck2014-a30}
{\sorthelp{Planck Collaboration 2015X}}{Planck Collaboration XXIV},
  {\textit{Planck} 2015 results. XXIV. Cosmology from Sunyaev-Zeldovich cluster
  counts}. 2016, \aap, submitted, \eprint{1502.01597}

\bibitem[{{\sorthelp{Planck Collaboration 2015Y}}{Planck Collaboration
  XXV}(2016)}]{planck2014-a31}
{\sorthelp{Planck Collaboration 2015Y}}{Planck Collaboration XXV},
  {\textit{Planck} 2015 results. XXV. Diffuse, low-frequency Galactic
  foregrounds}. 2016, \aap, submitted, \eprint{1506.06660}

\bibitem[{{\sorthelp{Planck Collaboration 2015ZA}}{Planck Collaboration
  XXVI}(2016)}]{planck2014-a35}
{\sorthelp{Planck Collaboration 2015ZA}}{Planck Collaboration XXVI},
  {\textit{Planck} 2015 results. XXVI. The Second Planck Catalogue of Compact
  Sources}. 2016, \aap, submitted, \eprint{1507.02058}

\bibitem[{{\sorthelp{Planck Collaboration 2015ZB}}{Planck Collaboration
  XXVII}(2016)}]{planck2014-a36}
{\sorthelp{Planck Collaboration 2015ZB}}{Planck Collaboration XXVII},
  {\textit{Planck} 2015 results. XXVII. The Second Planck Catalogue of
  Sunyaev-Zeldovich Sources}. 2016, \aap, in press, \eprint{1502.01598}

\bibitem[{{\sorthelp{Planck Collaboration 2015ZC}}{Planck Collaboration
  XXVIII}(2016)}]{planck2014-a37}
{\sorthelp{Planck Collaboration 2015ZC}}{Planck Collaboration XXVIII},
  {\textit{Planck} 2015 results. XXVIII. The Planck Catalogue of Galactic Cold
  Clumps}. 2016, \aap, in press, \eprint{1502.01599}

\bibitem[{{\sorthelp{Planck Collaboration IntG}}{Planck Collaboration Int.
  VII}(2013)}]{planck2012-VII}
{\sorthelp{Planck Collaboration IntG}}{Planck Collaboration Int. VII},
  {\textit{Planck} intermediate results. VII. Statistical properties of
  infrared and radio extragalactic sources from the Planck Early Release
  Compact Source Catalogue at frequencies between 100 and 857\,GHz}. 2013,
  \aap, 550, A133, \eprint{1207.4706}

\bibitem[{Press {et~al.}(1992)Press, Teukolsky, Vetterling, \&
  Flannery}]{Press:1992:NRC:148286}
Press, W.~H., Teukolsky, S.~A., Vetterling, W.~T., \& Flannery, B.~P. 1992,
  Numerical Recipes in C (2Nd Ed.): The Art of Scientific Computing (New York,
  NY, USA: Cambridge University Press)

\bibitem[{{Rigby} {et~al.}(2011){Rigby}, {Maddox}, {Dunne}, {Negrello},
  {Smith}, {Gonz{\'a}lez-Nuevo}, {Herranz}, {L{\'o}pez-Caniego}, {Auld},
  {Buttiglione}, {Baes}, {Cava}, {Cooray}, {Clements}, {Dariush}, {de Zotti},
  {Dye}, {Eales}, {Frayer}, {Fritz}, {Hopwood}, {Ibar}, {Ivison}, {Jarvis},
  {Panuzzo}, {Pascale}, {Pohlen}, {Rodighiero}, {Serjeant}, {Temi}, \&
  {Thompson}}]{rigby11}
{Rigby}, E.~E., {Maddox}, S.~J., {Dunne}, L., {et~al.}, {Herschel-ATLAS: first
  data release of the Science Demonstration Phase source catalogues}. 2011,
  \mnras, 415, 2336, \eprint{1010.5787}

\bibitem[{{Ter{\"a}sranta} {et~al.}(2004){Ter{\"a}sranta}, {Achren}, {Hanski},
  {Heikkil{\"a}}, {Holopainen}, {Joutsamo}, {Juhola}, {Karlamaa}, {Katajainen},
  {Kein{\"a}nen}, {Koivisto}, {Koskimies}, {K{\"o}n{\"o}nen}, {Lainela},
  {L{\"a}htenm{\"a}ki}, {M{\"a}kinen}, {Niemel{\"a}}, {Nurmi}, {Pursimo},
  {Rekola}, {Savolainen}, {Tornikoski}, {Torppa}, {Valtonen}, {Varjonen},
  {Vilenius}, {Virtanen}, \& {Wiren}}]{Terasranta04}
{Ter{\"a}sranta}, H., {Achren}, J., {Hanski}, M., {et~al.}, {Twenty years
  monitoring of extragalactic sources at 22, 37 and 87 GHz}. 2004, \aap, 427,
  769

\bibitem[{{Trippe} {et~al.}(2010){Trippe}, {Neri}, {Krips}, {Castro-Carrizo},
  {Bremer}, {Pi{\'e}tu}, \& {Fontana}}]{trippe10}
{Trippe}, S., {Neri}, R., {Krips}, M., {et~al.}, {The first IRAM/PdBI
  polarimetric millimeter survey of active galactic nuclei. I. Global
  properties of the sample}. 2010, \aap, 515, A40, \eprint{1003.3205}

\bibitem[{{Valiante et al.}(2015)}]{valiante15}
{Valiante et al.} 2015, in prep.

\bibitem[{{Wang} {et~al.}(2014){Wang}, {Rowan-Robinson}, {Norberg}, {Heinis},
  \& {Han}}]{wang14}
{Wang}, L., {Rowan-Robinson}, M., {Norberg}, P., {Heinis}, S., \& {Han}, J.,
  {The Revised IRAS-FSC Redshift Catalogue (RIFSCz)}. 2014, \mnras, 442, 2739,
  \eprint{1402.4991}

\bibitem[{{Weiland} {et~al.}(2011){Weiland}, {Odegard}, {Hill}, {Wollack},
  {Hinshaw}, {Greason}, {Jarosik}, {Page}, {Bennett}, {Dunkley}, {Gold},
  {Halpern}, {Kogut}, {Komatsu}, {Larson}, {Limon}, {Meyer}, {Nolta}, {Smith},
  {Spergel}, {Tucker}, \& {Wright}}]{Weiland11}
{Weiland}, J.~L., {Odegard}, N., {Hill}, R.~S., {et~al.}, {Seven-year Wilkinson
  Microwave Anisotropy Probe (WMAP) Observations: Planets and Celestial
  Calibration Sources}. 2011, \apjs, 192, 19, \eprint{1001.4731}

\end{thebibliography}
\appendix
\section{PSF photometry}
\label{sec:psfflux}

The flux density is obtained by fitting a model of the PSF at the position of the source. The model has four free parameters: the amplitude of the source, a background offset, and  two coordinates for the location of the source. The PSF is obtained from the effective beam ~\citep{planck2013-p02, planck2013-p03}. The model of the source is
\begin{equation}
\vec{m} = A \vec{P} + C,
\end{equation}
where $\vec{P}$ is the PSF at the position of the source (the
integrated response to a point-like source), $A$ is the amplitude of the source, and $C$ is the (constant) background.  The PSF at the position of the source is obtained
from the effective beam, which is defined only at the centre of each map pixel, by means of a bicubic interpolation between adjacent pixels. This step is new: in the previous version of the PCCS, the PSF was built from the effective beam at the centre of the pixel
associated with the location of the source. The PSF model $\vec{P}$ depends therefore on the position of the source, $\vec{P}=\vec{P}(x_s,y_s)$.

The best-fit values of the parameters $\beta = (A, C, x_s, y_s)$ are found by minimizing the $\chi^2$ between the model and the data, $\vec{d}$,
\begin{equation}
\chi^2(\beta) = \sum (\vec{d} - \vec{m})^{\rm T} \tens{N^{-1}} (\vec{d} - \vec{m}),
\label{eqn:phot_chisq}
\end{equation}
where $\tens{N}$ is the covariance matrix of the noise. The noise is assumed to be uncorrelated between pixels. 
The overall normalization of the noise is adjusted by setting $\chi^2=1$ at the best-fit
value of the parameters. We also include the uncertainty of the background flux in the error of the flux density estimation. This has the effect of
inflating the uncertainties to account for any mismatch between the modelled PSF and the true shape of the source and the background in the map. The uncertainties on
the parameters are computed from the curvature of the $\chi^2$. The best-fit amplitude and its uncertainty are converted to units of flux density using the area of the PSF and the unit conversion from K$_{\rm CMB}$ to MJy\,${\rm{sr}^{-1}}$ for each \textit{Planck} channel.

\section{Gaussian fitting method}
\label{sec:gauflux}

\begin{figure}
\includegraphics[width=0.53\textwidth]{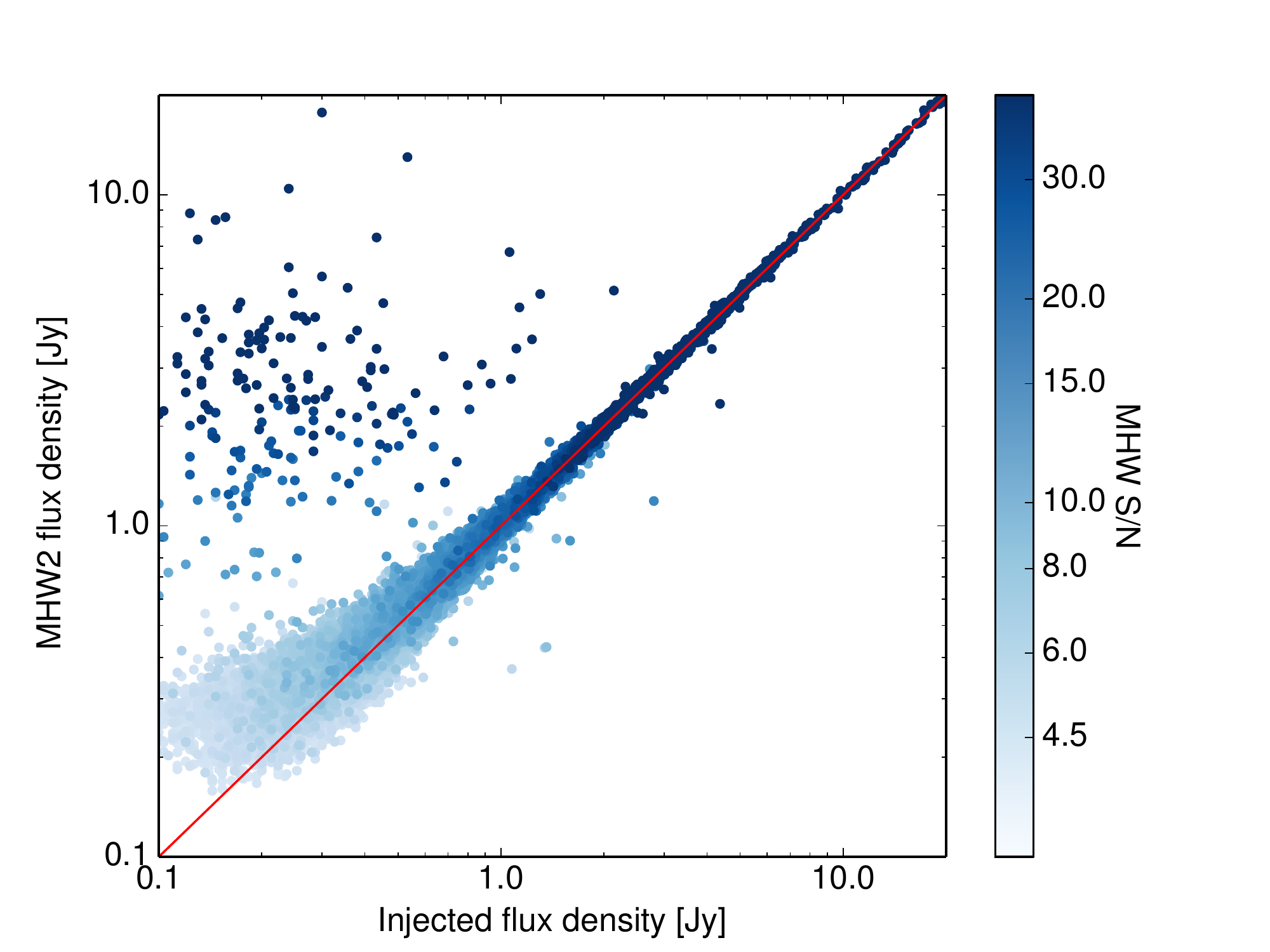}
\caption{Comparison of injected and recovered MHW2 flux densities for point sources at 100\,GHz.
The saturation of points is proportional to the \snr\ of the MHW2 detection. The outliers towards the upper left  are associated with faint
sources with injected flux  densities $<1$\,Jy located in areas with complicated backgrounds close to the Galactic plane. It is known that in these regions the MHW2 algorithm may give biased flux-density estimates.}
\label{TrueVsMHWPS}
\end{figure}

\begin{figure}
\includegraphics[width=0.53\textwidth]{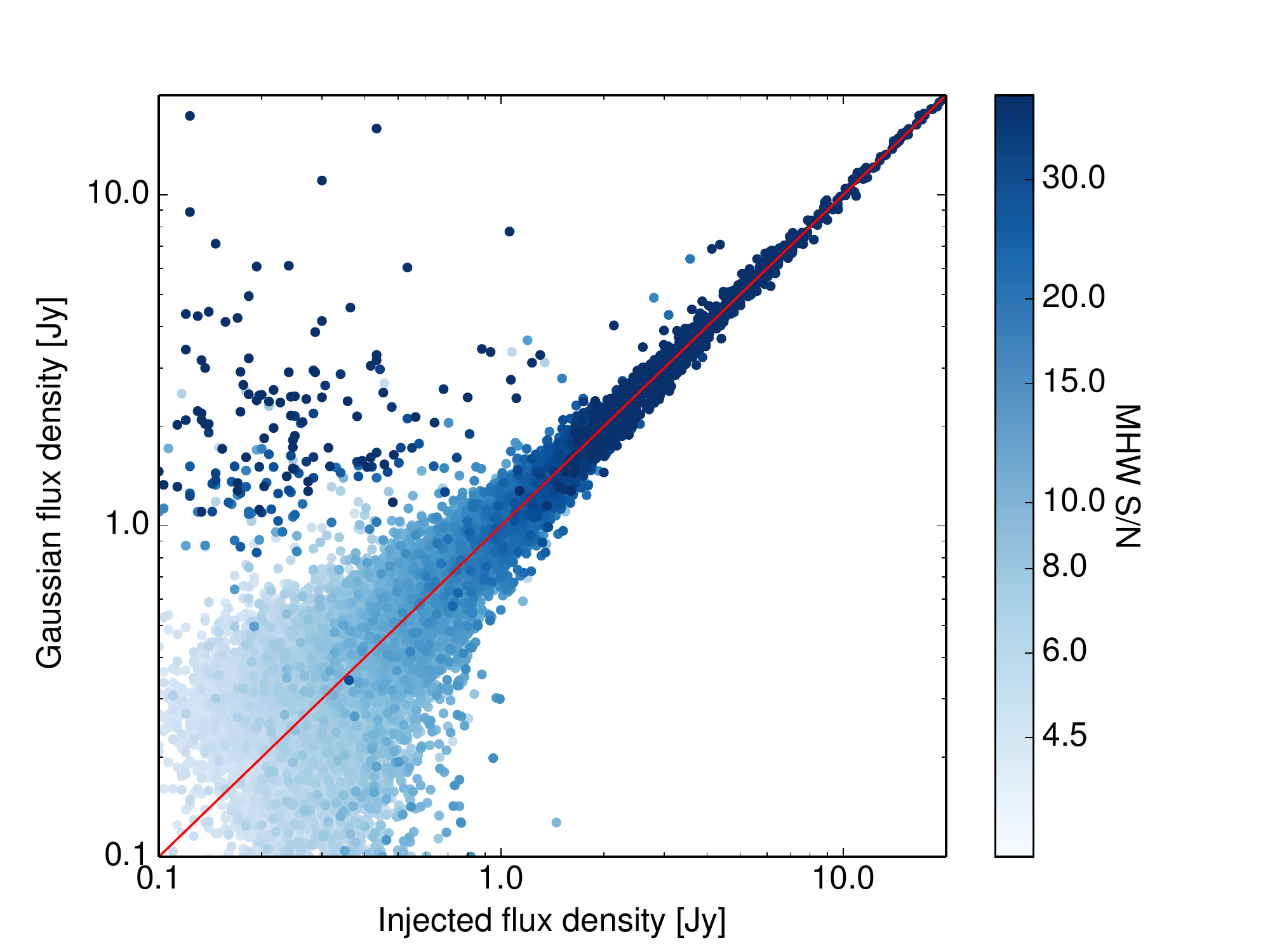}
\caption{Comparison of injected and recovered Gaussian flux densities for point sources at 100\,GHz.
The saturation of points is proportional to the \snr\ of the MHW2 detection. Gaussian fitting improves the outlier flux densities slightly,
but the flux-density estimates are still biased.}
\label{TrueVsGFPS}
\end{figure}

\begin{figure}
\includegraphics[width=0.53\textwidth]{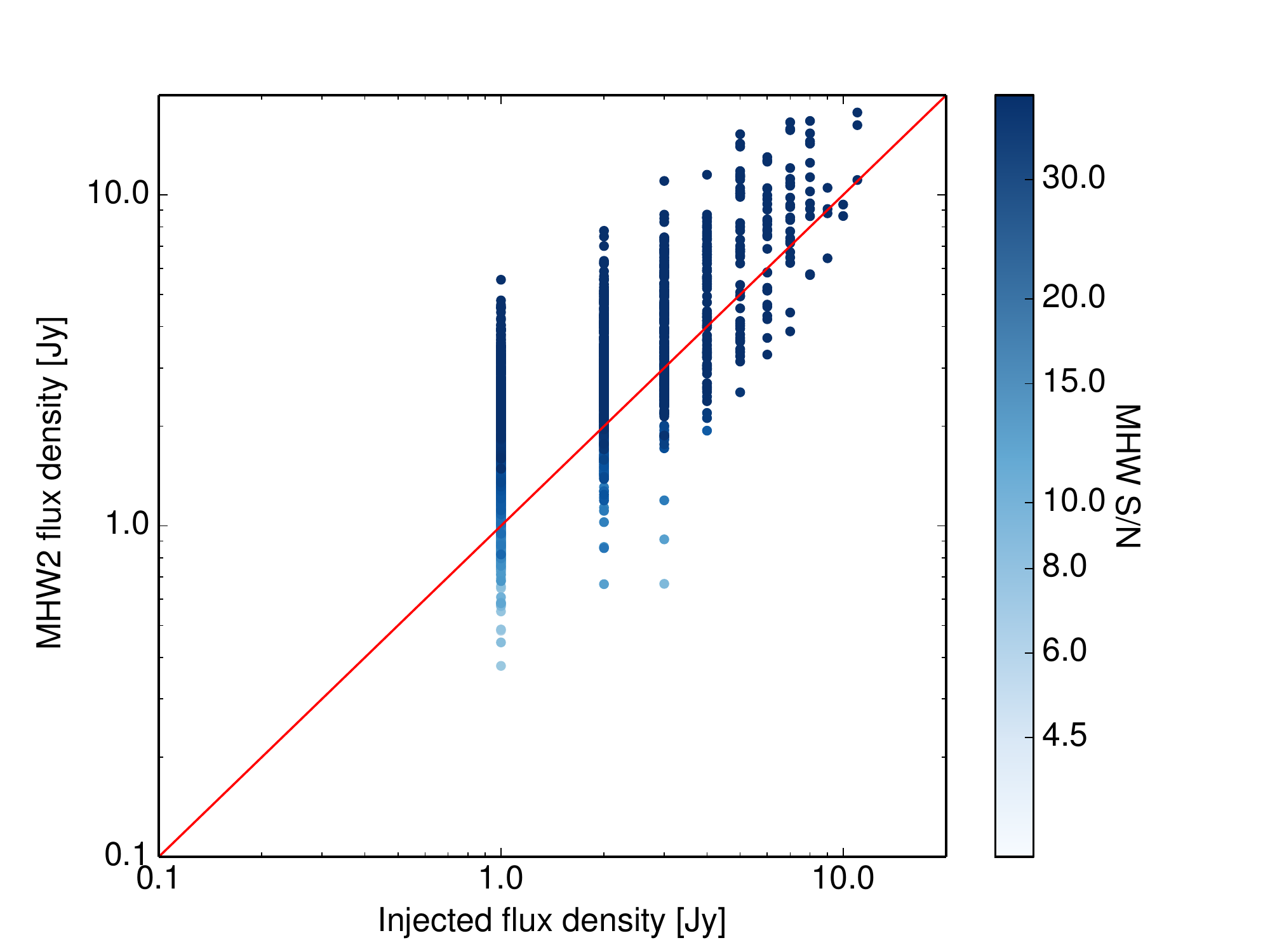}
\caption{Comparison of injected and recovered MHW2 flux densities for cold cores at 100\,GHz.
The saturation of points is proportional to the \snr\ of the MHW2 detection.  The visible digitization is because only integer values were used for the flux densities of injected sources.}
\label{TrueVsMHWCC}
\end{figure}

\begin{figure}
\includegraphics[width=0.53\textwidth]{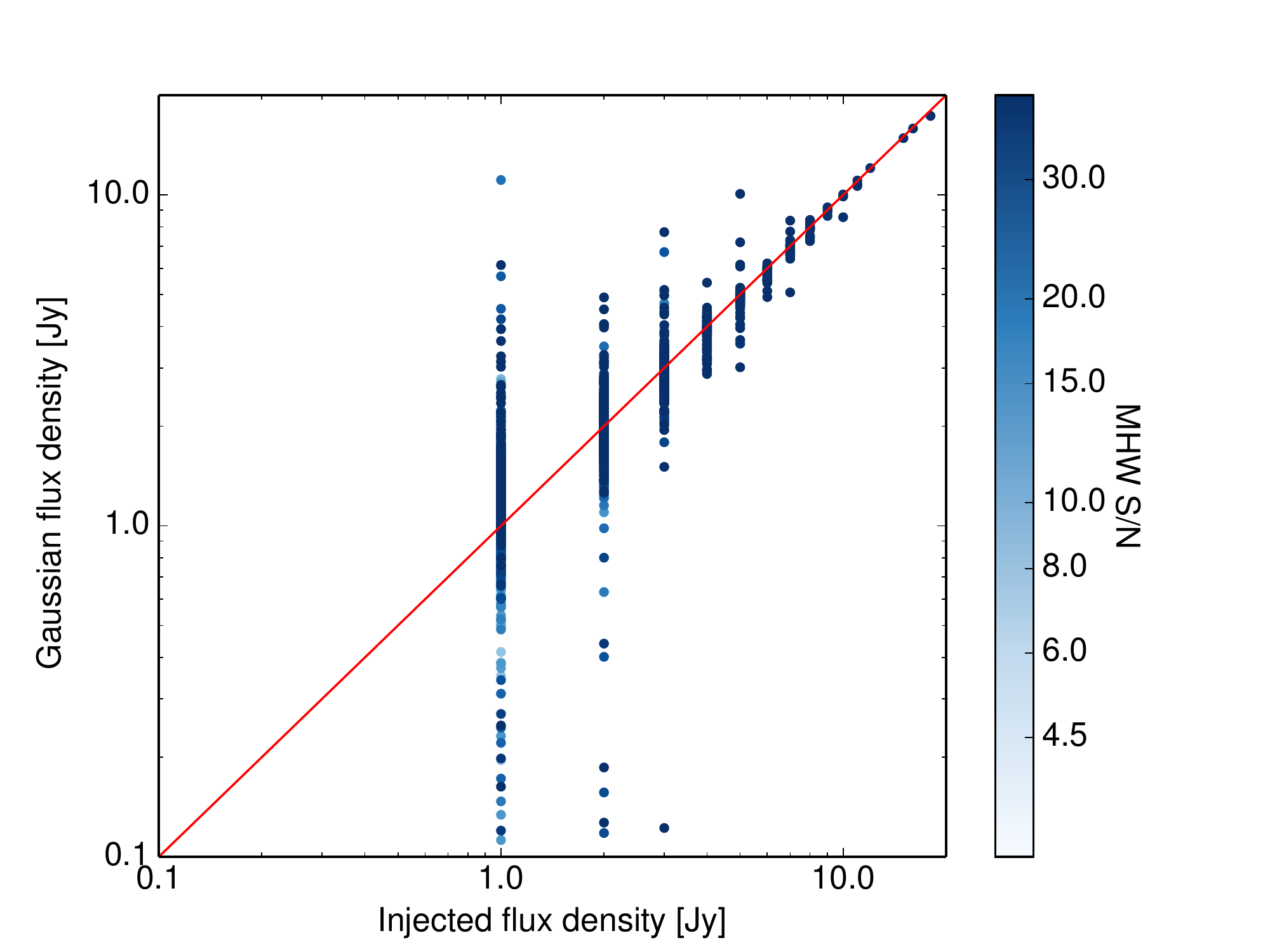}
\caption{Comparison of injected and recovered Gaussian flux densities for cold cores at 100\,GHz.
The saturation of points is proportional to the \snr\ of the MHW2 detection.  The visible digitization is because only integer values were used for the flux densities of injected sources.} 
\label{TrueVsGFCC}
\end{figure}

To recover the shape of the source and its orientation, and to improve on the MHW2 flux-density estimate, Gaussian fitting is used. Taking point source parameters from the MHW2 catalogue as input, a 2D Gaussian fit is performed at the location of each catalogue source. Six parameters are fitted: Galactic coordinates of the source $(l,b)$; the flux density; the major and minor semi-axes; and an orientation angle. The source semi-axes and orientation angle are used to flag elongated sources.

\subsection{Downhill simplex in multidimensions}

The downhill simplex method in multidimensions, i.e. the Nelder--Mead or ``amoeba'' method \citep{Press:1992:NRC:148286}, which is useful for problems where the derivatives are not known, is used for optimization in multiparameter space.
The functional to optimize is based on
the reduced log-likelihood with some prior regularization for the size
of the source defined by the effective beam at the given frequency.
The algorithm starts optimization at the MHW2 source location, assuming initially a circular Gaussian source with the FHWM of the PSF. Optimization of all six parameters converges in a reasonable number of iterations, usually less than 1000.

\subsection{Error estimation}

The downhill simplex method does not provide any information on the
uncertainties in the optimized parameters. We estimate errors using a Markov Chain
Monte Carlo (MCMC) method to sample the posterior distribution. In our case,
the burn-in time can be very small since to start the chain the simplex-optimized value of the parameter from the previous step can be used, and
it is usually very close to the peak of the distribution. The length of the MCMC chain is set to 1000 samples by default.

To obtain the proper sampling in the MCMC chain it is necessary to define
the optimal step size in the parameter space. This determination is complex since there are more than 100\,000 sources in the catalogue to process; it would be impossible to adjust every MCMC chain, since it would take an excessive amount of time to try several values of the sampling step using an iterative approach.
Hence we decided to use an analytical approximation for the errors of most of the Gaussian profile parameters, following \citet{hagen08}.
 We used an analytical approximation for the separable 2D Gaussian profile model; the errors are derived for five parameters; namely position coordinates, flux density, and two non-equal width values. The calculated errors were used to find the value of the step size in the parameter space to start the MCMC sampler, which delivered the error on the tilt angle. This combined technique allowed us to avoid most of the problems related to undersampled or oversampled MCMC chains.

\subsection{Validation of the method using model data}

The results of the Gaussian fit were compared with the MHW2 positions and flux densities for a test set of sources with known parameters. The PR2  100\,GHz map with injected sources and cold cores was used to check the 2D Gaussian fitting algorithm. The cold cores were modelled as 2D Gaussian profiles with uniformly distributed random orientation angles. The major FWHM was in the range from 4\farcm5 to 19\arcmin, and the ellipticity parameter, $e={\rm FWHM}_{\rm major}/{\rm FWHM}_{\rm minor}$, varied from 1 (circular source) to 7 (highly elliptical source). All sources were uniformly distributed across the sky.

Plots comparing the flux density values for MHW2 and Gaussian fitting for this test set
 are shown in Figs.~\ref{TrueVsMHWPS}--\ref{TrueVsGFCC}.
Both methods give almost the same point source flux-density values
(Figs.~\ref{TrueVsMHWPS} and \ref{TrueVsGFPS}), but MHW2 gives less scatter for the fainter point sources. For the cold cores,
Gaussian fitting works better and gives much less bias, especially
for the bright sources (Figs.~\ref{TrueVsMHWCC} and \ref{TrueVsGFCC}).
The visible digitization of the flux densities reflects the fact that  only integer values were used for the flux densities of injected sources.\

The group of outliers in the upper left part of Figs.~\ref{TrueVsMHWPS}~and~\ref{TrueVsGFPS} correspond to faint point sources ($<1$\,Jy) in regions with complicated backgrounds, near the Galactic plane. 
Both MHW2 and Gaussian fitting fail to recover their flux densities accurately and give biased estimates.

It is clear that Gaussian fitting is preferable for extended
objects like cold cores, giving more accurate flux densities as well as recovering the actual source shape and orientation.

\section{Bandpass mismatch and polarization measures}
\label{sec:mismatch}
For sources with spectra differing from the CMB spectrum, bandpass mismatch causes leakage of temperature to polarization (see Sect.~\ref{subsec:bandpass_mismatch}).  To correct for this leakage, we require a model of the source spectrum.  That requirement was handled differently by LFI and HFI.

\subsection{LFI-specific details}
 In the case of LFI  maps corrected for bandpass-mismatch leakage,  the spectral model is based on the diffuse component separation analysis \citep{planck2014-a03}. These models are not particularly
accurate for compact sources; moreover, the analysis is done at a common resolution of 1\degr\,FWHM, whereas
for compact sources it is best to work at full resolution. Therefore the \plccs\ polarization measurements were evaluated
by extracting the $Q$ and $U$ flux densities from the full-resolution uncorrected maps, and correcting for leakage
at the flux density level, using
\begin{equation}
\lcv Q \\ U \rcv_{\rm corrected} = \lcv Q \\ U \rcv_{\rm raw} - \lcv P_Q \\ P_U \rcv(\alpha - \alpha_{\rm CMB}) I,
\end{equation}
where $I$ is the total intensity, $\alpha = d \ln I/d \ln \nu$ is the source spectral index in the relevant frequency band, and $\alpha_{\rm CMB}$ is the spectral index of the CMB fluctuations ($1.96, 1.90$, and  $1.75$, at 28.4, 44.1, and 70.4\,GHz, 
respectively). $P_{Q,U}$ is the projection factor derived in \citet{planck2014-a03}, which is evaluated at
each map pixel. For the point-source correction
we averaged the projection factors over the source pixels using the same weights as were used to extract the $Q$ and
$U$ fluxes.

\subsubsection{Evaluation of LFI polarization measurements}
The polarized flux density $P$ is calculated using Eq.~(\ref{eqn:P_def}), where the $\hat{Q}$ and $\hat{U}$ maps are obtained by applying the ``filtered fusion'' maximum-likelihood estimator \citep{Argueso09,LopezCaniego09} to the original Stokes $Q$ and $U$ maps.
The errors in $P$ are calculated using Eq.~(\ref{eqn:P_err}), adding in quadrature the error estimate obtained from the Monte Carlo simulations and calculating the 1\,$\sigma$ asymmetric errors for a Rayleigh distribution in the intervals [$0,0.159$] and [$0.841,1$]. The errors in the polarization angle, $\psi$, are calculated using Eq.~(\ref{eqn:P_ang_err}).

\subsection{HFI-specific details}
\label{subsec:polHFI}

\begin{figure}
\begin{center}
\includegraphics[width=0.48\textwidth]{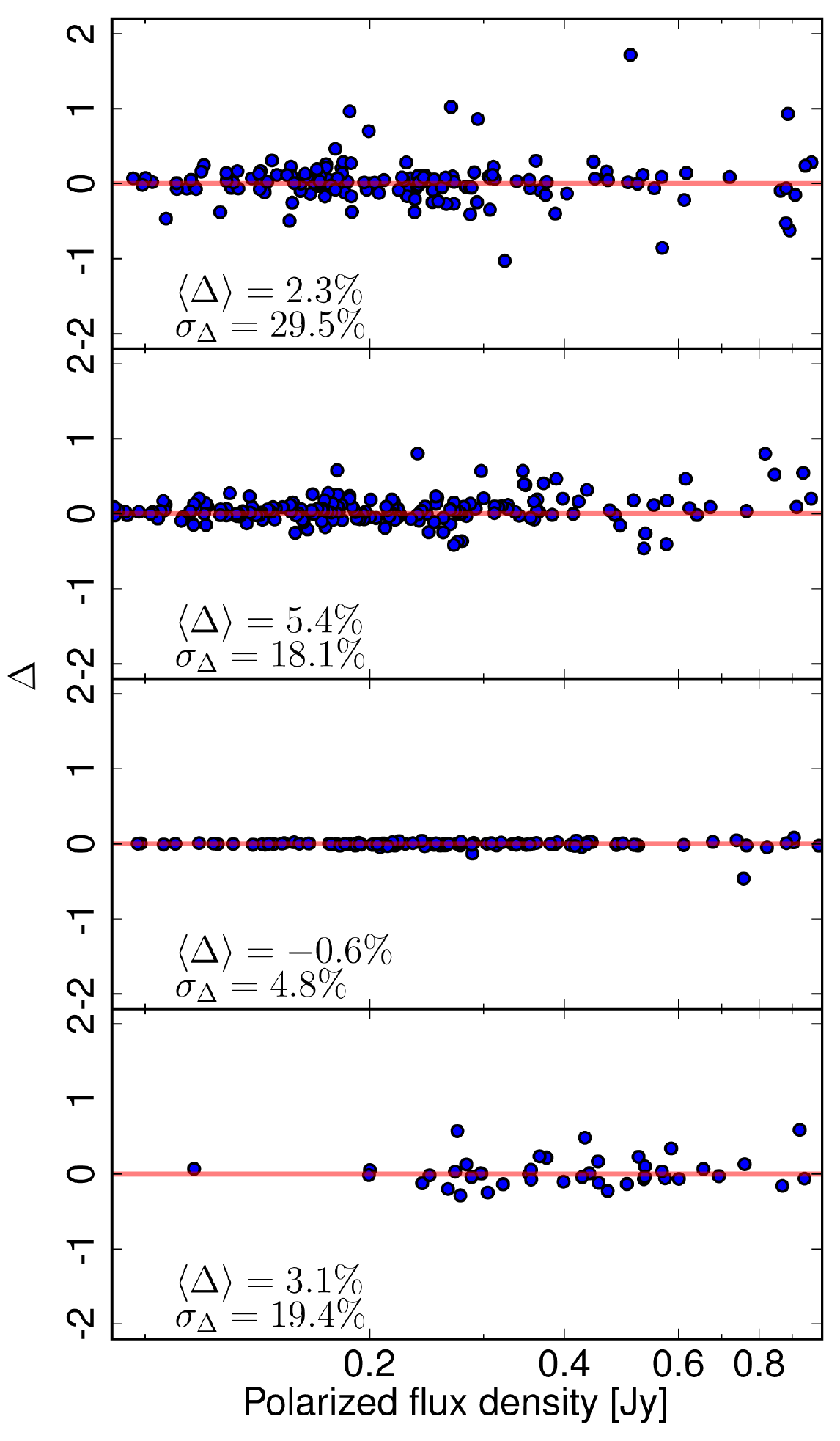}
\caption{Comparison of polarized flux densities from FFP8 simulations
  with ($S_{\mathrm{bpm}}$) and without ($S_{\mathrm{nobpm}}$)
  bandpass mismatch leakage. The difference between the polarized flux
  densities normalized by the uncertainty on the polarized flux
  density, $\Delta = (S_{\mathrm{bpm}}-S_{\mathrm{nobpm}}) / \sigma$,
  is plotted against the input polarized flux density. The errors in
  the polarized flux density due to leakage are always subdominant
  with respect to the uncertainties due to the noise.}
\label{fig:Pol_nbpm_bpm_PCCS2_RELEASE}
\end{center}
\end{figure}

Data from a number of bolometers are combined to make the HFI
polarized frequency maps.  Mismatch between the bandpasses of the
bolometers causes leakage from intensity to polarization for any
source of emission which has a non-CMB spectrum.  Small uncertainties
in the measured transmission for each bolometer may lead to large
uncertainties in the estimates of the bandpass mismatch
leakage~\citep{planck2014-a09}, so it is difficult to make an accurate
prediction of the leakage for compact sources.  Instead we use the
FFP8 simulations~\citep{planck2014-a14} to assess the effect of the
leakage.

Two sets of FFP8 maps have been generated.  One set was simulated with
the measured bandpasses for each HFI bolometer, so it contains the
bandpass mismatch leakage.  The other set was generated using the
average frequency channel bandpass for all bolometers in a channel.
In this idealized case, there is no mismatch between the bandpasses, so
no leakage is produced.  We compare the polarized flux densities of
sources from the two sets of maps.  A suitable quantity for assessing
the size of the leakage is the difference between the polarized flux
density from the maps with the leakage, $S_{\mathrm{bpm}}$, and that
from the maps without the leakage, $S_{\mathrm{nobpm}}$, normalized by
the uncertainty due to the noise, $\sigma$,
\begin{equation}
\Delta = \frac{(S_{\mathrm{bpm}}-S_{\mathrm{nobpm}})}{\sigma}.
\end{equation}
Figure \ref{fig:Pol_nbpm_bpm_PCCS2_RELEASE} shows this quantity
plotted against the input polarized flux density of the source for the
100--353\,GHz channels.  
The size of the effect depends on the differences between the individual bolometer bandpasses and the average frequency channel bandpass; they are smallest for the
217\,GHz channel.  For all channels the effect of the leakage is smaller than 1\,$\sigma$ for most sources.  The mean value of $\Delta$ gives the average bias on the polarized flux density measurements.  It is smaller than $0.06\,\sigma$ for all four channels.  Therefore we conclude that the effect is small and can be safely ignored.
\subsubsection{Evaluation of HFI polarization measurements}
We use the $Q$ and $U$ maps that have been corrected for the leakage of the diffuse temperature components into polarization. This does not correct for any leakage from temperature due to the compact sources themselves. However, as shown above, this effect is subdominant when compared with the statistical uncertainty of the measurements.

The polarized flux density estimator, $\hat{P}$, is evaluated using Eq.~(\ref{eqn:P_def}) and the maximum-likelihood estimates, $\hat{Q}$ and $\hat{U}$,  are extracted from the corresponding $Q$ and $U$ maps by fixing the position to that found from the intensity map and strictly assuming that the source is point-like.
This selection of the estimator is justified, because maximum-likelihood estimators produce minimum variance and unbiased estimates,  which follow a Gaussian distribution with a variance given by
\begin{equation}
\label{eq:MaxLikeEstimateSigma}
\frac{1}{\sigma^2} = \sum_{\boldsymbol \eta}  \widetilde{{\boldsymbol \psi}}^t({\boldsymbol \eta}) \mathcal{N}^{-1}({\boldsymbol \eta}) {\boldsymbol \psi}({\boldsymbol \eta}),
\end{equation}
where $\mathcal{N}^{-1}$ is the inverse power-spectrum of the background of the patch-map, and ${\boldsymbol \psi}({\boldsymbol \eta})$ is the beam transfer function as function of the bidimensional spatial frequency vector, ${\boldsymbol \eta}$.
The maximum-likelihood estimator employed for the extraction of the $Q$ and $U$ signals was {\tt PowellSnakes}  \citep[PwS;][]{PwSII,PwSI}.
The PwS likelihood assumes the background is a realization of a homogeneous Gaussian random field, where a power-spectrum is known, which is a good assumption for small, flat patches cut from the $Q$ and $U$ maps.
The PwS package has been extensively tested and used inside the \Planck\ Collaboration and is known to deliver robust and accurate estimates \citep{planck2011-1.10, planck2011-5.1a, planck2013-p05a}.
The flux densities in the $Q$ and $U$ maps  can be either positive or negative.
In order to reduce systematic effects, which could be induced by the ancillary steps of the likelihood evaluation (such as the estimation of the background power-spectrum of the patch-map), we perform the same estimation procedure on both the patch-map and its negative, and the average of the two flux-density estimates is taken.
This procedure helps in stabilizing the polarization signal, especially when tackling regions with complex backgrounds.
Equivalent estimates of $\hat{Q}$ and $\hat{U}$, and their uncertainties, can be obtained using an aperture-photometry estimator.
The aperture-photometry estimator is  robust to deviations from the likelihood data model -- e.g. extended sources, background deviations from Gaussianity, or variations in the beam shape -- at the cost of slightly larger error bars.
For both estimators the criterion for the acceptance of a putative detection was that $\hat{P}$ had to be 99.99\,\% significant with regard to the null-hypothesis, which is well described by a Rayleigh distribution when the assumption is made that $\sigma_Q \approx \sigma_U$, and that these errors are uncorrelated.
In the case of acceptance, the polarization angle estimate, $\hat{\psi}$, is evaluated using Eq.~(\ref{eqn:Pangle_def}).
The $\hat{Q}$ and $\hat{U}$ uncertainties are propagated onto the $\hat{P}$ and $\hat{\psi}$ estimates.
This is done by assuming the uncertainties are normally distributed and the
error bars are small compared with the measured quantities.
This approximation holds very well given the high significance threshold we have chosen.
In the case where a putative detection is rejected because it does not reach the required significance, we provide the 99\,\% upper limit.

\subsection{Methods to determine polarization properties of marginal detections}
\label{sec:marginal_pol}

For the four polarization-sensitive HFI channels, we provide another set
of polarized flux-density and polarization angle estimates for sources
with marginal detections of polarization.  The extraction of the
polarization signal from the $Q$ and $U$ maps follows the same
procedure described above and in Sect.~\ref{sec:data}.  We then proceed by assessing the
probability of obtaining a given measurement of polarization, given a
true value of polarization~$P$.

\label{sec:AppPolMarginalMathFrame}
\begin{figure}
\begin{center}
\includegraphics[width=0.5\textwidth]{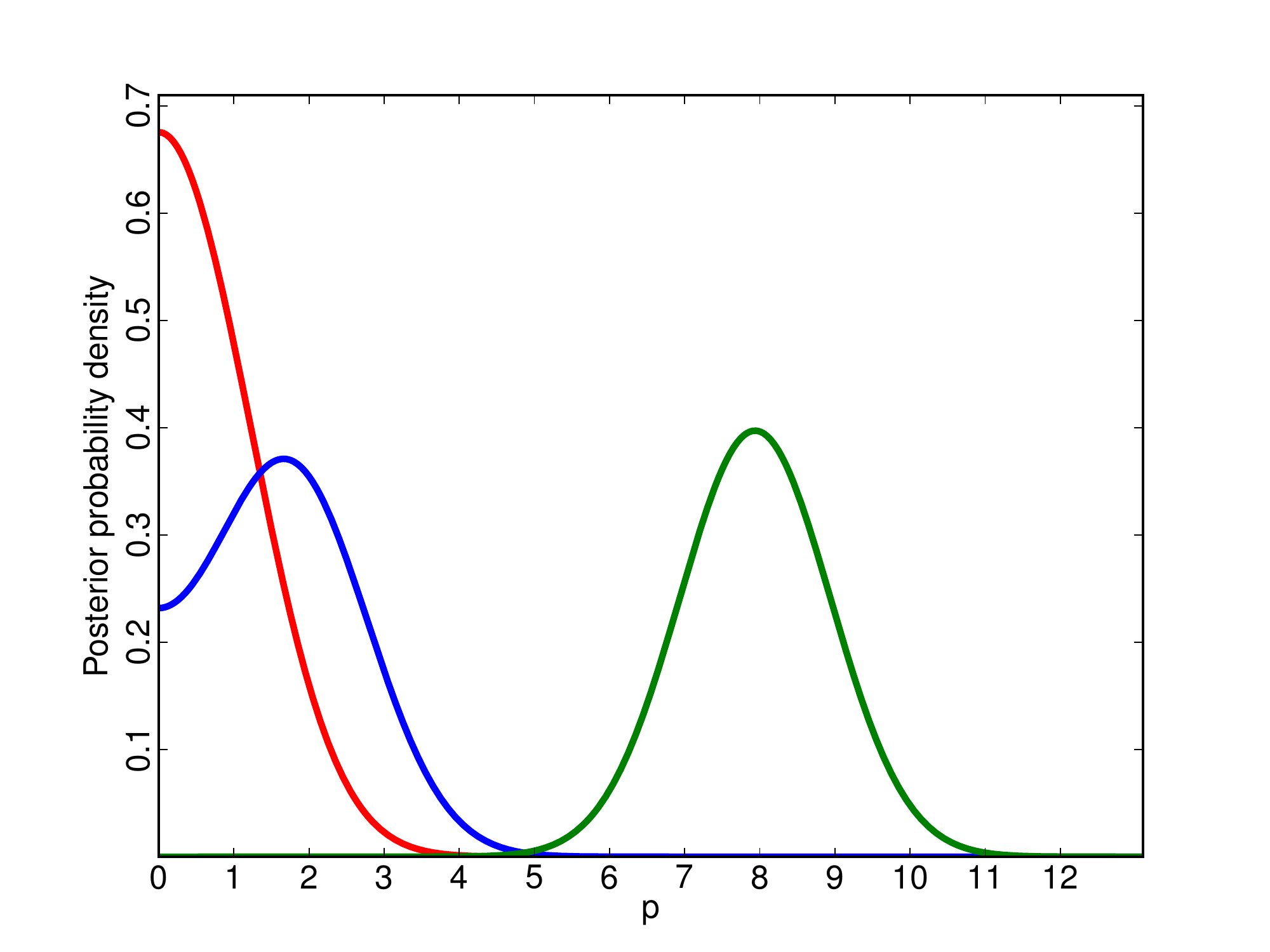}
\caption{Three different cases of polarized flux density posterior distributions (Eq.~\ref{eq:PolPosterior}). The red curve shows a non-detection ($p \lesssim 1.21$), the blue curve shows a marginal detection ($1.21 \lesssim p  \lesssim 3.0$), and the green curve shows a significant detection ($p \gtrsim 3.0$).
\label{fig:MargPolBayCases}}
\end{center}
\end{figure}

Assuming $\sigma_Q \approx \sigma_U \approx \sigma$, the probability of drawing $p = \sqrt{q^2 + u^2}/\sigma$ given the \textit{true} value of the polarized flux density $p_0 = P_0/\sigma$, is
\begin{equation}
\mathcal{L}(p ~ | ~ \sigma) \equiv \textrm{Pr}(p ~|~ p_0, \sigma)= p \, \exp \left( {-\frac{p^2 + p_0^2}{2}} \right) \, \mathcal{I}_0(p p_0),
\end{equation}
where $\mathcal{I}_0(x)$ is the zero-order modified Bessel function of the first kind \citep{PolLikelihood}.
Using Bayes' theorem, the posterior distribution of $p_0$ (the true polarized flux density) given the measurement $p$ is
\begin{equation}
\label{eq:PolPosterior}
\textrm{Pr}(p_0 ~|~ p) \propto  \exp \left( {-\frac{p^2 + p_0^2}{2}} \right) \, \mathcal{I}_0(p p_0) \, \Phi(p_0),
\end{equation}
where $\Phi(p_0)$ is the Heaviside step function. 
We have used the Heaviside step function as a prior, since the flux-density parameter is intrinsically positive, and by taking the asymptotic form of $\mathcal{I}_0(p p_0)$ it  may be shown that $p_0$ acts as a location parameter for large $p$.
Examples of three different posterior distributions for the polarized flux density are shown in Fig.~\ref{fig:MargPolBayCases}.

The polarization angle distribution was derived using Eq. (\ref{eqn:Pangle_def}).
It has been assumed that $U$ and $Q$ are independent Gaussian-distributed random variables with means $\mu_U = \hat{u}$,  $\mu_Q = \hat{q}$, where $\hat{u}, \hat{q}$ are maximum-likelihood estimates of $U$ and $Q$, and $\sigma_U$ and $\sigma_Q$ are given by Eq.~(\ref{eq:MaxLikeEstimateSigma}).
Then using the equality for changing variables in a probability distribution $\textrm{Pr}(\theta)\, \textrm{d}\theta = \textrm{Pr}(\zeta)\, \textrm{d}\zeta$, where $\zeta = U / Q$, the probability distribution for $\theta$ is
\begin{equation}
\label{eq:PolAngle}
\textrm{Pr}(\theta) = - 2 \cos(2\theta)^{-2} f\left(- \tan(2\theta)\right),
\end{equation}
where $f(w)$ is a function defined in \citet[equation~1]{AngleRatio}.
The best-fit value is the mode of the distribution and the asymmetric error-bars were computed using the 95\,\% highest probability density (HPD) region of the posterior distribution.
HPDs are discussed by \cite{BoxTiao} as a general method for compressing the information contained within a probability distribution. Each HPD is uniquely defined by the amount of probability it encloses and is constructed such that there exists no probability density value outside the HPD that is greater than any value contained within it.
In other words the 95\,\% HPD contains 95\,\% of the total probability under the posterior distribution, such that there is no point outside of this area with a higher probability than any point inside the area.
This approach allows us to provide a best-fit value and the asymmetric 2\,$\sigma$ error-bars for the marginal polarization entries in the catalogue.

\begin{figure}
\begin{center}
\includegraphics[width=0.52\textwidth]{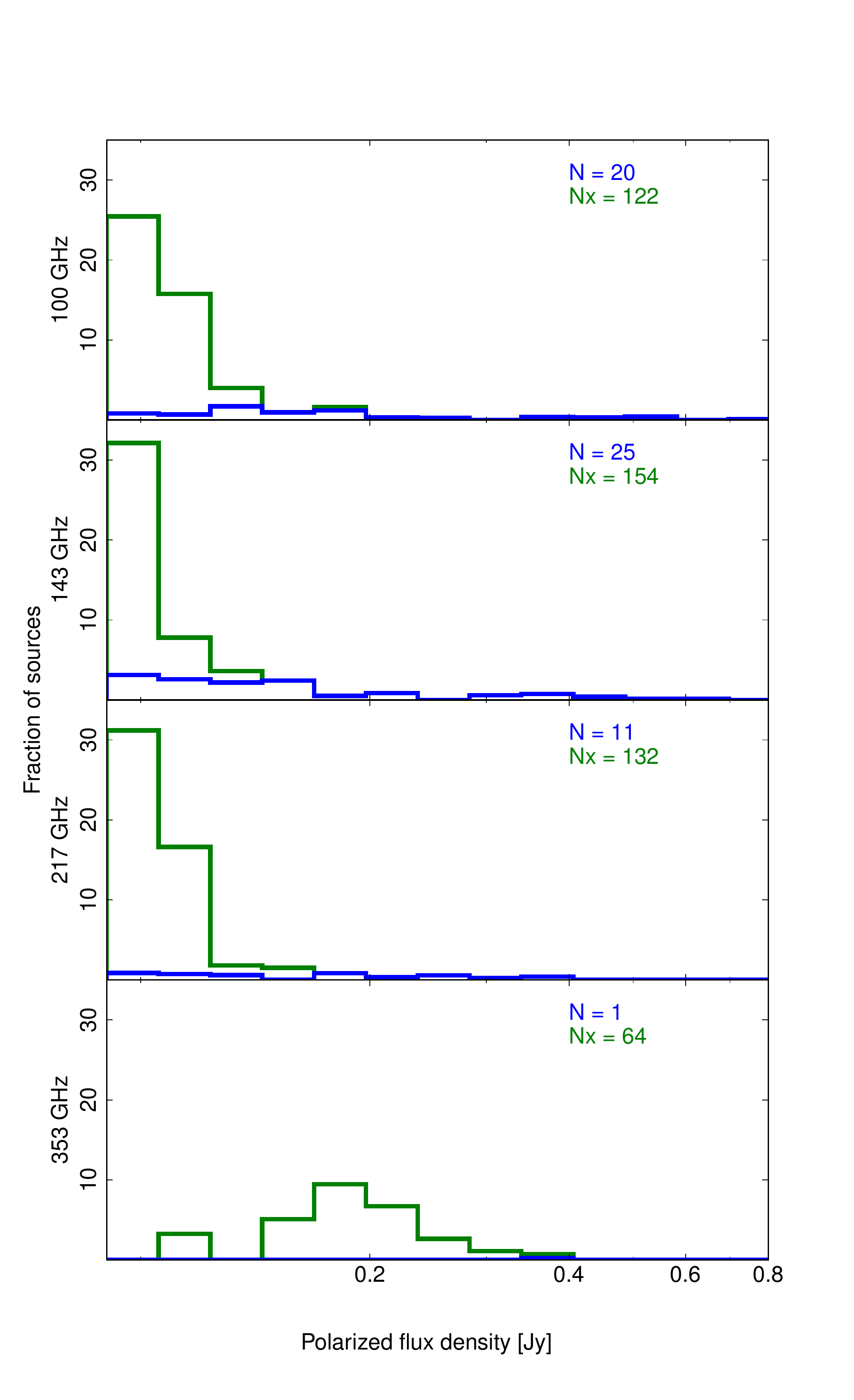}
\caption{Normalized histograms of the polarized flux density from the \plccs\, catalogues for  100 to 353\,GHz. The blue line shows the detections obtained using the common procedure used by both LFI and HFI, whereas the marginal detections,  which are produced only for HFI, are shown in green. The number of sources in each group is shown in the top right of each panel.
\label{fig:HFI_DX11c_FLUX_hist}}
\end{center}
\end{figure}

We shall refer to this set of measurements as the PowellSnakes marginal polarization (PwSPOL) data set.
PwSPOL permits a proper statistical characterization of fainter polarization signals and is therefore able to provide a deeper and more complete catalogue without any loss of reliability, as is shown in Sect.~\ref{sec:qa_polarization}.
An additional benefit of PwSPOL is that it provides a qualitative assessment about an even fainter population of polarized sources which could be valuable as targets for follow-up observations. It does this by splitting the non-detections into two separate groups: clear non-detections, where the polarized flux density posterior peaks at no-signal, and marginal detections, where the posterior does not peak at no-signal, but this possibility is still inside the 95\,\% HPD region of the posterior distribution.
In Fig.~\ref{fig:HFI_DX11c_FLUX_hist} we compare the histograms of the polarized flux-density measurements in the \plccs\  found with each approach.
The distribution of polarized flux densities in the 353\,GHz channel does not follow the same pattern as the other HFI channels. This is due to the much shallower completeness (see Sect.~\ref{sec:qa_polarization}) of this channel as compared with the others.
Although reliable, the polarized flux-density measurements of this marginal set are expected to be biased high as result of Eddington-type bias and are therefore not suitable for any statistical analysis.

\end{document}